%
%
%
%
%
%
%
\hsize=6.5truein
\vsize=9.0truein
\catcode`\@=11 

\def\nolabels{\def\wrlabel##1{}\def\eqlabel##1{}\def\reflabel##1{}}
\def\writelabels{\def\wrlabel##1{\leavevmode\vadjust{\rlap{\smash%
{\line{{\escapechar=` \hfill\rlap{\sevenrm\hskip.03in\string##1}}}}}}}%
\def\eqlabel##1{{\escapechar-1\rlap{\sevenrm\hskip.05in\string##1}}}%
\def\thlabel##1{{\escapechar-1\rlap{\sevenrm\hskip.05in\string##1}}}%
\def\reflabel##1{\noexpand\llap{\noexpand\sevenrm\string\string\string##1}}}
\nolabels
\global\newcount\secno \global\secno=0
\global\newcount\meqno \global\meqno=1
\global\newcount\mthno \global\mthno=1
\global\newcount\mexno \global\mexno=1
\global\newcount\mquno \global\mquno=1
\global\newcount\tblno \global\tblno=1
\def\newsec#1{\global\advance\secno by1 
\global\subsecno=0\xdef\secsym{\the\secno.}\global\meqno=1\global\mthno=1
\global\mexno=1\global\mquno=1\global\figno=1\global\tblno=1

\bigbreak\medskip\noindent{\bf\the\secno. #1}\writetoca{{\secsym} {#1}}
\par\nobreak\medskip\nobreak}
\xdef\secsym{}
\global\newcount\subsecno \global\subsecno=0
\def\subsec#1{\global\advance\subsecno by1 \global\subsubsecno=0
\xdef\subsecsym{\the\subsecno.}
\bigbreak\noindent{\bf\secsym\the\subsecno. #1}\writetoca{\string\quad
{\secsym\the\subsecno.} {#1}}\par\nobreak\medskip\nobreak}
\xdef\subsecsym{}
\global\newcount\subsubsecno \global\subsubsecno=0
\def\subsubsec#1{\global\advance\subsubsecno by1
\bigbreak\noindent{\it\secsym\the\subsecno.\the\subsubsecno.
                                   #1}\writetoca{\string\quad
{\the\secno.\the\subsecno.\the\subsubsecno.} {#1}}\par\nobreak\medskip\nobreak}
\global\newcount\appsubsecno \global\appsubsecno=0
\def\appsubsec#1{\global\advance\appsubsecno by1 \global\subsubsecno=0
\xdef\appsubsecsym{\the\appsubsecno.}
\bigbreak\noindent{\it\secsym\the\appsubsecno. #1}\writetoca{\string\quad
{\secsym\the\appsubsecno.} {#1}}\par\nobreak\medskip\nobreak}
\xdef\appsubsecsym{}
\def\appendix#1#2{\global\meqno=1\global\mthno=1\global\mexno=1
\global\figno=1\global\tblno=1
\global\subsecno=0\global\subsubsecno=0
\global\appsubsecno=0
\xdef\appname{#1}
\xdef\secsym{\hbox{#1.}}
\bigbreak\bigskip\noindent{\bf Appendix #1. #2}
\writetoca{Appendix {#1.} {#2}}\par\nobreak\medskip\nobreak}
%
%
\def\eqnn#1{\xdef #1{(\secsym\the\meqno)}\writedef{#1\leftbracket#1}%
\global\advance\meqno by1\wrlabel#1}
\def\eqna#1{\xdef #1##1{\hbox{$(\secsym\the\meqno##1)$}}
\writedef{#1\numbersign1\leftbracket#1{\numbersign1}}%
\global\advance\meqno by1\wrlabel{#1$\{\}$}}
\def\eqn#1#2{\xdef #1{(\secsym\the\meqno)}\writedef{#1\leftbracket#1}%
\global\advance\meqno by1$$#2\eqno#1\eqlabel#1$$}
%
%
\def\thm#1{\xdef #1{\secsym\the\mthno}\writedef{#1\leftbracket#1}%
\global\advance\mthno by1\wrlabel#1}
\def\exm#1{\xdef #1{\secsym\the\mexno}\writedef{#1\leftbracket#1}%
\global\advance\mexno by1\wrlabel#1}
%
%
\def\tbl#1{\xdef #1{\secsym\the\tblno}\writedef{#1\leftbracket#1}%
\global\advance\tblno by1\wrlabel#1}
%
\newskip\footskip\footskip14pt plus 1pt minus 1pt 
\def\f@@t{\baselineskip\footskip\bgroup\aftergroup\@foot\let\next}
\setbox\strutbox=\hbox{\vrule height9.5pt depth4.5pt width0pt}
\global\newcount\ftno \global\ftno=0
\def\foot{\global\advance\ftno by1\footnote{$^{\the\ftno}$}}
%
\newwrite\ftfile
\def\footend{\def\foot{\global\advance\ftno by1\chardef\wfile=\ftfile
$^{\the\ftno}$\ifnum\ftno=1\immediate\openout\ftfile=foots.tmp\fi%
\immediate\write\ftfile{\noexpand\smallskip%
\noexpand\item{f\the\ftno:\ }\pctsign}\findarg}%
\def\footatend{\vfill\eject\immediate\closeout\ftfile{\parindent=20pt
\centerline{\bf Footnotes}\nobreak\bigskip\input foots.tmp }}}
\def\footatend{}
%
%
\global\newcount\refno \global\refno=1
\newwrite\rfile
\def\ref{\the\refno\nref}
\def\bref{\nref}
\def\nref#1{\xdef#1{\the\refno}\writedef{#1\leftbracket#1}%
\ifnum\refno=1\immediate\openout\rfile=refs.tmp\fi
\global\advance\refno by1\chardef\wfile=\rfile\immediate
\write\rfile{\noexpand\item{[#1]\ }\reflabel{#1\hskip.31in}\pctsign}\findarg}
\def\findarg#1#{\begingroup\obeylines\newlinechar=`\^^M\pass@rg}
{\obeylines\gdef\pass@rg#1{\writ@line\relax #1^^M\hbox{}^^M}%
\gdef\writ@line#1^^M{\expandafter\toks0\expandafter{\striprel@x #1}%
\edef\next{\the\toks0}\ifx\next\em@rk\let\next=\endgroup\else\ifx\next\empty%
\else\immediate\write\wfile{\the\toks0}\fi\let\next=\writ@line\fi\next\relax}}
\def\striprel@x#1{} \def\em@rk{\hbox{}}
\def\lref{\begingroup\obeylines\lr@f}
\def\lr@f#1#2{\gdef#1{\ref#1{#2}}\endgroup\unskip}

\def\addref#1{\immediate\write\rfile{\noexpand\item{}#1}} 
\def\startrefs#1{\immediate\openout\rfile=refs.tmp\refno=#1}
\def\xref{\expandafter\xr@f}\def\xr@f[#1]{#1}
\def\refs#1{[\r@fs #1{\hbox{}}]}
\def\r@fs#1{\edef\next{#1}\ifx\next\em@rk\def\next{}\else
\ifx\next#1\xref #1\else#1\fi\let\next=\r@fs\fi\next}
%

%
 \newwrite\ffile\global\newcount\figno \global\figno=1
%
%
\def\fig{\the\figno\nfig}
\def\nfig#1{\xdef#1{\secsym\the\figno}%
\writedef{#1\leftbracket \noexpand~\the\figno}%
\ifnum\figno=1\immediate\openout\ffile=figs.tmp\fi\chardef\wfile=\ffile%
\immediate\write\ffile{\noexpand\medskip\noexpand\item{Figure\ \the\figno. }
\reflabel{#1\hskip.55in}\pctsign}\global\advance\figno by1\findarg}
\def\xfig{\expandafter\xf@g}\def\xf@g \penalty\@M\ {}
\def\figs#1{figs.~\f@gs #1{\hbox{}}}
\def\f@gs#1{\edef\next{#1}\ifx\next\em@rk\def\next{}\else
\ifx\next#1\xfig #1\else#1\fi\let\next=\f@gs\fi\next}
%
%
\newwrite\lfile
\def\tmpfile#1{\def\xtmpfile{#1}}
{\escapechar-1\xdef\pctsign{\string\%}\xdef\leftbracket{\string\{}
\xdef\rightbracket{\string\}}\xdef\numbersign{\string\#}}

\def\writestop{\def\writestoppt{\immediate\write\lfile{\string\pageno%
\the\pageno\string\startrefs\leftbracket\the\refno\rightbracket%
\string\def\string\secsym\leftbracket\secsym\rightbracket%
\string\secno\the\secno\string\meqno\the\meqno}\immediate\closeout\lfile}}
\def\writestoppt{}\def\writedef#1{}

\def\seclab#1{\xdef #1{\the\secno}\writedef{#1\leftbracket#1}\wrlabel{#1=#1}}
\def\applab#1{\xdef #1{\appname}\writedef{#1\leftbracket#1}\wrlabel{#1=#1}}
\def\subseclab#1{\xdef #1{\secsym\the\subsecno}%
\writedef{#1\leftbracket#1}\wrlabel{#1=#1}}
\def\appsubseclab#1{\xdef #1{\secsym\the\appsubsecno}%
\writedef{#1\leftbracket#1}\wrlabel{#1=#1}}
\def\subsubseclab#1{\xdef #1{\secsym\the\subsecno.\the\subsubsecno}%
\writedef{#1\leftbracket#1}\wrlabel{#1=#1}}
\newwrite\tfile \def\writetoca#1{}
\def\leaderfill{\leaders\hbox to 1em{\hss.\hss}\hfill}
\def\writetoc{\immediate\openout\tfile=toc.tmp
   \def\writetoca##1{{\edef\next{\write\tfile{\noindent ##1
   \string\leaderfill {\noexpand\number\pageno} \par}}\next}}}

%
\catcode`\@=12 
%
%
%
%
%
\def\dbend{{\manual\char127}}
\def\d@nger{\medbreak\begingroup\clubpenalty=10000
    \def\par{\endgraf\endgroup\medbreak} \noindent\hang\hangafter=-2
    \hbox to0pt{\hskip-\hangindent\dbend\hfill}\ninepoint}
\outer\def\danger{\d@nger}

\def\p{\partial}

\def\vev#1{\langle #1 \rangle}

\def\darr#1{\raise1.5ex\hbox{$\leftrightarrow$}\mkern-16.5mu #1}
\def\half{{\textstyle{1\over2}}} 

%
%
\def\al{\alpha}
\def\be{\beta}
\def\ga{\gamma}  \def\Ga{\Gamma}
\def\de{\delta}  \def\De{\Delta}
\def\ep{\epsilon}  

\def\et{\eta}
\def\th{\theta}  \def\Th{\Theta}

\def\la{\lambda} \def\La{\Lambda}
\def\rh{\rho}
\def\si{\sigma}

\def\ph{\phi}  \def\Ph{\Phi}  

  \def\Ps{\Psi}
\def\om{\omega}  \def\Om{\Omega}
%
%

\def\bPh{{\bf \Ph}}
%

%
%
\def\cA{{\cal A}} 
\def\cC{{\cal C}} \def\cD{{\cal D}}
\def\cE{{\cal E}} 

\def\cI{{\cal I}} 
\def\cL{{\cal L}}
\def\cM{{\cal M}}

\def\cO{{\cal O}}
\def\cP{{\cal P}}
\def\cR{{\cal R}} \def\cS{{\cal S}}
\def\cU{{\cal U}} 
\def\cV{{\cal V}}
\def\cW{{\cal W}}

\def\mapright#1{\smash{\mathop{\longrightarrow}\limits^{#1}}}
\def\vev#1{\langle #1 \rangle}

\def\ie{{\it i.e.}}
\def\eg{{\it e.g.}} \def\cf{{\it cf.}}
\def\proof{\noindent {\it Proof:}\ }
\def\Box{\hbox{$\rlap{$\sqcup$}\sqcap$}}

\def\Hom{{\rm Hom}}
\def\End{{\rm Hom}}
\def\vps{^{\vphantom{*}}}
\def\pxiu#1{\hbox{$\RSr\partial\over\LWrr{\partial x^#1}$}}
\def\pxid#1{\hbox{$\RSr\partial\over\LWrr{\partial x_#1}$}}

\def\str#1#2#3{f{}_{#1}{}_{#2}{}^{#3}}
\def\pr{\partial}
\def\Ep{{\cal E}}
\def\cgM{{\overline M}}
%
%
\def\amsyes{y }

\def\answ{y }
\ifx\answ\amsyes

\input amssym.def

\def\CC{{\Bbb C}}
\def\ZZ{{\Bbb Z}}
\def\NN{{\Bbb N}}

\def\RR{{\Bbb R}}
\def\bfg{{\frak g}}
\def\bfh{{\frak h}}

\def\bfnp{{\frak n}_+}
\def\bfnm{{\frak n}_-}
\def\hg{{\widehat{\frak g}}}

\def\fC{{\frak C}} \def\fQ{{\frak Q}}
\def\fH{{\frak H}}
\def\sln{\frak{sl}_N}   
\def\sltw{\frak{sl}_2}  \def\hsltw{\widehat{\frak{sl}_2}}
\def\slth{\frak{sl}_3}  \def\hslth{\widehat{\frak{sl}_3}}

\def\sosi{\frak{so}_6}
\def\soei{\frak{so}_8}
\def\sotwon{\frak{so}_{2N}} \def\sotwontwo{\frak{so}_{2N+2}}
\def\gltwon{\frak{gl}_{2N}}
\def\frde{\hbox{$\frak{d}$}}
\def\slN{\frak{sl}_N}
\def\uone{\frak{u}_1}
\def\fA{\frak{A}} \def\bga{\slth\oplus(\uone)^2}
\def\fV{\frak{V}} \def\fP{{\frak P}} \def\fD{{\frak D}}
\def\fH{\frak{H}} \def\fM{{\frak M}}
\def\fC{\frak{C}} \def\fI{{\frak I}}
\else
\def\ZZ{{Z\!\!\!Z}}
\def\CC{{I\!\!\!\!C}}
\def\NN{{I\!\!N}}

\def\RR{I\!\!R}
\def\bfg{{\bf g}}
\def\bfh{{\bf h}}
\def\bfnm{{\bf n}_-}
\def\bfnp{{\bf n}_+}
\def\hg{\hat{\bf g}}

\def\sln{s\ell_n}   
\def\sltw{s\ell_2}  \def\hsltw{\widehat{s\ell_2}}
\def\slth{s\ell_3}  \def\hslth{\widehat{s\ell_3}}

\def\sosi{so_6}
\def\soei{so_8}
\def\sotwon{so_{2N}}  \def\sotwontwo{so_{2N+2}}
\def\gltwon{gl_{2N}}
\def\slN{s\ell_N}
\def\uone{u_1}

\def\frde{\de}
\def\cA{{\cal A}} \def\bga{{\bf a}}
\def\cR{{\cal R}}
\fi
\def\cG{{\cR}}
\def\homhom{{\rm Hom}_{\bfnp} (\cL(\La),\hbox{$\bigwedge$}{}^n (\bfnp
\backslash \bfg))}
\def\homhomt{{\rm Hom}_{\bfnp} (\cL(\La),T)}
\def\homhomtn{{\rm Hom}_{\bfnp} (\cL(\La),T^n)}

%
%
\def\comdiag{
\def\normalbaselines{\baselineskip15pt
\lineskip3pt \lineskiplimit3pt }}

\def\mapright#1{\smash{\mathop{\longrightarrow}\limits^{#1}}}

\def\mapdown#1{\Big\downarrow\rlap{$\vcenter{
                \hbox{$\scriptstyle#1$}}$}}

%
%
\def\vM{v_M} \def\vF{v_F} \def\vMb{\bar{v}_{M}} \def\vMp{v_{M'}}
\def\oom{\overline{\om}} \def\vFp{v_{F'}} \def\vV{v_V}
\newsymbol\ltimes 226E
\newsymbol\rtimes 226F
%
%
%
\def\AdM#1{Adv.\ Math.\ {\bf #1}}
\def\AnM#1{Ann.\ Math.\ {\bf #1}}

\def\CMP#1{Comm.\ Math.\ Phys.\ {\bf #1}}

\def\FAP#1{Funct.\ Anal.\ Appl.\ {\bf #1}}
\def\IJMP#1{Int.\ J.\ Mod.\ Phys.\ {\bf #1}}
\def\InM#1{Inv.\ Math.\ {\bf #1}}
\def\JGP#1{J.\ Geom.\ Phys.\ {\bf #1}}

\def\LMP#1{Lett.\ Math.\ Phys.\ {\bf #1}}
\def\LNM#1{Lect.\ Notes in Math.\ {\bf #1}}
\def\MPL#1{Mod.\ Phys.\ Lett.\ {\bf #1}}
\def\NPB#1{Nucl.\ Phys.\ {\bf B#1}}
\def\PLB#1{Phys.\ Lett.\ {\bf {#1}B}}
\def\PNAS#1{Proc.\ Natl.\ Acad.\ Sci. USA {\bf #1}}
\def\PRep#1{Phys.\ Rep.\ {\bf #1}}

\def\PTP#1{Prog.\ Theor.\ Phys.\ Suppl.\ {\bf #1}}
\def\SMD#1{Sov.\ Math.\ Dokl.\ {\bf {#1}}}

\def\TMP#1{Theor.\ Math.\ Phys.\ {\bf {#1}}}

%

%
%
\def\SMu{\hbox{\lower 3pt\hbox{ \epsffile{su10.eps}}}}
\def\SMs{\hbox{\lower 3pt\hbox{ \epsffile{ss10.eps}}}}
\def\SMd{\hbox{\lower 3pt\hbox{ \epsffile{sd10.eps}}}}
\def\NOsmile{
    \def\SMu{\Box} \def\SMs{\Box} \def\SMd{\Box}}
\def\SMS{\leavevmode\vadjust{\rlap{\smash%
{\line{{\escapechar=` \hfill\rlap{\hskip.3in%
                 \hbox{\lower 2pt\hbox{\epsffile{sd10.eps}}}}}}}}}}
\def\SMH{\leavevmode\vadjust{\rlap{\smash%
{\line{{\escapechar=` \hfill\rlap{\hskip.3in%
                 \hbox{\lower 2pt\hbox{\epsffile{su10.eps}}}}}}}}}}
%
%
\def\LW#1{\lower .5pt \hbox{$\scriptstyle #1$}}
\def\LWr#1{\lower 1.5pt \hbox{$\scriptstyle #1$}}
\def\LWrr#1{\lower 2pt \hbox{$\scriptstyle #1$}}
\def\RSr#1{\raise 1pt \hbox{$\scriptstyle #1$}}
\def\cWth{\cW_3}
\def\vdeg{\kappa}
\def\cWth{{\cal W}_3}
\def\cWthp{{\cal W}_{3,+}} \def\cWthm{{\cal W}_{3,-}}
\def\cWthz{{\cal W}_{3,0}} \def\cWthpm{{\cal W}_{3,\pm}}
\def\ker{{\rm Ker}}
\def\txt#1{$\hbox{$#1$}$}
\def\crM#1{{\cal M}^{(#1)}}  \def\crI#1{{\cal I}^{(#1)}}
\NOsmile

\input tables.tex
\input epsf
\tmpfile{filem.xtmp}
\def\JMI{1}
\def\genint{1.1}
\def\Physmot{1.1.1}
\def\Mathmot{1.1.2}
\def\Out{1.2}
\def\Not{1.3}
\def\PBW{2}
\def\Walg{2.1}
\def\Wintro{2.1.1}
\def\Wth{2.1.2}
\def\eqPAaa{(2.1)}
\def\eqPAab{(2.2)}
\def\eqPAac{(2.3)}
\def\eqPAad{(2.4)}
\def\eqPAae{(2.5)}
\def\eqPAaf{(2.6)}
\def\eqPAag{(2.7)}
\def\Wthmod{2.2}
\def\WOcat{2.2.1}
\def\thPBaa{2.1}
\def\eqPBaab{(2.8)}
\def\eqPBaac{(2.9)}
\def\eqPBaaa{(2.10)}
\def\thPBaaa{2.2}
\def\thJHa{2.3}
\def\eqJHa{(2.11)}
\def\thJHb{2.4}
\def\eqJHb{(2.12)}
\def\thPBae{2.5}
\def\eqPBac{(2.13)}
\def\eqPBaca{(2.14)}
\def\thPBaf{2.6}
\def\eqPBad{(2.15)}
\def\thPBag{2.7}
\def\WVer{2.2.2}
\def\thPBbab{2.8}
\def\eqPBbaa{(2.16)}
\def\eqPBbab{(2.17)}
\def\thPBbac{2.9}
\def\eqPBbac{(2.18)}
\def\thPBba{2.10}
\def\thPBbad{2.11}
\def\eqPBbae{(2.19)}
\def\thPBbae{2.12}
\def\eqPBCa{(2.20)}
\def\eqBPa{(2.21)}
\def\eqBPb{(2.22)}
\def\eqBPc{(2.23)}
\def\eqBPd{(2.24)}
\def\thPBuni{2.13}
\def\eqPBuni{(2.25)}
\def\thPBbaa{2.14}
\def\thPBaea{2.15}
\def\eqPBbgabc{(2.26)}
\def\eqPBbga{(2.27)}
\def\eqBPe{(2.28)}
\def\eqPBbh{(2.29)}
\def\eqPBba{(2.30)}
\def\thPBbc{2.16}
\def\eqPBbj{(2.31)}
\def\eqPBbb{(2.32)}
\def\eqPBbc{(2.33)}
\def\eqPBbd{(2.34)}
\def\thPBbd{2.17}
\def\eqBPg{(2.35)}
\def\eqPBbi{(2.36)}
\def\eqPBbfa{(2.37)}
\def\eqPBbfb{(2.38)}
\def\thPBbe{2.18}
\def\thPBbf{2.19}
\def\eqPBbf{(2.39)}
\def\thPBbfa{2.20}
\def\eqPBbg{(2.40)}
\def\eqPBzc{(2.41)}
\def\eqPBzb{(2.42)}
\def\eqPBza{(2.43)}
\def\eqPBbk{(2.44)}
\def\thPBbi{2.21}
\def\eqPBbla{(2.45)}
\def\eqPBblb{(2.46)}
\def\eqPBbm{(2.47)}
\def\eqPBbn{(2.48)}
\def\eqPBbo{(2.49)}
\def\eqBPn{(2.50)}
\def\WFock{2.2.3}
\def\eqPBca{(2.51)}
\def\eqPBcab{(2.52)}
\def\eqPBcac{(2.53)}
\def\eqPBcad{(2.54)}
\def\eqPBcae{(2.55)}
\def\eqPBcaa{(2.56)}
\def\eqPBcb{(2.57)}
\def\eqJMphi{(2.58)}
\def\eqJMstop{(2.59)}
\def\eqPBcc{(2.60)}
\def\eqPbcdko{(2.61)}
\def\eqPBcdc{(2.62)}
\def\eqPBcd{(2.63)}
\def\eqPBcda{(2.64)}
\def\eqPBcdb{(2.65)}
\def\eqPBce{(2.66)}
\def\thPBcb{2.22}
\def\eqPBcdh{(2.67)}
\def\eqPBcdkt{(2.68)}
\def\eqPBcdi{(2.69)}
\def\eqPBcdj{(2.70)}
\def\thPBcab{2.23}
\def\thPBca{2.24}
\def\eqPBcdd{(2.71)}
\def\eqPBcdda{(2.72)}
\def\eqPBcde{(2.73)}
\def\eqPBcdf{(2.74)}
\def\thPBcaa{2.25}
\def\eqPBcca{(2.75)}
\def\thPBcc{2.26}
\def\eqBPh{(2.76)}
\def\eqPBcg{(2.77)}
\def\eqBPi{(2.78)}
\def\eqPBcga{(2.79)}
\def\eqPBcgb{(2.80)}
\def\eqPBcgc{(2.81)}
\def\thPBcd{2.27}
\def\eqBPj{(2.82)}
\def\thPBcda{2.28}
\def\eqBPk{(2.83)}
\def\eqBPl{(2.84)}
\def\thPBce{2.29}
\def\Wctwo{2.3}
\def\Wctwoa{2.3.1}
\def\thPBdaa{2.30}
\def\eqPBdaa{(2.85)}
\def\thPBda{2.31}
\def\eqPBda{(2.86)}
\def\eqPBDa{(2.87)}
\def\eqPBDb{(2.88)}
\def\eqPBdc{(2.89)}
\def\eqPBdb{(2.90)}
\def\eqPBdd{(2.91)}
\def\eqBPm{(2.92)}
\def\eqPBde{(2.93)}
\def\thPBdb{2.32}
\def\thPBdc{2.33}
\def\eqPBEa{(2.94)}
\def\Wctwob{2.3.2}
\def\fiPBaa{~1}
\def\fiPBab{~2}
\def\fiPBac{~3}
\def\eqPBdh{(2.95)}
\def\eqPBdg{(2.96)}
\def\eqPBdi{(2.97)}
\def\thPBdd{2.34}
\def\Wres{2.4}
\def\thPCa{2.35}
\def\thJMaa{2.36}
\def\eqJMea{(2.98)}
\def\WVerres{2.4.1}
\def\eqPBRc{(2.99)}
\def\eqPBRa{(2.100)}
\def\eqPBRd{(2.101)}
\def\tbBPa{2.1}
\def\eqPBRe{(2.102)}
\def\fiPBba{~4}
\def\fiPBbb{~5}
\def\fiPBbc{~6}
\def\thJMab{2.37}
\def\eqPCga{(2.103)}
\def\eqPCg{(2.104)}
\def\eqPCh{(2.105)}
\def\eqJMb{(2.106)}
\def\NSbrst{3}
\def\Scomplex{3.1}
\def\SSghosts{3.1.1}
\def\ghostope{(3.1)}
\def\modes{(3.2)}
\def\comrel{(3.3)}
\def\strenergy{(3.4)}
\def\vaccum{(3.5)}
\def\ghbasis{(3.6)}
\def\vaccumtwo{(3.7)}
\def\antiinv{(3.8)}
\def\ghform{3.1}
\def\SSbrst{3.1.2}
\def\existbrst{3.2}
\def\brstcurr{(3.9)}
\def\brstdd{(3.10)}
\def\dlead{(3.11)}
\def\defsemcoh{3.3}
\def\brstact{(3.12)}
\def\Swcohprob{3.2}
\def\prodinL{(3.13)}
\def\ontheL{3.4}
\def\defofl{(3.14)}
\def\weighofl{(3.15)}
\def\profpa{(3.16)}
\def\lattice{3.5}
\def\sumcom{(3.17)}
\def\genformofpsi{(3.18)}
\def\phvac{(3.19)}
\def\optostate{(3.20)}
\def\opegen{(3.21)}
\def\cocycless{3.6}
\def\eqJMcoc{(3.22)}
\def\phcoc{(3.23)}
\def\anzcoc{(3.24)}
\def\eqsforxi{(3.25)}
\def\ohthose{(3.26)}
\def\SSpreliminaries{3.3}
\def\SSrelative{3.3.1}
\def\totops{(3.27)}
\def\vanishw{3.7}
\def\virvanish{(3.28)}
\def\vanishhhh{(3.29)}
\def\decompp{(3.30)}
\def\punchw{(3.31)}
\def\relsubs{(3.32)}
\def\SSreduction{3.3.2}
\def\redthm{3.8}
\def\genwec{(3.33)}
\def\defoff{(3.34)}
\def\degrees{(3.35)}
\def\ghostdeg{(3.36)}
\def\dzero{(3.37)}
\def\suppofeone{(3.38)}
\def\donediff{(3.39)}
\def\condonwgts{(3.40)}
\def\secterm{(3.41)}
\def\moreredthm{3.9}
\def\bounding{(3.42)}
\def\vancohm{(3.43)}
\def\vancohcm{(3.44)}
\def\newdzero{(3.45)}
\def\neweone{(3.46)}
\def\fockfiltr{3.10}
\def\defofffock{(3.47)}
\def\SSslth{3.3.3}
\def\slthonfc{(3.48)}
\def\energlev{(3.49)}
\def\slthac{3.11}
\def\SSbilinear{3.3.4}
\def\conjdiff{(3.50)}
\def\lemform{3.12}
\def\cohduality{3.13}
\def\dualcoh{(3.51)}
\def\SSfundamental{3.4}
\def\SSHfirst{3.4.1}
\def\jctw{(3.52)}
\def\jcth{(3.53)}
\def\fundcoh{3.14}
\def\condforweg{(3.54)}
\def\primest{(3.55)}
\def\dimnsns{(3.56)}
\def\fstspsq{(3.57)}
\def\othsctrm{(3.58)}
\def\jzz{(3.59)}
\def\restate{3.15}
\def\primecoh{(3.60)}
\def\fullvsprime{(3.61)}
\def\SHgen{3.4.2}
\def\setofshwe{3.16}
\def\fockcoh{3.17}
\def\dimfockcoh{(3.62)}
\def\triples{3.18}
\def\cohfweyl{3.19}
\def\decomfweyl{(3.63)}
\def\polytips{3.1}
\def\isomorph{(3.64)}
\def\isomorphpr{(3.65)}
\def\SSgeneral{3.5}
\def\SSgenintro{3.5.1}
\def\SSvanishh{3.5.2}
\def\ghnumlim{3.20}
\def\lowestsect{(3.66)}
\def\conjone{3.21}
\def\SSgencoh{3.5.3}
\def\gencones{3.2}
\def\twistedco{3.22}
\def\shiftsss{3.23}
\def\twlengthbig{~1}
\def\twlengexs{~2}
\def\twconess{3.24}
\def\fullcoh{3.25}
\def\bigdecomp{(3.67)}
\def\thBa{3.26}
\def\eqBa{(3.68)}
\def\ennns{(3.69)}
\def\BValgebra{4}
\def\SGBValg{4.1}
\def\SSdefini{4.1.1}
\def\BVghal{4.1}
\def\eqBPza{(4.1)}
\def\eqBPzb{(4.2)}
\def\eqBPzc{(4.3)}
\def\BVscor{(4.4)}
\def\BVbv{4.2}
\def\thCd{4.3}
\def\eqCc{(4.5)}
\def\eqCd{(4.6)}
\def\triplec{4.4}
\def\derofa{(4.7)}
\def\eqPBexa{(4.8)}
\def\LLaut{4.5}
\def\SSpolyderaa{4.1.2}
\def\BVleib{(4.9)}
\def\defpolvs{4.6}
\def\BVderv{(4.10)}
\def\derivas{4.7}
\def\prodpol{(4.11)}
\def\prodex{(4.12)}
\def\defsch{4.8}
\def\schbrack{(4.13)}
\def\schbrtwo{(4.14)}
\def\jab{(4.15)}
\def\bvschal{4.9}
\def\SBVpolder{4.1.3}
\def\polasder{(4.16)}
\def\notforget{(4.17)}
\def\coeff{(4.18)}
\def\product{(4.19)}
\def\bracket{(4.20)}
\def\evop{(4.21)}
\def\BVoppl{(4.22)}
\def\bvpolvs{4.10}
\def\eqJMvol{(4.23)}
\def\SSpolyder{4.1.4}
\def\defpi{(4.24)}
\def\gmorphism{4.11}
\def\indprod{(4.25)}
\def\fststep{(4.26)}
\def\genind{(4.27)}
\def\bvhomscr{4.12}
\def\condforpi{(4.28)}
\def\gencasebv{(4.29)}
\def\Smodules{4.2}
\def\gmodule{4.13}
\def\gmAA{(4.30)}
\def\gmAB{(4.31)}
\def\gmAC{(4.32)}
\def\defbvmod{4.14}
\def\BVscormod{(4.33)}
\def\gmAE{(4.34)}
\def\gmAF{(4.35)}
\def\gmAG{(4.36)}
\def\SSpolydermod{4.2.1}
\def\gmodder{4.15}
\def\gmBA{(4.37)}
\def\gmBB{(4.38)}
\def\gmBR{(4.39)}
\def\gmBQ{(4.40)}
\def\bvpolyderivations{4.3}
\def\SSgrrinal{4.3.1}
\def\vanishideal{(4.41)}
\def\metric{(4.42)}
\def\sosixder{(4.43)}
\def\strgrr{4.16}
\def\basisss{(4.44)}
\def\decgrr{(4.45)}
\def\hiddensymmetry{4.3.2}
\def\sotwonpluss{4.17}
\def\sotwonnn{(4.46)}
\def\polyderofcgn{4.3.3}
\def\poldercond{4.18}
\def\conscond{(4.47)}
\def\formder{(4.48)}
\def\realcond{(4.49)}
\def\prfcond{(4.50)}
\def\endbasis{(4.51)}
\def\tenspr{(4.52)}
\def\speccs{(4.53)}
\def\basione{(4.54)}
\def\actionone{(4.55)}
\def\firstdec{(4.56)}
\def\genasymtns{(4.57)}
\def\actonp{(4.58)}
\def\genact{(4.59)}
\def\gendec{(4.60)}
\def\basisextwo{(4.61)}
\def\baseistwo{(4.62)}
\def\gendec{(4.63)}
\def\genderbasone{(4.64)}
\def\genderbastwo{(4.65)}
\def\generalact{(4.66)}
\def\genasymtnsbig{(4.67)}
\def\eqvolel{(4.68)}
\def\gendec{(4.69)}
\def\bigcand{(4.70)}
\def\schid{(4.71)}
\def\tblpolyvec{4.1}
\def\polyvects{4.19}
\def\dblgrd{(4.72)}
\def\generat{4.20}
\def\conone{(4.73)}
\def\contwo{(4.74)}
\def\conthr{(4.75)}
\def\confour{(4.76)}
\def\confive{(4.77)}
\def\products{(4.78)}
\def\galgstr{4.3.4}
\def\expolbr{4.21}
\def\homogen{4.22}
\def\brackmap{(4.79)}
\def\sophi{(4.80)}
\def\sophigen{(4.81)}
\def\twopcom{(4.82)}
\def\commcphi{4.23}
\def\Cph{(4.83)}
\def\genproduct{4.24}
\def\bigproduct{(4.84)}
\def\bvpolyvectors{4.3.5}
\def\uniquebv{4.25}
\def\firstvan{(4.85)}
\def\firstone{(4.86)}
\def\exxxc{(4.87)}
\def\seconex{(4.88)}
\def\resoneo{(4.89)}
\def\simplelemma{4.26}
\def\deloncph{(4.90)}
\def\candph{(4.91)}
\def\bvopmth{4.27}
\def\exformbv{(4.92)}
\def\morerestr{(4.93)}
\def\firstcomm{(4.94)}
\def\sopigenn{(4.95)}
\def\trvl{(4.96)}
\def\psxi{(4.97)}
\def\xidexi{(4.98)}
\def\bvgener{4.28}
\def\fstcond{(4.99)}
\def\scncond{(4.100)}
\def\newelem{(4.101)}
\def\newdelc{(4.102)}
\def\trplid{(4.103)}
\def\cohofdel{4.29}
\def\chiral{4.3.6}
\def\slnalg{(4.104)}
\def\holthm{4.30}
\def\holommult{(4.105)}
\def\SSthree{4.4}
\def\SSpirthree{4.4.1}
\def\branch{(4.106)}
\def\modelspace{4.31}
\def\bigbrone{(4.107)}
\def\defofc{(4.108)}
\def\defofp{(4.109)}
\def\defoflam{(4.110)}
\def\deffod{(4.111)}
\def\hidsym{4.4.2}
\def\decomada{(4.112)}
\def\decomadb{(4.113)}
\def\extrathr{(4.114)}
\def\twistfor{(4.115)}
\def\twostforA{(4.116)}
\def\twistedmodules{4.4.3}
\def\twmodls{4.32}
\def\twpairs{(4.117)}
\def\wactring{(4.118)}
\def\twistoprs{4.2}
\def\plvmodule{4.33}
\def\suthgenp{(4.119)}
\def\SStwplcl{4.4.4}
\def\conforplvct{(4.120)}
\def\gonevec{4.3}
\def\genvecone{4.1}
\def\exofgph{(4.121)}
\def\psets{(4.122)}
\def\forfg{(4.123)}
\def\grringact{(4.124)}
\def\redcondfo{(4.125)}
\def\pigenpol{4.34}
\def\gpldecom{(4.126)}
\def\pltwcnsi{4.4}
\def\ofspom{(4.127)}
\def\freegen{4.35}
\def\quargvf{4.5}
\def\twplsumm{4.36}
\def\defoftwdel{(4.128)}
\def\twistbvop{(4.129)}
\def\Sbvalgebra{5}
\def\SBVintro{5.1}
\def\SBVgen{5.1.1}
\def\cftbv{5.1}
\def\product{(5.1)}
\def\prodaslim{(5.2)}
\def\bvopac{(5.3)}
\def\barckop{(5.4)}
\def\slthaut{5.2}
\def\SSmnotation{5.1.2}
\def\notforst{(5.5)}
\def\Ssurvey{5.2}
\def\SSground{5.2.1}
\def\unoppsi{(5.6)}
\def\grringgenx{(5.7)}
\def\grconst{5.3}
\def\grconstraint{(5.8)}
\def\cohringgen{5.4}
\def\bigprod{(5.9)}
\def\hwopsex{(5.10)}
\def\pisdep{(5.11)}
\def\normalhigh{(5.12)}
\def\phiopes{(5.13)}
\def\groundring{5.5}
\def\defofpi{(5.14)}
\def\SSautom{5.2.2}
\def\slthreeop{(5.15)}
\def\bonslth{(5.16)}
\def\higslth{(5.17)}
\def\slthalgebra{5.6}
\def\tebnspr{(5.18)}
\def\pionpiss{(5.19)}
\def\expciipieces{(5.20)}
\def\biioncc{(5.21)}
\def\ohhatccc{(5.22)}
\def\pioncccc{(5.23)}
\def\techlemmacp{5.7}
\def\ctwothph{(5.24)}
\def\ctwoththph{(5.25)}
\def\ctwothphp{(5.26)}
\def\booncis{5.8}
\def\boonciss{(5.27)}
\def\intertwns{5.9}
\def\equivproof{(5.28)}
\def\SSmorehone{5.2.3}
\def\hgsklth{(5.29)}
\def\ohthosepi{(5.30)}
\def\goodproj{(5.31)}
\def\ohomegas{(5.32)}
\def\SSextens{5.2.4}
\def\fullpiss{(5.33)}
\def\noslfour{5.10}
\def\twopistoom{(5.34)}
\def\kinofslf{(5.35)}
\def\prodofomr{(5.36)}
\def\SSsummary{5.2.5}
\def\summofone{5.11}
\def\decomofhone{(5.37)}
\def\genofhone{5.12}
\def\Spolybv{5.3}
\def\SSmain{5.3.1}
\def\maintheorem{5.13}
\def\exactseq{(5.38)}
\def\SSpi{5.3.2}
\def\volumeop{(5.39)}
\def\thepisss{(5.40)}
\def\comtox{(5.41)}
\def\projofX{(5.42)}
\def\pisonpis{5.14}
\def\pissonpiss{(5.43)}
\def\piofhonp{5.15}
\def\piofhonP{(5.44)}
\def\SSpihom{5.3.3}
\def\piisbv{5.16}
\def\piandbv{(5.45)}
\def\ohxprodpi{(5.46)}
\def\SSembed{5.3.4}
\def\imathdef{5.17}
\def\defiotaaone{(5.47)}
\def\defiotaatwo{(5.48)}
\def\ctimesx{(5.49)}
\def\JJeon{(5.50)}
\def\JJetw{(5.51)}
\def\JJeth{(5.52)}
\def\invthepiss{(5.53)}
\def\nicecomm{(5.54)}
\def\comwithzero{(5.55)}
\def\firstpartt{(5.56)}
\def\JJind{(5.57)}
\def\toohigh{(5.58)}
\def\exceptionone{(5.59)}
\def\exceptionone{(5.60)}
\def\Scomstr{5.4}
\def\SStwmod{5.4.1}
\def\tipoftwc{(5.61)}
\def\genmodcoh{5.18}
\def\grgenonom{(5.62)}
\def\onebdry{(5.63)}
\def\thdees{(5.64)}
\def\brackonom{(5.65)}
\def\othbry{(5.66)}
\def\comofom{(5.67)}
\def\basisinrone{(5.68)}
\def\honeoneact{5.19}
\def\SSgenpol{5.4.2}
\def\genproj{5.20}
\def\generpolof{(5.69)}
\def\asympisom{(5.70)}
\def\specst{(5.71)}
\def\spstdb{5.21}
\def\Scomplstr{5.5}
\def\SScomplstrint{5.5.1}
\def\cohofbo{5.22}
\def\tachquar{(5.72)}
\def\tachdbl{(5.73)}
\def\crsubsp{(5.74)}
\def\SSdualdec{5.5.2}
\def\bronid{5.23}
\def\brmmms{(5.75)}
\def\vanishomeg{(5.76)}
\def\mappip{(5.77)}
\def\dothomie{(5.78)}
\def\gmodofi{5.24}
\def\brahomie{(5.79)}
\def\conjfori{5.25}
\def\projofga{(5.80)}
\def\decofdblt{(5.81)}
\def\bvminus{5.26}
\def\imofga{(5.82)}
\def\expliii{(5.83)}
\def\decomofh{(5.84)}
\def\SSconject{5.5.3}
\def\sqspst{(5.85)}
\def\profomeg{(5.86)}
\def\appPBA{A}
\def\appPBAa{\hbox {A.}1}
\def\tbPBaa{\hbox {A.}1}
\def\tbPBab{\hbox {A.}2}
\def\tbPBac{\hbox {A.}3}
\def\tbPBad{\hbox {A.}4}
\def\appPBAb{\hbox {A.}2}
\def\tbPBae{\hbox {A.}5}
\def\tbPBaf{\hbox {A.}6}
\def\appPBAc{\hbox {A.}3}
\def\tbPBag{\hbox {A.}7}
\def\tbPBah{\hbox {A.}8}
\def\tbPBai{\hbox {A.}9}
\def\appPBC{B}
\def\eqCoa{(\hbox {B.}1)}
\def\eqJMphase{(\hbox {B.}2)}
\def\eqCoc{(\hbox {B.}3)}
\def\eqCod{(\hbox {B.}4)}
\def\eqCoe{(\hbox {B.}5)}
\def\eqCof{(\hbox {B.}6)}
\def\eqCog{(\hbox {B.}7)}
\def\eqCoh{(\hbox {B.}8)}
\def\eqCoi{(\hbox {B.}9)}
\def\eqCoj{(\hbox {B.}10)}
\def\eqCok{(\hbox {B.}11)}
\def\eqCol{(\hbox {B.}12)}
\def\eqCom{(\hbox {B.}13)}
\def\eqCon{(\hbox {B.}14)}
\def\eqCoo{(\hbox {B.}15)}
\def\eqCop{(\hbox {B.}16)}
\def\eqCoq{(\hbox {B.}17)}
\def\eqCor{(\hbox {B.}18)}
\def\eqCos{(\hbox {B.}19)}
\def\eqCot{(\hbox {B.}20)}
\def\appPBB{C}
\def\tbPBba{\hbox {C.}1}
\def\tbPBbb{\hbox {C.}2}
\def\tbPBbc{\hbox {C.}3}
\def\tbJMbd{\hbox {C.}4}
\def\APexplicit{D}
\def\SSapecintro{\hbox {D.}1}
\def\levelll{(\hbox {D.}1)}
\def\partfunct{(\hbox {D.}2)}
\def\cohdimmm{(\hbox {D.}3)}
\def\zerozero{\hbox {D.}1}
\def\Appthetables{\hbox {D.}2}
\def\hhmfour{\hbox {D.}2}
\def\hhmthree{\hbox {D.}3}
\def\hhmtwo{\hbox {D.}4}
\def\hhmone{\hbox {D.}5}
\def\hhmzero{\hbox {D.}6}
\def\primeplots{E}
\def\plotttsss{~1}
\def\Apolder{F}
\def\Apprelim{\hbox {F.}1}
\def\ttensor{(\hbox {F.}1)}
\def\masterlemma{\hbox {F.}1}
\def\tracee{(\hbox {F.}2)}
\def\tracefactors{(\hbox {F.}3)}
\def\coefficients{(\hbox {F.}4)}
\def\dfactor{(\hbox {F.}5)}
\def\piciex{\hbox {F.}2}
\def\pisandcis{(\hbox {F.}6)}
\def\notone{(\hbox {F.}7)}
\def\multlemma{\hbox {F.}3}
\def\identthree{(\hbox {F.}8)}
\def\identmulti{(\hbox {F.}9)}
\def\ASproof{\hbox {F.}2}
\def\prodofes{(\hbox {F.}10)}
\def\commone{(\hbox {F.}11)}
\def\identone{(\hbox {F.}12)}
\def\newproduct{(\hbox {F.}13)}
\def\productpis{(\hbox {F.}14)}
\def\productpispr{(\hbox {F.}15)}
\def\piscomm{(\hbox {F.}16)}
\def\geeee{(\hbox {F.}17)}
\def\newprrr{(\hbox {F.}18)}
\def\twopispr{(\hbox {F.}19)}
\def\auxlemma{\hbox {F.}4}
\def\auxident{(\hbox {F.}20)}
\def\letsgo{(\hbox {F.}21)}
\def\etimesp{(\hbox {F.}22)}
\def\gencase{(\hbox {F.}23)}
\def\commalg{(\hbox {F.}24)}
\def\Appproof{\hbox {F.}3}
\def\cpdelta{(\hbox {F.}25)}
\def\realbigstuff{(\hbox {F.}26)}
\def\oneextra{(\hbox {F.}27)}
\def\Saffine{G}
\def\SSaffine{\hbox {G.}1}
\def\eqAa{(\hbox {G.}1)}
\def\commutrel{(\hbox {G.}2)}
\def\thAb{\hbox {G.}1}
\def\peterweyl{(\hbox {G.}3)}
\def\thBa{\hbox {G.}2}
\def\modeldec{(\hbox {G.}4)}
\def\genvectf{(\hbox {G.}5)}
\def\eqAd{(\hbox {G.}6)}
\def\eqAf{(\hbox {G.}7)}
\def\xisfrompw{(\hbox {G.}8)}
\def\constraint{(\hbox {G.}9)}
\def\SSaffpol{\hbox {G.}2}
\def\polyvectors{\hbox {G.}3}
\def\polyvecdef{(\hbox {G.}10)}
\def\polyinf{(\hbox {G.}11)}
\def\ghostsntt{(\hbox {G.}12)}
\def\fdghvac{(\hbox {G.}13)}
\def\pirepr{(\hbox {G.}14)}
\def\smallspace{\hbox {G.}4}
\def\tttt{(\hbox {G.}15)}
\def\nplusrep{(\hbox {G.}16)}
\def\comofx{(\hbox {G.}17)}
\def\bigrepr{(\hbox {G.}18)}
\def\inside{\hbox {G.}5}
\def\polyes{(\hbox {G.}19)}
\def\frobenius{\hbox {G.}6}
\def\frobdec{(\hbox {G.}20)}
\def\condonph{(\hbox {G.}21)}
\def\crucial{(\hbox {G.}22)}
\def\diffph{(\hbox {G.}23)}
\def\infdiff{(\hbox {G.}24)}
\def\homscond{\hbox {G.}7}
\def\condforim{(\hbox {G.}25)}
\def\coness{\hbox {G.}8}
\def\connes{(\hbox {G.}26)}
\def\compexp{(\hbox {G.}27)}
\def\ghostvecexp{(\hbox {G.}28)}
\def\suthreepv{\hbox {G.}3}
\def\nplusmodone{~1}
\def\condforimth{(\hbox {G.}29)}
\def\derofg{(\hbox {G.}30)}
\def\liuvvec{(\hbox {G.}31)}
\def\hmsandplvc{\hbox {G.}1}
\def\polystable{\hbox {G.}9}
\def\poldecom{(\hbox {G.}32)}
\def\polydisaa{(\hbox {G.}33)}
\def\polydisba{(\hbox {G.}34)}
\def\polydisbb{(\hbox {G.}35)}
\def\polydisca{(\hbox {G.}36)}
\def\polydiscb{(\hbox {G.}37)}
\def\polydiscc{(\hbox {G.}38)}
\def\polydisda{(\hbox {G.}39)}
\def\polydisdb{(\hbox {G.}40)}
\def\polydisdc{(\hbox {G.}41)}
\def\polydisea{(\hbox {G.}42)}
\def\polydisea{(\hbox {G.}43)}
\def\SSbvaff{\hbox {G.}4}
\def\prodinghst{(\hbox {G.}44)}
\def\bvghost{(\hbox {G.}45)}
\def\bvprop{\hbox {G.}10}
\def\bvplgh{(\hbox {G.}46)}
\def\expanofd{(\hbox {G.}47)}
\def\bvoponpa{\hbox {G.}11}
\def\commofvec{(\hbox {G.}48)}
\def\brackofde{\hbox {G.}12}
\def\bvonhoms{(\hbox {G.}49)}
\def\donvectf{(\hbox {G.}50)}
\def\cohofdelpr{\hbox {G.}13}
\def\cohofdel{(\hbox {G.}51)}
\def\comofdandc{(\hbox {G.}52)}
\def\homotop{(\hbox {G.}53)}
\def\twodeleq{\hbox {G.}14}
\def\freemodules{H}
\def\gengr{\hbox {H.}1}
\def\holgen{(\hbox {H.}1)}
\def\relone{(\hbox {H.}2)}
\def\reltwo{(\hbox {H.}3)}
\def\arelfree{(\hbox {H.}4)}
\def\broper{(\hbox {H.}5)}
\def\relthree{(\hbox {H.}6)}
\def\relfour{(\hbox {H.}7)}
\def\relfive{(\hbox {H.}8)}
\def\firstmonom{(\hbox {H.}9)}
\def\typeone{(\hbox {H.}10)}
\def\typethree{(\hbox {H.}11)}
\def\typetwo{(\hbox {H.}12)}
\def\typefour{(\hbox {H.}13)}
\def\qiuscones{\hbox {H.}2}
\def\APexstates{I}
\def\SSgrring{\hbox {I.}1}
\def\exone{(\hbox {I.}1)}
\def\exdotthree{(\hbox {I.}2)}
\def\SSident{\hbox {I.}2}
\def\identqu{(\hbox {I.}3)}
\def\stctwo{(\hbox {I.}4)}
\def\stcthree{(\hbox {I.}5)}
\def\identctwoth{(\hbox {I.}6)}
\def\SShonegen{\hbox {I.}3}
\def\pissenone{(\hbox {I.}7)}
\def\pissenonec{(\hbox {I.}8)}
\def\ponemonetwo{(\hbox {I.}9)}
\def\slthreeopp{(\hbox {I.}10)}
\def\ptwoonemtwo{(\hbox {I.}11)}
\def\pisstwo{(\hbox {I.}12)}
\def\pisstwotwo{(\hbox {I.}13)}
\def\FGaa{(\hbox {I.}14)}
\def\FGab{(\hbox {I.}15)}
\def\FGad{(\hbox {I.}16)}
\def\FGac{(\hbox {I.}17)}
\def\FGae{(\hbox {I.}18)}
\def\FGaf{(\hbox {I.}19)}
\def\pissthreeoh{(\hbox {I.}20)}
\def\FGag{(\hbox {I.}21)}
\def\FGah{(\hbox {I.}22)}
\def\FGak{(\hbox {I.}23)}
\def\FGal{(\hbox {I.}24)}
\def\ohhiiijj{(\hbox {I.}25)}
\def\FGam{(\hbox {I.}26)}
\def\FGan{(\hbox {I.}27)}
\def\ohfive{(\hbox {I.}28)}
\def\volumeom{(\hbox {I.}29)}
\def\SStwgen{\hbox {I.}4}
\def\omegaone{(\hbox {I.}30)}
\def\omegarone{(\hbox {I.}31)}
\def\omegartwo{(\hbox {I.}32)}
\def\omegaonetwo{(\hbox {I.}33)}
\def\omegatwoone{(\hbox {I.}34)}
\def\omegathree{(\hbox {I.}35)}
\def\AVir{J}
\def\AVcoh{\hbox {J.}1}
\def\vircones{\hbox {J.}1}
\def\virreltbl{~1}
\def\sltcoh{\hbox {J.}1}
\def\vircohdec{(\hbox {J.}1)}
\def\virabstbl{~2}
\def\AVbv{\hbox {J.}2}
\def\strthsltw{\hbox {J.}2}
\def\plembvir{~3}
\def\genofvirbv{\hbox {J.}3}
\def\extenofvir{\hbox {J.}4}
\def\doublets{(\hbox {J.}2)}
\def\nonvanishpr{(\hbox {J.}3)}


\bref\Sunnyb{
O.~Aharony, O. Ganor, J. Sonnenschein and  S. Yankielowicz,
{\it $C=1$  string as a topological $G/G$ model}, \PLB{305} (1993)
35 {\tt (hep-th/9302027)}.}

\bref\Sunny{
O. Aharony, J. Sonnenschein and  S. Yankielowicz,
{\it $G/G$ models and $\cW_N$ strings}, \PLB{289} (1992) 309
({\tt hep-th/9206063}).}

\bref\BBSS{
F.A.~Bais, P.~Bouwknegt, K.~Schoutens and M.~Surridge,
{\it Extensions of the Virasoro algebra constructed
from Kac-Moody algebras using higher order Casimir invariants},
\NPB{304} (1988) 348.}

\bref\BV{
I.~Batalin and G.~Vilkovisky, {\it Quantization of gauge theories with
linearly dependent generators}, Phys.\ Rev.\ {\bf D28} (1983) 2567.}

\bref\BPZ{
A.A.~Belavin, A.M.~Polyakov and A.B.~Zamolodchikov,
{\it Infinite conformal symmetry in two dimensional quantum field
theory}, \NPB{241} (1984) 333.}

\bref\BDDT{
E.~Bergshoeff, J.~de~Boer, M.~de~Roo and T.~Tjin,
{\it The cohomology of the noncritical $\cW$-string},
\NPB{420} (1994) 379 {\tt (hep-th/9312185)}.}

\bref\BergSev{
E.~Bergshoeff, A.~Sevrin and  S.~Shen, {\it A derivation of the BRST
operator for non-critical $\cW$ strings}, \PLB{296} (1992) 95
{\tt (hep-th/9209037)}.}

\bref\BGG{
I.N.~Bernstein, I.M.~Gel'fand and S.I.~Gel'fand, {\it Differential
operators on the base affine space and a study of $\bfg$-modules}, in
``Lie groups and their representations,'' Proc.\ Summer School in
Group Representations, Bolyai Janos Math.\ Soc., Budapest 1971, pp.\
21, (Halsted, New York, 1975).}

\bref\BLNW{
M.~Bershadsky, W.~Lerche, D.~Nemeschansky and N.P.~Warner,
{\it A BRST operator for noncritical W strings},
\PLB{292} (1992) 35 {\tt (hep-th/9207067)}.}

\bref\BLNWnp{
M.~Bershadsky, W.~Lerche, D.~Nemeschansky and N.P.~Warner,
{\it Extended N=2 superconformal structure of gravity and W gravity
coupled to matter}, \NPB{401} (1993) 304 {\tt (hep-th/9211040)}.}

\bref\BeOo{
M.~Bershadsky and H.~Ooguri,
{\it Hidden $SL(n)$ symmetry in conformal field theories},
\CMP{139} (1991) 71.}

\bref\BdFl{
L.C.~Biedenharn and D.E.~Flath, {\it On the structure of tensor
operators in SU(3)}, \CMP{93} (1984) 93.}

\bref\ABil{
A.~Bilal, {\it What is $W$ geometry?}, \PLB{249} (1990) 56.}

\bref\Bor{
R.E.~Borcherds,
{\it Vertex algebras, Kac-Moody algebras, and the monster},
\PNAS{83}\ (1986) 3068.}

\bref\BT{
R. Bott and L.W.\ Tu, {\it Differential forms  in algebraic
topology} (Springer-Verlag, New York, 1982).}

\bref\Bou{
P.~Bouwknegt, {\it Extended conformal algebras from Kac-Moody
algebras}, in ``Infinite-dimensional Lie algebras and Lie
groups,'' {\it ed.} V.G.~Kac, Adv.\ Ser.\ Math.\ Phys.\ {\bf 7}
(1988) 527 (World Scientific, Singapore, 1989).}

\bref\BMPqg{
P.~Bouwknegt, J.~McCarthy and K.~Pilch, {\it Quantum group
structure in the Fock space resolutions of $\widehat{\frak{sl}}(n)$
representations}, \CMP {\bf 131} (1990) 125.}

\bref\BMPstr{
P.~Bouwknegt, J.~McCarthy and K.~Pilch,
{\it Some spects of free field resolutions in $2D$ CFT with
application to the quantum Drinfel'd-Sokolov reduction},
in the proceedings of ``Strings and Symmetries 1991,''
{\it eds.} N.~Berkovitz et al., (World Scientific, Singapore, 1991),
{\tt (hep-th/9110007)}.}

\bref\BMPvir{
P.~Bouwknegt, J.~McCarthy and K.~Pilch, {\it BRST analysis of physical
states for 2d gravity coupled to $c\leq 1$ matter}, \CMP{145} (1992)
541.}

\bref\BMPkar{
P.~Bouwknegt, J.~McCarthy and K.~Pilch,
{\it Semi-infinite cohomology in conformal field theory and $2D$ gravity},
\JGP{11}\ (1993) 225 {\tt (hep-th/9209034)}.}

\bref\BMPa{
P.~Bouwknegt, J.~McCarthy and K.~Pilch, {\it Semi-infinite
cohomology of $\cW$-algebras}, \LMP{29} (1993) 91
{\tt (hep-th/9302086)}.}

\bref\BMPb{
P.~Bouwknegt, J.~McCarthy and K.~Pilch, {\it On the BRST structure of
$\cW_3$ gravity coupled to $c=2$ matter},
in ``Perspectives in Mathematical
Physics,'' Vol.\ III, {\it eds.} R.~Penner and S.T.~Yau, pp.\ 77,
(International Press, Boston, 1994) {\tt (hep-th/9303164)}.}

\bref\BMPc{
P.~Bouwknegt, J.~McCarthy and K.~Pilch, {\it On the $\cW$-gravity spectrum
and its $G$-structure}, in the proceedings of the workshop
``Quantum Field Theory and String Theory,'' Carg\`ese 1993,
{\it eds.\ } L.~Baulieu et al., p.\ 59
(Plenum Press, New York, 1995) {\tt (hep-th/9311137)}.}

\bref\BMPstring{
P.~Bouwknegt, J.~McCarthy and K.~Pilch, {\it Operator algebra of the
 $4D$ $\cWth$ string}, to appear in the Proceedings of ``Strings '95:
 Future Perspectives in String Theory,'' USC-95/24, ADP-95-47/M39 {\tt
 (hep-th/9509nnn)}.}

\bref\BPist{
P.~Bouwknegt and K.~Pilch, {\it The BV-algebra structure of $\cWth$
 cohomology}, in ``G\"ursey Memorial Conference I: Strings and
 Symmetries,'' Lect.\ Notes in Phys.\ {\bf 447}, {\it eds.}
 G.~Akta\c{s} et al. (Springer Verlag, Berlin, 1995) {\tt
 (hep-th/9509nnn)}.}

\bref\BS{
P.~Bouwknegt and K.~Schoutens, {\it $\cW$ symmetry in conformal field
theory}, \PRep{223} (1993) 183 {\tt (hep-th/9210010)}.}

\bref\BSb{
P.~Bouwknegt and K.~Schoutens, {\it $\cW$-symmetry},
Adv.\ Series in Math.\ Phys.\ {\bf 22}, (World Scientific,
Singapore, 1995).}

\bref\Da{
F.~David, {\it Conformal field theories coupled to 2-D gravity
in the conformal gauge}, \MPL{A3} (1988) 1651.}

\bref\dBGb{
  J. de Boer and J. Goeree, {\it $\cW$-gravity from Chern-Simons theory},
  \NPB{381} (1992) 329 {\tt  (hep-th/9112060)}.}

\bref\DeBoer{
J. de Boer and  J. Goeree,
{\it KPZ analysis for $\cW_3$ gravity}, \NPB{405} (1993) 669
{\tt  (hep-th/9211108)}.}

\bref\Sevetal{
A. Deckmyn, R.  Siebelink, W. Troost and  A. Sevrin,
{\it On the Lagrangian realization of noncritical $\cW$ strings},
Phys.\ Rev.\ {\bf D51} (1995) 6970 ({\tt hep-th/9411221}).}

\bref\vDdV{
K.~de~Vos and P.~van~Driel, {\it The Kazhdan-Lusztig conjecture
for finite $\cW$-algebras},
\LMP{} to appear, {\tt (hep-th/9312016)}.}

\bref\vDdVb{
K.~de~Vos and P.~van~Driel, {\it The Kazhdan-Lusztig conjecture
for $\cW$-algebras}, {\tt (hep-th/9508020)}.}

\bref\DK{
J.~Distler and H.~Kawai, {\it Conformal field theory and 2-D
quantum gravity}, \NPB{321} (1989) 509.}

\bref\Dixmier{
J.\ Dixmier, {\it Enveloping algebras} (North Holland, Amsterdam, 1977).}

\bref\FLa{
V.A.~Fateev and S.L.~Lukyanov,
{\it The models of two dimensional conformal quantum field theory
with $\ZZ_n$ symmetry}, \IJMP{A3} (1988) 507.}

\bref\FLb{
V.A.~Fateev and S.L.~Lukyanov,
{\it Additional symmetries and exactly soluble models in two-dimensional
conformal field theory},
Sov.\ Sci.\ Rev.\ A.\ Phys.\ {\bf 15} (1990) 1.}

\bref\FZ{
V.A.~Fateev and A.B.~Zamolodchikov,
{\it Conformal quantum field theory models in two dimensions
having $\ZZ_3$ symmetry},  \NPB{280} (1987) 644.}

\bref\Feigin{
B.~Feigin, {\it The semi-infinite cohomology of the Virasoro
and Kac-Moody Lie algebras},
Usp.\ Mat.\ Nauk {\bf 39} (1984) 195.}

\bref\FeFrtw{
B.L.~Feigin and E.V.~Frenkel,
{\it Affine Kac-Moody algebras and semi-infinite flag manifolds},
\CMP{128}\ (1990) 161.}

\bref\FeFr{
B.~Feigin and E.~Frenkel,
{\it Quantization of the Drinfeld-Sokolov reduction},
\PLB{246} (1990) 75.}

\bref\FeFrloc{
B.~Feigin and E.~Frenkel, {\it Affine Kac-Moody algebras at the
critical level and Gelfand-Dikii algebras}, \IJMP{A7}
Suppl.\ A1 (1992) 197.}

\bref\FF{
B.L.~Feigin and D.~Fuchs,
in {\it Representation of Lie groups and related topics},
{\it eds.}\ A.M.~Vershik and D.P~Zhelobenko
(Gordon and Breach, New York, 1990).}

\bref\Fig{
J.M.~Figueroa-O'Farrill,
{\it On the homological construction of Casimir algebras},
\NPB{343} (1990) 450.}

\bref\FKW{
E.~Frenkel, V.G.~Kac and M.~Wakimoto, {\it Characters and fusion
rules for $\cW$-algebras via quantized Drinfeld-Sokolov reductions},
\CMP{147} (1992) 295.}

\bref\FGZ{
I.B. Frenkel, H. Garland and G.J. Zuckerman, {\it Semi-infinite
cohomology and string theory}, Proc. Natl. Acad. Sci. USA {\bf 83}
(1986) 8442.}

\bref\FHL{
I.B.~Frenkel, Y.-Z.~Huang and J.~Lepowsky,
{\it On axiomatic approaches to vertex operator algebras and modules},
Memoirs of the Amer.\ Math.\ Soc.\ {\bf 104}\ (1993) no.~494.}

\bref\FK{
I.B.~Frenkel and V.G.~Kac,
{\it Basic representations of affine Lie algebras and dual resonance
models}, \InM{62} (1980) 23.}

\bref\FLM{
I.B.~Frenkel, J.~Lepowsky and A.~Meurman, {\it Vertex operator
algebras and the monster}, Pure and Appl.\ Math.\ {\bf 134},
(Academic Press, San Diego, 1988).}

\bref\FNij{
A.~Fr\"ohlicher and A.~Nijenhuis, {\it Theory of vector-valued
differential forms I}, Indag.\ Math. {\bf 18} (1956) 338.}

\bref\GKa{
I.M.~Gel'fand and A.A.~Kirillov, {\it On the structure of the field of
quotients of the enveloping algebra of a semisimple Lie algebra},
\SMD{9} (1968) 669.}

\bref\GKb{
I.M.~Gel'fand and A.A.~Kirillov,
{\it The structure of the Lie field connected with a split semisimple
Lie algebra}, \FAP{3} (1969) 6.}

\bref\GeZl{I.M. Gel'fand and A.V. Zelevinskij, {\it Models of
representations of classical groups and their hidden symmetries},
Funct.\ Anal.\ Appl.\  {\bf 18} (1984) 183.}

\bref\Ge{
M.~Gerstenhaber, {\it The cohomology structure of an associative
ring},  \AnM{78} (1962) 267;
{\it On the deformation of rings and
algebras}, \AnM{79} (1964) 59.}

\bref\Gerv{
J.-L.~Gervais, {\it $W$ geometry from chiral embeddings},
\JGP{11} (1993) 293.}

\bref\Wgeom{
J.-L.~Gervais and Y.~Matsuo,
{\it $\cW$-geometries}, \PLB{274} (1992) 309 {\tt (hep-th/9110028)};
{\it Classical $A_n$ $\cW$-geometry},
\CMP{152} (1993) 317 {\tt (hep-th/9201026)}.}

\bref\Get{
E.~Getzler, {\it Batalin-Vilkovisky algebras and two-dimensional
topological field theories}, \CMP{159} (1994) 265
{\tt (hep-th/9212043)}. }

\bref\Go{
P.~Goddard, {\it Meromorphic conformal field theory}, in `Infinite
dimensional Lie algebras and Lie groups,' ed.\ V.~Kac, Proc.\ CIRM-Luminy
conf., 1988, Adv.\ Series in Math.\ Phys.\ {\bf 7} (1988) 556
(World Scientific, Singapore, 1989).}

\bref\GNOS{
P.~Goddard, W.~Nahm, D.~Olive and A.~Schwimmer,
{\it Vertex operators for non-simply-laced algebras},
\CMP{107} (1986) 179.}

\bref\Gu{
V.~Gurarie, {\it Logarithmic operators in conformal field theory},
\NPB{410} (1993) 535 {\tt (hep-th/9303160)}.}

\bref\Hullgeom{
C.~Hull, {\it The geometry of $W$ gravity},
\PLB{269} (1991) 257;  {\it $W$ geometry},
\CMP{156} (1993) 245.}

\bref\Hullgr{
C.~Hull, {\it Lectures on $W$ gravity, $W$ geometry and $W$ strings}, in the
proceedings of ``Trieste Summer School on High Energy Physics and
Cosmology,'' 1992 {\tt (hep-th/9302110)}.}

\bref\Jan{
J.C.~Jantzen, {\it Moduln mit einem h\"ochsten Gewicht},
Lect.\ Notes in Math.\ {\bf 750} (1979).}

\bref\Ka{
V.G.~Kac, {\it Infinite dimensional Lie algebras}, (Cambridge
University Press, Cambridge, 1985).}

\bref\KP{
V.G.~Kac and D.H.~Peterson, {\it Infinite dimensional Lie
algebras, theta functions and modular forms}, \AdM{53} (1984) 125.}

\bref\KR{
V.G.~Kac and A.K.~Raina, {\it Highest weight representations
of infinite dimensional Lie algebras}, Adv.\ Series in Math.\
Phys.\ {\bf 2}, (World Scientific, Singapore, 1987).}

\bref\Kleb{
I.~Klebanov, {\it Ward identities in two-dimensional string theory},
\MPL{A7} (1992) 723 {\tt (hep-th/9201005)}.}

\bref\Knapp{
A.W. Knapp, {\it Lie groups, Lie algebras, and cohomology},
(Princeton University Press, Princeton, 1988).}

\bref\KosSch{
Y.\ Kosmann-Schwarzbach, {\it Exact Gerstenhaber algebras and Lie
bialgebroids}, U.R.A. 169 CNRS preprint.}

\bref\Ko{
J.-L.~Koszul, {\it Crochet de Schouten-Nijenhuis et cohomologie},
Ast\'erisque (hors s\'erie) (1985) 257.}

\bref\Kr{
I.S.~Krasil'shchik, {\it Hamiltonian cohomology of canonical
algebras}, Sov.\ Math.\ Dokl.\ {\bf 21} (1980) 625;
{\it Schouten bracket
and canonical algebras}, \LNM{1334} (1988) 79.}

\bref\Kutas{
D. Kutasov,  E. Martinec and N. Seiberg, {\it Ground rings and their
modules in 2-d gravity with $c<1$}, \PLB{276} (1992) 437
{\tt (hep-th/9111048)}.}

\bref\Lerche{
W. Lerche and  A. Sevrin, {\it On the Landau-Ginzburg realization of
topological gravities}, \NPB{428} (1994) 259 {\tt (hep-th/9403183)}.}

\bref\LZvir{
B.H. Lian and G.J. Zuckerman, {\it New selection rules and physical
states in 2-d gravity: conformal gauge}, Phys.\ Lett.\ {\bf 254B}
(1991) 417;
{\it 2-d gravity with $c=1$ matter},
Phys.\ Lett.\ {\bf 266B} (1991) 21;
{\it Semi-infinite homology and
2-d gravity}, \CMP {\bf 145} (1992) 561.}

\bref\LZbv{
B.H.~Lian and G.J.~Zuckerman, {\it New perspectives on the
BRST-algebraic structure of string theory}, \CMP{154} (1993) 613
{\tt (hep-th/9211072)}.}

\bref\LPWX{
  H. Lu, C.N. Pope, X.J. Wang and K.W. Xu,
  {\it The complete cohomology of the $\cW_3$ string},
  Class.\ Quant.\ Grav. {\bf 11} (1994) 967; {\tt (hep-th/9309041)}.}

\bref\MMMO{
A.~Marshakov, A.~Mironov, A.~Morozov and M.~Olshanetsky,
{\it $c=r_G$ theories of $W_G$-gravity: the set of observables as a
model of simply laced $G$}, \NPB{404} (1993) 427.}

\bref\Mats{
Y.~Matsuo, {\it Classical $W$ geometry in conformal and light cone gauges},
\PTP{114} (1993) 243.}

\bref\Mfest{
J.~McCarthy, {\it Operator algebra of the $D=2+2$ $\cWth$ string}, in
proceedings of the ``H.S.~Green and A.~Hurst Festschrift:
Confronting the Infinite,'' Adelaide, March 1994,
 Mod.\ Phys.\ Lett.\ {\bf A} (1995),
 to appear {\tt (hep-th/9509nnn)}.}

\bref\MS{
M.~Penkava and A.~Schwarz, {\it On some algebraic structures arising
in string theory}, in ``Perspectives in Mathematical Physics,'' Vol.\
III, {\it eds.} R.~Penner and S.T.~Yau, (International Press, 1994)
{\tt (hep-th/9212072)}.}

\bref\Popb{
  C.N. Pope, {\it $\cW$-strings 93}, in the proceedings of ``Strings
  '93,'' {\it eds.} M.\ Halpern et al. (World Scientific, Singapore, 1995)
 {\tt (hep-th/ 9309125)}.}

\bref\Pope{
C.N.~Pope, E.~Sezgin, K.S.~Stelle and X.J.~Wang, {\it Discrete states
in the $W_3$ string}, \PLB{299} (1993) 247
{\tt (hep-th/9209111)}.}

\bref\RS{
L.~Rozansky and H.~Saleur, {\it Quantum field theory for the
multivariable Alexander-Conway polynomial}, \NPB{376} (1992) 461.}

\bref\Sadov{
V. Sadov, {\it On the spectra of $SL(N)_k/SL(N)_k$ cosets and $\cW_N$
gravities, I},
\IJMP{A8} (1993) 5115 {\tt (hep-th/9302060)}.}

\bref\Sch{
J.A.~Schouten, {\it Uber differentialkomitanten zweier kontravarianter
groessen}, Proc.\ Ser.\ {\bf A43} (1940) 449.}

\bref\SSvNg{
  K. Schoutens, A. Sevrin and P. van Nieuwenhuizen, {\it Induced gauge
  theories and $\cW$-gravity}, in Proc. of ``Strings and Symmetries 1991,''
  Stony Brook, May 1991, {\it eds.} \ N. Berkovits  et al.
  (World Scientific, Singapore, 1992).}

\bref\Th{
C.~Thielemans, {\it A Mathematica package for computing operator
product expansions}, \IJMP{C2} (1991) 787.}

\bref\TM{
J.~Thierry-Mieg,  {\it BRS analysis of Zamolodchikov's spin two and
three current algbera},  \PLB{197} (1987) 368.}

\bref\EV{
E.~Verlinde, {\it The master equation of 2-D string theory},
\NPB{381} (1992) 141 {\tt (hep-th/9202021)}.}

\bref\Wat{
G.M.T.~Watts, {\it
Determinant formulae for extended algebras in two
  dimensional conformal field theory},
\NPB{326} (1989) 648; erratum, \NPB{336} (1990) 720.}

\bref\Wa{
G.M.T.~Watts, {\it \cW-algebras and coset models}, \PLB{245} (1990) 65.}

\bref\Wia{
E.~Witten, {\it A note on the antibracket formalism},
\MPL{A5} (1990) 487.}

\bref\Wi{
E.~Witten, {\it Ground ring of two-dimensional string theory},
\NPB{373} (1992) 187 {\tt (hep-th/9108004)}.}

\bref\WiZw{
E.~Witten and B.~Zwiebach, {\it Algebraic structures and differential
geometry in 2d string theory}, \NPB{377} (1992) 55
{\tt (hep-th/9201056)}.}

\bref\WY{
Y.-S.~Wu and C.-J.~Zhu, {\it The complete structure of the cohomology
ring and associated symmetries in $d=2$ string theory},
\NPB{404} (1993) 245 {\tt (hep-th/9209011)}.}

\bref\Za{
A.B.~Zamolodchikov, {\it Infinite additional symmetries in two
dimensional conformal quantum field theory}, \TMP{65} (1985) 1205.}

\bref\CJZhu{
C.-J.~Zhu, {\it The structure of the ground ring in critical
$\cW_3$ gravity}, {\tt (hep-th/9508125)}.}


\hfuzz=32pt

\nopagenumbers
\pageno=0
%
%
%
%
\line{}
\vskip1cm
\centerline{\bf THE $\cWth$ ALGEBRA:}\smallskip

\centerline{\bf MODULES, SEMI-INFINITE COHOMOLOGY AND BV-ALGEBRAS}
\vskip1cm

\centerline{Peter BOUWKNEGT$\,^{1}$, Jim McCARTHY$\,^1$ and
Krzysztof PILCH$\,^2$}
\bigskip

\centerline{\sl $^1$ Department of Physics and Mathematical Physics}
\centerline{\sl University of Adelaide}
\centerline{\sl Adelaide, SA~5005, Australia}
\bigskip

\centerline{\sl $^2$ Department of Physics and Astronomy }
\centerline{\sl  University of Southern California}
\centerline{\sl Los Angeles, CA~90089-0484, USA}
\medskip
\vskip1.5cm

\centerline{\bf ABSTRACT}\medskip
{\rightskip=1cm
\leftskip=1cm
\noindent
The noncritical $D=4$ $\cW_3$ string is a model of $\cW_3$ gravity
coupled to two free scalar fields.  In this paper we discuss its BRST
quantization in direct analogy with that of the $D=2$ (Virasoro)
string.  In particular, we calculate the physical spectrum as a
problem in BRST cohomology.  The corresponding operator cohomology
forms a BV-algebra. We model this BV-algebra on that of the
polyderivations of a commutative ring on six variables with a
quadratic constraint, or, equivalently, on the BV-algebra of
(polynomial) polyvector fields on the base affine space of
$SL(3,\CC)$. \hfil\break
\indent
In deriving this picture over the past few years, a number of results
on the representation theory of $\cW$-algebras were developed.  In
this paper we attempt to present a complete summary of the progress
made in these studies.  Thus, the paper consists of three main parts.
In Part I we develop the machinery required to study $\cW$-modules,
and apply it, in particular, to Verma modules and Fock modules of the
$\cWth$ algebra at central charge $c=2$.  The main results of this
part are a formula for the Jordan-H\"older multiplicities of the $c=2$
Verma modules and a resolution of $c=2$ irreducible modules in terms
of (generalized) Verma modules.  In Part II we use the results of Part
I to compute the semi-infinite cohomology of the $\cWth$ algebra with
values in the tensor product of a $c=2$ Fock module and a $c=98$ Fock
module.  In Part III, after developing some general results about
BV-algebras, their modules and discussing some examples, we show how
the corresponding operator cohomology $\fH$ can be given the structure
of a BV-algebra.  This algebra is $\ZZ$-graded by ghost number and
nonvanishing in degrees $0$ to $8$. The dot product is induced from
the normal ordered product in the vertex operator algebra.  The
abelian ring at ghost number $0$ is shown to be a model space of
$\slth$, isomorphic to the polynomial ring $\cR$ generated by
$\{x_\si,x^{\si}\}_{\si=1,2,3}$ divided out by the ideal generated by
the constraint $x_\si x^\si = 0$.  In our general discussion the set
of polyderivations, $\fP$,  of the ring, $\cR$,  is shown to form a
BV-algebra.  By construction, we find a BV-homorphism of $\fH$ to the
BV-algebra $\fP$.  Conversely, we show how $\fP$ embeds into
$\fH$ as a dot algebra.  Using these results the remainder of $\fH$ is
partially understood in terms of specific $\fP$-modules.}

\vfil
\line{USC-95/18 \hfil}
\line{ADP-95-46/M38 \hfil}
\line{{{\tt hep-th/9509119}}\hfil September 1995}

\eject

%
\baselineskip=1.2\baselineskip
\footline{\hss \tenrm -- \folio\ -- \hss}

%
%
\centerline{\bf CONTENTS}\medskip

\centerline{Section \JMI. INTRODUCTION AND PRELIMINARIES}\bigskip

\item{\genint.} General introduction  \dotfill  4
\itemitem{\Physmot.} Physical context and motivation \dotfill 4
\itemitem{\Mathmot.} Mathematical context and motivation \dotfill 5
\item{\Out.} Outline and summary of results \dotfill 6
\item{\Not.} Glossary of notation \dotfill 9
\bigskip

\centerline{Section \PBW. $\cW$-ALGEBRAS AND THEIR MODULES}\bigskip

\item{\Walg.} $\cW$-algebras \dotfill  11
\itemitem{\Wintro.} Introduction to $\cW$-algebras \dotfill 11
\itemitem{\Wth.} The $\cW_3$ algebra \dotfill 11
\item{\Wthmod.} $\cW_3$ modules \dotfill 12
\itemitem{\WOcat.} The category $\cO$ \dotfill 12
\itemitem{\WVer.} (Generalized) Verma modules \dotfill 15
\itemitem{\WFock.} Fock spaces \dotfill 23
\item{\Wctwo.} Verma modules and Fock modules at $c=2$ \dotfill 27
\itemitem{\Wctwoa.} Generalities \dotfill 27
\itemitem{\Wctwob.} Explicit examples \dotfill 30
\item{\Wres.} Resolutions \dotfill 33
\itemitem{\WVerres.} Verma module resolutions of $c=2$ irreducible $\cWth$
modules \dotfill 34

\bigskip

\centerline{Section \NSbrst.
BRST COHOMOLOGY OF THE $4D$ $\cWth$ STRING}
\bigskip

\item{\Scomplex.} Complexes of  semi-infinite cohomology of the
$\cW_3$ algebra \dotfill 41
\itemitem{\SSghosts.} The $\cW_3$ ghost system \dotfill 41
\itemitem{\SSbrst.} The BRST current and the differential \dotfill 42
\item{\Swcohprob.} The $\cW_3$ cohomology problem for $4D$ $\cWth$
gravity \dotfill 43
\item{\SSpreliminaries.} Preliminary results  \dotfill 46
\itemitem{\SSrelative.} A comment on the relative cohomology \dotfill
46
\itemitem{\SSreduction.} Reduction theorems \dotfill 47
\itemitem{\SSslth.} The $\bga$ symmetry of $H(\cW_3,\fC)$ \dotfill 49
\itemitem{\SSbilinear.} A bilinear form on $\fC$ and $H(\cW_3,\fC)$
        \dotfill 50
\item{\SSfundamental.} The cohomology in the ``fundamental Weyl
chamber'' \dotfill 50
\itemitem{\SSHfirst.} $ H(\cW_3,L(\La,0)\otimes F(\La^L,2i))$ with
$-i\La^L+2\rh\in P_+$ \dotfill 51
\itemitem{\SHgen.} $H(\cW_3,F(\La^M,0)\otimes F(\La^L,2i))$ with
$-i\La^L+2\rh\in P_+$ \dotfill 54
\item{\SSgeneral.} The  conjecture for $H(\cW_3,\fC)$  \dotfill 56
\itemitem{\SSgenintro.} Introduction  \dotfill 56
\itemitem{\SSvanishh.} A vanishing theorem  \dotfill 57
\itemitem{\SSgencoh.} $H(\cW_3,F(\La^M,0)\otimes F(\La^L,2i))$
\dotfill 57
\vfill\eject

\centerline{Section \BValgebra. BATALIN-VILKOVISKY ALGEBRAS}
\bigskip
\item{\SGBValg.} G-algebras and BV-algebras \dotfill 61
\itemitem{\SSdefini.} Definitions   \dotfill 61
\itemitem{\SSpolyderaa.} The G-algebra of polyderivations of an Abelian
    algebra   \dotfill 62
\itemitem{\SBVpolder.} Example: the BV-algebra of polyvectors on a
    free algebra $\cC_N$    \dotfill 64
\itemitem{\SSpolyder.} Algebra of polyderivations associated with a
    BV-algebra \dotfill 65
\item{\Smodules.} G-modules  and  BV-modules  \dotfill 66
\itemitem{\SSpolydermod.} Natural G-modules for the G-algebra
    $(\cP(\cR),\,\cdot\,,[-,-]_S)$ \dotfill 67
\item{\bvpolyderivations.} The BV-algebra of polyderivations of the
    ground ring algebra $\cG_N$ \dotfill 67
\itemitem{\SSgrrinal.} The ``ground ring'' algebra $\cG_N$  \dotfill 68
\itemitem{\hiddensymmetry.} A ``hidden symmetry'' of $\cG_N$ \dotfill 68
\itemitem{\polyderofcgn.} Polyderivations of $\cG_N$   \dotfill 69
\itemitem{\galgstr.} The G-algebra structure of $\cP(\cG_N)$ \dotfill 73
\itemitem{\bvpolyvectors.} The BV-algebra structure of $\cP(\cG_N)$
\dotfill 74
\itemitem{\chiral.} ``Chiral'' subalgebras of $\cP(\cG_N)$   \dotfill 76
\item{\SSthree.} The $N=3$ case  \dotfill 77
\itemitem{\SSpirthree.} The algebra $\cP(\cR_3)$ \dotfill 77
\itemitem{\hidsym.} The hidden symmetry structure \dotfill 78
\itemitem{\twistedmodules.} ``Twisted'' modules of $\fP$  \dotfill 79
\itemitem{\SStwplcl.} A classification of twisted polyderivations \dotfill 80

\bigskip

\centerline{Section \Sbvalgebra. THE BV-ALGEBRA OF THE $\cW_3$ STRING}
\bigskip
\item{\SBVintro.}   Introduction \dotfill 84
\itemitem{\SBVgen.} General results \dotfill 84
\itemitem{\SSmnotation.} More notation \dotfill 85
\item{\Ssurvey.} A preliminary survey of $\fH$  \dotfill 85
\itemitem{\SSground.} The ground ring $\fH^0$   \dotfill 85
\itemitem{\SSautom.} $\fH^1$: the $\bga$ symmetry of $\fH$ revisited
\dotfill 87
\itemitem{\SSmorehone.} More $\fH^1$  \dotfill 89
\itemitem{\SSextens.} An extension of $\sosi$  \dotfill 90
\itemitem{\SSsummary.} A summary for $\fH^1$  \dotfill 90
\item{\Spolybv.} The relation between $\fH$ and $\fP$   \dotfill 90
\itemitem{\SSmain.} The main theorem   \dotfill 90
\itemitem{\SSpi.} $\pi(\fH)=\fP$  \dotfill 91
\itemitem{\SSpihom.} $\pi$ is a BV-algebra homomorphism  \dotfill 91
\itemitem{\SSembed.} An embedding $\imath:\fP\rightarrow \fH$
\dotfill 92
\item{\Scomstr.} The bulk structure of $\fH$ \dotfill 93
\itemitem{\SStwmod.} Twisted modules of $\fH^0$ \dotfill 94
\itemitem{\SSgenpol.} Interpretation of $\fH$ in terms of
twisted polyderivations \dotfill 95
\item{\Scomplstr.} Towards the complete structure of $\fH$ \dotfill 96
\itemitem{\SScomplstrint.} The BV-operator $b_0$ \dotfill 96
\itemitem{\SSdualdec.} The dual decomposition of $\fH$ \dotfill 97
\itemitem{\SSconject.} Concluding remarks and open problems \dotfill 99

\vfill\eject
\centerline{APPENDICES}
\bigskip

\item{\appPBA.} Verma modules at $c=2$ \dotfill 101
\itemitem{\appPBAa.} Primitive vectors \dotfill 101
\itemitem{\appPBAb.} Irreducible modules\dotfill 103
\itemitem{\appPBAc.} Verma modules\dotfill 104
\item{\appPBC.} Vertex Operator Algebras associated to root lattices
\dotfill 106
\item{\appPBB.} Tables for resolutions of $c=2$ irreducible
modules\dotfill 108
\item{\APexplicit.} Summary of explicit computations \dotfill 110
\itemitem{\SSapecintro.} Introduction \dotfill 110
\itemitem{\Appthetables.} The tables  \dotfill 111
\item{\primeplots.} A graphical representation of
        $H_{pr}(\cW_3,\fC)$ \dotfill 114
\item{\Apolder.} Polyderivations $\cP(\cG_N)$ \dotfill 116
\itemitem{\Apprelim.} Preliminary results \dotfill 116
\itemitem{\ASproof.}  Proof of Theorem \genproduct \dotfill 117
\itemitem{\Appproof.} Proof of Theorem \bvopmth \dotfill 118
\item{\Saffine.} BV-algebra of polyvectors on the base affine space
$A(G)$ \dotfill 119
\itemitem{\SSaffine.} The base affine space $A(G)$  \dotfill 119
\itemitem{\SSaffpol.} Polyvectors on $A(G)$ \dotfill 120
\itemitem{\suthreepv.} Example of $\slth$ \dotfill 124
\itemitem{\SSbvaff.} The BV-algebra structure of $\cP(A)$ \dotfill 127
\item{\freemodules.} Free modules of $\fP_\pm$ \dotfill 130
\item{\APexstates.} Some explicit cohomology states  \dotfill 133
\itemitem{\SSgrring.} The ground ring generators   \dotfill 133
\itemitem{\SSident.} The identity quartet   \dotfill 133
\itemitem{\SShonegen.} Generators of $\fH_1^n$, $ n\geq 1$  \dotfill 134
\itemitem{\SStwgen.}  Twisted modules of the ground ring  \dotfill 136
\item{\AVir.} The BV-algebra of $2D$ $\cW_2$ string  \dotfill 137
\itemitem{\AVcoh.} The cohomology problem \dotfill 137
\itemitem{\AVbv.} The BV-algebra and structure theorems \dotfill 138
\bigskip

\item{} REFERENCES \dotfill 140
\vfil\eject
%
%
\newsec{INTRODUCTION AND PRELIMINARIES}
\seclab\JMI

\subsec{General introduction}
\subseclab\genint

This paper is an exposition of our work, over the past few years, on the
$\cW_3$ algebra: the representation theory; the corresponding
semi-infinite cohomology for special modules; the
operator algebra of related $\cW_3$ string models and its
BV-algebra interpretation.
This study has several motivations from
different directions, which we will briefly indicate before
returning to an outline of the results.\smallskip

\subsubsec{Physical context and motivation}
\subsubseclab\Physmot

In $\cW$-gravities%
\foot{For a general review see, \eg, [\Hullgr,\Popb,\SSvNg,\dBGb] and
references therein.} the vector-generated diffeomorphism symmetry of
ordinary gravity is extended by higher tensor structures.  The
resulting gauge theory then involves massless higher-spin tensor
fields in addition to the spin two field corresponding to metric
deformations.  It is an intriguing possibility that there should be
some corresponding generalization of geometry which will allow a
natural description of these $\cW$ generalizations.  A number of
groups have made preliminary studies of the subject (see, \eg,
[\Hullgeom,\Wgeom,\ABil,\Gerv,\Mats]),
but as yet it has not been developed to an elegant theory.
One can hope that a better understanding of the models themselves will
aid in this development.  Recent progress in constructing
$\cW$-gravities in two dimensions, where the quantization may be
carried through, suggests this as the promising avenue for exploration.
Given that a first-quantized description of a propagating string must
be independent of the parametrization of the two-dimensional string
world-sheet, a study of 2-d $\cW$-gravity is further motivated as a
possible extension of string theory.  We will follow the string
motivation here by restricting our attention to models for which the
matter content is a conformal field theory.

 Let us briefly recall the corresponding situation for ordinary 2-d
gravity.  In a conformal gauge quantization using the DDK ansatz
[\Da,\DK], a well-defined BRST quantization of the model exists for a
restricted range of the central charge of the matter CFT; namely,
$c^M\leq 1$.  The model then splits into almost-decoupled left- and
right-moving ``chiral'' sectors, and the physical states can be computed
[\LZvir,\BMPvir] from the BRST cohomology (also called, semi-infinite
cohomology)
of the Virasoro algebra with values in a
tensor product of two scalar field Fock modules.
In [\Wi], Witten instigated the study of the algebra of the
corresponding physical operators for the case of a single free matter
scalar ($c^M = 1$), the $2D$ (Virasoro) string.  He found a rich
structure partially described by an interesting geometrical
interpretation.  Further study showed how this structure could be
implemented in deriving the physical consequences of the model
[\WiZw,\Kleb,\EV], and how it extends to $c^M \leq 1$ [\Kutas].
Finally, Lian and Zuckerman [\LZbv] identified the underlying
mathematical structure as a Batalin-Vilkovisky (BV-) algebra -- a
special case of a Gerstenhaber (G-) algebra (see also
[\MS,\Get]).  They found that 2-d gravity models generically have this
BV-algebra structure, and they further showed how the geometrical nature
of Witten's results arise through a homomorphism from the BV-algebra of
operator cohomology to the BV-algebra of regular polyvector fields on
the (complex) plane. The remainder of the physical operator algebra was
then interpreted, through a nondegenerate pairing, as a module
of this BV-algebra.
\smallskip

 To describe $\cW$-gravity one must quantize a gauge theory based on a
nonlinear algebra of constraints; namely, a $\cW$-algebra extension of
the Virasoro algebra (see [\BS,\BSb] and references therein). It is
this algebraic structure which allows a definition of certain
$\cW$-gravity models through BRST quantization even though the
associated $\cW$-geometry is not yet well understood in general.  In
fact, working by analogy with ordinary 2-d gravity, there exists a
well-motivated BRST quantization of $\cW$-gravity coupled to conformal
matter with a restricted range for the central charge in the matter
CFT [\BLNW].  For technical reasons, the most complete discussion of
non-critical $\cW$-string models to date has been given for $c^M \leq
2$ in $\cWth$ gravity, \ie, $\cW$-gravity based on the $\cWth$ algebra
(see, \eg, [\BLNW,\BLNWnp,\BMPa,\BMPb,\LPWX,\BMPc,\BDDT,\Lerche]).  A
convenient representation\foot{This representation is directly
relevant to the situation in which all ``cosmological constant'' terms
are tuned to zero.  Indirect arguments make these results relevant to
the generic case as well.} for this class of models is the system of
$\cWth$ gravity coupled to a matter sector consisting of two free
scalar fields.  It is presumed that a similar treatment may be given
in the general rank $N$ case for $c^M \leq N$, and indeed many of the
results we present are completely general with this in mind.  The
models with $c^M < 2$ are obtained by choosing appropriate background
charge sources in the free field matter system, and a rather
well-known projection maps these to the minimal models of the $\cW_3$
algebra [\FKW,\BMPstr].  The case $c^M=2$ is the limit without
background charge insertions. \smallskip

 The analysis of the $2D$ string recalled above has been extended to
$\cWth$ gravity models over a number of years [\BMPa,\BMPb,\BMPc].  In
this paper we will discuss the special case of $c^M=2$.  In the
corresponding string interpretation the matter scalar fields would
embed the world sheet of the string into a two-dimensional space-time.
But, moreover, since this is a non-critical theory there are dynamical
gravitational degrees of freedom -- under the DDK-type ansatz these
are described by a pair of scalar fields of ``wrong sign'' with a
background charge source, the so-called Liouville sector.  Thus, in
this string language, the model describes a ($2+2$)-dimensional string
in non-trivial background fields.  We call it the $4D$ $\cW_3$
string. \smallskip

\subsubsec{Mathematical context and motivation}
\subsubseclab\Mathmot

$\cW$-algebras are non-linear extensions of the Virasoro algebra (see
[\BS,\BSb] and references therein).  In the class of $\cW$-algebras
the simplest one is the so-called $\cWth$ algebra, which possesses --
apart from the Virasoro generators $L_n,\, n\in \ZZ$ -- one additional
(infinite) set of generators $W_n,\, n\in \ZZ$.  The $\cWth$ algebra,
being the simplest infinite dimensional algebra with non-linear
defining relations, is a useful laboratory to see which properties,
constructions and techniques from the Lie algebra case extend, or do
not extend as the case may be, to the non-linear case.  Specifically,
in this paper, we study the structure theory of a suitable category of
$\cWth$ modules (\eg, composition series) and various aspects of
homological algebra (\eg, resolutions, semi-infinite
cohomology).\smallskip

There are two immediately obvious differences with the Lie algebra case.
First, the adjoint action of the Cartan subalgebra on $\cWth$
is not diagonalizable, and similarly its action on most interesting
$\cWth$ modules is not diagonalizable.%
\foot{This is very reminiscent of the Lie superalgebra case.}
As a consequence, we are led to incorporate so-called
generalized Verma modules, \ie, modules induced
from an indecomposable module of the Cartan subalgebra by the negative root
operators, into our framework.
Secondly, the tensor product of two $\cWth$ modules does not, in general,
carry the structure of a $\cWth$ module.  This necessitates
a generalization of what we mean by semi-infinite cohomology
of an algebra with values in a tensor product of modules.  Moreover, it
prevents (at least a straightforward) application of many standard
techniques in calculating such a cohomology.\smallskip

Besides the interest from a purely technical point of view, studying
the semi-infinite cohomology of $\cWth$ algebras is interesting
because it provides us with beautiful, yet highly nontrivial, examples
of so-called BV-algebras [\Ko,\LZbv,\MS,\Get].  BV-algebras are
$\ZZ$-graded, supercommutative, associative algebras with a second
order derivation $\De$ of degree $-1$, satisfying $\De^2=0$.  They
naturally possess the structure of a G-algebra [\Ge], \ie, a
$\ZZ$-graded, supercommutative, associative algebra under a product,
as well as a $\ZZ$-graded Lie superalgebra under a bracket, and such
that the bracket acts as a superderivation of the product.  G-algebras
show up in many different areas of mathematics.  An important example
of a G-algebra is the set of polyderivations $\cP(\cR)$ of a
commutative algebra $\cR$.  It is an interesting question to
determine for which algebras $\cR$, the set of polyderivations
$\cP(\cR)$ possesses, in fact, the structure of a BV-algebra.  In this
paper we present some examples where this turns out to be the case,
namely, the free algebra $\cC_{N}$ on $N$ generators and the algebra
$\cR_{N}$ obtained from $\cC_{2N}$ by dividing out the ideal generated
by a quadratic relation.

A special class of BV-algebras are those for which the cohomology
of the BV-operator $\De$ vanishes.  It turns out that the
semi-infinite cohomology of the $\cWth$ algebra with coefficients in
the tensor product of two $\cWth$ Fock modules can be equipped
with the structure of a BV algebra and
contains, in a sense, the algebra of polyderivations $\cP(\cR_3)$ in
exactly such a way as to make the cohomology of $\De$ trivial.
In fact, at least superficially, the cohomology looks like a specific
``patching'' of a set of G-modules $\cP(\cR_3,M_w)$, \ie, polyderivations
with coefficients in an $\cR_3$ module $M_w$, labeled by elements
of the Weyl group of $\slth$ ($\cP(\cR_3,M_{w=1}) \cong \cP(\cR_3)$).
Since $\cP(\cR_3)$ can, in fact, be identified with the set of polyvector
fields (with polynomial coefficients) on the so-called base affine space
of $SL(3,\CC)$, we also get an interesting lead
to studying geometrical aspects of $\cW$-algebras.

\subsec{Outline and summary of results}
\subseclab\Out

A semi-infinite cohomology may be defined for the $\cWth$
algebra by analogy with that for the Virasoro case.  Corresponding to
the two sets of generators $L_m$ and $W_m$, $m\in \ZZ$, introduce two
sets of ghost oscillators $(b^{[i]}_m,c^{[i]}_m)\,,i=2,3$ , generating
the Fock space $F^{\rm gh}$. For any two positive energy\foot{A
positive energy module has $L_0$ diagonalizable with finite
dimensional eigenspaces, and with the spectrum bounded from below.
The category $\cO$ of relevant modules is defined more precisely in
Section \PBW.}  $\, \cWth$-modules $V^M$ and $V^L$, such that $c^M +
c^L = 100$, there exists a complex $(V^M \otimes V^L\otimes F^{{\rm
gh}}, d)$, graded by ghost number, and with a nilpotent differential
(BRST operator) $d$ of degree $1$ [\TM,\BLNW], with leading terms
$$
d = \sum_{m}\big( c^{[3]}_{-m}(\hbox{$1\over\sqrt{\be^M}$} W^M_m -
\hbox{$i\over\sqrt{\be^L}$} W^L_m) +
		c^{[2]}_{-m}(L^M_m+L^L_m) \big) + \dots \, ,
$$
where $\be=16/(22+5c)$.
The cohomology of $d$ at degree $n$ will be denoted by
$H^{n}(\cWth,V^M\otimes V^L)$ and called the BRST cohomology of the
$\cWth$ algebra on $V^M\otimes V^L$.
\smallskip

 The central problem motivating the present study was the computation of
this cohomology for Fock modules, particularly the case $c^M=2$ which is
the case of interest for the $4D$ $\cWth$ string. The result is given
in Section \NSbrst.  In turn, this problem spawned several other studies
of mathematical and possibly physical  interest.  In particular, the
calculations of Section \NSbrst\ require a detailed knowledge of the
representation theory of $\cWth$, which is discussed in Section \PBW.
General results show that the corresponding operator cohomology, $\fH$,
forms a BV-algebra [\LZbv].   In Section \BValgebra\ we discuss models
of BV-algebras which allow us to  develop  a detailed understanding of
$\fH$ in Section \Sbvalgebra.  The results there go a long way
towards verifying those of Section \NSbrst.
\smallskip

 In the remainder of this introduction we summarize the main results of
each section.
\smallskip

  The modules of interest in Section \PBW\ are Fock modules and
(generalized) Verma modules.  The structure of a given module
$V\in\cO$ can be exhibited in part through its composition series,
JH$(V)$ (Theorem \thJHa). In fact (Lemma \thPBaea) there is a 1--1
correspondence between primitive vectors in $V$ of (generalized)
weight $(h,w)$ and irreducible modules $L(h,w,c)$ appearing in
JH$(V)$.  Denote the multiplicity with which a given irreducible
module $L$ appears in JH$(V)$ by $(V:L)$.
Linear independence of the
characters of irreducible modules (Theorem
\thPBaaa), gives the result (Theorem \thPBce) that
$$
(M(\La,\al_0) : L) = (F(\La,\al_0) : L) \, .
$$
Thus by studying Fock modules one learns detailed information about
related Verma modules, and {\it vice versa}.  The crucial result here is
Corollary \thPBcda, which shows that generically the two are
isomorphic as $\cWth$ modules.  In particular, one finds that if
$-i(\La + \al_0\rho) \in D_+$,
with $\al_0{}^2<-4$, then $F(\La,\al_0) \cong \overline{M}(\La,\al_0)$.
\smallskip

For $c=2$ ($\al_0=0$),
the case of most interest in this paper, the  Fock module is
unitary with respect to a Hermitian inner product and we show in Theorem
\thPBda\ that $F(\La,0)$ is completely reducible, and in fact, for
$\La\in P$
$$
(F(\La,0) : L(\La',0)) = m_\La^{\La'} \, ,
$$
where $m_\La^{\La'}$ is the multiplicity of the weight $\La$ in the
finite-dimensional
irreducible $\slth$ module $\cL(\La')$.  This is a proof of the
Kazhdan-Lusztig conjecture for this special case (for a general
discussion of the Kazhdan-Lusztig conjecture for $\cW$-algebras, see
[\vDdV,\vDdVb]).  Further, for $c=2$, we apply this understanding to derive
(generalized) Verma module resolutions for irreducible $\cWth$ modules.
\smallskip

 The calculation of the semi-infinite cohomology is detailed in
Section $\NSbrst$.  We begin, in Section \Swcohprob, by finding a
lattice of momenta for the matter Fock modules so that the cohomology
complex lifts to a Vertex Operator Algebra, $\fC$, on which the
differential acts as the charge of a spin-1 current.  By means of the
usual state/operator mapping of conformal field theory, the main
results of this section are summarized as the cohomology of the
complex $(\fC,d)$ , denoted $H(\cWth,\fC)$ (with a slight abuse of notation).
When considered as an algebra with the induced VOA structure, we
will denote this operator cohomology by $\fH$.  There
exists a bilinear form on $\fC$ which induces a nondegenerate pairing
between the cohomology at ${\rm gh}=n$ and ${\rm gh}=8-n$ (Theorem \lemform).

 The fundamental result which allows the techniques developed in
Section \PBW\ to be applied is the reduction theorem (Theorem
\redthm):

\smallskip
\noindent
{\it For an arbitrary generalized Verma module
$M^{(\vdeg)}(\La^M,\al_0^M)$ and a contragredient Verma module
$\cgM(\La^L,\al_0^L)$, $c^M+c^L=100$, the cohomology
$H(\cW_3,M^{(\vdeg)}(\La^M,\al_0^M)\otimes \cgM(\La^L,\al_0^L))$ is
nonvanishing if and only if
$$
-i(\La^L+\al_0^L\rh)=w(\La^M+\al^M_0\rh)\,,
$$
for some $w\in W$, in which case it is spanned by the states
$$
v_0 \, ,\quad c_0^{[2]}v_0 \, , \quad c_0^{[3]}v_{\vdeg-1}\, , \quad
c_0^{[3]}c_0^{[2]}v_{\vdeg-1} \, ,
$$
where $v_i=v_i^M\otimes \overline v^L\otimes |0\rangle_{gh}$,
$i=0,\ldots,\vdeg-1$, span the highest weight space.}
\smallskip

\noindent For $c^M=2$ and $-i\La^L + 2\rho \in P_+$,
combining this theorem with the Verma module resolutions of
Section \PBW\ computes $H(\cW_3,L(\La,0)\otimes F(\La^L,2i))$ (Theorem
\fundcoh).  The Fock space decomposition theorem (Theorem \thPBda) then
computes (Theorem \fockcoh) the desired cohomology,
$H^n(\cW_3,F(\La^M,0)\otimes F(\La^L,2i))$.  The resulting operator
cohomology for this sector of Liouville momenta may
be decomposed under $\bga$ into cones of finite-dimensional
irreducible modules
at different ghost numbers (Theorem \cohfweyl).  For the remaining
sectors, \ie, for $w(-i\La^L+2\rho) \in P_+$, $w\in W$, we are able to
derive the full result from the assumption of a kind of Weyl group
symmetry -- the result being that up to ghost number shifts the cones
are essentially reflected to the
other Weyl chambers.  Thus, the complete
cohomology for the $c^M=2$ case is summarized in Theorem
\fullcoh.  \smallskip

 General results [\LZbv] imply that the operator cohomology forms a
BV-algebra, $(\fH,\,\cdot\,,b_0)$, graded by ghost number,
with the dot product given by the operator product expansion and with the
BV operator identified with $b_0^{[2]}$.  To prepare for a
thorough analysis of the operator algebra, we
develop in Section \BValgebra\
some general machinery as well as explicit examples
of G- and BV-algebras.
\smallskip

Given an Abelian ring, $\cR$, the archetypal example of a G-algebra
is the algebra of polyderivations of the ring $\cR$,
$(\cP(\cR),\,\cdot\,,[-,-]_S)$, equipped with the Schouten bracket.
For any BV-algebra, $(\fA,\,\cdot\,,\De)$, the subspace $\fA^0$ is an
Abelian ring with respect to the dot product.  In fact, there is a
natural G-algebra homomorphism, $\pi$, from $(\fA,\,\cdot\,,\De)$ to
$(\cP(\fA^0),\,\cdot\,,[-,-]_S)$.  If there is a compatible
BV-operator on the space of polyderivations, then in
we show under what conditions $\pi$ lifts to a
BV-homomorphism (Theorem \bvhomscr).
There is a similar development for the modules of
these algebras.  Given an $\cR$ module, $M$, one may construct the
polyderivations, $\cP(\cR,M)$, of $\cR$ with values in $M$.  Then in
Theorem \gmodder\ we show under what conditions $\cP(\cR,M)$ will be a
G-module of $\cP(\cR)$.
\smallskip

The examples we build are based on the ring $\cG_N=\cC_{2N}/\cI$, where
$\cC_{2N}\cong\CC\,[ x^1,\ldots, x^{2N}]$ is a free Abelian algebra,
and $\cI$ is the ideal generated by a quadratic vanishing relation
(Section \SSgrrinal).  The natural $\sotwon$ action by derivations
of $\cC_{2N}$ descends to $\cG_N$.  Using it, we construct an explicit
basis for the space of polyderivations, $\cP(\cG_N)$, which is
summarized in Theorem \polyvects.  Moreover, in Theorem \generat, we
find a finite set of generators and relations which characterize
$\cP(\cG_N)$ as a dot algebra.  These results allow us to demonstrate
that $\cP(\cG_N)$ is actually a BV-algebra (Theorem \bvopmth), and
to explicitly calculate the homology of the corresponding BV-operator
(Theorem \cohofdelpr).
\smallskip

 For comparison with the operator cohomology the most relevant case is
$N=3$.  The
ground ring in this case, $\cR_3$, is a model space for $\slth
\subset \sosi$.  As a
remnant of the $\sosi$ (or, in fact, a ``hidden" $\soei$) structure,
we see in Section \twistedmodules\ that there are six natural $\cR_3$
module structures, $M_w$, $w \in W$.  The main result of this section
is Theorem \plvmodule, where we show that for each $w\in W$, the space
of twisted polyderivations $\fP_w \equiv \cP(\cG_3,M_w)$ is a G-module
of $\fP\equiv\cP(\cR_3)$.  In Section \SStwplcl\ we are then able to
give quite explicit $\bga$ decompositions for these G-modules.
\smallskip

In Section \Sbvalgebra\ we put the above results together to
obtain a description of the BV-algebra of operator cohomology, $\fH$.
We first observe in Theorem \cohringgen\ that $\fH^0 \cong
\cR_3$, thus bringing into play all the results of Section
\BValgebra.%
\foot{The description of the ground ring of $\cW_N$ gravity in terms
of the $\sln$ model space was anticipated in [\MMMO].}
In particular, we prove (Theorem \maintheorem):

\smallskip
\noindent
\item{i.} {\it There exists a  natural map
$\pi:\fH\rightarrow \fP$ that  is  a BV-algebra homomorphism between
$(\fH,\,\cdot\,,b_0)$ and $(\fP,\,\cdot \,,\De)$.}
\item{ii.} {\it Let $\frak{I} \equiv {\rm Ker\,}\pi$ be a BV-ideal of $\fH$.
We have an exact sequence of BV-algebras
$$
\matrix{
0&\mapright{}&\frak{I}&\mapright{}&\fH&\mapright{\pi}&\fP&\mapright{}&
0\,.\cr}
$$
There exists a dot algebra homomorphism $\imath:\fP\rightarrow \fH$,
such that $\pi\circ\imath={\rm id}$, \ie, the sequence splits as a
sequence of $\imath(\fP)$ dot modules.}
\smallskip

\noindent
Similarly, we show that there are $\fH^0$ modules, $\widehat M_w\subset
\fI$, which are isomorphic to $M_w$ as twisted $\cR_3$ modules
(Theorem \genmodcoh).  Indeed, up to some subtleties at the boundaries
of the different Weyl chambers, we deduce that the bulk of $\fI$
admits a description in terms of the twisted polyderivations
$\fP_w$ (Theorem \genproj).
\smallskip

To be more precise, one can try to understand how these different
sectors of twisted polyderivations are patched together.  We first
show in Theorem \cohofbo\ that the cohomology of $b_0$ on $\fH$ is
trivial.  Thus $(\fH,\,\cdot\,,b_0)$ is an extension of
$\fP$ which is acyclic with respect to the BV-operator.  Next, we
construct a further projection on $\fI$, $\pi': \fI^n \longrightarrow
\cP^{n-1}(\cR_3, M_{r_1}\oplus M_{r_2})$, which is the identity on
$\fI^1\cong \widehat M_{r_1}\oplus \widehat M_{r_2}$.  Then (Theorem
\gmodofi) the map $\pi'$ is a G-morphism between the G-module $\fI$ of
$\fH$ and the G-module $\fP_{r_1}\oplus \fP_{r_2}$ of $\fP$ (and
conjecturally a BV-morphism).  Together with the nondegenerate
pairing on
cohomology, this gives a relatively explicit description of the
dot-module structure of $\fH$ over $\imath(\fP)$.
\smallskip

Finally, there are a number of appendices which typically either
record results of explicit calculations, or give technical details of
particular derivations.  The two exceptions are: Appendix \Apolder,
where we discuss the BV-algebra of polyvectors on a base affine space
of a group $G$; Appendix \AVir, where we summarize -- in the notations
and with the insights of present paper -- the corresponding
understanding of the $c^M=1$ Virasoro case, the $2D$ string.\smallskip

For short summaries of some of the results the reader may examine
[\BPist,\Mfest,\BMPstring].


\bigskip\bigskip

\leftline{\bf Acknowledgements}

The authors would like to thank each others' physics department, as
well as the Theory Division at CERN for hospitality at various times
during this work.  We have enjoyed discussions with
I.~Bars, K.~de Vos, E.~Frenkel, E.~Getzler,
W.~Lerche, M.~Varghese, G.~Moore, I.~Penkov, C.~Pope,
A.~Schwarz, P.~van Driel, N.~Warner, C.-J.~Zhu and especially
G.~Zuckerman.
P.B.\ acknowledges the
support of the Packard Foundation during his time at the University of
Southern California when most of this work was done.  P.B.\ and J.M.\
acknowledge the support of the Australian Research Council, while
K.P.\ is supported in part by the U.S.\ Department of Energy Contract
\#DE-FG03-84ER-40168.

\bigskip\bigskip

\subsec{Glossary of notation}
\subseclab\Not

\noindent Throughout the paper we will use the following notations:
\bigskip

\settabs 6\columns

\+ $\cO$    &   category of modules, see Section \WOcat\cr
\+ JH$(V)$ & Jordan-H\"older composition series for $V\in\cO$.\cr
\+ $U(\,\cdot\,)$   &   universal enveloping algebra functor\cr
\+ $U(\,\cdot\,)_{\rm loc}$ &  corresponding local completion\cr\bigskip

\+ $\bfg$ &  complex simple Lie algebra\cr
\+ $\bfg\cong\bfnp\oplus\bfh\oplus\bfnm$ & Cartan decomposition \cr
\+ $\bfh$ &  Cartan subalgebra with dual $\bfh^*$\cr
\+ $(\ \, ,\ \, )$ & bilinear form on $\bfh$ or $\bfh^*$,
   sometimes also denoted by $\,\cdot$\cr
\+ $\rh$ & element of $\bfh^*$
   such that $\rh(h_i)=1, i=1,\dots,\ell$\cr
\+ $\ell$ &  rank of $\bfg$\cr
\+ $\De$, $\De_\pm$ & roots, positive/negative roots of $\bfg$ \cr
\+ $Q =\ZZ\cdot\De_+$ & root lattice of $\bfg$\cr
\+ $Q_+ = \ZZ_{\geq0}\cdot \De_+$ \cr
\+ $P$, $P_+$, $P_{++}$ & set of integral, dominant integral, strictly
   dominant integral weights, respectively\cr
\+ $\al_i\,, i=1,\dots,\ell$ & simple roots of $\bfg$\cr
\+ $\La_i\,, i=1,\dots,\ell$ & fundamental weights of $\bfg$\cr
\+ $D_+$ & fundamental Weyl chamber, \ie, $D_+ = \{ \la\in\bfh^*_{\RR}
   \,|\, (\la,\al_i)\geq0,\, \forall i=1,\ldots,\ell\}$\cr
\+ $\cL(\La)$ & finite dimensional irreducible representation of
   $\bfg$ with highest weight $\La\in P_+$\cr
\+ $m^{\La'}_{\La}$ & multiplicity of the weight $\La$ in ${\cal L}(\La')$\cr
\+ $W$ & Weyl group of $\bfg$\cr
\+ $w_0$ & Coxeter element of $W$, \ie, longest element in $W$\cr
\+ $r_i$ &   reflection in simple root $\al_i$; for $\slth$ $i=1,2$,
   $r_{ij} = r_ir_j$, $r_3 = r_1r_2r_1 = r_2r_1r_2 = w_0$\cr
\+ $\widetilde W=W\cup\{\si_1,\si_2\}$ & \ where $\si_i$, $i=1,2$, act by
   zero on all weights $\la\in\bfh^*$\cr
\+ $\ell(\si)$ & the length of $\si\in \widetilde W$\cr
\+ $\ell_w(\si),\,w\in W$ & twisted length of $\si\in\widetilde W$\cr
\+ $\si\circ\La = \La+\rh-\si\rh$, $\si \in \widetilde W$\cr
\+ $\hg$ & affine Lie algebra with underlying finite-dimensional Lie algebra
   $\bfg$\cr
\+ $\La_i\,, i=0,\ldots,\ell$ & fundamental weights of $\hg$\cr
\+ $\widehat{W}\cong W \ltimes T$ & Weyl group of $\hg$\cr
\+ $T\equiv \{ t_\al\,|\, \al\in Q\}$ & translation subgroup of
   $\widehat W$\cr
\+ $G$, $N_+$, $H$, $N_-$ & complex groups generated by
   $\bfg$, $\bfnp$, $\bfh$, $\bfnm$ respectively\cr
\+ $A = N_+\backslash G$ & base affine space\cr
\+ $\Ep(G)$ & regular functions on $G$\cr
\+ $\Ep(A)$ & regular functions on $A$\cr\bigskip

\+ $\cA$ & Heisenberg algebra\cr
\+ $F(\La,\al_0)$ &  Fock space ($\cA$-module) with weight $\La$ and
   background charge $\al_0 \in \CC$, see Section \WFock\cr
\+ $F^{\rm gh}$ & ghost Fock space, see Section \SSghosts\cr
\+ $w \cdot \La ~=~ w(\La+\al_0\rh) - \al_0\rh$, for $w\in {W}$\cr
\+ $M(h,w,c)$ & Verma module of $\cWth$, see Definition \thPBbac\cr
\+ $M^{(\vdeg)}(h,w,c)$ & generalized Verma module of $\cWth$, see
   Definition \thPBbi\cr
\+ $M^{(\vdeg)}(\La,\al_0) = M(h(\La,\al_0),w(\La,\al_0),c(\al_0))$ as
   specified in \eqPBbb\cr
\+ $M^{(\vdeg)}[s_1,s_2] = M^{(\vdeg)}(s_1\La_1 + s_2\La_2,0)$
   as found below Theorem \thPBdc\cr
\+ $M(S) \equiv M(v_1,v_2,\ldots)$ && submodule of $M(\La,\al_0)$
   generated by $S = \{ v_1,v_2,\ldots\}$\cr
\+ $[\ ,\ ]$ & graded commutator (\eg, anti-commutator for ghost fields)\cr
\bigskip

\+ $\fC$ & chiral algebra specified in Theorem \lattice\cr
\+ $H(\cW_3,\fC)$ & cohomology of the complex $(\fC,d)$ with differential
   $d$ given by \brstdd\ acting as in \brstact\cr
\+ $\fH = H(\cW_3,\fC)$ & considered as an operator algebra\cr
\+ $\fH_w = H(\cW_3,\fC_w)$ & cohomology of the subcomplex $(\fC_w,d)$ of
   operators with $-i\La^L+2\rh\in w^{-1}P_+$\cr\bigskip

\+ $(\fA,\,\cdot\,,[-,-])$ & Gerstenhaber algebra, see Definition \BVghal\cr
\+ $(\fA,\,\cdot\,,\De)$ & BV-algebra with BV-operator $\De$, see
   Definition \BVbv\cr
\+ $\cP^n(\cR,M)$ & order $n$ polyderivations of an algebra
   $\cR$ with values in $M$\cr
\+ $\cD(\cR,M) \equiv \cP^1(\cR,M)$\cr
\+ $\cP^n(\cR)\equiv \cP^n(\cR,\cR)$\cr
\+ $\fP_w$ & twisted polyderivations, see Section \twistedmodules\cr\bigskip

\vfil\eject
%
\def\vrh{\varrho}
%
%
\secno=1
\newsec{${\cW}$-ALGEBRAS AND THEIR MODULES}
\seclab\PBW

\subsec{${\cW}$-algebras}
\subseclab\Walg

\subsubsec{Introduction to $\cW$-algebras}
\subsubseclab\Wintro

$\cW$-algebras are certain nonlinear, higher spin extensions of the
2-dimensional conformal algebra, \ie, the Virasoro algebra. They were
first introduced by Zamolodchikov [\Za] and have subsequently been
investigated by many people (see, \eg, [\BS,\BSb], and references
therein).  The proper mathematical setting is that of ``Vertex
Operator Algebras'' (VOAs) [\FLM] or, equivalently, ``Meromorphic
Conformal Field Theory'' [\Go].\smallskip

The simplests $\cW$-algebras are the algebras ${\cW}[\bfg]$ associated
to some simple, simply-laced Lie algebra $\bfg$ either by
Drinfel'd-Sokolov reduction or by a coset-construction. They have
generators of conformal dimension equal to the orders of the
independent Casimir operators of $\bfg$. In particular ${\cW}_N \equiv
{\cW}[\sln]$ has $N-1$ generators of dimension
$2,3,\ldots,N$. \smallskip

In this paper we restrict our attention to the simplest nonlinear
$\cW$-algebra, namely $\cWth$, although most of the results
continue to hold for the more general algebras ${\cW}[\bfg]$. We
formulate many of our results using generic Lie algebra notation so
that the generalization to ${\cW}[\bfg]$ should be obvious.\smallskip


\subsubsec{The $\cWth$ algebra}
\subsubseclab\Wth

The $\cWth$ algebra with central charge $c\in\CC$
can be defined as the quotient of the
universal enveloping algebra of the free
Lie algebra generated by $L_m, W_m,\, m\in \ZZ$, by the ideal generated
by the following commutation relations
\eqn\eqPAaa{ \eqalign{
[L_m,L_n]  ~=~ & (m-n) L_{m+n} + {\textstyle {c\over12}} m (m^2 -1 )
  \de_{m+n,0} \,,\cr
[L_m,W_n]  ~=~ & (2m-n) W_{m+n} \,,  \cr
[W_m,W_n]  ~=~ & (m-n) ( {\textstyle {1\over15}}(m+n+3)(m+n+2)
 - {\textstyle{1\over6}} (m+2)(n+2) ) L_{m+n} \cr
 & + \be (m-n) \La_{m+n} + {\textstyle {c\over360}} m (m^2-1)(m^2-4)
  \de_{m+n,0}\,,\cr}
}
where $\be = 16/(22+5c)$ and
\eqn\eqPAab{
\La_m ~=~ \sum_{n\leq -2} L_n L_{m-n} + \sum_{n>-2} L_{m-n} L_n -
  {\textstyle{3\over10}} (m+3)(m+2) L_m \,.
}\smallskip

Equivalently, one can introduce fields $T(z)$ and $W(z)$ (\ie, formal
power series in $\cWth [[z,z^{-1}]]$) by
\eqn\eqPAac{
T(z) ~=~ \sum_{m\in\ZZ} \ L_m z^{-m-2}\,,\qquad W(z) ~=~ \sum_{m\in \ZZ}
W_m z^{-m-3}\,,
}
in terms of which \eqPAaa\ can be translated into so-called ``Operator
Product Expansions'' (OPEs) (see, \eg,
[\BPZ] for an early discussion of the
use of OPEs in conformal field theory, and [\Bor,\FLM,\FHL] for the
mathematical theory)
\eqn\eqPAad{ \eqalign{
T(z) T(w) ~=~ & { c/2 \over (z-w)^4} +
  { 2T(w) \over (z-w)^2} + {\p T(w) \over z-w} + \ldots\,, \cr
T(z) W(w)  ~=~  & { 3 W(w) \over (z-w)^2} + {\p W(w) \over z-w}
  + \ldots\,,\cr
W(z) W(w)  ~=~ & { c/3 \over (z-w)^6} +
 {2 T(w) \over (z-w)^4} + {\p T(w) \over (z-w)^3} +
  {1\over (z-w)^2} ( 2\be\La(w) + {\textstyle{3\over10}} \p^2 T(w)) \cr
  & + {1\over (z-w)} ( \be\p\La(w) + {\textstyle{1\over15}} \p^3 T(w))
  + \ldots \,,\cr}
}
where
\eqn\eqPAae{
\La(z) ~=~ \sum_{m\in \ZZ} \La_m z^{-m-3}
  ~=~ (TT)(z) - {\textstyle{3\over10}} \p^2 T(z)\,.
}\smallskip

It is useful to split the $\cWth$-generators into three groups
according to their modings.  Let
\eqn\eqPAaf{
\cWthpm ~\equiv~ \{ L_m, W_m ~|~ \pm m>0 \}\,,\qquad
\cWthz ~\equiv~ \{ L_0, W_0 \} \,.
}
Note that, while the generators in $\cWthz$ form an (Abelian) subalgebra
of $\cWth$, the so-called Cartan subalgebra, the generators in $\cWthpm$
do not form a subalgebra.
The action of $\cWthz$ on the generators of $\cWth$ is explicitly given by
\eqn\eqPAag{ \eqalign{
[L_0, L_n] & ~=~ -n\, L_n \,, \cr
[L_0, W_n] & ~=~ -n\, W_n \,, \cr
[W_0, L_n] & ~=~ -2n\, W_n \,, \cr
[W_0, W_n] & ~=~ - \textstyle{1\over 15} n(n^2-4) L_n - \be n \La_n \,.\cr}
}
{}From here we easily see that it is impossible to diagonalize the
action of $\cWthz$ on $\cWth$, \ie, we do not have a root space
decomposition with respect to the Cartan subalgebra.  It should, however,
be possible to find a generalized root space decomposition, \ie, a basis
of $\cWth$ in which the action of $\cWthz$ is in Jordan normal form.
To our knowledge no such basis has been explicitly constructed yet,
and we consider this an important open problem.
Later, when we discuss modules of $\cWth$, we encounter the same
problem of nondiagonalizability of the Cartan subalgebra, and we will
discuss this issue in more detail.
Here it suffices to remark that this nondiagonalizability
is one of the major differences with, for example, affine Lie algebras
or the Virasoro
algebra, and is reflected in a much more subtle and complicated
submodule structure.  As a consequence, both the construction of resolutions
and the calculation of semi-infinite cohomology are correspondingly more
difficult.

\subsec{$\cWth$ modules}
\subseclab\Wthmod

\subsubsec{The category $\cO$}
\subsubseclab\WOcat

To prove properties of $\cWth$ modules in some generality, we first
need to define a proper category of modules -- henceforth referred to
as the category $\cO$.
This category should be
small enough to allow for certain ``nice'' properties, \eg,
the existence of Jordan-H\"older series and
the existence of a semi-infinite cohomology. On the
other hand the category should be
big enough to incorporate the (physically) interesting
modules, such as free field Fock spaces and Verma modules, and to
allow for the existence of certain homological constructions,
\eg, resolutions of irreducible modules, within the category.
In addition one usually requires that the category is closed under
certain basic operations such as taking direct sums, tensor products
and contragredients.\smallskip

For (affine) Lie algebras $\bfg$ the
category $\cO$ is, loosely speaking,
the category of $\bfh$-diagonalizable modules with finite-dimensional
weight spaces and weights bounded from above [\BGG,\Ka].
To require $\cWthz$-diagonalizability for $\cWth$ modules
is too strong a requirement, however. As we will see later,
as a direct consequence of \eqPAag,
in general not even Verma modules are $W_0$-diagonalizable. We
thus only require the modules to be $L_0$-diagonalizable.%
\foot{Modules with a nondiagonalizable action of the Virasoro
generator $L_0$ also exist (see, \eg, [\RS,\Gu]), and have some
important applications.  However, we do not need them for the
purpose of this paper, so we do not include them in the definition
of the category $\cO$.}  Moreover, since $L_0$ is identified with the
energy operator, we require the $L_0$ eigenvalues to be bounded from
below. For an $L_0$-diagonalizable module $V$, let $V_{(h)} = \{ v\in
V\,|\, L_0\, v = h\, v\}$ be its eigenspaces, such that $V =
\coprod_{h\in\CC}\, V_{(h)}$.  Let $P(V) = \{ h\in \CC\,|\, V_{(h)}
\neq 0\}$ denote the set of $L_0$-eigenvalues of $V$.  We refer
to the $L_0$-eigenvalue of a state $v\in V$ as its ``$L_0$-level.''

\thm\thPBaa
\proclaim Definition \thPBaa\ [\FKW]. The category $\cO$, of ``positive
energy $\cW$-modules'', is the set of $L_0$-diagonalizable
modules $V$ such that each $L_0$-eigenspace $V_{(h)}$ is finite-dimensional,
and for which there exist a finite set
$h_1,\ldots,h_s\in \CC$ ($h_i\neq h_j\,{\rm mod\,}1$ for all $i,j$),
such that $P(V) \,\subset\, \cup_{i=1}^s
\cup_{N\in \ZZ_{\geq0}} \{ h\in \CC\,|\, h = h_i + N \}$.\par\smallskip

It is clear that (finite) direct sums, submodules and quotients of
modules in $\cO$ are again in $\cO$.
Generally, however, contrary to the Lie
algebra case, tensor products of modules in $\cO$ are not
in $\cO$ for the simple reason that -- due to the nonlinear nature of
the $\cWth$ algebra -- they do not carry the structure of a $\cWth$
module.\smallskip

Despite the fact that $W_0$ need not be
diagonalizable on $V\in \cO$,
we do of course have a generalized eigenspace
decomposition (Jordan normal form) of $W_0$ on each $V_{(h)}$.
We have $V_{(h)} = \coprod_{w\in \CC} V_{(h,w)}$, where we have
denoted by $V_{(h,w)} = \{ v \in V_{(h)}\,|\, \exists N\in\NN:
(W_0 - w)^N\, v \,=\, 0 \ {\rm for\ some\ } w\in\CC\,\}$ the generalized
eigenspaces of $W_0$.\smallskip

Within each Jordan block of $V_{(h,w)}$ we may choose a basis
$\{ v_0,\ldots,v_{\vdeg-1}\}$ such that $(W_0-w)\, v_i\, =\, v_{i-1}$
for $i=1,\ldots,\vdeg-1$ and $(W_0-w)\, v_0 \,=\, 0$.  That is,
with respect to this basis
\eqn\eqPBaab{
W_0 ~=~ \left( \matrix{
  w & 1 & &&\cr
    & w & 1 && \cr
    &   & \ddots & \ddots\cr
    &   &  & w & 1 \cr
    &   &  & & w \cr}  \right) \,.
}
For such a basis, we also use the notation
\eqn\eqPBaac{
\matrix{
v_{\vdeg-1} & \mapright{W_0-w} & v_{\vdeg-2} & \mapright{W_0-w} & \ldots &
  \mapright{W_0-w} & v_0 \cr} \,.
}\smallskip

The (generalized) character ${\rm ch\,}_V$
of a module $V\in\cO$ is now defined as
\eqn\eqPBaaa{
{\rm ch\,}_V(q,p) ~=~ {\rm Tr\,}_V \left(q^{L_0} p^{W_0} \right)
 ~=~ \sum_{(h,w) \in \CC^2} \  {\rm dim}_{\CC} \, \left( V_{(h,w)} \right)\,
 q^h p^w \,.
}
We have

\thm\thPBaaa
\proclaim Theorem \thPBaaa. The characters ${\rm ch\,}_L$, where $L$
runs over the set of irreducible modules in $\cO$ are linearly
independent over $\CC$, in particular ${\rm ch\,}_L = {\rm ch\,}_{L'}$ iff
$L ~\cong~ L'$.\par\smallskip

The proof of this theorem is given after Theorem \thPBbaa.\smallskip

Unfortunately, no explicit expression for the generalized
character is known for any $\cWth$ module (with the exception of the
trivial module, of course).  For most purposes it suffices, however, to
consider the specialization ${\rm ch\,}_V(q,1)$ of the character.
Expressions for these specialized characters are known for most
interesting modules in $\cO$. \smallskip

Clearly, since $\cWth$ is $\ZZ$-graded, any $V\in\cO$
decomposes into a direct sum $V = \bigoplus_i V(h_i)$
of $\cWth$ modules, where $h_i$ are
the elements in Definition \thPBaa, and the $L_0$-eigenvalues of
$V(h_i)$ are concentrated on strings $h_i + N,\, N\in \ZZ_{\geq0}$.
For all practical purposes,
we may thus equally well consider modules built on a single $h_i$ only.
Let us denote this subcategory by $\cO_{(h_i)}$.\smallskip

An important ingredient in unraveling the
structure of a $V\in \cO$ is to consider filtrations of
$V$ by submodules, in particular so-called ``composition series''
or ``Jordan-H\"older series'' that are characterized by the condition
that quotients of subsequent terms in the filtration should be
irreducible.
As opposed to the finite-dimensional setting, such composition series
do not, in general, exist for infinite-dimensional algebras.
We need a slight modification of the usual construction; namely,
a ``cut-off'' which renders the filtrations finite.
In complete analogy with the affine Lie algebra case
we have (see [\Ka])

\thm\thJHa
\proclaim Theorem \thJHa.  Every module $V\in \cO_{(h)}$ possesses
a composition series (or Jordan-H\"older series) JH$(V)$, \ie,
for all $N\in \ZZ_{\geq0}$, there exists a (finite) filtration by
submodules of $V$ (denoted by JH$_N(V)$)
\eqn\eqJHa{
\matrix{
V = V_0 & \supset & V_1 & \supset & \ldots & \supset & V_s \,=\, 0\,, \cr}
}
and a subset $I\,\subset\,\{0,\ldots,s-1\}$,
such that
\item{i.}
$V_i/V_{i+1}$ is irreducible for $i\in I$,
\item{ii.} $\coprod_{M\leq N}\,(V_i /
V_{i+1})_{(h+M)} \,=\, 0$ for $i\not\in I$.\par\smallskip

\proof
The proof (which requires some results on highest
weight modules to be discussed in Section \WVer)
parallels the one given in [\Ka] with one minor
modification; namely, the maximal element has to be chosen
to be an eigenvector of $W_0$.  That this can always be done is
obvious.  We refer to [\Ka] for more details.
\Box\hfil\smallskip

For any $V\in \cO_{(h)}$, only irreducible modules
$L\in \cO_{(h)}$ appear as quotients in the
composition series of $V$.  So, for any such $L\in\cO_{(h)}$, choose
$M\in\ZZ_{\geq0}$ such that $P(L)\,\subset\, h+M+\ZZ_{\geq0}$ and
$V_{(h+M)} \,\neq\,0$.  Then, choose  any $N\geq M$ and
denote by $(V\!:\!L)$ the multiplicity with which the irreducible
module $L$ appears in the composition series JH$_N(V)$.  Clearly,
$(V\!:\! L)$ is independent of the choice of $N\geq M$.  Moreover,

\thm\thJHb
\proclaim Theorem \thJHb. We have
\eqn\eqJHb{
{\rm ch\,}_V ~=~ \sum_L\ (V\!:\!L)\ {\rm ch\,}_L \,.
}
\par \smallskip

\proof As in [\Ka]. \Box\hfil\smallskip

\noindent {\it Remarks:}\hfil
\item{i.} The characters ${\rm ch\,}_L$ of the irreducible
modules $L$ are not only independent (Theorem \thPBaaa), but they also
span the space of characters ${\rm ch\,}_V,\, V\in\cO$.
\item{ii.} The fact that we would like quotients of subsequent terms in
the composition series to be irreducible forced us to introduce
additional terms whose quotients do not contribute states up to the
level we are interested in.  This can be avoided by making
the following modification to Theorem \thJHa\
which is easily seen to be equivalent to
the original.  Let us say that a $\cWth $ module
$V\in \cO_{(h)}$ is ``irreducible up to level $N$'' if for all proper
submodules $W \subset V$ we have $W \,\bigcap\, ( \coprod_{k\leq N}
V_{(h+k)} ) \,=\, 0$.  Then, for every $V\in\cO_{(h)}$ and
$N\in\ZZ_{\geq0}$, we have a filtration \eqJHa\ of $V$
such that each $V_i/V_{i+1}$ is irreducible up to degree $N$.  We
give examples of such filtrations in Section \Wctwob.

\thm\thPBae
\proclaim Definition \thPBae. Let $V\in\cO$.
\item{i.} A vector $v\in V$ is called ``primitive'' if there exists
a proper submodule $U\subset V$ such that $\cWthp \cdot v \subset U$
while $v\not\in U$.
\item{ii.} A vector $v\in V$ in called ``pseudo-singular'' (or
p-singular, for short) if $\cWthp\cdot v=0$.
\item{iii.} A vector $v\in V$ is called ``singular'' if it is
p-singular and a $\cWthz$ eigenvector.
\par\smallskip

Let us denote the set of singular, p-singular and
primitive vectors in $V\in \cO$ by
${\rm Sing}(V)$,  ${\rm pSing}(V)$ and ${\rm Prim}(V)$,
respectively, and let $SV$ be the module generated by all
p-singular vectors in $V$.
Clearly, ${\rm Prim}(V) \supset {\rm pSing}(V)\supset {\rm Sing}(V)$.
Also, a primitive
vector $v\in V$ becomes p-singular in the quotient module $V/U$.
\smallskip

While the notion and use of a singular vector is probably well-known
from their Virasoro analogue, the notion of a primitive vector
might be less familiar.  It is, nevertheless, rather useful.
In particular, after examining in more detail the collection
of irreducible modules in $\cO$, we establish a 1--1
correspondence between primitive vectors in a module $V\in \cO$ and
the set of irreducible modules $L$ occuring in the composition
series JH$(V)$ of $V$ (see Lemma \thPBaea).\smallskip

The last property of modules in $\cO$ that we discuss before proceeding
to more explicit examples is that to every module $V\in\cO$ there
is associated a contragredient module $\overline{V}\in\cO$.  Hereto,
let $\om_{\cW}$ be the $\CC$-linear anti-involution of $\cWth$ defined by%
\foot{In later sections we also need the anti-linear
anti-involution defined by
these relations.  We denote it by $\oom_{\cW}$.}
\eqn\eqPBac{
\om_{\cW}(L_n) ~=~ L_{-n} \,, \qquad
\om_{\cW}(W_n) ~=~ W_{-n} \,.
}
For future use, note that \eqPBac\ is
equivalent to the anti-involution acting on fields in $\cWth[[z,z^{-1}]]$
defined by
\eqn\eqPBaca{
\om_{\cW}(T(z)) ~=~  z^{-4}\,T(z^{-1})\,,\qquad
   \om_{\cW}(W(z)) ~=~ z^{-6}\,W(z^{-1})\,.
}

\thm\thPBaf
\proclaim Definition \thPBaf.
Consider $V\in \cO$. The ``contragredient module'' $\overline{V}$
is defined, as a vector space, to be $\overline{V} = \coprod_{N\in
\ZZ} {\rm Hom\,}_{\CC}(V_{(h+N)},\CC)$.  The $\cWth$
module structure is given by
\eqn\eqPBad{
x\, f (v) \equiv f (\om_{\cW}(x)\, v)\,,
}
where $f\in \overline{V},\, v\in V$ and $x\in\cWth$. \par\smallskip

Clearly, for $V\in \cO$, the contragredient module $\overline V$ is
again in $\cO$. Note, however, that the module dual to $V$
has $L_0$-eigenvalues bounded from {\it above} and thus is not in
$\cO$ [\FF].

\thm\thPBag
\proclaim Lemma \thPBag. Let $V,W\in \cO$ and suppose $\pi \in
{\rm Hom\,}_{\cW}(V,W)$. Then, there exists a $\overline{\pi} \in
{\rm Hom\,}_{\cW}(\overline{W},\overline{V})$ defined by
$\overline{\pi}(\bar{w})(v) = \bar{w}(\pi(v))$ for all $v\in V,
\bar{w}\in \overline{W}$.\par

\subsubsec{(Generalized) Verma modules}
\subsubseclab\WVer

Important examples of modules in $\cO$ are the so-called
``Verma modules'' or, more generally, ``highest weight modules.''
In this section we recall their definitions and some
important properties.  In fact, it turns out that we need
modules which are slightly more general than Verma modules,
the so-called ``generalized Verma modules.'' However, their structure
theory can be developed along the same lines as for
Verma modules, so we just restrict ourselves to stating the analogous
theorems.\smallskip

\thm\thPBbab
\proclaim Definition \thPBbab. A $\cWth$ module $V \in \cO$ is
called a highest weight module with highest weight $(h,w)\in\CC^{\,2}$
if there exists a nonzero vector $\vV\in V$, the
so-called ``highest weight vector,'' such that
\eqn\eqPBbaa{
\cWthp \cdot \vV = 0\,,\qquad L_0\, \vV = h\,\vV
  \,,\qquad W_0\, \vV = w\,\vV\,.
}
and
\eqn\eqPBbab{
V ~\cong~ \cWth \cdot \vV\,.
}\par\smallskip

\noindent {\it Remark:} Note that a ``module with highest weight'' is
a module that satisfies \eqPBbaa, but not necessarily \eqPBbab.
For example, Fock spaces -- to be discussed in Section \WFock\ -- are
modules with highest weight, but are {\it not}
highest weight modules in general.\smallskip

Verma modules are highest weight modules which are, in a sense, maximal.
Namely,

\thm\thPBbac
\proclaim Definition \thPBbac.  A Verma module $M(h,w,c)$ is the
module ``induced'' by the action of $\cWthm$ from a highest weight vector $\vM$
of highest weight $(h,w)$, \ie, the $\cWth$ module
$\cWth\cdot \vM$, divided by the relations
\eqn\eqPBbac{
\cWthp \cdot \vM ~=~0\,,\qquad L_0\, \vM ~=~ h\, \vM \,,\qquad
  W_0\, \vM ~=~ w\, \vM\,.
}
The action of $\cWth$ on \eqPBbac\ is determined by the
commutation relations \eqPAaa\ and equation \eqPBbac.\par\smallskip

One of the most important properties of Verma modules is their
co-universality

\thm\thPBba
\proclaim Lemma \thPBba. Let $V \in \cO$ and let $v_0\in V$ be
a singular vector of weight $(h,w)$. Then there exists a
unique $\cWth$ homomorphism $\pi\in {\rm Hom\,}_{\cWth}
(M(h,w,c),V)$ such that $\pi(\vM) = v_0$ where $\vM$ is the highest weight
vector of $M(h,w,c)$.\par\smallskip

\noindent {\it Proof:}
Clearly, $\pi$ is uniquely defined by $\pi(x\, \vM)  \,=\,
x\,\pi(\vM) \,=\, x\, v_0$
for all $x \in \cWth$. \Box\hfil\smallskip

And, as an immediate consequence, we have

\thm\thPBbad
\proclaim Corollary \thPBbad. Every highest weight module $V\in\cO$
with highest weight $(h,w)\in \CC^{\,2}$ is a quotient of the Verma
module $M(h,w,c)$.\par\smallskip

\proof Let $\vM$ be the highest weight vector of $M(h,w,c)$, and let
$\vV$ be the highest weight vector of $V$.  Lemma \thPBba\ provides
us with a unique  $\cWth$ homomorphism $M(h,w,c) ~\mapright{\pi}~ V$ such
that $\pi(\vM)=\vV$.  Clearly, because of \eqPBbab, $\pi$ is actually
an epimorphism.  Let $K = {\rm Ker\,}\pi$, then we have
$V \,\cong\, M(h,w,c)/K$. \Box\hfil\smallskip

For many purposes it is useful to have an explicit basis of the Verma
module.  Clearly, the set of all monomials
\eqn\eqPBbae{
e_{m_1 \ldots m_K;n_1 \ldots n_L} ~=~
L_{-m_1}\cdots L_{-m_K} W_{-n_1}\cdots W_{-n_L}\, v\,,\qquad
m_i, n_j \in \ZZ_{>0}\,,\ K,L\in\ZZ_{\geq0}\,,
}
form a basis of $M(h,w,c)$, but this basis is overcomplete.
One may, instead, find a linearly independent set.

\thm\thPBbae
\proclaim Theorem \thPBbae\ [\Wa]. (Poincar\'e-Birkhoff-Witt)
The vectors $e_{m_1 \ldots m_K;n_1\ldots n_L}$ with
$m_1\geq \ldots \geq m_{K-1}\geq m_K >0$, $n_1\geq \ldots \geq
n_{L-1} \geq n_L >0$ constitute a set of  independent basis
vectors of the Verma module $M(h,w,c)$.\par\smallskip

\noindent {\it Remark:}  Similarly, the vectors
\eqn\eqPBCa{
\overline e_{m_1\ldots m_K;n_1\ldots n_L}\,,\quad
  m_1\geq \ldots \geq m_{K-1}\geq m_K >0\,,\quad n_1\geq \ldots \geq
  n_{L-1} \geq n_L >0\,,
}
dual to $e_{m_1 \ldots m_K;n_1\ldots n_L}$, constitute a basis
for the contragredient Verma module $\overline M(h,w,c)$.\smallskip

The idea of the proof is quite standard.  Note thereto that the
theorem would be straightforward if the algebra was Abelian.  The idea
is thus to try to reduce the problem to that of an Abelian algebra
and show that the correction terms are immaterial.  We
furnish the details of the proof since similar ideas are most
crucial in the computation of a certain semi-infinite cohomology in
later sections.  Namely, this cohomology can be computed by taking the
cohomology, and corresponding complex (the so-called ``Koszul complex''),
of the Abelianized algebra as a starting point.\smallskip

\noindent {\it Proof of Theorem \thPBbae:}
We define a grading of $\cWth$ by ${\rm deg}(L_n)=1$
and ${\rm deg}(W_n)=2$.%
\foot{Evidently, other choices of degree are possible.  A choice
that  works for any $\cW$-algebra is to put
${\rm deg}\,(W^{(\De)}_n) = \De$, where $W^{(\De)}(z)$ is a
$\cW$-generator of conformal dimension $\De$ (see, \eg, [\Wa]).}
Note that, with respect to this grading,
the degrees of terms on the right hand side of the commutator
\eqPAaa\ are strictly less than the degree of the left hand side.
Similarly, we associate a degree to the monomials \eqPBbae\ by
${\rm deg}\,(e_{m_1\ldots m_K;n_1\ldots n_L}) = K + 2 L$.  Now
observe that for any permutation $\si\in S_K$ and $\si' \in S_L$ we
have
\eqn\eqBPa{
e_{m_{\si(1)}\ldots m_{\si(K)};n_{\si'(1)}\ldots n_{\si'(L)}} ~=~
   e_{m_1 \ldots m_K;n_1\ldots n_L} + \dots\,,
}
where the dots stand for a (finite) sum of monomials \eqPBbae\ of
degrees strictly less than $K+2 L$.  We can choose $\si$ and
$\si'$ such that we obtain the lexicographical ordering $m_1\geq
\ldots \geq m_{K-1}\geq m_K >0$, $n_1\geq \ldots \geq n_{L-1}
\geq n_L >0$.  The theorem is now proved by induction on the
degree. \Box\hfil\smallskip

The above proof can be formalized.
Note that, upon defining $M_d = \{ v \in M ~|~ \deg\, v \leq d\}$ we
obtain an increasing filtration
\eqn\eqBPb{ \matrix{
\CC\cdot v ~\equiv~ M_0 & \subset & M_1 & \subset & M_2 & \subset &
\cdots & \subset & M \,.\cr}
}
Similarly, we obtain an increasing filtration of $\cWth$.
Now, obviously, the associated graded space
\eqn\eqBPc{
{\rm Gr}\,M ~\equiv~ \coprod_{d\geq0} \ (M_{d+1}/M_{d})\,,
}
becomes a module of the Abelian algebra ${\rm Gr\,}\cWth$.  In fact,
\eqn\eqBPd{
e_{m_{\si(1)},\ldots,m_{\si(K)};n_{\si'(1)},\ldots,n_{\si'(L)}}
  ~=~ e_{m_1,\ldots,m_K;n_1,\ldots,n_L}\,,
}
in ${\rm Gr}\,M$.\smallskip

A useful observation is the following.  Since every highest weight
module $V$ is the image of a Verma module $M$ under a $\cWth$
homomorphism (see Corollary \thPBbad), $V$ inherits the increasing
filtration of $M$ upon defining $V_d = \pi(M_d)$.  Under this grading
${\rm Gr}\,V$ becomes a ${\rm Gr}\,\cWth$ module.\smallskip

Even though $\cW_{3,\pm}$ does not define a subalgebra of $\cWth$, and
thus, strictly speaking, the universal enveloping algebra
$U(\cW_{3,\pm})$ is not defined, for practical purposes
and motivated by Theorem \thPBbae\ it is useful
to define $U(\cW_{3,\pm})$ not as an algebra but merely as a subspace
of $\cWth$ as follows

\thm\thPBuni
\proclaim Definition \thPBuni.  The universal envelope
$U(\cWthm)$ of $\cWthm$ is defined to be the
subspace of $\cWth$ spanned by the vectors
\eqn\eqPBuni{
L_{-m_1}\cdots L_{-m_K} W_{-n_1}\cdots W_{-n_L}\,,\qquad
m_1\geq \ldots \geq m_{K-1}\geq m_K >0\,,n_1\geq \ldots \geq
n_{L-1} \geq n_L >0\,,
}
and similarly for $U(\cWthp)$.
\par\smallskip

{}From Theorem \thPBbae\ it follows that every $v\in M(h,w,c)$ can be
written as $v = u\, \vM$ for some $u\in U(\cWthm)$, and that for $v =
u_1 u_2\, \vM$ with $u_1,u_2\in U(\cWthm)$ we can find $u\in
U(\cWthm)$ such that $v= u\, \vM$ by using the $\cWth$ commutation
relations \eqPAaa\ and the defining relations \eqPBbaa\ for
$\vM$. \smallskip

We now return to the study of Verma modules and a particular
class of quotient modules, namely the irreducible modules.

\thm\thPBbaa
\proclaim Theorem \thPBbaa.
\item{i.} $M(h,w,c)$ contains a maximal submodule $I(h,w,c)$ and
$L(h,w,c) \,\equiv\, M(h,w,c)/I(h,w,c)$ is irreducible.  Conversely,
every irreducible module $L\in\cO$ is isomorphic to some $L(h,w,c)$.
\item{ii.} $I(h,w,c) \,\cong\, P\!M(h,w,c)$, where $P\!M(h,w,c)$ is the
submodule of $M(h,w,c)$ generated by all (proper) primitive
vectors in $M(h,w,c)$.
\item{iii.} Every (nonzero) $\cWth$ homomorphism of Verma modules
is injective.\par\smallskip

\noindent {\it Proof:} \hfill\break
(i) Standard. \hfil\break
(ii) Clearly, by maximality of $I(h,w,c)$ we have $PM(h,w,c) \,\subset\,
I(h,w,c)$.  Now suppose $I(h,w,c) \,\not\cong\, PM(h,w,c)$.  Take a
vector $v \in I(h,w,c)
\backslash PM(h,w,c)$ of minimal level.  Since  $\cWthp$ lowers the level,
 $\cWthp \cdot v$ vanishes in $I(h,w,c) \backslash PM(h,w,c)$, so
$\cWthp \cdot v \in PM(h,w,c)$.  But this implies that
$v\in PM(h,w,c)$, which is a contradiction.
\hfil\break
(iii) Suppose $\pi \in {\rm Hom\,}_{\cWth}(M,M')$.  Denote by $\vM$ and
$\vMp$ the highest weight vectors of the Verma modules $M$ and $M'$,
respectively.  Clearly, $\pi(\vM) = u_2\, \vMp$ for some $u_2 \in
U(\cWthm)$.  To show injectivity of $\pi$ we
have to prove that $u_1 u_2\, \vMp = 0$ with $u_1\in U(\cWthm)$
implies $u_1=0$ or $u_2=0$.  This is obvious for Abelian algebras,
so we use the fact that the $\cWth$ homomorphism of Verma modules
induces a ${\rm Gr\,}\cWth$-homomorphism of the associated graded
Verma modules (see the discussion after Theorem \thPBbae). \Box
\hfil\smallskip

\noindent {\it Proof of Theorem \thPBaaa:} Follows from the fact that,
because of Theorem \thPBbaa\ (i), $L$ is determined
up to isomorphism by $(h,w,c)$, \ie, by the leading term in the
character ${\rm ch\,}_L(q,p)$. \Box\hfil\smallskip

After having determined that all irreducible modules in the category
$\cO$ are of the type $L(h,w,c)$ (Theorem \thPBbaa\ (i)) we now have

\thm\thPBaea
\proclaim Lemma \thPBaea. There is a 1--1 correspondence
between elements in ${\rm Prim}(V)$ of generalized weight $(h,w)$
and irreducible modules $L(h,w,c)$ occurring in JH($V$).\par\smallskip

\noindent {\it Proof:} Let $v\in {\rm Prim}(V)$
and have generalized weight $(h,w)$. By definition there exists a
proper submodule $U\subset V$ such that $v\not\in U$ while $\cWthp\cdot v
\subset U$. Consider the module $V/U$. Clearly $L(h,w,c)$ occurs in
JH($V/U$). By merging the JH-series for $V/U$ with the JH-series for $U$
we immediately obtain that $L(h,w,c) \in$ JH($V$).\hfil\break
Conversely, suppose $L(h,w,c)\in$ JH($V$), then $L(h,w,c) \cong
V_s/V_{s+1}$ for some $s$ (and $N$ sufficiently large). Let $v$ be
a representative of the highest weight vector of $L(h,w,c)$ in $V_s\subset V$.
Clearly, $\cWthp\cdot v\in V_{s+1}$ while $v\not\in V_{s+1}$, \ie,
$v\in {\rm Prim}(V)$. \Box\hfil\smallskip

A very convenient ingredient in the study of the submodule structure
of Verma modules is the determinant of a certain bilinear form
defined on Verma modules, the so-called Shapovalov form.  Let us first
briefly recall the definition and properties of the Shapovalov form.
\smallskip

First, applying Lemma \thPBba\ to the highest weight vector $\vMb$
of the contragredient Verma module $\overline{M}(h,w,c)$
yields a (unique) $\cWth$ homomorphism
\eqn\eqPBbgabc{
\imath\in {\rm Hom\,}_{\cWth}(M(h,w,c),\overline{M}(h,w,c))\,,
}
such that $\imath(\vM) = \vMb$.  This, in turn, immediately provides
us with a bilinear form
$\vev{-|-}_M\,:\,M(h,w,c) \,\times\, M(h,w,c) ~\mapright{}~ \CC$ by
\eqn\eqPBbga{
\vev{u|v}_M ~=~ \imath(u)(v)\,,\qquad u,v \in M(h,w,c)\,.
}
which is such that $\vev{\vM|\vM}_M = \imath(\vM)(\vM) = \vMb(\vM) = 1$.
Moreover, the fact that $\imath$ is a $\cWth$ homomorphism translates
in the property that
the form \eqPBbga\ is contravariant with respect to $\om_{\cW}$,
namely, for all $u,v\in M(h,w,c)$,
\eqn\eqBPe{
\vev{x\,u|v}_M ~=~ \imath(x\, u)(v) ~=~ (x\, \imath(u))(v)
  ~=~ \imath(u)(\om_{\cW}(x)\, v) ~=~ \vev{u|\om_{\cW}(x)\, v}_M \,.
}
Conversely, every contravariant bilinear form $\vev{-|-}$
on $M(h,w,c) \,\times\, M(h,w,c)$
such that $\vev{\vM|\vM} = 1$ leads to a $\cWth$ homomorphism
$\bar{\imath} \in {\rm Hom\,}_{\cWth}(M(h,w,c),\overline{M}(h,w,c))$,
satisfying $\bar{\imath}(\vM) = \vMb$, by
defining $\bar{\imath}(v) = \vev{v|-}_M$.  Moreover,
because of the uniqueness of $\imath$, we necessarily have $\bar{\imath} =
\imath$.  \smallskip

To get a more explicit form for $\vev{-|-}_M$, define
for any $v\in M(h,w,c)$, the ``vacuum expectation value''
$\vev{v}$ as the coefficient of the highest weight
vector $\vM$ of $M(h,w,c)$ in $v$.
Now, for $u,v \in M(h,w,c)$ and $u=x\, \vM$
for some $x\in U(\cWthm)$, the formula
\eqn\eqPBbh{
\vev{u|v} \equiv \vev{ \om_{\cW}(x) v } \,,
}
clearly defines such a contravariant bilinear form and hence equals
$\vev{-|-}_M$ above.

Now, upon recalling that
the ``radical'' ${\rm Rad\,}\langle - | - \rangle_M$ of the
form $\langle - | - \rangle_M$ is defined as
\eqn\eqPBba{
{\rm Rad\,}\langle - | - \rangle_M ~\equiv~
\{ u\in M(h,w,c)~|~\vev{u|v}_M = 0,\ \forall\,v\in M(h,w,c)\}\,,
}
we can formulate the main properties of the Shapovalov form

\thm\thPBbc
\proclaim Theorem \thPBbc.
\item{i.} $M(h,w,c)$ carries a unique contravariant bilinear form
$\vev{-|-}_M$ such that $\vev{\vM |\vM }_M = 1$, where
$\vM$ is the highest weight vector of $M(h,w,c)$.  This form is
symmetric.
\item{ii.} The generalized eigenspaces of $\cWthz \,=\, \{ L_0,W_0\}$ are
pairwise orthogonal.
\item{iii.} ${\rm Ker\ } \imath
\,\cong\, {\rm Rad\,}\langle - | - \rangle_M \,\cong\,
I(h,w,c)$, hence $L(h,w,c) \,\cong\, M(h,w,c)/I(h,w,c)$ carries a
unique contravariant
bilinear form $\vev{-|-}_L$ such that $\langle v_L|v_L\rangle_L =1$, where
$v_L$ is the highest weight vector of $L(h,w,c)$.
This form is nondegenerate.\par\smallskip

\noindent {\it Proof:}\hfil\break
(i) It remains to show that $\vev{-|-}_M$ is symmetric.
This is, however, evident from \eqPBbh,
and the fact that $\om_{\cW}{}^2=1$.  \hfil\break
(ii) Consider two generalized eigenspaces spanned by
$\{u_0,\ldots,u_{\vdeg-1}\}$ and
$\{v_0,\ldots,v_{\vdeg'-1}\}$ corresponding to eigenvalues
$(h,w)$ and $(h',w')$, respectively.
That is, $L_0\, u_i = h\, u_i$ for all $i$, and $L_0\, v_j = h'\,
v_j$ for all $j$.  Further, $(W_0 - w)^M\, u_i = 0$ for all $M\geq i+1$ and
$(W_0 - w')^M\, v_j = 0$ for all $M\geq j+1$.  Then, as usual,
$(h-h')\vev{v_j|u_i}_M\, = \,
\vev{v_j|L_0\, u_i}_M - \vev{L_0\, v_j| u_i}_M\, =\, 0$,
and so $\vev{v_j|u_i}_M \,=\, 0$
for $h\neq h'$.  Moreover, for $M\geq i+1$, we have
\eqn\eqPBbj{
0 = \vev{ v_j | (W_0 - w)^M \, u_i}_M  = \sum_{k=0}^{M}
\left( \matrix{ M \cr k \cr} \right) (w'-w)^{M-k}
\vev{  (W_0 - w')^k \, v_j | u_i}_M
}
Consider \eqPBbj\ for $j=0$ and arbitrary $i$. It follows from \eqPBbj\ that
$\vev{ v_0 | u_i }_M = 0$ for $w\neq w'$. Now proceed by induction to
$j$ to conclude that $\vev{ v_j | u_i }_M = 0$ for
all $i,j$ if $w\neq w'$.\hfil\break
(iii) By definition, ${\rm Ker\ } \imath
\,\cong\, {\rm Rad\,}\vev{-|-}_M$.  Furthermore, ${\rm Rad\,}\vev{-|-}_M$
is clearly a (proper) submodule of $M(h,w,c)$, so it remains to
be shown that $I(h,w,c) \subset {\rm Rad\,}\vev{-|-}_M$.  By
Theorem \thPBbaa\ (ii) we have $I(h,w,c) \cong PM(h,w,c)$, so
suppose $v\in PM(h,w,c)$.  Clearly, $\vM \not\in U(\cWthp)\cdot v$ so,
in view of \eqPBbh\ this immediately implies that $v\in {\rm Rad\,}
\vev{-|-}_M$. \Box\hfil\smallskip

It turns out to be convenient to parametrize the Verma modules
$M(h,w,c)$  by an $\slth$ weight $\La\in \bfh^*_{\CC}$
and a complex scalar $\al_0 \in \CC$ (called the ``background charge'')
as follows
\eqn\eqPBbb{ \eqalign{
c(\al_0)     & = 2 - 24 \al_0{}^2 \,,\cr
h(\La,\al_0) & = - (\th_1\th_2 + \th_1\th_3 + \th_2\th_3) =
  \half (\La,\La+2\al_0\rh) \,, \cr
w(\La,\al_0) & = \sqrt{3\be} \th_1\th_2\th_3 \,,\cr}
}
where
\eqn\eqPBbc{
\th_i = (\La + \al_0\rh, \ep_i)\,,
}
and $\ep_i,\, i=1,2,3$ are the weights of the
$3$-dimensional representation of $\slth$
with highest weight $\La_1$, \ie,
\eqn\eqPBbd{
\ep_1 = \La_1\,,\qquad
\ep_2 = \La_2 - \La_1 \,,\qquad
\ep_3 = -\La_2 \,.
}
The origin of this parametrization will become apparent in Section \WFock.
Clearly,

\thm\thPBbd
\proclaim Lemma \thPBbd. We have
\eqn\eqBPg{
h(\La,\al_0) ~=~ h(\La',\al_0)\,,\qquad
w(\La,\al_0) ~=~ w(\La',\al_0) \,,
}
if and only if $\La' = w(\La+\al_0\rh) - \al_0\rh$ for some Weyl group
element $w\in W$.\par\smallskip

For convenience, we define, for fixed background
charge $\al_0$, the shifted (dotted) action of the
Weyl group ${W}$ of $\bfg$ on $P$ by
\eqn\eqPBbi{
w \cdot \La ~=~ w(\La+\al_0\rh) - \al_0\rh \,,\qquad
w\in {W}\,.
}
Henceforth, we simply write $M(\La,\al_0)$ for $M(h(\La,\al_0),
w(\La,\al_0),c(\al_0))$, and similarly $L(\La,\al_0)$, for the irreducible
quotient.  Since we will mostly restrict our attention to a specific
value of the central charge $c$ (or background charge $\al_0$) we
will, in fact, often write $M(\La)$ etc., if no confusion can
arise.  Note that, because of Lemma \thPBbd, we have
$M(\La,\al_0) \cong M(\La',\al_0)$ if $\La' = w\cdot \La$ for some
$w\in W$, \ie, to parametrize Verma modules one may choose $\La + \al_0\rh$
to lie in a specific Weyl chamber if one so desires. \smallskip

We denote the subspace of $M(\La,\al_0)$ of $L_0$-eigenvalue
$h(\La,\al_0) + N$ by $M(\La,\al_0)_N$.%
\foot{Note that $M(\La,\al_0)_N \,=\, M(\La,\al_0)_{(h(\La)+N)}$ in the
notation introduced in Section \WOcat.}
Now,
let $\imath_N\,:\,M(\La,\al_0)_N ~\mapright{}~ \overline{M}(\La,\al_0)_N$
denote the restriction of $\imath$ to $M(\La,\al_0)_N$.  Clearly,
$\imath_N$ is a linear map between two vector spaces of equal dimension,
so, after choosing bases for $M(\La,\al_0)_N$ and
$\overline M(\La,\al_0)_N$, the determinant of $\imath_N$,
the so-called Kac determinant, is well-defined.  Put
\eqn\eqPBbfa{
\cS(\La,\al_0)_N ~\equiv~ {\rm det\ }\imath_N\,.
}
Equivalently, for any basis $\{ u_i\}$ of $M(\La,\al_0)_N$
\eqn\eqPBbfb{
\cS(\La,\al_0)_N ~\sim~ {\rm det}( \vev{u_i|u_j}_M )\,,
}
where $\sim$ means proportionality with a factor independent of $\La$
and $\al_0$.
Obviously,  ${\rm Ker\,}\imath_N = 0$ if and only if
$\cS(\La,\al_0)_N \neq 0$.  Moreover,  $\cS(\La,\al_0)_N = 0$ clearly
implies $\cS(\La,\al_0)_k = 0$ for all $k\leq N$.
So, we conclude from Theorem \thPBbc,

\thm\thPBbe
\proclaim Lemma \thPBbe. The Verma module $M(\La,\al_0)$ is irreducible
up to level $N$ if and only if ${\cal S}(\La,\al_0)_N \neq 0$.\par
\smallskip

The following explicit result for the Kac determinant is well-known
(see, \eg, [\BS] and references therein)

\thm\thPBbf
\proclaim Theorem \thPBbf.
\eqn\eqPBbf{
{\cal S}(\La,\al_0)_N ~\sim~ \prod_{\al\in \De} \
\prod_{\scriptstyle{r,s\in \NN} \atop \scriptstyle{rs\leq N}}
\Big( (\La+\al_0\rh,\al) - (r\al_+ + s\al_-) \Big) ^{p_2(N-rs)} \,,
}
where we have introduced $\al_\pm$ such that $\al_0 = \al_+ + \al_-,\,
\al_+\al_- = -1$. The (2-colour) partition function $p_2(N)$ is
defined by
$\prod_{n\geq1} (1-q^n)^{-2} \,=\, \sum_{N\geq0}\, p_2(N)\, q^N$. \par
\smallskip

\proof  The proof can be given either by constructing
a sufficient number of singular vectors explicitly (for example
through a free field construction) or, as in [\KR], by determining
a sufficient number of vanishing lines using the fact that we have
a realization of $\cWth$ on the coset module $(\hslth)_1 \oplus
(\hslth)_k / (\hslth)_{k+1}$ (see, \eg, [\BS]). \Box\hfil\smallskip

An immediate consequence of Theorem \thPBbf\ is the $\cWth$ analogue
of the so-called Kac-Kazhdan condition

\thm\thPBbfa
\proclaim Corollary \thPBbfa.  The Verma module $M(\La,\al_0)$ is
irreducible if and only if
$(\La+\al_0\rh,\al) \not\in (\NN\al_+ + \NN \al_-)$ for
all roots $\al\in \De$.\par\smallskip

In the case of the Virasoro algebra the Kac determinant \eqPBbf\
suffices to determine the complete structure of submodules of
a Verma module [\FF]. In particular, one finds that ${\rm Prim}(M) =
{\rm Sing}(M)$, and that the weights of all singular vectors
are concentrated on the orbit of the highest weight under the affine
Weyl group of $\hsltw$. For $\cWth$, however, knowledge of
the Kac determinant
\eqPBbf\ is by itself not enough to ascertain the submodule structure
of a Verma module, essentially because \eqPBbf\ only carries information
about the $L_0$-weight.  The full submodule structure could probably be
deduced if, for instance, we could find the (nonzero)
generalized eigenspaces of $\cWthz$ and were able to compute the
determinant of the bilinear form on these generalized eigenspaces.
It is generally believed that, also for the $\cWth$ algebra,
the weights of all primitive vectors in a Verma module
are concentrated on a certain orbit of the highest weight vector
under the affine Weyl group of $\hslth$.  To our knowledge,
no general proof of this ``linkage principle'' is available,
but there is certainly ample evidence for it coming from the
Quantum Drinfel'd-Sokolov reduction.
Remarkably, in the case of our immediate interest, namely $c=2$ Verma
modules, it can indeed be proved -- see, Theorem
\thPBdc.
Clearly, knowing the weights of all primitive vectors in Verma
modules is of utmost importance, since these not only
determine the possible irreducible modules in the composition series
of a Verma module (by Lemma \thPBaea), but also the possible nontrivial
$\cWth$ homomorphisms between Verma modules.
\smallskip

We have already emphasized that, for generic modules $V\in\cO$, the
action of the Cartan subalgebra $\cWthz$ need not be diagonalizable.
In fact, this phenomenon already occurs in Verma modules.  Consider
thereto the following easy computation (see [\Wat]).
The level $h+1$ eigenspace
in the Verma module $M(h,w,c)$ is
two-dimensional and spanned by $\{L_{-1}\,\vM, W_{-1}\,\vM\}$.
The action of $W_0$ on this two-dimensional space is given by
\eqn\eqPBbg{ \eqalign{
W_0 ( L_{-1}\,\vM ) & ~=~ (w\,L_{-1} + 2\,W_{-1}) \,\vM\,, \cr
W_0 ( W_{-1} \,\vM) & ~=~  \left(\textstyle{1\over5}
  \left( -1 + 2\be(5h+1) \right) \, L_{-1} + w\,W_{-1} \right)\,\vM\,. \cr}
}
Clearly, $W_0$ is not diagonalizable iff the following
equation holds
\eqn\eqPBzc{
2\be(5h+1) -1 ~=~ 0\,.
}
In that case, we have a two-dimensional generalized eigenspace
corresponding to a $W_0$ eigenvalue $w$.  Upon defining%
\foot{Note that $v_1$ is determined by \eqPBza\ up to the
addition of an arbitrary multiple of $v_0$.}
\eqn\eqPBzb{ \eqalign{
v_0 & ~=~  W_{-1} \,\vM\,, \cr
v_1 & ~=~  \textstyle{1\over2} L_{-1}\,\vM\,,\cr}
}
we find from \eqPBbg\
\eqn\eqPBza{ \eqalign{
W_0\ v_1 & ~=~ w\,v_1 + v_0\,, \cr
W_0\ v_0 & ~=~ w\,v_0\,, \cr}
}
\ie, the vectors $\{ v_0,v_1 \}$ form a Jordan basis for
$M(h,w,c)$ at $L_0$-level $h+1$.  \smallskip

For example, \eqPBzc\ is satisfied for
$h=w=0$ and $c=2$ (\ie, $\La=0,\, \al_0 =0$).
In this specific case, it is easily seen that, in fact,
$v_0$ is a singular vector, whilst $v_1$ is p-singular.  Thus,
the module generated by $v_1$, which
we denote by $M(v_1)$, is a proper
submodule of $M(0,0,2)$ (note that
$v_0 = W_0\, v_1 \in M(v_1)$).  Moreover, $M(v_1)$
is entirely contained
in the maximal ideal, and therefore has to be projected out
in (the first step of) a resolution of the irreducible module $L(0,0,2)$
(see Section \Wres\ for more details).  We elaborate on this example
in later sections.\smallskip

To summarize, we have seen in the simple example above that modules
where the highest weight space corresponds to some
indecomposable representation of the Cartan subalgebra $\cWthz$,
such as $M(v_1)$, naturally occur as submodules of Verma
modules and that, moreover, these modules are required in the construction
of (Verma module type) resolutions for the irreducible modules.\smallskip

After this lengthy discussion, let us now introduce generalized Verma
modules.\smallskip

Let $V^{(\vdeg)}$ denote a $\vdeg$-dimensional
indecomposable representation of
$\cWthz$ of generalized weight $(h,w)$.  As explained in Section \WOcat,
we may choose a basis $\{ v_0,\ldots,v_{\vdeg-1}\}$ of $V^{(\vdeg)}$ such that
(see  \eqPBaac)
\eqn\eqPBbk{
\matrix{
v_{\vdeg-1} & \mapright{W_0-w} & \ldots & \mapright{W_0-w} & v_1 &
  \mapright{W_0-w} & v_0 & \mapright{W_0-w} & 0 \,.\cr}
}

\thm\thPBbi
\proclaim Definition \thPBbi.  The generalized Verma module
$M^{(\vdeg)}(h,w,c)$,
with generalized highest weight $(h,w)$, is defined as the $\cWth$ module
``induced'' from $V^{(\vdeg)}$ by the action of $\cWthm$, \ie, the
$\cWth$ module $\cWth \cdot V^{(\vdeg)}$ modded out by the relations
\eqn\eqPBbla{
\cWthp \cdot V^{(\vdeg)}~=~ 0 \,,}
as well as
\eqn\eqPBblb{ \eqalign{
L_0 \, v_i & ~=~ h\, v_i \quad {\rm for\ } i=0,\ldots ,\vdeg-1\,, \cr
W_0 \, v_i & ~=~ \cases{ w\,v_i + v_{i-1} & for $i=1,\ldots,\vdeg-1$ \cr
                          w\,v_0 & for $i=0\,.$ \cr} \cr}
}
The action of $\cWth$ on $M^{(\vdeg)}(h,w,c)$ is
defined by means of the commutation relations \eqPAaa\
and the relations \eqPBbla\ and \eqPBblb.\par\smallskip

Most of the theorems as well as their proofs
that have been discussed for Verma modules have a
straightforward analogue for generalized Verma modules.  We
refrain from going into details.  Let us just remark that a basis
of $M^{(\vdeg)}(h,w,c)$ is provided by the vectors
\eqn\eqPBbm{
e^{(i)}_{m_1\ldots m_K;n_1\ldots n_L} ~=~
L_{-m_1}\cdots L_{-m_K} W_{-n_1}\cdots W_{-n_L}\, v_i\,,
}
where, $m_1\geq \ldots \geq m_{K-1} \geq m_K >0$ and
$n_1\geq \ldots \geq n_{L -1} \geq n_{L} > 0$, while
$i=0,\ldots,\vdeg-1$.\smallskip

There exists a unique bilinear symmetric form $\vev{-|-}_M$ on
$M^{(\vdeg)}(h,w,c)$, contravariant with respect to $\om_{\cW}$,
and such that $\vev{v_i|v_j}_M = \de_{ij}$.  \smallskip

Since a generalized Verma module $M^{(\vdeg)}(h,w,c)$ is generated
(over $\cWth$) by the vector $v_{\vdeg-1}$, every $\pi
\in {\rm Hom\,}_{\cWth}(M^{(\vdeg)}(h,w,c), V)$ with $V\in \cO$ is
uniquely determined by the image of $v_{\vdeg-1}$ under $\pi$
(\cf, Lemma \thPBba).
Clearly, $\pi(v_{\vdeg-1}) \in {\rm pSing}(V)$ and has generalized
weight $(h,w)$.  Moreover, $(W_0 -w)^{\vdeg}\, \pi(v_{\vdeg-1}) = 0$.
Conversely, for every $v\in {\rm pSing}(V)$ of generalized
weight $(h,w)$ such that $(W_0 -w)^{\vdeg}\, v = 0$ the map
$\pi( v_{\vdeg-1}) = v$ uniquely extends to a $\pi
\in {\rm Hom\,}_{\cWth}(M^{(\vdeg)}(h,w,c), V)$.  \smallskip

In particular, if we apply the above to the case that $V$ itself is a
generalized Verma module, we find a sequence of $\cWth$ homomorphisms
\eqn\eqPBbn{
\matrix{
0 & \mapright{} &
M(h,w,c) & \mapright{\pi_0} & \ldots & \mapright{\pi_{\vdeg-3}} &
M^{(\vdeg-1)}(h,w,c) & \mapright{\pi_{\vdeg-2}} & M^{(\vdeg)}(h,w,c)\,, \cr}
}
defined by $\pi_i(v_{i}) = v_i$ for $i=0,\ldots,\vdeg-2$.
Even more so, since every $\pi_i$ is injective, we obtain
a decreasing filtration of $M^{(\vdeg)}(h,w,c)$ by generalized
Verma modules, \ie,
\eqn\eqPBbo{
\matrix{
M^{(\vdeg)}(h,w,c) & \supset & M^{(\vdeg-1)}(h,w,c) & \supset &
\ldots & \supset & M(h,w,c)\,, \cr}
}
such that all quotients are isomorphic to $M(h,w,c)$.
This filtration
is very useful in relating properties of generalized Verma modules to
those of ordinary Verma modules and, in particular, for relating
generalized Verma module cohomology to Verma module cohomology by
means of the spectral sequence associated to the filtration.
As an example, it follows from the filtration \eqPBbo\ that
the weights of primitive vectors in $M^{(\vdeg)}(\La,\al_0)$ coincide
with the weights of those in $M(\La,\al_0)$.\smallskip

More generally, for $\vdeg_1 < \vdeg_2$, we have an
injective $\pi \in
{\rm Hom\,}_{\cWth}(M^{(\vdeg_1)}(h,w,c), M^{(\vdeg_2)}(h,w,c))$
such that the quotient is isomorphic to $M^{(\vdeg_2-\vdeg_1)}(h,w,c)$, \ie,
we have exact sequences
\eqn\eqBPn{ \matrix{
0 & \mapright{} & M^{(\vdeg_1)}(h,w,c) & \mapright{\pi} &
M^{(\vdeg_2)}(h,w,c) & \mapright{} &
M^{(\vdeg_2 - \vdeg_1)}(h,w,c) & \mapright{} & 0\,. \cr}
}\smallskip

Although it is still true that any $\cWth$ homomorphism of a Verma module
to a generalized Verma module is injective (\cf,
Theorem \thPBbaa\ (iii)), this
property does not hold for $\cWth$ homomorphisms between arbitrary
generalized Verma modules.
Consider, for example, the $\cWth$ automorphism $\pi$ of
$M^{(2)}(h,w,c)$ defined by $\pi(v_1) = v_0$.  Clearly
$\pi$ is not injective.  For a more complicated example, consider
the Verma module $M = M(0,0,2)$.  We have already seen that there
exists a $\cWth$ homomorphism $M^{(2)}(1,0,2) ~\mapright{\pi}~
M(0,0,2)$ defined by $\pi(v'_1) = v_{1}$ (with $v_1$ as defined
in \eqPBzb), whose image is
$M(v_{1})$.  Explicit computation shows that $M(v_{1}) \,\not\cong\,
M^{(2)}(1,0,2)$.  We come back to this example in more detail
in Section \Wctwob.

\subsubsec{Fock spaces}
\subsubseclab\WFock

In this section we define an extremely useful realization
of the $\cW_3$ algebra known as the ``free field realization.''
The corresponding modules are known as Fock modules or
Feigin-Fuchs modules.\smallskip

Let $\cA$ be the oscillator algebra (Heisenberg algebra) with
basis $\{ \al_m^i\,|\, m\in \ZZ,\, i=1,2\}$ and commutation
relations
\eqn\eqPBca{
[ \al_m^i , \al_n^j ] ~=~ m \de_{m+n,0} \de^{ij} \,.
}
In terms of formal power series
\eqn\eqPBcab{
i\p\ph^i(z) ~=~ \sum_{n\in\ZZ} \ \al_{n}^i\, z^{-n-1}\,\quad
   \in \cA[[z,z^{-1}]]\,,
}
also referred to as ``free scalar fields,''
the commutation relations \eqPBca\ are encoded in the following OPEs
\eqn\eqPBcac{
i\p\ph^i(z)\ i\p\ph^j(w) ~=~  { \de^{ij} \over (z-w)^2 }+ \ldots\,.
}
Also, for convenience, we often use vector notation by introducing
an orthonormal basis (with respect to the Euclidean inner product)
$\{\vec{e}_1, \vec{e}_2 \}$ of $\CC^{\,2}$, \eg,
\eqn\eqPBcad{
\vec{\al}_n ~\equiv~ \sum_i \ \al_n^i\,\vec{e}_i\,,
}
and identify $\CC^{\,2}$ with the weight space $\bfh^*_\CC$ of $\slth$.
The algebra $\cA$ has a Cartan decomposition $\cA ~\cong~ \cA_- \oplus
\cA_0 \oplus \cA_+$, where
\eqn\eqPBcae{
\cA_\pm ~=~ \{ \al_n^i ~|~ \pm n>0 \} \,,\qquad
\cA_0 ~=~ \{ \al_0^i \} \,.
}
The universal enveloping algebra $U(\cA)$, as well as its local
completion $U(\cA)_{\rm loc}$ [\FeFrloc] are defined as usual.\smallskip

For any $\La\in \bfh^*_\CC$ and $\al_0\in\CC$ we have an (irreducible)
$\cA$-module
$F(\La,\al_0)$, \ie, a Fock module, with basis
\eqn\eqPBcaa{
f_{m_1,\ldots,m_M; n_1,\ldots n_N} ~=~ \al_{-m_1}^2 \cdots
\al_{-m_M}^2 \al_{-n_1}^1 \cdots \al_{-n_N}^1 |\La\rangle\,,
}
where $n_1\geq n_{2} \geq \ldots \geq n_N,\,
m_1\geq m_{2} \geq \ldots \geq m_M$ and the ``vacuum vector''
$|\La\rangle$ satisfies
\eqn\eqPBcb{ \eqalign{
\vec{\al}_m\, |\La\rangle ~=~ &  0\,,\qquad {\rm for}\quad m>0 \,,\cr
\vec{\al}_0 \,|\La\rangle  ~=~ & \La\, |\La\rangle \,. \cr}
}
The action of $\cA$ on $F(\La,\al_0)$ is defined by \eqPBca\ and
\eqPBcb.%
\foot{Clearly, at this point, the parameter $\al_0$ does
not play a role, since all modules $F(\La,\al_0)$ for different values
of $\al_0$ are isomorphic as $\cA$-modules.  The reader should easily
distinguish, in context, the complex number $\al_0$ from the vector of
``momentum'' operators $\vec{\al}_0$.}  It is usual to extend the
representation by the operator $\vec{q}$ such that $[ q^i , \al_n^j ]
{}~=~ i \de_{n,0} \de^{ij}$.  In canonical quantization, $\vec{q}$ is
simply the zero mode of the free scalar field,
\eqn\eqJMphi{
\ph^i(z) ~=~ q^i
-i \al_0^i \, \log z + i \sum_{n\neq 0} \ {\al_{n}^i\over n} \, z^{-n}\, .
}
Then the vacuum vector of different Fock modules, \ie,
$|\La\rangle$ for different $\La$, can be generated from $|0\rangle$
(the so-called $\frak{sl}(2,\CC)$ invariant vacuum) via
$|\La \rangle = e^{i\La \cdot q}|0\rangle$.\smallskip

It is convenient to state more precisely the relation between
operators and states.  First observe that for fixed $\La$ there is an
isomorphism between the states in \eqPBcaa\ and the space of fields
obtained by a finite number of normal products of a finite number of
derivatives of the basic field $i \p\phi^j(z)$.  Moreover, using the
normal ordering prescription, we have ${\rm lim}_{z\rightarrow 0} e^{i\La
\cdot \phi(z)}|0\rangle = |\La\rangle$.  So, the isomorphism can be
extended to arbitrary $\La$: Introduce the space, $\fV$, of
normal-ordered fields of the form $P(i\p\phi^j(z))
e^{i\La\cdot\phi(z)}$, where $P$ is a polynomial in $i\p\phi^j(z)$
and its derivatives, and $\La \in \bfh^*_\CC$.  Then for any
state $|\cO\rangle \in F(\La,\al_0)$, there is a corresponding field
$\cO(z)\in\fV$ such that
\eqn\eqJMstop{
|\cO\rangle=\lim_{z\rightarrow 0} \cO(z)|0\rangle\, .  }  When the
space of allowed $\La$ in $\fV$ is restricted to a lattice $L$ such
that the OPEs of all fields are meromorphic, we call the space
$\fV$ a chiral algebra.  Under certain further conditions we call
the chiral algebra a Vertex Operator Algebra (VOA).  The strongest of
these is the requirement that for any two fields in the chiral algebra
the OPE in one order is related to that in the other order by analytic
continuation.  For further discussion, and a complete list of the
defining relations for a VOA, see, \eg, [\Bor,\FLM,\FHL].
To impose
these conditions generally requires that we extend the
construction of operator fields to include phase-cocycles.  An example
which will be required later is discussed in Appendix \appPBC.\smallskip

For any given $\al_0\in\CC$, let $\om_{\cA}$ be the $\CC$-linear
anti-involution of $\cA$ defined by
\eqn\eqPBcc{
\om_{\cA}(\vec{\al}_n) ~=~ w_0 (\vec{\al}_{-n})  - 2 \al_0 \rh
  \de_{n,0}\,,
}
which we may equivalently specify on fields, \ie, $\cA[[z,z^{-1}]]$, by
\eqn\eqPbcdko{
\om_{\cA} ( i\p\vec{\ph}(z))
  ~=~ z^{-2} \,w_0 (i\p\vec{\ph}(z^{-1})) - 2\al_0 \rh
  \,z^{-1}  \, .
}
Clearly, $\om_{\cA}$ extends to a $\CC$-linear anti-involution
on $U(\cA)_{\rm loc}$.%
\foot{There exist many other anti-involutions on
$\cA$ and consequently many different bilinear forms contravariant
with respect to the chosen anti-involution.
The one we have chosen here is the most natural with
regards to the $\cWth$ module structure of $F(\La,\al_0)$
(see Theorem \thPBca).  We will, however, introduce a different
anti-involution and the associated form, needed for the proof of
Theorem \thPBda, shortly.}
Since,
\eqn\eqPBcdc{
\om_{\cA}(\vec{\al}_0) |\La\rangle ~=~ (w_0(\La) - 2\al_0\rh) |\La\rangle
  ~=~ (w_0\cdot \La) |\La\rangle \,,
}
the anti-involution $\om_{\cA}$ provides us with a map
\eqn\eqPBcd{
F(w_0\cdot\La,\al_0) \,\times\, F(\La,\al_0) ~\mapright{}~ F(\La,\al_0)\,,
}
defined by
\eqn\eqPBcda{
( x \, \vFp  , y \, \vF)  ~\mapsto~
  \om_{\cA} (x) y \, \vF\,,
}
where $x,y \in U(\cA_-)_{\rm loc}$, and we have denoted, for convenience,
$\vF = |\La,\al_0\rangle$ and $\vFp = |w_0\cdot\La,\al_0\rangle$.
\smallskip

Furthermore, we define the ``vacuum expectation value''
$\vev{-}\,:\, F(\La,\al_0) ~\mapright{}~ \CC$
as the coefficient of $\vF$ in the expansion
of $v\in F(\La,\al_0)$ in the basis \eqPBcaa.  \smallskip

By combining the bilinear map \eqPBcd\ with $\vev{-}$ we obtain
a bilinear form
\eqn\eqPBcdb{
\vev{-|-}_F ~:~ \matrix{
F(w_0\cdot\La,\al_0) \,\times\, F(\La,\al_0) & \mapright{} & \CC \,,\cr}
}
\ie,
\eqn\eqPBce{
\vev{v|w} ~\equiv~ \vev{ \om_\cA(x) w } \,.
}
where $v = x \, \vFp$ with $x\in U(\cA_-)_{\rm loc}$.
We have

\thm\thPBcb
\proclaim Theorem \thPBcb.
There exists a unique bilinear
form $\vev{-|-}_F\,:\, F(w_0\cdot\La,\al_0) \,\times\, F(\La,\al_0)
{}~\mapright{}~ \CC$, contravariant with respect to $\om_{\cA}$,
such that $\vev{\vFp |\vF}_F = 1$, where $\vF$ and $\vFp$
are the highest weight vectors of $F(\La,\al_0)$ and $F(w_0\cdot\La,\al_0)$,
respectively.  This form is nondegenerate.\par\smallskip

Another useful anti-involution of $\cA$,
to be denoted by $\oom_{\cA}$, is defined by
\eqn\eqPBcdh{
\oom_{\cA} ( \al_n^i) ~=~ \al_{-n}^i \,,
}
or, equivalently, by
\eqn\eqPBcdkt{
\oom_{\cA} ( i\p\ph^i(z))  ~=~ z^{-2} \, i\p\ph^i(z^{-1}) \,.
}
We extend $\oom_{\cA}$ to an anti-linear anti-involution
on $U(\cA)_{\rm loc}$.
In complete analogy with eq.\ \eqPBce\ we now obtain a sesquilinear
form
\eqn\eqPBcdi{
(-|-)_F~:~  F(\La,\al_0) \,\times\, F(\La,\al_0) ~\mapright{}~ \CC\,,
}
by
\eqn\eqPBcdj{
(x\, \vF | y \, \vF)_F ~=~ \vev{ \oom_{\cA}(x) y \, v_F } \,.
}
In fact,

\thm\thPBcab
\proclaim Theorem \thPBcab. There exists a unique sesquilinear form
$(-|-)_F~:~  F(\La,\al_0) ~\times~ F(\La,\al_0) ~\mapright{}~ \CC$,
contravariant with respect to $\oom_{\cA}$, such that $(\vF|\vF)_F=1$.
This form is Hermitian, \ie, $(v|w)_F = \overline{(w|v)}_F$, and
positive definite.  The basis \eqPBcaa\ is orthogonal with respect
to $(-|-)_F$.\par\smallskip

\proof Standard (see, \eg, [\KR]). \Box\hfil\smallskip

\thm\thPBca
\proclaim Theorem \thPBca. For any $\al_0\in \CC$ such that
$c = 2 - 24\al_0{}^2$,
we have a homomorphism of algebras
$\vrh \,:\, \cWth \to U(\cA)_{\rm loc}$ defined by
\eqn\eqPBcdd{
\vrh(T(z)) ~=~  -\half \p\vec{\ph}\cdot \p\vec{\ph} - i\al_0 \rh\cdot
  \p^2\vec{\ph} \,,
}
\eqn\eqPBcdda{ \eqalign{
\vrh(W(z)) ~=~  & \hbox{${-i\sqrt{\be}\over {3 \sqrt{2}}}$}\,
  (\p\ph^1\p\ph^1\p\ph^2 - \p\ph^2\p\ph^2\p\ph^2)
  + \hbox{$\al_0\sqrt{\be}$}
  (\hbox{${\sqrt{3}\over 2}$} \p\phi^1 \p^2\phi^1 +
   \p\phi^2 \p^2\phi^1 -
   \hbox{${\sqrt{3}\over 2}$} \p\phi^2 \p^2\phi^2) \cr
& + \hbox{${i\sqrt{3 \be}\over {2 \sqrt{2}}}$} \al_0^2\,
(\p^3\phi^1 - \hbox{${1\over \sqrt{3}}$} \p^3\phi^2) \, .  \cr}
}
Furthermore,
\eqn\eqPBcde{
\vrh\,\om_{\cW} ~=~ \om_{\cA}\,\vrh\,, \qquad {\rm for\ all\ } \al_0\in\CC\,,
}
and
\eqn\eqPBcdf{
\vrh\,\oom_{\cW} ~=~ \oom_{\cA}\,\vrh\,, \qquad {\rm for}\ \al_0=0\,,
}
such that, in particular, the form $\vev{-|-}_F$ is contravariant
with respect to $\om_{\cW}$ for all $\al_0\in\CC$, and
$(-|-)_F$ is contravariant with respect to $\oom_{\cW}$ for $\al_0=0$.
\par\smallskip

\proof By a straightforward, albeit tedious, calculation. \Box\hfil\smallskip

\noindent {\it Remark:} The homomorphism $\vrh$ was first discussed
in [\FZ].  In [\FLa,\FLb] a systematic method to derive
$\vrh$, the so-called Quantum Drinfel'd-Sokolov (QDS) reduction, was
first presented.  In \eqPBcdda\
we have chosen an orthonormal basis with respect to which
the simple roots of $\slth$ are $\al_1 = (\sqrt{2},0)\, , \,\,
\al_2 = (-{1\over{\sqrt{2}}},{\sqrt{3}\over{\sqrt{2}}})$.
\smallskip

By means of the homomorphism $\vrh$ we can equip the $\cA$-module
$F(\La,\al_0)$ with the structure of a $\cW_3$ module.  We denote
this module by $F(\La,\al_0)$ as well.  Clearly, $F(\La,\al_0) \in
\cO$; the highest weight space is one-dimensional and spanned by
$|\La,\al_0\rangle$.  The central charge of this representation, along
with the weight of $|\La,\al_0\rangle$, are parametrized exactly as in
\eqPBbb\ by the background charge $\al_0$.
The module contragredient to $F(\La,\al_0)$, as a $\cW_3$ module, is
determined by the following theorem

\thm\thPBcaa
\proclaim Theorem \thPBcaa. We have a $\cWth$ isomorphism
\eqn\eqPBcca{
\imath_F~:~  \matrix{ F( w_0\cdot \La ,\al_0)  & ~\mapright{\cong}~   &
\overline{F} (\La,\al_0)\,, \cr}
}
where $\imath_F$ is explicitly given by $\imath_F(v) = \langle v | -
\rangle_F$.
\par\smallskip

\proof The fact that $\imath_F \in {\rm Hom}_{\cWth}(F( w_0\cdot \La ,\al_0),
\overline{F} (\La,\al_0))$ follows from the contravariance of
$\langle -|- \rangle_F$ with respect to $\om_{\cA}$ and \eqPBcde.
That $\imath_F$ is, in fact, an isomorphism follows from the
fact that $\langle -|- \rangle_F$
is nondegenerate (see Theorem \thPBcb). \Box\hfil\smallskip

To determine the structure of $F(\La,\al_0)$ as a $\cW_3$ module
it turns out to be useful to ``compare'' $F(\La,\al_0)$ to a
(contragredient-) Verma module. We have

\thm\thPBcc
\proclaim Theorem \thPBcc.  Let $\vM, \vF$ and $\vMb$ denote
the highest weight vectors of $M(\La,\al_0)$, $F(\La,\al_0)$ and
$\overline{M}(\La,\al_0)$, respectively.
\item{i.} There exist unique $\cWth$ homomorphisms
\eqn\eqBPh{ \matrix{
M(\La,\al_0) & \mapright{\imath'} & F(\La,\al_0) & \mapright{\imath''}
 & \overline{M}(\La,\al_0)\,, \cr}
}
such that $\imath'(\vM) = \vF$ and $\imath''(\vF) = \vMb$.
\item{ii.} $\imath = \imath{''} \, \imath' $.
\item{iii.} $\imath'(x\, \vM) ~=~ \vrh(x)\, \vF$ and
$\overline{\imath''}(x\, \vM) ~=~ \vrh(x)\, \vFp$.
\item{iv.} We have a commutative diagram
\eqn\eqPBcg{  \comdiag{  \matrix{
M(\La,\al_0) \,\times\, M(\La,\al_0) & \mapright{\overline{\imath''}
  ~\times~ \imath'}
  & F(w_0\cdot \La,\al_0) \,\times\, F(\La,\al_0) \cr
&&\cr
\mapdown{ \langle - | - \rangle_M} &&  \mapdown{ \langle - | - \rangle_F}\cr
&&\cr
\CC & \mapright{{\rm id}} & \CC  \cr}
}}
\par\smallskip

\noindent {\it Proof:} \hfil\break
(i) The existence (and definition) of $\imath'$ follows immediately from
Lemma \thPBba.  Similarly, Lemma \thPBba\ gives a $\cWth$ homomorphism
$\overline{\imath''}\,:\, M(\La,\al_0)
{}~\mapright{}~ F(w_0\cdot\La,\al_0) ~\cong~ \overline{F}(\La,\al_0)$
which, by Lemma \thPBag, is contragredient to the map $\imath''$ sought
for.\hfil\break
(ii) Follows from the uniqueness of $\imath$. \hfil\break
(iii) Follows from the uniqueness of $\imath'$ and $\imath''$.\hfil\break
(iv) Let $x,y\in U(\cWthm)$, then
\eqn\eqBPi{ \eqalign{
\langle \overline{\imath''}(x \, \vM) , \imath'(y\, \vM) \rangle_F & ~=~
  \langle (\om_{\cA}\, \rh) (x) \rh(y) \, \vF \rangle_F ~=~
  \langle (\rh\, \om_{\cW})(x) \rh(y) \, \vF \rangle_F \cr
  & ~=~ \langle \rh( \om_{\cW}(x) y) \, \vF \rangle_F  ~=~
  \langle \om_{\cW}(x) y \, \vM \rangle_M ~=~ \langle x \, \vM,
  y\, \vM \rangle_M \,.\cr}
}
\Box\hfil\smallskip

Let, $\imath'_N$ and $\imath''_N$ denote the restrictions of
$\imath'$ and $\imath''$ to $M(\La,\al_0)_N$ and $F(\La,\al_0)_N$,
respectively.  Since $\imath'_N$ and $\imath''_N$ are linear maps between
vector spaces of equal dimension we can define
\eqn\eqPBcga{
\cS'(\La,\al_0)_N ~\equiv~ {\rm det\,} \imath'_N\,,\qquad
\cS{''}(\La,\al_0)_N ~\equiv~ {\rm det\,} \imath''_N \,.
}
where the determinants are defined by means
of bases $\{ v_i\}$ and $\{ w_i\}$ of
$F(\La,\al_0)_N$ and $M(\La,\al_0)_N$, respectively.  We have
\eqn\eqPBcgb{
\cS'(\La,\al_0)_N ~\sim~ {\rm det\,}( \vev{v_i|\imath'(w_j)}_F )\,.
}
In addition,
it follows from the proof of Theorem \thPBcc\ (i) that
\eqn\eqPBcgc{
\cS{''}(\La,\al_0)_N ~=~ \cS'(w_0\cdot\La,\al_0)_N\,.
}
We have

\thm\thPBcd
\proclaim Theorem \thPBcd\ [\BMPa].
\eqn\eqBPj{ \eqalign{
\cS'(\La,\al_0)_N & ~\sim~ \prod_{\al\in \De_+}
\prod_{\scriptstyle{ r,s\in \NN} \atop \scriptstyle{rs\leq N}}
\Big( (\La+\al_0\rh,\al) - (r\al_+ + s\al_-) \Big) ^{p_2(N-rs)} \,,\cr
\cS{''}(\La,\al_0)_N & ~\sim~ \prod_{\al\in \De_+}
\prod_{\scriptstyle{ r,s\in \NN} \atop \scriptstyle{rs\leq N}}
\Big( (\La+\al_0\rh,\al) + (r\al_+ + s\al_-) \Big) ^{p_2(N-rs)} \,.\cr}
}\par\smallskip

\noindent {\it Sketch of proof:} The proof is based on the explicit
construction of a sufficient number of singular vectors in $F(\La,\al_0)$
in terms of multi-contour integrals over products of screeners.
Note that, by Theorem \thPBcc\ (ii), the Kac determinant $\cS(\La,\al_0)$
of Theorem \thPBbfa\ factorizes, up to
a proportionality factor, as $\cS'(\La,\al_0) \cS''(\La,\al_0)$.
\Box\hfil\smallskip

As an immediate consequence of Theorem \thPBcd\ we have

\thm\thPBcda
\proclaim Corollary \thPBcda\ [\BMPa].
\item{i.}
\eqn\eqBPk{
F(\La,\al_0) ~\cong~ \cases{  M(\La,\al_0)  & if $(\La+\al_0\rh,\al) \not\in
  (\NN \al_+ + \NN \al_-)$ for all $\al\in\De_+\,,$ \cr
                  \overline{M}(\La,\al_0)  & if $(\La+\al_0\rh,\al) \not\in
  -(\NN \al_+ + \NN \al_-)$ for all $\al\in\De_+\,.$ \cr}
}
In particular, if $(\La+\al_0\rh,\al) \not\in (\NN \al_+ + \NN \al_-)$ for
all $\al\in\De$, then $M(\La,\al_0) ~\cong~ F(\La,\al_0) ~\cong~
\overline{M}(\La,\al_0)$ are irreducible.
\item{ii.} For $\al_0{}^2\in \RR$ with $\al_0{}^2 < -4$ (or,
equivalently, $c\geq c_{\rm crit}
 - 2 = 98$), we have
\eqn\eqBPl{
F(\La,\al_0) ~\cong~ \cases{  M(\La,\al_0)  & for $i(\La+\al_0\rh) \in
  \et D_+\,,$ \cr
                  \overline{M}(\La,\al_0)  & for $-i(\La+\al_0\rh) \in
  \et D_+\,,$ \cr}
}
where $D_+ ~=~ \{ \la \in \bfh^*_\RR ~|~ (\la,\al) \geq 0,\,
\forall \al\in \De_+ \}$ denotes the fundamental Weyl chamber, and
$\et = {\rm sign}(-i\al_0)$.
\par\smallskip

It immediately follows from Corollary \thPBcda\ that for
almost all $\La\in P$ we have an isomorphism $M(\La,\al_0) \,\cong\,
F(\La,\al_0)$ of $\cWth$ modules. Note, however, that since the
(generalized) eigenvalues of $\cWthz$ are algebraic in (the components
of) $\La$, they are in fact equal (and have the same multiplicity)
on $M(\La,\al_0)$ and $F(\La,\al_0)$ for {\it all} $\La \in \bfh^*_\CC$.
This in turn implies the equality of the characters and since the
characters ${\rm ch\,}L$ are algebraically independent
(see Theorem \thPBaaa) it follows
immediately from \eqJHb\ and Lemma \thPBaea\ that

\thm\thPBce
\proclaim Theorem \thPBce.
\item{i.} For all $\La\in \bfh^*_\CC$ and all irreducible
modules $L$ we have $(M(\La,\al_0)\!:\! L) =
(F(\La,\al_0)\!:\! L)$.
\item{ii.} There is a 1--1 correspondence between
primitive vectors in $M(\La,\al_0)$ and in $F(\La,\al_0)$.
\par

\subsec{Verma modules and Fock modules at $c=2$}
\subseclab\Wctwo

\subsubsec{Generalities}
\subsubseclab\Wctwoa

In this section we study in more detail the structure
of Verma modules, irreducible modules and Fock spaces for central
charge $c=2$.  This is the case of most interest for the rest of this
paper, where we study the $4D$ $\cWth$ string -- \ie, the off-critical
$\cWth$ string with two flat embedding coordinates.  These embedding
coordinates correspond to the ``matter'' free fields in the above for
$c^M = 2$, thus motivating the interest in such Fock modules.  The
results for the remaining modules are required to obtain a framework
in which calculations for $c=2$ Fock spaces are feasible.  This
becomes more clear below, and in the following section.  \smallskip

Remarkably, the $\hslth$ structure which appears for $c=2$ allows
us to derive strong results; in particular, we obtain the weights
and multiplicities of primitive vectors in $c=2$ Verma modules.
Although the construction
of a level-1 representation of $\hslth$ on $c=2$ Fock spaces
is standard, we give a brief review in Appendix \appPBC.  This serves
to set conventions, as well as to introduce the concept of
Vertex Operator Algebra (VOA) associated with a given lattice of Fock
spaces. \smallskip

A preliminary result is the character of irreducible representations
at $c=2$.

\thm\thPBdaa
\proclaim Theorem \thPBdaa\ [\Bou,\BS].  For the irreducible modules
$L(\La,0)$ at $c=2$ with $\La\in P_+$  we have
\eqn\eqPBdaa{\eqalign{
{\rm ch}_{L(\La,0)}(q) & ~=~ {1\over \prod_{n\geq1} (1-q^n)^2}\,
  \sum_{w\in W}\, \ep(w)\, q^{ {1\over2} |w(\La+\rh) - \rh|^2} \cr
& ~=~
  { q^{ {1\over2} |\La|^2}  \over \prod_{n\geq1} (1-q^n)^2}\,
  \prod_{\al\in \De_+} \, (1-q^{(\la+\rh, \al)})\,. \cr}
}\par\smallskip

\thm\thPBda
\proclaim Theorem \thPBda. Consider $c=2$, \ie, $\al_0=0$, and $\La\in
\bfh^*_\CC$.
\item{i.} The Fock space $F(\La,0)$ is completely reducible for
all $\La\in \bfh^*_\CC$.
\item{ii.} For all $\La\in P$ we have
\eqn\eqPBda{
F(\La,0) ~\cong~ \bigoplus_{\La' \in P_+} \
  m^{\La'}_{\La} L(\La',0)\,.
}
where $m^{\La'}_{\La}$ is equal to the multiplicity of the weight
$\La$ in the irreducible finite dimensional representation
${\cal L}(\La')$ of $\slth$ with highest weight $\La'$.
\item{iii.} $(F(\La,0):L(\La',0)) = m^{\La'}_{\La}$.
\par\smallskip

\noindent {\it Proof:} \hfil\break
(i) By Theorems \thPBcab\ and \thPBca\ we have a positive definite
Hermitian form $(-|-)_F$, contravariant with respect to $\oom_{\cW}$,
on the Fock space $F(\La,0)$, \ie, the $\cW_3$ module $F(\La,0)$
is unitary with respect to $(-|-)_F$.  As in, \eg,
Prop.\ 3.1 of [\KR]
this immediately implies the complete reducibility of $F(\La,0)$.\hfil\break
(ii) From the Frenkel-Kac-Segal vertex operator construction
it follows that $\bigoplus_{\La\in P}\, F(\La,0)$ is an $\hslth$ module
at level $1$.  In fact, it is known that
\eqn\eqPBDa{
F ~\equiv~ \bigoplus_{\La\in P}\ F(\La,0) ~\cong~
  L^{\hslth}(\La_0) \oplus L^{\hslth}(\La_1) \oplus L^{\hslth}(\La_2)\,,
}
where $L^{\hslth}(\La_i),\,i=0,1,2$, denotes the integrable $\hslth$
highest weight module at level-$1$ with highest weight $\La_i$.
Under the horizontal algebra $\slth$, $F$ decomposes as
\eqn\eqPBDb{
F ~\cong~ \bigoplus_{\La'\in P_+} \ \left( \cL(\La') \otimes
  V(\La') \right) \,,
}
But, since $\cWth$
is in the commutant of $\slth$ [\BBSS], it acts
on the ``multiplicity spaces'' $V(\La')$.  In fact, by comparing the
characters on each side of \eqPBDb\ and using
Theorem \thPBdaa, one easily verifies that
$V(\La') \cong L(\La',0)$ (see, \eg, [\KP]).
Now, decomposing $F$ under $\bfh$
immediately gives \eqPBda. \hfil\break
(iii) Follows directly from (ii). \Box\hfil\smallskip

\noindent {\it Remarks:}
\item{i.} Note that since the weight multiplicities $m^{\La'}_{\La}$
are Weyl invariant, \ie, $m^{\La'}_{w\La} =  m^{\La'}_{\La}$, for
all $w\in W$, we have an isomorphism
\eqn\eqPBdc{
F(\La,0) ~\cong~ F(w\La,0) \,.
}
Similar isomorphisms do {\it not} hold for $\al_0 \neq 0$.
\item{ii.} All the results given to this point directly
extend to $c = \ell$ representations of $W_{\ell + 1}$,
where $\slth$ is replaced by $\frak{sl}_{\ell+1}$.
\item{iii.} If $\La\in P_+$
we have the following explicit formula -- specific to $\slth$ -- for
the weight multiplicities $m^{\La'}_{\La}$
\eqn\eqPBdb{ \eqalign{
\sum_{\be\in Q_+} \ m_{\La}^{\La + \be} e^\be ~=~ &
 {1\over (1-e^{\al_1})(1-e^{\al_2})(1-e^{\al_3})} -
 {e^{(\La+\rh,\al_1)\al_2}\over (1-e^{\al_2})(1-e^{\al_3})(1-e^{\al_1+2\al_2})}
 \cr
& -{e^{(\La+\rh,\al_2)\al_1}\over
 (1-e^{\al_1})(1-e^{\al_3})(1-e^{2\al_1+\al_2})} \,.\cr}
}\smallskip

Part (ii) of Theorem \thPBda\ can also
be argued more heuristically, along the lines of the
decomposition theorem for Virasoro Fock modules at $c=1$ (see, \eg,
[\KR]). One proceeds by explicitly constructing a standard set of
singular vectors in the Fock space.  Then, by comparing the character
of the irreducible modules built on those singular vectors with the
character of the Fock space, one concludes that this set exhausts all
possible singular vectors.
The standard set of singular vectors
is naturally determined by the screening
operator construction, which we briefly recall. Consider, for $\La\in P$,
the ``screening operators'' $ Q_i ~:~ F(\La+\al_i,0) ~\to~ F(\La,0)$
associated to the simple roots $\al_i,\, i=1,\ldots,\ell$ of $\bfg$,
defined by
\eqn\eqPBdd{
Q_i ~=~ \oint\ {dz\over 2\pi i} \
  e^{ -i \al_i\cdot \ph} \,.  } It is straightforward to
check that, for each $i$, $Q_i$ is a $\cW$-homomorphism. Also, the
$Q_i$ satisfy the Serre relations of $U_q(\bfnm)$ for $q=-1$ [\BMPqg].
Clearly then, provided it is nonvanishing, the image of the highest
weight vector $|\La+\al_i\rangle \in F(\La + \al_i,0)$ under $Q_i$ is
a singular vector in $F(\La,0)$.  More generally, the image of
$|\La+\be\rangle$ under the composite operator $Q_\be = Q_{i_1} \cdots
Q_{i_n}$, $\be = \al_{i_1} + \ldots + \al_{i_n}$, yields a singular
vector in $F(\La,0)$, provided this image is nonvanishing.
{}From the inequality
\eqn\eqBPm{
h(\La +\be + r\al_i) \geq h(\La+\be) \qquad {\rm iff\ }\quad
r\leq (\La+\be,\al_i)\,,
}
it follows trivially that
\eqn\eqPBde{
(Q_i)^r |\La + \be \rangle = 0 \qquad {\rm if} \ r\geq
(\La+\be+\rh,\al_i)\,.  } Further, because of the algebra of the
$Q_i$, we may identify $Q_\be$ with a state at weight $\La$ in the
$\slth$ Verma module with highest weight $\La+\be$.  Then \eqPBde\
implies that the combinations of screening operators which act
nontrivially on $|\La+\be\rangle$ can be, at most, identified with the
weight $\La$ subspace of the irreducible quotient of the Verma module
$M_{\La+\be}$.  In other words, the number of nonvanishing singular
vectors in $F(\La,0)$ of type $Q_\be |\La+\be\rangle,\, \be \in
Q,\, \La+\be \in P$, is at most equal to $m^{\La+\be}_\La$, the
multiplicity of the weight $\La$ in the irreducible $\bfg$-module with
highest weight $\La+\be$.  The proof would now be complete if we
could show that the number of nonvanishing singular vectors
is exactly {\it equal} to $m^{\La+\be}_\La$.  For one can
easily sum the characters of the irreducible modules built on these
singular vectors, using Theorem \thPBdaa, and finds that the result is
exactly equal to the character of the Fock space $F(\La,0)$.
Thus it would follow that these singular vectors in fact exhaust the set of
all singular vectors in $F(\La,0)$.  Hence the last step for the proof
along these lines involves a careful study of the integral representations
of the singular vectors constructed above.  We have not carried out this
step.  However, it seems that the proof presented earlier could be interpreted
exactly as a ``nonvanishing theorem'' for these integrals.\smallskip

This concludes our discussion of the $c=2$ Fock spaces. We now
turn our attention to $c=2$ (\ie, $\al_0=0$) Verma modules.  Unfortunately,
the precise submodule structure of Verma modules is unknown.  We can,
however, conclude a lot from the known structure of the Fock modules.
First of all

\thm\thPBdb
\proclaim Theorem \thPBdb. Let $M(\La,0) ~\mapright{\imath'}~
F(\La,0) ~\mapright{\imath''}~ \overline{M}(\La,0)$
be the $\cW$-homomorphisms of Theorem \thPBcc.
We have $\imath'(M(\La,0)) ~\cong~ L(\La,0)$ or, in other words,
$\imath'(I(\La,\al_0)) = 0$. Similarly, $\imath''(F(\La,0))
{}~\cong~ L(\La,0)$.\par\smallskip

\noindent {\it Proof:} Follows from the complete reducibility of $F(\La,0)$
and the fact that $M(\La,0)$ is generated by $\cWthm$. \Box\hfil\smallskip

Furthermore, since the composition series for $c=2$ Fock modules is
now completely known, and Verma modules have the same composition
factors, we have

\thm\thPBdc
\proclaim Theorem \thPBdc.  Let $\La,\La'\in P$ and let $w'\in W$
such that $w'\La'\in P_+$, then
\item{i.} $(M(\La,0)\!:\!L(\La',0)) ~=~ m_\La^{w'\La'}$.
\item{ii.} ${\rm Prim}(M^{(\vdeg)}(\La,0)) ~\subset~
\coprod_{\La'\in P_+ \,;\,
m^{\La'}_\La \neq 0}\ M^{(\vdeg)}(\La,0)_{(h(\La'),w(\La'))}\,, $.
\item{iii.} $\pi \in {\rm Hom}_{\cWth}(M^{(\vdeg')}(\La',0),
M^{(\vdeg)}(\La,0))$ is nontrivial only if $m^{w'\La'}_\La \neq 0$.
\par\smallskip

\noindent {\it Proof:} \hfil\break
(i) Follows from Theorems \thPBce\ (i) and \thPBda\ (iii).\hfil\break
(ii) Follows from (i), Lemma \thPBaea\
(or Theorem \thPBce\ (ii)) and the filtration \eqPBbo.\hfil\break
(iii) Follows from (ii), Lemma \thPBbd\ (recall that $\al_0=0$ for $c=2$)
and the fact that the
image of of $v_{\vdeg' -1} \in M^{(\vdeg')}(\La',0)$ is a (nontrivial)
p-singular vector in $M^{(\vdeg)}(\La,0)$.\Box\hfil\smallskip

\noindent {\it Remark:}
Obviously, for $\La'\in P_+,\, \La\in P$, we have $m^{\La'}_{\La}\neq0$
only if $\La' - \La \in Q_+$.  For $c=2$ it therefore makes sense to
extend the action $W$ on $\bfh^*$ to $\widehat W$ by defining
\eqn\eqPBEa{
t_\al \La ~=~ \La + \al \,,\quad \al\in Q\,,\La\in \bfh^*\,,
}
where we have used that $\widehat W \cong W \ltimes T$, \ie, every
$\widehat w\in\widehat W$ can be (uniquely) decomposed as $\widehat w
= w t_{\al}$ for some $w\in W,\, \al\in Q$.  Using this affine Weyl
group action, Theorem \thPBdc\ (ii) can now be formulated as the
statement that the weights
of primitive vectors in a generalized Verma module $M^{(\vdeg)}(\La,0)$
are on the orbit of $\La$ under $\widehat W$.

\subsubsec{Explicit examples}
\subsubseclab\Wctwob

Let us introduce some more notation.  For any set of vectors $S=\{v_1,v_2,
\ldots\}\subset
M(\La,\al_0)$ we denote by $M(S) = M(v_1,v_2,\ldots)$ the submodule
of $M(\La,\al_0)$ generated by $\{v_1,v_2,\ldots\}$.
Further, in the remainder of this chapter
we generically use the symbol $w$ for a primitive vector which is
not p-singular, $v$ for a p-singular vector which is not singular
and $u$ for a singular vector.
By Theorem \thPBdc, the weights of the primitive vectors
in $M(\La,0)$ are concentrated on the orbit of $\La$ under
the coset $\widehat{W}/W$, so we find it convenient to
label primitive vectors by the Dynkin labels of the corresponding weight,
\ie, we
use the notation $w_{s_1 s_2}$ for a primitive vector of weight
$(h(\La),w(\La))$ where $\La = s_1 \La_1 + s_2 \La_2$.  This notation
is adopted for the $u$ and $v$ vectors also.
Moreover, we label $c=2$ (generalized) Verma modules by the
Dynkin indices of their highest weight (inside square brackets),
\ie, we use the notation
$M[s_1,s_2]$ for $M(s_1\La_1+s_2\La_2,0)$ etc.\ ($\al_0=0$ is
implicitly understood in this notation).\smallskip

\noindent {\it Example:}
Let us discuss in more detail the example of Section \WVer, \ie, we
consider the Verma module $M[0,0]$.  Its highest weight vector is,
conforming to the conventions above, denoted by $u_{00}$.  We have
already seen that $M[0,0]$, at $L_0$-level $1$, consists of a
two-dimensional Jordan block under $W_0$ corresponding to weight
$(h=1, w=0)$ ($\La = \La_1 + \La_2$); \ie, at this weight there is
a singular vector $u_{11}$ and a p-singular vector $v_{11}$.  Note
that since $m^{\La_1+\La_2}_0=2$ this is consistent with Theorem
\thPBdc\ (ii).  Next, at energy level $h=3$ we find two singular
vectors $u_{30}$ and $u_{03}$ in accordance with $m^{3\La_1}_0 =
m^{3\La_2}_0 = 1$.  \smallskip

At energy level $h=4$ something interesting happens.  Explicit
computation shows that there are only two p-singular vectors, while on
the other hand $m^{2\La_1 + 2\La_2}_0 = 3$.  The resolution of this
paradox is that besides the singular and p-singular vectors $u_{22}$
and $v_{22}$, respectively, there is also a primitive vector $w_{22}$.
In fact, the generalized eigenspace corresponding to $\La = 2\La_1 +
2\La_2$ (\ie, $h=4,\, w=0$) has dimension four and decomposes into
$3+1$ dimensional Jordan blocks.  The remaining vector, \ie, the vector
in the $1$-dimensional block, is in the irreducible module.  As far as
the content of the submodules generated by these primitive vectors is
concerned, explicit computation shows that $u_{30}, u_{03}, v_{22} \in
M(u_{11})$, $w_{22} \in M(v_{11})$, $u_{22} \in M(u_{30})$, $u_{22}
\in M(u_{03})$, but $w_{22} \not\in M(u_{11})$ and $v_{22} \not\in
M(u_{30},u_{03})$.  Combining the fact that $w_{22}\in M(v_{11})$ but
$w_{22}\not\in M(u_{11})$, with the fact that $w_{22}$ is primitive,
leads, in particular, to the conclusion that $\cWthp\cdot w_{22}
\subset M(u_{11})$; \ie, $w_{22}$ becomes singular in the quotient
module $M[0,0]/M(u_{11})$.  We have checked this by explicit
calculation as well. \smallskip

\nfig\fiPBaa{}
\nfig\fiPBab{}
\nfig\fiPBac{}

All of this information is summarized in Figure \fiPBaa.  The figure
contains all primitive vectors up to level 6 (the level increases
going down), the horizontal arrows between the primitive vectors
refer to the action of $W_0 -w$.  The cones built on a set of vectors
$S$ depict the module generated by $S$, \ie, $M(S)$.
\smallskip

\vbox{\epsfysize=250pt \epsfbox{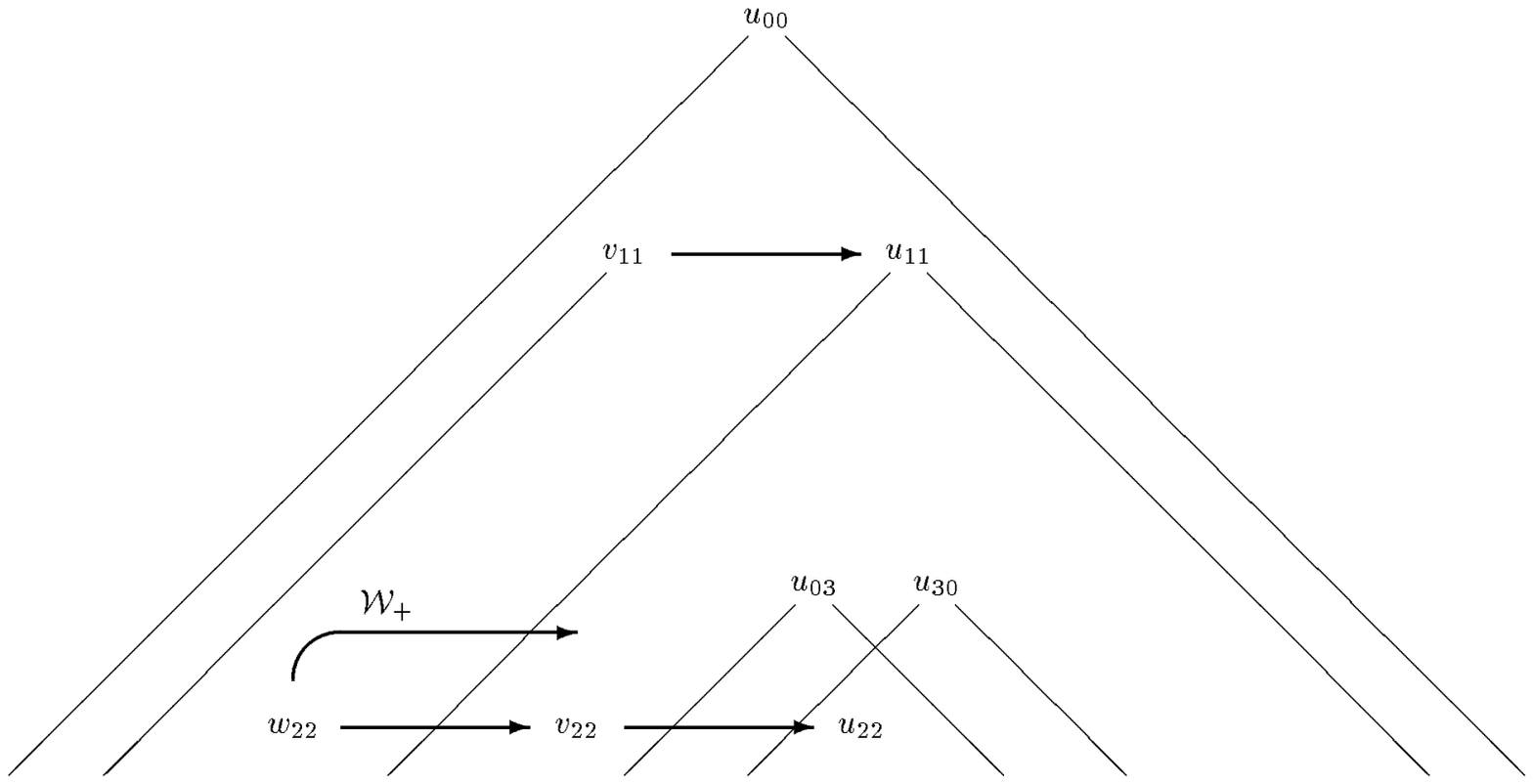}}\smallskip

\centerline{Figure~\fiPBaa. Embedding structure for $M[0,0]$}\medskip

One may deduce from the above that (a possible choice for) the
Jordan-H\"older series JH$_N(M[0,0]),\, N\leq3$ is given by
(see Remark (ii) after Theorem \thJHb)
\eqn\eqPBdh{ \matrix{
M & \supset & M(v_{11}) & \supset & M(u_{11}) &
 \supset & M(u_{30}, u_{03}) & \supset &
 M(u_{30})\,. \cr}
}
The quotients are isomorphic (up to $N=3$) with
$L[0,0], L[1,1], L[1,1], L[0,3]$ and $L[3,0]$, respectively.
For $N\geq4$, though, the quotient $M(v_{11})/M(u_{11})$ is no
longer irreducible due to the appearance of the primitive vector
$w_{22}$.  The following is, however, a viable
Jordan-H\"older series for $N\leq6$
\eqn\eqPBdg{ \comdiag{ \matrix{
M & \supset & M(v_{11}) & \supset & M(w_{22},u_{11}) & \supset &
 M(u_{11}) & \supset & M(u_{30}, u_{03}, v_{22}) & \supset &
 M(u_{30},v_{22}) \supset \cr
& \supset & M(v_{22}) & \supset & M(u_{22}) \,.\cr}
}}\smallskip

In Appendix \appPBA\ we have summarized some explicit computations
regarding the submodule structure of $c=2$ Verma modules.
In these tables we have labelled the Verma modules as well as
irreducible modules
by the the Dynkin indices of their highest weights, \eg,
$M[s_1,s_2]$, as before.  The triality of $\La$
is defined, as usual, by $(s_1+2s_2)\, {\rm mod\,}3$.\smallskip

Tables \tbPBaa--\tbPBad\ provide a list of primitive vectors
(arranged in Jordan blocks) for (generalized) Verma modules of
low lying highest weights and levels.%
\foot{For $h\geq9$, Tables \tbPBaa--\tbPBad, do not necessarily give the
entire Jordan blocks, \ie, it is possible that the Jordan
blocks contain additional non-primitive vectors.  Also, there often
exist additional Jordan blocks at the same weight $(h,w)$ as the ones
in the table, \eg, $M[0,0]$ has an additional $1$-dimensional
Jordan block at $(h,w)=(4,0)$ -- the corresponding vector $\tilde u_{22}$
is in $L[0,0]$.}
A prime on a primitive vector in $M^{(2)}[s_1,s_2]$ indicates
that this vector is in the kernel of the natural homomorphism
$M^{(2)}[s_1,s_2] \to M[s_1-1,s_2-1]$.
Tables \tbPBae\ and \tbPBaf\ list the
dimensions of the level $h$ subspaces of irreducible $c=2$ modules
and Tables \tbPBag--\tbPBai\ list the dimensions for some
submodules of, respectively, $M[0,0], M[1,0]$ and $M[1,1]$,
generated by primitive vectors.  All computations were done
with the help of Mathematica$^{\rm TM}$,%
\foot{We thank L.~Romans for supplying routines for working with conformal
fields at the level of modes.}
except for those in Tables
\tbPBae\ and \tbPBaf\ which follow from Theorem \thPBdaa, and some cases
for which the submodule is known to be isomorphic to a Verma module
(see the discussion in Section \WVer).\smallskip

With the help of the tables in Appendix \appPBA\ one can verify
that, for example, the quotients in the JH-series \eqPBdh\ and \eqPBdg\
are indeed irreducible up to the asserted level.  Additional examples,
like the one discussed above, can be worked out using the tables.
For illustrational purposes we give the embedding structures
of $M[1,0]$ and $M[1,1]$ below.\smallskip

\vbox{\epsfysize=250pt \epsfbox{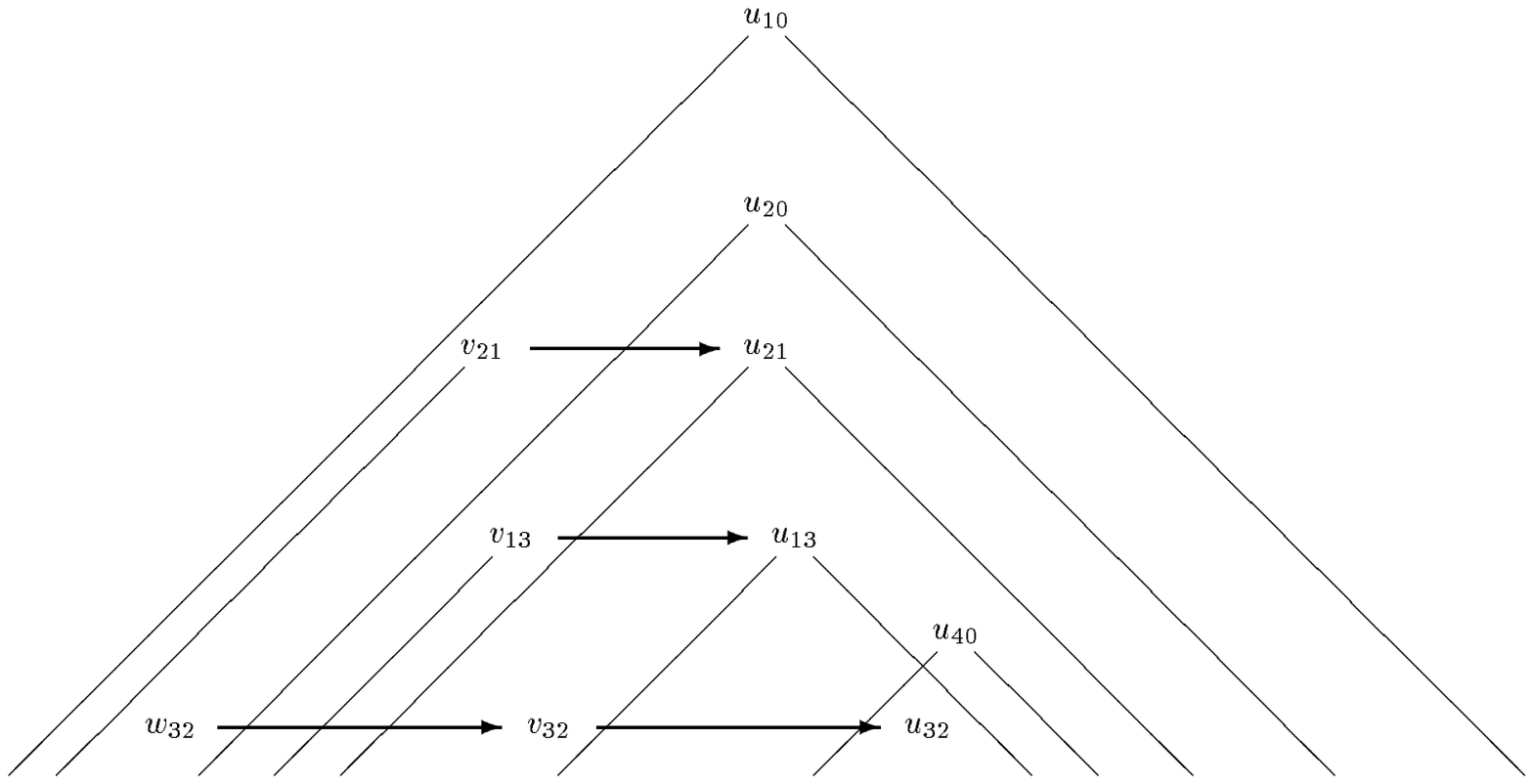}}\smallskip

\centerline{Figure~\fiPBab. Embedding structure for $M[1,0]$}\medskip

As another example of a JH-series, that can be read off from the tables,
we give $JH_N(M[1,1])$ valid for $N\leq8$
\eqn\eqPBdi{
\comdiag{ \matrix{
M[1,1] & \supset & M(u_{30},u_{03},v_{22}) &
 \supset & M(u_{30},u_{03},w_{33}) & \supset &
 M(u_{03},w_{33},v_{41}) & \supset & M(u_{22},v_{41},v_{14},w_{33}) \supset \cr
 & \supset & M(v_{41},v_{14},w_{33}) & \supset & M(v_{41},v_{14})
 & \supset & \ldots \,.\cr}
}}
\smallskip

We conclude this section with the following observation.
While for the Virasoro algebra all submodules of Verma modules are generated
by singular vectors (see, \eg, [\FF]), here we have

\thm\thPBdd
\proclaim Corollary \thPBdd.  Not every submodule of a $\cWth$ Verma
module is generated by p-singular vectors.\par\smallskip

\proof The submodule $M(w_{22},u_{11})$ of $M[0,0]$ in the example
above provides a counterexample. \Box\hfil\smallskip

The Corollary above is another manifestation that the $\cWth$ algebra
behaves, in many respects, as a rank $3$ Lie algebra (in fact as
$\hslth$), while the Virasoro algebra is a rank $2$ Lie algebra whose
submodule structure is considerably simpler [\FF].

\vbox{\epsfysize=250pt \epsfbox{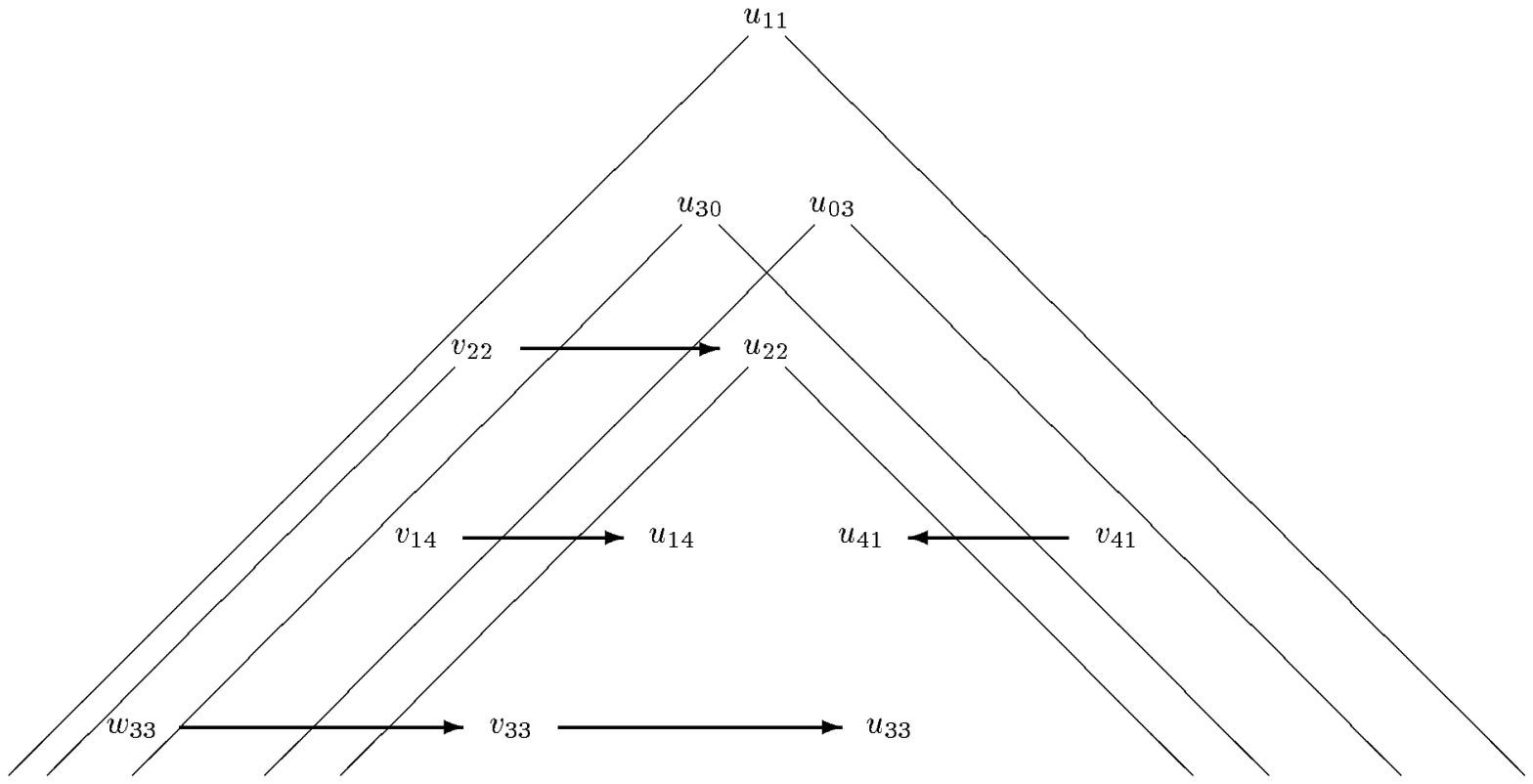}}\smallskip

\centerline{Figure~\fiPBac. Embedding structure for $M[1,1]$}\medskip

\subsec{Resolutions}
\subseclab\Wres

 An important construction in homological algebra is that of a
``resolution'' of a module.  Its utility lies in the fact that, through
a resolution of a module $V$, many computations involving the module
$V$ can be reduced to computations involving the modules in the
resolution of $V$, \eg, by means of spectral sequence techniques.  By
choosing the modules in the resolution to have certain simple
properties --
exactly which properties should be considered simple depends on the
problem under investigation -- the latter computations might become
tractable.  In the physical problem of the $D=4$ $\cW_3$ string we are
required to work with free fields, since these are the embedding
coordinates of the string into spacetime.  Thus, in the context of
this paper, the need for resolutions follows from the complicated
nature of the free field realization in Theorem \thPBca.  Given the
Fock space decomposition in Theorem \thPBda \ for $c=2$, it is enough
to understand resolutions of irreducible modules.

\thm\thPCa
\proclaim Definition \thPCa.  A resolution of a
$\cWth$ module $V\in \cO$ is a $\ZZ$-graded complex $(\cC,\de)$
of $\cWth$ modules
with a differential $\de$ of degree $1$, \ie, $\de \,:\, \cC^{(n)}
{}~\mapright{}~ \cC^{(n+1)},\ \de^2= 0$,
such that $H^{n}(\de,\cC) ~\cong~ \de_{n,0}\, V$.\par\smallskip

As an example of a resolution, consider%
\foot{More generally, Fock space resolutions for the $\cWth$ irreducible
modules with a completely degenerate highest weight (``minimal models'')
were constructed in [\FKW,\BMPstr] by applying the Quantum
Drinfel'd-Sokolov reduction to the
Fock space resolutions of admissible $\hslth$ modules.}

\thm\thJMaa
\proclaim Theorem \thJMaa.
There exists a resolution $\cC^{(n)}$ for the $c=2$ irreducible module
$L(\La,0),\, \La\in P_+$, in terms of Fock spaces.  Here
\eqn\eqJMea{
\cC^{(n)}\ \cong\ \bigoplus_{ \{w\in W\,|\, \ell(w)=n\} }
\ F(w(\La+\rh) - \rh,0)\,.
}
\par\smallskip

\noindent {\it Proof:} Follows directly from the Fock space
decomposition \eqPBda. \Box\hfil\smallskip

Interestingly, the resolution is of finite length -- it contains a
finite number of Fock spaces, here labelled by the Weyl group of
$\slth$.  This result is clearly consistent with the character formula
\eqPBdaa.  In fact, with the differential constructed from screening
charges as discussed below the proof of Theorem \thPBda, the structure
of the resolution reflects those of the (dual to the) BGG resolutions
for irreducible finite dimensional $\slth$ modules.

\subsubsec{Verma module resolutions of $c=2$ irreducible $\cWth$ modules}
\subsubseclab\WVerres

Verma modules, and also generalized Verma modules, have the ``simple
property'' that they are, in a sense, free over $\cWthm$.  This is
the main reason that Verma
module resolutions (also called BGG resolutions) are important
homological constructions.  As we have already seen in Section \Wctwob,
resolutions of $c=2$ irreducible $\cWth$ modules in terms of Verma modules
will in general not exist, \eg, the kernel of the canonical
projection $M[0,0] \to L[0,0]$ is isomorphic with the image
of a {\it generalized} Verma module, namely $M^{(2)}[1,1]$, in
$M[0,0]$.
However, in this section we present, for any given $c=2$ irreducible
$\cWth$ module $L(\La,0),\, \La\in P_+$, the
construction of a resolution, to be denoted by $(\cM(\La,0),\de)$,
in terms of generalized Verma modules, \ie, a resolutions
where each of the terms $\cM^{(n)}(\La,0)$ is the direct sum of a (finite)
number of generalized Verma modules of $\cWth$.  By construction
we have $\cM^{(n)}(\La,0) = 0$ for $n>0$.  It turns out
that, in fact, $\cM^{(n)}(\La,0) = 0$ for $n$ sufficiently negative,
namely $n<-4$, as well, so that the resolutions are of ``finite length.''
It should be remarked that the fact that such resolutions exist in
the first place is rather remarkable, since, as we have seen in
Corollary \thPBdd, not every submodule of a $\cWth$ Verma module is
generated by p-singular vectors.\smallskip

Let us now, assuming their existence,
try to construct such generalized Verma module resolutions, by combining
the various results of the previous sections. \smallskip

By Theorem \thPBdc\ (iii), nontrivial homomorphisms
$M^{(\vdeg')}(\La',0) \to M^{(\vdeg)}(\La,0)$ with $\La,\La'\in  P$
exist only
if $m^{w'\La'}_\La \neq 0$, where $w'\in W$ is such that $w'\La'\in P_+$.
Using the redundancy in parametrization by $\La$
(Lemma \thPBbd\ with $\al_0=0$), it follows that,
in order to build resolutions of an irreducible module
$L(\La,0)$, it suffices for the various terms $\cM^{(n)}(\La,0)$
to consider sums of generalized Verma
modules $M^{(\vdeg)}(\La',0)$, with $\La'\in P_+$ such that
$m^{\La'}_{\La} \neq 0 $, only.

Furthermore, since
\eqn\eqPBRc{
w^{-1} ( w (\La + \rh) - \rh) = \La + \rh - w^{-1}\rh\,,
}
one might think that, in analogy with Theorem \thJMaa, only
(generalized) Verma modules with highest weights $\La' = \La +
\rh - w^{-1}\rh,\,
w\in W$ -- corresponding to translations $t_{\rh - w^{-1}\rh}$
in \eqPBEa\ -- will enter the resolution. This turns out to be false.
In addition, as we will see later, weights corresponding to the
translation $t_{\rh}$ will arise.%
\foot{Unfortunately, we have no intrinsic understanding why
exactly this particular subset of $T\equiv \{ t_\al\,|\, \al\in Q\}$
occurs in the generalized Verma module resolutions at $c=2$.}
\smallskip

As will become clear in Section \SSHfirst\ it is useful to
introduce an extension $\widetilde W$
of the Weyl group $W$ of $\slth$, by $\widetilde W \equiv
W \cup \{\si_1,\si_2\}$
and extend the length function on $W$ to $\widetilde W$ by
assigning $\ell(\si_1)=1$ and $\ell(\si_2)=2$.
Similarly, we can extend the ``twisted length'' $\ell_w(\si)
\equiv \ell(w^{-1}\si) - \ell(w^{-1}),\, w,\si\in W$ to $\si\in
\widetilde W$ by defining the multiplications
\eqn\eqPBRa{
w\si_i  ~=~ \si_i\,,\qquad i=1,2\,,\quad w\in W\,.
}
Furthermore, $\widetilde W$ acts on $\bfh^*$ by $\si_i \la =0,\,
i=1,2$.  Note that this action is consistent with the multiplications
\eqPBRa.
Then, motivated by \eqPBRc, we define the ``circle action''
of $\widetilde W$ on $\bfh^*$ by
\eqn\eqPBRd{
\si \circ \La ~=~ \La + \rh - \si \rh \,,\qquad \si\in\widetilde W\,.
}

To denote the weights in the resolution we will
use both the notation $\si\circ\La$ as well as
their Dynkin labels.  Below we provide a translation
table for quick reference.\bigskip

\tbl\tbBPa
\begintable
$\quad\si\quad$ | $\si\circ\La$ | $[\si\circ\La]$ | $\quad\ell(\si)\quad$ \cr
$1$     |    $\La$                       | $[s_1,s_2]$     | $0$ \nr
$r_1$   |  $\La+\al_1$                | $\quad [s_1+2,s_2-1]\quad$ | $1$ \nr
$r_2$   |  $\La+\al_2$                   | $[s_1-1,s_2+2]$ | $1$ \nr
$\si_1$ | $\La+\al_1+\al_2$              | $[s_1+1,s_2+1]$ | $1$ \nr
$r_{12}$|  $\La+2\al_1+\al_2$            | $[s_1+3,s_2]$   | $2$ \nr
$r_{21}$|  $\La+\al_1+2\al_2$            | $[s_1,s_2+3]$   | $2$ \nr
$\si_2$ |  $\La+\al_1+\al_2$             | $[s_1+1,s_2+1]$ | $2$ \nr
$r_3$   |  $\quad\La+2\al_1+2\al_2\quad$ | $[s_1+2,s_2+2]$ | $3$
\endtable
\smallskip
\centerline{Table \tbBPa.  The circle action of $\widetilde W$.}
\medskip

As we will see, not all $\si\in\widetilde W$ enter the
(generalized) Verma module resolution of $L(\La,0)$.%
\foot{This is, of course, intimately related to the fact that
$m^{\La}_{\La-\al_i}=0$ iff $(\La,\al_i)=0$.}
It proves useful to define a subset $W(\La) \subset \widetilde W$ for all
$\La\in P_+$ as follows
\eqn\eqPBRe{
W(\La) ~=~ \cases{
  \widetilde W  &  if $\La\in P_{++}\,,$ \cr
  \{ 1,r_i,\si_1,r_{12},r_{21},r_3 \} & if $(\La,\al_i)=0,\, \La\neq0\,,$\cr
  \{ 1,\si_1,r_{12},r_{21},r_3\} & if $\La=0\,.$\cr}
}

The resolutions $(\cM(\La,0), \de)$ of $L(\La,0)$
have the following structure:  Only generalized Verma modules
$M^{(\vdeg)}(\si\circ\La,0)$ with $\si\in W(\La)$ occur.
For any given $\si\in W(\La)$ a (generalized) Verma
module with either $\vdeg=1$ or $\vdeg=2$
and highest weight $\si\circ\La$ occurs as a direct
summand of $\cM^{(n)}(\La,0)$ for $n = -\ell(\si)$,
and, if $M^{(2)}(\si\circ\La,0)$ occurs as
a direct summand of $\cM^{(-\ell(\si))}(\La,0)$,
then $M(\si\circ\La,0)$ occurs as a direct
summand of $\cM^{(-\ell(\si)-1)}(\La,0)$ provided $\La\neq0$.
Otherwise, \ie, if $M(\si\circ\La,0)$ occurs as
a direct summand of $\cM^{(-\ell(\si))}(\La,0)$
and/or $\La=0$, then
generalized Verma modules with highest weight $\si\circ\La$
will not occur as a summand of $\cM^{(n)}(\La,0)$ for
$n\neq -\ell(\si)$.\smallskip

A more precise statement is contained in the following

\nfig\fiPBba{}
\nfig\fiPBbb{}
\nfig\fiPBbc{}

\thm\thJMab
\proclaim Conjecture \thJMab.
The resolution, $(\cM(\La,0),\de)$, of an
irreducible $\cWth$ module $L(\La,0),\, \La\in P_+$
is one of three types, depending on whether $\La\in P_{++}$,
$\La\in P_+\backslash P_{++}$ but $\La\neq0$, or $\La=0$.
The resolutions are depicted in Figures \fiPBba--\fiPBbc.
In these pictures, each $\cM^{(n)}$
($n$ decreases going downward) is the direct sum of the generalized
Verma modules on the same horizontal line, and the differentials
$\de^{(n)} \,:\, \cM^{(n)} \to \cM^{(n+1)}$ are given by the collection
of homomorphisms represented by the arrows.  The homomorphisms are fully
determined by the image of the lowest vector,
\ie, $v_{\vdeg -1}$,  in each highest weight Jordan block of
the generalized Verma modules $M^{(\vdeg)}(\La',0)$.  These are
given in equations \eqPCga--\eqPCh.\par\smallskip

\vfil\eject

\vbox{\epsfysize=310pt \epsfbox{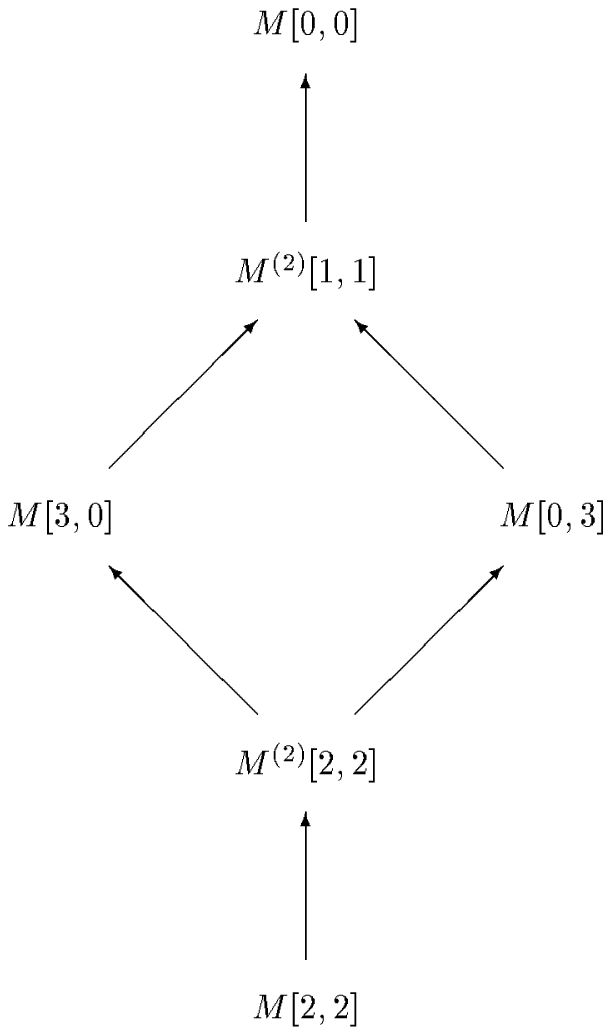}}\smallskip

\eqn\eqPCga{ \eqalign{
\de^{(-1)} (v_{1\,1}) & ~=~ v_{1\,1} \cr
\de^{(-2)} ((u_{3\,0},0)) & ~=~ u_{3\,0} \cr
\de^{(-2)} ((0,u_{0\,3})) & ~=~ u_{0\,3} \cr
\de^{(-3)} (v_{2\,2}) & ~=~ (u_{2\,2},0) - (0,u_{2\,2}) \cr
\de^{(-4)} (u_{2\,2}) & ~=~ u_{2\,2} \cr}
}\bigskip

\centerline{Figure~\fiPBba. Resolution of $L[0,0]$}\medskip

\vfil\eject

\vbox{\epsfysize=310pt \epsfbox{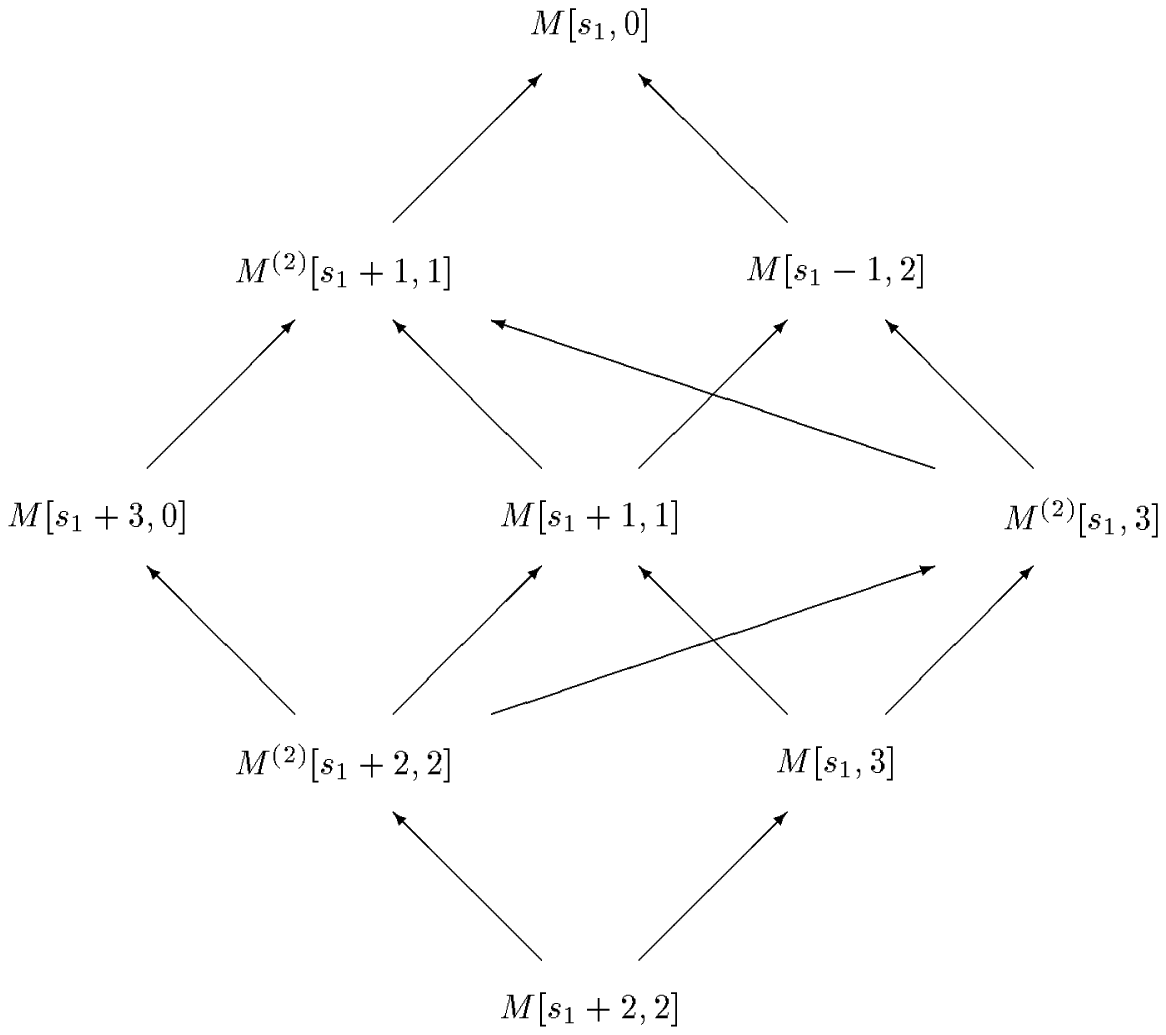}}\smallskip

\eqn\eqPCg{ \eqalign{
\de^{(-1)}((v_{s_1+1\, 1},0)) & ~=~ v_{s_1+1\, 1}\cr
\de^{(-1)}((0,u_{s_1-1\, 2})) & ~=~ u_{s_1-1\, 2}\cr
\de^{(-2)}((u_{s_1+3\, 0},0,0)) & ~=~  (u'_{s_1+3\, 0},0) \cr
\de^{(-2)}((0,u_{s_1+1\, 1},0)) & ~=~
  (u_{s_1+1\, 1},0) - (0,u_{s_1+1\, 1}) \cr
\de^{(-2)}((0,0,v_{s_1\, 3})) & ~=~
  - (v_{s_1\, 3},0) + (0,v_{s_1\, 3}) \cr
\de^{(-3)}((v_{s_1+2\, 2},0)) & ~=~
  \hbox{ $\textstyle{1\over12}$} (u_{s_1+2\, 2},0,0) +
  (0,v_{s_1+2\, 2},0) + (0,0,v_{s_1+2\, 2}) \cr
\de^{(-3)}((0,u_{s_1\, 3})) & ~=~ (0,u_{s_1\, 3},0) + (0,0,u_{s_1\, 3}) \cr
\de^{(-4)}(u_{s_1+2\, 2}) & ~=~
  - (u_{s_1+2\, 2},0) + (0,u_{s_1+2\, 2})\cr}
}\bigskip

\centerline{Figure~\fiPBbb. Resolution of $L[s_1,0],\ s_1>0$}\medskip

\vfil\eject

\vbox{\epsfysize=310pt \epsfbox{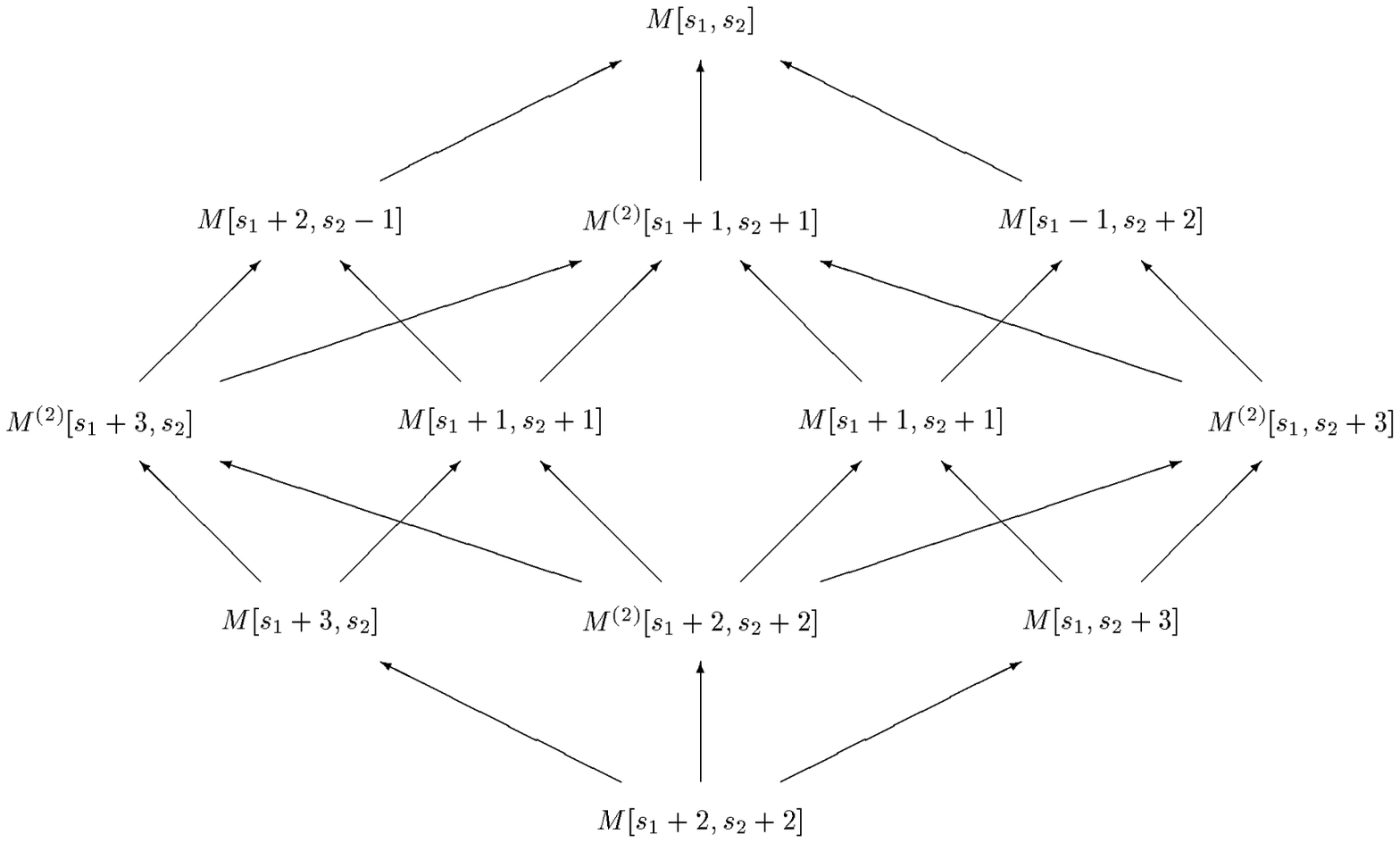}}\smallskip

\eqn\eqPCh{ \eqalign{
\de^{(-1)}((u_{s_1+2\, s_2-1},0,0)) & ~=~ u_{s_1+2\, s_2-1}\cr
\de^{(-1)}((0,v_{s_1+1\, s_2+1},0)) & ~=~ v_{s_1+1\, s_2+1}\cr
\de^{(-1)}((0,0,u_{s_1-1\, s_2+2})) & ~=~ u_{s_1-1\, s_2+2}\cr
\de^{(-2)}((v_{s_1+3\, s_2},0,0,0)) & ~=~
  -(v_{s_1+3\, s_2},0,0) + (0,v_{s_1+3 \,s_2},0) \cr
\de^{(-2)}((0,u_{s_1+1\, s_2+1},0,0)) & ~=~
  -(u_{s_1+1\, s_2+1},0,0) + (0,u_{s_1+1\, s_2+1},0) \cr
\de^{(-2)}((0,0,u_{s_1+1\, s_2+1},0)) & ~=~
   (0,u_{s_1+1\, s_2+1},0) - (0,0,u_{s_1+1\, s_2+1}) \cr
\de^{(-2)}((0,0,0,v_{s_1\, s_2+3})) & ~=~
  (0,v_{s_1\, s_2+3},0) - (0,0,v_{s_1\, s_2+3}) \cr
\de^{(-3)}((u_{s_1+3\, s_2},0,0)) & ~=~
  -(u_{s_1+3\, s_2},0,0,0) + (0,u_{s_1+3\, s_2},0,0)\cr
\de^{(-3)}((0,v_{s_1+2\, s_2+2},0)) & ~=~
  -(v_{s_1+2\, s_2+2},0,0,0) + (0,v_{s_1+2\, s_2+2},0,0)
  - (0,0,v_{s_1+2\, s_2+2},0)
  + (0,0,0,v_{s_1+2\, s_2+2}) \cr
\de^{(-3)}((0,0,u_{s_1\, s_2+3})) & ~=~
  (0,0,u_{s_1\, s_2+3},0) - (0,0,0,u_{s_1\, s_2+3}) \cr
\de^{(-4)}(u_{s_1+2\, s_2+2}) & ~=~
  -(u_{s_1+2\, s_2+2},0,0) + (0,u_{s_1+2\, s_2+2},0) +
  (0,0,u_{s_1+2\, s_2+2})\cr}
}\bigskip

\centerline{Figure~\fiPBbc. Resolution of $L[s_1,s_2],\ s_1,s_2>0$}\medskip

\vfil\eject

\noindent {\it The evidence for Conjecture \thJMab:} \hfil\break
We have explicitly carried out the program of constructing and
checking the resolution in four different cases; namely, for $L[0,0]$,
$L[1,0]$, $L[2,0]$, and $L[1,1]$.
This is done as follows:
First we examine the primitive vector structure of $M[s_1,s_2]$,
and of the (generalized) Verma modules with the same highest
weights as the primitive vectors, and so on.  This information is
given, down to a finite $L_0$-level, in Tables \tbPBaa--\tbPBad\ of
Appendix \appPBA\ (see also the discussion in Section \Wctwob).
At the first step in the resolution, we take $\cM^{(-1)}[s_1,s_2]$
in such a way that the image of $\cM^{(-1)}[s_1,s_2]$ in $M[s_1,s_2]$
is precisely the maximal ideal $I[s_1,s_2]$.  From, again, the
multiplicities in
Tables \tbPBaa\ and \tbPBab\ one now concludes that the previously
constructed homomorphism has a nontrivial kernel, \ie, that the
various summands of $\cM^{(-1)}[s_1,s_2]$ have some ``overlap'' in
$M[s_1,s_2]$.  This is taken care of by a proper choice of
$\cM^{(-2)}[s_1,s_2]$, and so on.  This reasoning by itself leads to
the ``minimal Ansatz'' for the resolutions as depicted in the figures.
Secondly, we fix the normalization of all homomorphisms constituting
the differential by imposing the condition that
$ \de^{(n+1)} \de^{(n)} = 0$ on the highest weight vectors.
The last step, the actual verification of the resolution, now comes
down to the explicit calculation of the dimensions of the images $\cI^{(n)}
\subset \cM^{(n)}$
of the homomorphisms $\de^{(n-1)}\,:\,\cM^{(n-1)} \to \cM^{(n)}$
at each step in the resolution.  Then we must prove, for all $n\leq0$,
that at each  $L_0$-level $h$,
\eqn\eqJMb{
{\rm dim\,} \cM^{(n)}_{(h)} ~=~ {\rm dim\,}\cI^{(n)}_{(h)} + {\rm
dim\,}\cI^{(n+1)}_{(h)}\,,  }
where we have defined, for convenience, $\cI^{(1)} \equiv L[s_1,s_2]$.
\smallskip

To compute the dimension of the image at a specific
$L_0$-level in a given module is straightforward in
principle: We
calculate the action of the standard basis vectors -- as given in Theorem
\thPBbae\ or \eqPBbm\ -- on the p-singular vectors of interest, the
level of each basis vector being chosen so that the result is an
vector in the given module at $L_0$-level $h$.  Then we calculate
the rank of the matrix of coefficients of these vectors in the
standard $L_0$-level $h$ basis of the given module.
The computations are done using Mathematica$^{\rm TM}$.
The results for $L[0,0]$, $L[1,0]$, $L[1,1]$ and $L[2,0]$
are displayed in Tables \tbPBba, \tbPBbb, \tbPBbc\ and
\tbJMbd, respectively.
Clearly, the data collected in the tables provide
a verification of the resolutions down to
$L_0$-level at which the last space is expected to appear.
{}From the explicit examples at low lying highest weight we have
extrapolated to the general result.
\smallskip

Some comments are in order.
Superficially, the resolutions for $\La\in P_+\backslash P_{++}$
look like subdiagrams of the generic resolution, \ie, for $\La\in P_{++}$.
There are however important differences.  While in the generic resolution
the various generalized Verma modules at steps $n$ and $n+2$ are
connected by $0$, $1$ or $2$ squares, and the $\de^2=0$ condition
works through cancelation within each square, the boundary case $L[s_1,0]$
is more subtle.  First, there is no square originating at $M[s_1+3,0]$,
which means that $u_{s_1+3\,0}$ has to map to the singular vector
$u'_{s_1+3\,0}$ in $M^{(2)}[s_1+1,1]$ that is in the kernel of the
homomorphism $M^{(2)}[s_1+1,1] \to M[s_1,0]$.  Secondly, there are
three possible paths from $M^{(2)}[s_1+2,2]$ to $M^{(2)}[s_1+1,1]$.
The third path is crucial since without it,
and with the normalizations (uniquely) fixed from
the other squares in the diagram, $\de^2$ would not be zero on
$M^{(2)}[s_1+2,2]$.  \smallskip

\noindent {\it Remarks:}
\item{i.} An independent consistency check on the conjectured
resolutions is the fact that the resulting character of $L(\La,0)$, as
obtained from the Lefschetz principle, coincides with that of
Theorem \thPBdaa.  In fact, our belief that the resolutions are of
finite length is to a large extent based on the character
formula \eqPBdaa.
\item{ii.} Another, {\it a posteriori}, consistency check
is provided by the resulting semi-infinite cohomology and
its underlying BV-structure computed in Sections \NSbrst\ and
\Sbvalgebra.  This BV-structure is sufficiently rigid that potential errors
in the resolution are likely to lead to inconsistencies at this
stage.
\item{iii.} The fact that there are three different types of resolutions
depending on the type of $\La$, is
presumably related to the existence of three possible posets
at $c=2$, which determine the Kazhdan-Lusztig
polynomial that encodes the multiplicities of the irreducible
modules in the composition series of a Verma module [\vDdVb].
It is quite probable that the resolutions within each case
can be related by invoking a ``shift principle'' a la Jantzen [\Jan].
\item{iv.} It is an interesting open problem to derive the resolutions
of Conjecture \thJMab\ from analogous resolutions of $\hslth$ modules
by means of the Quantum Drinfel'd-Sokolov reduction.\smallskip

This concludes our discussion of the structure theory of $\cWth$ modules.
In the next section we discuss how to apply the above results in
the computation of the semi-infinite cohomology  of the $\cWth$ algebra.

\vfill\eject

%
%
\secno2
\newsec{BRST COHOMOLOGY OF THE $4D$ $\cW_3$ STRING}
\seclab\NSbrst

\subsec{Complexes of  semi-infinite cohomology of the $\cW_3$ algebra}
\subseclab\Scomplex

The notion of semi-infinite cohomology of the $\cW_3$ algebra with
values in a positive energy module was first introduced in [\TM]. In
this section we briefly summarize an extension of this construction to
the category of tensor products of {\it two} positive energy modules
[\BLNW].

\subsubsec{The $\cW_3$ ghost system}
\subsubseclab\SSghosts

The first step in the construction of a complex for the semi-infinite
cohomology of the $\cW_3$ algebra is the same as that for the case of
the Virasoro or affine Lie algebras (see, \eg,
[\Feigin,\FGZ]). Corresponding to the currents $T(z)$ and $W(z)$, we
introduce
two anticommuting
$bc$-ghost systems $(b^{[j]},c^{[j]})$, with $j=2$ and $j=3$,
respectively. The nonvanishing OPEs of the ghost fields are
\eqn\ghostope{
c^{[j]}(z) b^{[j']}(w)~\sim~ {\de^{j j'}\over z-w}\,,\quad b^{[j]}(z)
c^{[j']}(w)~\sim~ {\de^{j j'}\over z-w}\,,} so that the mode operators,
$c_n^{[j]}$ and $b_n^{[j]}$, defined by the expansions,
\eqn\modes{
c^{[j]}(z)~=~\sum_{n\in\ZZ} c_n^{[j]} z^{-n+j-1}\,,\quad
b^{[j]}(z)~=~\sum_{n\in\ZZ} b_n^{[j]} z^{-n-j}\,,}
satisfy the anticommutation relations of a Clifford algebra:
\eqn\comrel{
[c_m^{[j]},c_n^{[j']}]~=~0\,,\quad [b_m^{[j]},b_n^{[j']}]=0\,,\quad
[c_m^{[j]},b_n^{[j']}]~=~\de^{j j'}\de_{m+n,0}\,,\quad m,n\in\ZZ\,.}
The dimensions of the fields $b^{[j]}(z)$ and $c^{[j]}(z)$ are equal
to $j$ and $-j+1$, respectively, and follow from the stress-energy
tensor
\eqn\strenergy{
T^{gh\,[j]}(z)~=~
-(j-1)(\partial b^{[j]}c^{[j]})(z)
-j(b^{[j]}\partial c^{[j]})(z)\,, \quad j=2,3\,.}
\smallskip

Let $F^{gh}$ denote the ghost Fock space defined as the standard
positive energy module of the Clifford algebra \comrel. It is freely
generated by $c_{-n}^{[j]}$, $n\geq 0$, and $b_{-n}^{[j]}$, $n>0$, from the
``physical'' ghost vacuum $|0\rangle_{gh}$, satisfying
\eqn\vaccum{
c_n^{[j]} |0\rangle_{gh}~=~0\,,\quad n\geq 1\,,\qquad
b_n^{[j]}|0\rangle_{gh}~=~0\,,\quad n\geq 0\,; \quad j=2,3\,.}
A standard basis in $F^{gh}$ consists of the elements
\eqn\ghbasis{
g_{k_1\ldots k_K;\ell_1\ldots \ell_L;m_1\ldots m_M;n_1\ldots n_N}~=~
c_{-k_1}^{[2]}\ldots b_{-\ell_1}^{[2]}\ldots c_{-m_1}^{[3]} \ldots
b_{-n_1}^{[3]}\ldots |0\rangle_{gh}\,,}
where $k_1>\ldots>k_K\geq 0$, etc.
Exactly as discussed in Section \WFock, there is an isomorphism
between the states in \ghbasis\ and the chiral algebra $\fV^{gh}$ of
fields obtained by a finite number of normal products of a finite
number of derivatives of the basic fields $(b^{[j]},c^{[j]})$: for any
state $|\cO\rangle \in F^{gh}$, there is a corresponding field
$\cO(z)\in\fV^{gh}$ such that $|\cO\rangle=\lim_{z\rightarrow 0}
\cO(z)|0\rangle$, where $|0\rangle$ is the $\frak{sl}(2,\CC)$ invariant
vacuum (for the ghost system, $|0\rangle =
b_{-1}^{[2]}b_{-1}^{[3]}b_{-2}^{[3]} |0\rangle_{gh}$).  Both
$\fV^{gh}$ and $F^{gh}$ are graded by the ghost number ${\rm
gh}(\cdot)$, with the usual assignment ${\rm gh}(c^{[j]})=1$ and ${\rm
gh}(b^{[j]})=-1$, and normalized such that the ghost number of the
identity operator, \ie, the $\frak{sl}(2,\CC)$ vacuum, is equal to
zero.  In this normalization the ghost number of the physical ghost
vacuum, $|0\rangle_{gh}$, which corresponds to the operator
$c^{[2]}\partial c^{[3]}c^{[3]}(z)$, is equal to three.
\smallskip

It is convenient to also define the Fock space $\overline{F}^{gh}$,
precisely as $F^{gh}$ but now with a vacuum $| \bar{0}
\rangle_{gh}$, satisfying
\eqn\vaccumtwo{
c_n^{[j]} |\bar{0}\rangle_{gh} ~=~ 0 \,,\quad n\geq0\,,\qquad
b_n^{[j]} |\bar{0}\rangle_{gh} ~=~ 0 \,,\quad n\geq1\,;\quad
j=2,3\,.}
Clearly, we have a $\fV^{gh}$
isomorphism $\overline{F}^{gh} \cong F^{gh}$ by identifying
$|\bar{0}\rangle_{gh} = c_0^{[2]} c_0^{[3]} |0\rangle_{gh}$.
This isomorphism preserves ghost number if we assign ghost number
five to $|\bar{0}\rangle_{gh}$.
We may now introduce the linear anti-involution $\om_{gh}$ of $\fV^{gh}$
defined by
\eqn\antiinv{
\om_{gh} (c_n^{[j]}) ~=~ c_{-n}^{[j]}\,,\qquad
\om_{gh} (b_n^{[j]}) ~=~ b_{-n}^{[j]}\,.
}
Similarly to the discussion of Fock spaces in  Section \WFock, we have
\thm\ghform
\proclaim Theorem \ghform. There exists a unique bilinear form
$\vev{-|-}_{gh}\,:\,\overline{F}^{gh} \,\times~\,F^{gh} ~\mapright{}~
\CC$, contravariant with respect to $\om_{gh}$, such that
$\vev{\bar{0} | 0}_{gh} = 1$.  This form is non-degenerate on
$\overline{F}^{gh, 8-n} \,\times F^{gh,n}$.\par

\subsubsec{ The BRST current and the differential}
\subsubseclab\SSbrst

\thm\existbrst
\proclaim Theorem \existbrst\ [\BLNW]. Let $V^M$ and
$V^L$ be two arbitrary positive energy modules of the $\cW_3$
algebra. Consider the  current
\eqn\brstcurr{\eqalign{
J(z)~=~ &\ c^{[3]}(\hbox{$1\over\sqrt{\be^M}$} W^M -
  \hbox{$i\over\sqrt{\be^L}$} W^L) + c^{[2]}(T^M+T^L+\half T^{gh\,[2]} +
  T^{gh\,[3]}) \cr
& + (T^M-T^L) b^{[2]} c^{[3]}\partial c^{[3]}
  -\mu \, b^{[2]}\partial c^{[3]}
  \partial^2 c^{[3]}  +\hbox{$2\over 3$} \mu\, b^{[2]} c^{[3]}\partial^3
  c^{[3]} + \hbox{$3\over 2$} \partial^2 c^{[2]}\,,\cr}
}
where $\mu=(1-17\be^M)/(10\be^M)$ and $\be^{M,L}=
16/(22+5 c^{M,L})$ (see Section \Wth). Then the operator
\eqn\brstdd{
d~=~\oint {dz\over 2\pi i}\ J(z)\,,
}
acting on $V^M\otimes V^L\otimes F^{gh}$ satisfies $d^2=0$ if and only if
$c^M+c^L=100$.
\smallskip

The current \brstcurr\ is  a natural generalization of the BRST current
constructed in [\TM]. In particular,
the leading terms
\eqn\dlead{
J(z) ~=~ c^{[3]}\left (\hbox{$1\over\sqrt{\be^M}$} W^M -
\hbox{$i\over\sqrt{\be^L}$} W^L\right) + c^{[2]}
\left(T^M+T^L\right)+\ldots \,,}
have the form one would expect  if $\cW_3$ were a Lie algebra acting on
the tensor product of two modules. It has been shown in [\BLNW] that
the completion of \dlead\ by the higher order terms in
\brstcurr\ is unique, up to a total derivative, if one requires that
the corresponding charge $d$ is a differential of ghost number one,
\ie, $d^2=0$.  Thus the following definition is quite natural.

\thm\defsemcoh
\proclaim Definition \defsemcoh. Let $V^M$ and $V^L$ be positive energy
modules of the $\cW_3$ algebra with $c^M+c^L=100$. Then the complex
$(V^M\otimes V^L\otimes F^{gh},d)$ graded by the ghost number (degree),
and with the differential $d$ of ghost number  one, is the complex of
semi-infinite (BRST) cohomology of the $\cW_3$ algebra with values in
the tensor product $V^M\otimes V^L$.  The corresponding cohomology
will be denoted by $H(\cW_3,V^M\otimes V^L)$ and called the
non-critical $\cW_3$ cohomology.
\smallskip

\noindent
{\it Remarks:}
\item{i.} When $V^L$ is the trivial $\cW_3$ module,
the above complex reduces to the original complex introduced in
[\TM].  We will call the corresponding cohomology (with values in a
single $\cW_3$ module) the critical $\cW_3$ cohomology.
\item{ii.} Alternative derivations of the BRST current \brstcurr\
 were given in [\BergSev,\DeBoer].
\smallskip

One should note that the existence of an extension of the complex from
the critical to the non-critical case is by no means obvious, because,
unlike for Lie algebras, the tensor product of two $\cWth$ modules
does not have a natural $\cW_3$ module structure.  A more conceptual
explanation of the result in Theorem \existbrst\ has been given in
[\BLNWnp,\DeBoer] and, more recently, [\Sevetal], where it is
argued that non-critical complexes may be constructed from a suitable
complex of semi-infinite cohomology of an affine Lie algebra, using
the fact that the $\cW_3$ algebra itself is a (Quantum
Drinfel'd-Sokolov) reduction of an affine Lie algebra (see, \eg,
[\BeOo,\Fig,\FeFr]).  It seems, however, that in the cases we want to
study explicitly in this paper the precise relation between the two
cohomologies is to a large extent conjectural -- by extrapolating the
results for the $\cW_2$ (Virasoro) string (see,
\eg, [\Sunny,\Sunnyb,\Sadov]) -- and thus  we will not pursue this point
of view further.
\smallskip

In the following we will also need the ``operator version'' of the
cohomology, in which the modules are replaced by chiral algebras of
operators.  More precisely, let $\fV$, $\fV^M$ and $\fV^L$ be chiral
algebras that decompose as $\cW_3$ modules into direct sums of
positive energy $\cW_3$ modules, with the central charges $c=100$ and
$c^M+c^L=100$, respectively.  Then we have ``operator valued''
complexes given by $\fC=\fV\otimes \fV^{gh}$ in the critical case and
$\fC=\fV^M\otimes\fV^L\otimes
\fV^{gh}$ in the non-critical case.
Let $\cO(z)$ be a field in the
chiral algebra $\fC$. The action of the differential $d$ is given by
the OPE with the BRST current $J(z)$, namely
\eqn\brstact{
(d\cO)(z)=\oint_{C_z} {dw\over {2\pi i}} J(w) \, \cO(z) \, ,} where
the contour $C_z$ surrounds the point $w=z$ counterclockwise.  It is
straightforward to verify that $(\fC,d)$ is a complex. We will denote
the corresponding ``operator valued'' cohomology by $H(\cW_3,\fC)$.
\smallskip

One can use the relation between an operator $\cO(z)$ and the
corresponding state $|\cO\rangle$ at the level of the whole complex
given   by the analogue of \eqJMstop. This allows one to pass from an
``operator valued'' to a ``state valued'' complex. In cases where
there is an equivalence between the state and operator formulations --
as discussed above for the ghost system or in Section \WFock\ for the
Fock spaces -- we will switch freely between the two depending on
which one is more convenient.  One should remember, however, that for
certain classes of modules, \eg, Verma modules, the operator valued
counterpart of the complex may not exist.
\smallskip

A natural problem is to understand the algebraic structure on the
cohomology space $H(\cW_3,\fC)$ that is induced from the underlying
chiral algebra $\fC$. It turns out that if $\fC$ is a VOA, then
$H(\cW_3,\fC)$ has the structure of a BV-algebra. We will discuss this in
detail in Section \Sbvalgebra.  First, however, we need to define more
precisely what is the cohomology problem we want to solve.

\subsec{The $\cW_3$ cohomology problem for $4D$ $\cWth$ string }
\subseclab\Swcohprob

The spectrum of physical states of $4D$ $\cWth$ gravity is
computed as non-critical $\cW_3$ cohomology with values in the tensor
product of two Fock modules, $F(\La^M,0)\otimes F(\La^L,2i)$,
\ie, the background charges of the matter and the Liouville Fock spaces
are $\al^M=0$ and $\al^L=2i$, with the corresponding central charges
$c^M=2$ and $c^L=98$.  In principle the matter and the Liouville
momenta, $\La^M$ and $\La^L$, are arbitrary.  However, for reasons
that will be explained shortly, we will assume in addition that
$(\La^M,-i\La^L)$ are restricted to lie on a lattice
$L\subset \bfh^*_\RR\times \bfh^*_\RR$ characterized by the
following properties:
\smallskip

\item{i.} $\la\in P$ for all $(\la,\mu)\in L$.
\item{ii.} $(\La_i,\La_i)\in L$, $i=1,2$.
\item{iii.} $L$ is an integral lattice (of signature $(2,2)$), \ie,
\eqn\prodinL{
\la\cdot\la'-\mu\cdot\mu'\in\ZZ\,,}
for all $(\la,\mu)$, $(\la',\mu')\in L$.
\smallskip

\thm\ontheL
\proclaim Lemma \ontheL. The maximal lattice $L$ satisfying (i)-(iii)
consists of weights $(\la,\mu)$ such that
\eqn\defofl{
\la\,,\mu\in P\,,\quad \la -\mu \in  Q\,.}
\smallskip

\proof Set $\la'=\mu'=\La_i$ in \prodinL. Then for all $(\la,\mu)\in
L$ we find
\eqn\weighofl{
\La_i\cdot(\la-\mu)\in\ZZ\,,\quad i=1,2\,,}
which,  together with (i), implies \defofl.
Conversely, given a lattice satisfying  \defofl, and thus (i) and
(ii),  we may use the identity
\eqn\profpa{
\la\cdot\la'-\mu\cdot\mu'=
(\la-\mu)\cdot\la' +\mu\cdot(\la'-\mu')\in\ZZ\,,}
to deduce (iii). \SMu
\smallskip

The choice of the lattice $L$ is partly motivated by the following result.
\thm\lattice
\proclaim Theorem \lattice. {Let $\fC$ be the chiral algebra
corresponding to
\eqn\sumcom{
\cC=\bigoplus_{(\La^M,-i\La^L)\in L}F(\La^M,0)\otimes
F(\La^L,2i)\otimes F^{gh}\,.}
Then  $(\fC,\cC)$ can be equipped with a structure of a  VOA.}
\smallskip

\proof  An  operator
$\cO(z)\in\fC$ is of the form
\eqn\genformofpsi{
\cO(z)=P[\partial\ph^{M,i},\partial\ph^{L,i},c^{[j]},b^{[j]},\ldots\,]\,
\cV_{\La^M,-i\La^L}(z)\,,}
 where $P[\,\ldots\,]$ is a polynomial in the fields
$\partial\ph^{M,i}$, $\partial\ph^{L,i}$, $i=1,2$; $c^{[j]}$,
$b^{[j]}$, $j=2,3$, and their derivatives, while
$\cV_{\La^M,-i\La^L}(z)=V_{\La^M,\La^L}(z)\, c_{\La^M,\La^L}$,
the vertex operator
corresponding to the vacuum state
\eqn\phvac{
|\La^M,\La^L\rangle=|\La^M,0\rangle\otimes|\La^L,2i\rangle \otimes
b_{-1}^{[2]}b_{-1}^{[3]}b_{-2}^{[3]} |0\rangle_{gh}\,,} is, up to a
phase-cocycle $c_{\La^M,\La^L}$, the normal ordered exponent
\eqn\optostate{
V_{\La^M,\La^L}(z)= e^{i\La^M\cdot\ph^M+ i\La^L\cdot\ph^L}(z)\,.}  The
conformal dimension of the operator $P[\,\ldots\,]$ will be called the
operator-level of the operator $\cO$.
\smallskip

The OPE of two operators in $\fC$ with the momenta
$(\La^M_A,\La^L_A)$ and $(\La^M_B,\La^L_B)$, respectively, is
schematically of the form
\eqn\opegen{
\cO_{\La^M_A,\La^L_A}(z)\,\cO_{\La^M_B,\La^L_B}(w)=
\sum_{n\in\ZZ} \, (z-w)^{h_{AB}+n}\,\cO^{(n)}_{\La^M_A+\La^M_B,
\La^L_A+\La^L_B}\,,}
where $h_{AB}=\La^M_A\cdot\La^M_B+\La^L_A\cdot\La^L_B$. Here the
common factor $(z-w)^{h_{AB}}$ comes from the contraction of
exponentials, the remaining contractions clearly only modify this by
integer powers of $(z-w)$.  By setting the momenta of the operators to
lie on the lattice $L$, we find, using (iii), that all OPEs are
meromorphic.
\smallskip

To prove that $(\fC,\cC)$ is in fact a VOA, we must still show that it
is possible to choose the phase-cocycles, $c_{\La^M,\La^L}$, such that
the analytic continuation of the right hand side\ in \opegen\ is consistent
with the graded commutativity of the OPE determined by the ghost
number of operators. The existence of  the required phase-cocycles is
proved in the lemma below. \Box
\smallskip

\thm\cocycless
\proclaim Lemma \cocycless.
Let $\xi:Q\rightarrow P$ be a linear map satisfying
\eqn\eqJMcoc{
\xi(\al)\cdot\al' - \xi(\al')\cdot\al = \al \cdot \al' \quad
{\rm mod\, 2}\, ,
}
then
\eqn\phcoc{
c_{\La^M,\La^L}~=~e^{i\pi\left(\xi(\La^M+i\La^L)-i
\La^L\right)\cdot (p^M+ip^L)} \et(\La^M + i\La^L)\,,
}
are the required phase-cocycles turning
$(\fC,\cC)$ into a VOA.

\noindent
{\it Remark:} As reviewed in Appendix \appPBC, the map $\xi$ defines,
through \eqCol, a phase-cocycle in the vertex operator construction of
$\hslth$.  A particular choice for $\xi$ is given in \eqCom.
\smallskip

\proof Following \eqCol, let us set
\eqn\anzcoc{c_{\La^M,\La^L}=e^{i\pi \xi^M\left((\La^M,\La^L)\right)\cdot
p^M+i\pi\xi^L\left((\La^M,\La^L)\right) \cdot p^L}\,.}
Then the linear map $(\xi^M,\xi^L): L \rightarrow P\times iP$ must satisfy
\eqn\eqsforxi{
\xi^M_A\cdot\La^M_B+\xi^L_A\cdot\La^L_B -
\xi^M_B\cdot\La^M_A+\xi^L_B\cdot\La^L_A =
\La^M_A\cdot\La^M_B+\La^L_A\cdot\La^L_B\quad {\rm mod\ }\, 2\,,}
where $\xi^M_A=\xi^M\left((\La^M_A,\La^L_A)\right)$, etc. This may be
rewritten as
\eqn\ohthose{\eqalign{
\xi^M_A\cdot(\La_B^M+i\La^L_B)& -(\La_A^M+i\La^L_A)\cdot\xi^M_B
-i(\xi^M_A+i\xi^L_A)\cdot\La_B^L+i\La^L_A\cdot(\xi^M_B+i\xi^L_B)\cr
& =(\La_A^M+i\La^L_A)\cdot (\La_B^M+i\La^L_B)
-i\La_A^L\cdot(\La_B^M+i\La^L_B)-i(\La_A^M+i\La^L_A)\cdot \La_B^L
\quad {\rm mod\ }\, 2\,,\cr}}
which is  solved by  $\xi^L=i\xi^M$ and
$\xi^M\left((\La^M,\La^L)\right)=\xi(\La^M+i\La^L)-i\La^L$,
as one verifies immediately using \defofl\ and \eqJMcoc. \Box
\smallskip

Let us comment on the conditions (i)-(iii) on the lattice $L$.  One
expects that the most interesting subsector of the cohomology should
arise for maximally degenerate matter Fock modules of the $\cW_3$
algebra. For, if the Fock module is degenerate just along one root
direction then the calculation will reduce to an analogue of the
Virasoro case, and if it is irreducible we will obtain at most the
vacuum state as non-trivial cohomology -- a result that follows from
reduction theorems below.  As discussed in Section
\Wctwo, at $c=2$ the maximally degenerate Fock modules have integral
weights, which explains (i).  Condition (iii), via Theorem \lattice,
allows us to study the cohomology as a BV-algebra and thus is equally
natural. The remaining condition can be justified only {\it a
posteriori}, as by explicit cohomology computation we will find that
the ground ring of the theory, \ie, the ghost number zero subalgebra
of the full cohomology, has generators with weights $(\La_i,i\La_i)$,
$i=1,2$, as required by (ii).  However, we should stress that the
cohomology problem is well defined for any weights, and that at this
point our choice merely selects what should be the most interesting
subsector both from the mathematical and the physical point of view.
\smallskip

To summarize, let us formulate the main mathematical problem in the
quantization of the $4D$ $\cW_3$ string.

\smallskip
\noindent {\bf Problem.}
{\sl For the VOA, $(\fC,\cC)$, given in
\sumcom, compute the semi-infinite cohomology $H(\cW_3,\fC)$ and
determine explicitly its  BV-algebra structure.}
\smallskip

In the following sections we present a (partially conjectural) solution
to this problem. Given the length of the analysis and its reliance on
technical results, it may be useful at this point to outline the main
steps.
\smallskip

The problem clearly splits into two parts: a computation of the
cohomology, $H(\cW_3,\fC)$, and a study of its global structure. The
two steps are of course related, as the BV-algebra  structure of
$H(\cW_3,\fC)$ provides quite a lot of information on the cohomology
itself.
\smallskip

An explicit computation of the cohomology requires a rather detailed
understanding of the action of the $\cW_3$ algebra on the complex. In
this respect the results of Section \WVerres\ are adequate for the subcomplex
in which the shifted Liouville momentum $-i\La^L+2\rh$ is in the
fundamental Weyl chamber. Let us denote this subcomplex by
$\fC_1$.  More generally, we denote the subcomplex with $-i\La^L + 2 \rh
\in w^{-1}\, P_+$ by $\fC_w$.
(Note that $\fC_1$ is not closed under the OPE!).
A series of technical results in Section
\SSpreliminaries\ allows then a straightforward computation  of
$H(\cW_3,\fC_1)$ in Section \SSfundamental. The general form of the
result suggests an extension to the arbitrary Weyl chamber. This is
discussed in Section \SSgeneral. Then the complete
BV-algebra is studied in the last part of the paper.

\subsec{Preliminary results}
\subseclab\SSpreliminaries

\subsubsec{A comment on the relative cohomology}
\subsubseclab\SSrelative

We begin with some results on the general structure of the $\cW_3$
cohomology, in both the critical and the non-critical cases. We will
use  $V$ to denote either a single positive energy $\cW_3$ module
or a tensor product of two such modules.
\smallskip

Consider the operators
\eqn\totops{
L^{tot}_n~=~[d,b_n^{[2]}]\,,\quad W^{tot}_n~=~[d,b_n^{[3]}]\,, \quad
n\in\ZZ\,.
}
Then the $L_n^{tot}=L_n+L_n^{[2]}+L_n^{[3]}$ define a positive
energy representation of the ``total'' Virasoro algebra on
$V\otimes F^{gh}$, with vanishing central charge and
diagonalizable $L_0^{tot}$. The eigenspaces of
$L_0^{tot}$ yield a decomposition of the complex into finite
dimensional subcomplexes. By the
usual argument (see, \eg, [\FGZ]), the nontrivial cohomology can arise only
in the subcomplex annihilated by $L_0^{tot}$.
However, the operators  $L_n^{tot}$ and $W_n^{tot}$ do  not generate  a
``total'' $\cW_3$ algebra,%
\foot{It has been shown in [\BLNWnp] that
$T^{tot}(z)$, $W^{tot}(z)$, together with $J(z)$, $b^{[2]}(z)$ and
$b^{[3]}(z)$, form a subset of generators of the topological $N=2$
$\cW_3$ super-algebra.}  as would have been the case if
$\cW_3$ were a Lie algebra. Moreover, following the discussion in
Section \PBW, $W_0^{tot}$ is in general non-diagonalizable on the
complex. As a consequence, nontrivial cohomology states need not be
annihilated by $W_0^{tot}$.

\thm\vanishw
\proclaim Lemma \vanishw. A nontrivial cohomology may arise only in the
subcomplex whose elements  $|\Ph\rangle$ satisfy
\eqn\virvanish{L_0^{tot}|\Ph\rangle=0\,,}
and
\eqn\vanishhhh{(W_0^{tot})^N|\Ph\rangle=0\,,}
for some $N>0$.
\smallskip

\proof The first condition  \virvanish\ follows by diagonalizing
$L_0^{tot}$ on the complex and is the same as in the case of the
Virasoro algebra [\FGZ].  Since the subcomplex corresponding to $\ker\,
L_0^{tot}$ is finite dimensional, it can be decomposed into a direct
sum of generalized eigenspaces of $W_0^{tot}$ that are preserved by
$d$, since $[d,W_0^{tot}]=0$.  Thus we may assume that
\eqn\decompp{
(W_0^{tot}-w)^N|\Ph\rangle=0\,,}
for some $w\in\CC$ and $N>0$. Together with \totops, this implies
\eqn\punchw{
w^N|\Ph\rangle=db^{[3]}_0
\big(\sum_{n=1}^N (-1)^{n+1}\left(\hbox{$N\atop n$}\right)
w^{N-n}(W_0^{tot})^{n-1}\big)| \Ph\rangle\,,} provided
$d|\Ph\rangle=0$. Thus $|\Ph\rangle$ is a trivial cohomology state
whenever $w\not=0$. \Box
\smallskip

\noindent
One can define the complex of relative $\cW_3$ cohomology
with respect to the  ``Cartan subalgebra'' $\cW_{3,0}$ as
the intersection
\eqn\relsubs{{\rm Ker}\,W_0^{tot}\cap {\rm
Ker}\,L_0^{tot}\cap{\rm Ker}\,b_0^{[2]} \cap{\rm Ker}\,b_0^{[3]}\subset
V\otimes F^{gh}\,,} with the differential $d$, which clearly preserves
this subspace.  The corresponding cohomology will be called relative.
However, unlike in cases where the Cartan algebra acts semi-simply on
the complex, this relative cohomology is not only difficult to compute
(\eg, in most nontrivial examples considered below it is practically
impossible to determine the relative subcomplex explicitly) but also
cumbersome to relate to the full cohomology.
\smallskip

It turns out, however, that the description of $H(\cW_3,\fC)$ may
nevertheless be simplified, as the explicit results below suggest that
the $\cW_3$ cohomology carries a (non-canonical) quartet structure,
first recognized in the critical case in [\Pope]. The lowest ghost
number members of the quartets have been called the ``prime states,''
and, for enumeration purposes, they play the role analogous to the
relative cohomology states.

\subsubsec{Reduction theorems}
\subsubseclab\SSreduction

\def\vdeg{\kappa}
\thm\redthm
\proclaim Theorem \redthm\ [\BMPa]. For an arbitrary generalized Verma module
$M^{(\vdeg)}(\La^M,\al_0^M)$ and a contragredient Verma module
$\cgM(\La^L,\al_0^L)$, $c^M+c^L=100$, the cohomology
$H(\cW_3,M^{(\vdeg)}(\La^M,\al_0^M)\otimes
\cgM(\La^L,\al_0^L))$ is nonvanishing if and only if
\eqn\genwec{
-i(\La^L+\al_0^L\rh)=w(\La^M+\al^M_0\rh)\,,} for some $w\in W$, in
which case it is spanned by the states $$ v_0 \, ,\quad c_0^{[2]}v_0
\, , \quad c_0^{[3]}v_{\vdeg-1}\, , \quad c_0^{[3]}c_0^{[2]}v_{\vdeg-1} \, ,
$$
where $v_i=v_i^M\otimes \overline
v^L\otimes |0\rangle_{gh}$, $i=0,\ldots,\vdeg-1$ (see Definition \thPBbi).
\smallskip

\def\ep{\varepsilon} \def\fdeg{\nu}
\proof Consider  linear functions $\fdeg$, called the $\fdeg$-degree, on
$M^{(\vdeg)}(\La^M,\al_0^M)$, $\cgM(\La^L,\al_0^L)$, $F^{gh}$ with
values in $\CC\,[\,\ep^{-1},\ep\,]$, which map standard basis
elements \eqPBbm, \eqPBCa\ and
\ghbasis\ into powers of an indeterminate $\ep$,
\eqn\defoff{
\eqalign{
{\fdeg}\,(e_{m_1\ldots m_M;n_1\ldots n_N}^{(i)})&=\ep^{-M-2N}\,,\quad
i=0,\ldots , \vdeg - 1\,,\cr
\fdeg(\overline e_{m_1\ldots m_M;n_1\ldots n_N}) &=\ep^{M+2N}\,,\cr
\fdeg(g_{k_1\ldots k_K;\ell_1\ldots \ell_L;m_1\ldots m_M;n_1\ldots n_N})
&=\ep^{L-K+2N-2M}\,.\cr}}
Extend $\fdeg$ multiplicatively to the tensor product
$\cC=M^{(\vdeg)}(\La^M,\al_0^M)\otimes \cgM(\La^M,\al_0^M)\otimes
F^{gh}$. Let $\cC_{(m)}$ denote the linear span of elements of
$\fdeg$-degree $\ep^m$, \ie, $\cC_{(m)}=\fdeg^{-1}(\ep^m)$.%
\foot{This decomposition  generalizes  the filtration of
Verma modules introduced in the proof of Theorem \thPBbae.}
Then an
operator $A$ on $\cC$ has the $\fdeg$-degree equal $n$ if
$A\,\cC_{(m)}\subset
\cC_{(m+n)}$. In such case we will simply say that $A$ acts like
$\ep^n$. For an arbitrary $A$, let $A_{\ep^n}$ be its component of
degree $n$. The action of the generators of the $\cW_3$ algebra on the
basis vectors \eqPBbm\ and those of the corresponding contragredient
basis \eqPBCa\ can be studied explicitly using the commutation
relations \eqPAaa.  It is then straightforward to determine the
degrees present in the decomposition of each generator acting on
$\cC$. In schematic notation, we find
\eqn\degrees{
\eqalign{
L^M_{-n}&\sim{1\over \ep}+1+{\ep} + \cdots\,,\quad n>0\,,\cr
L^M_{n}&\sim 1 + \ep +\ep^2 +\cdots\,,\quad n\geq
0\cr
W^M_{-n}&\sim {1\over \ep^2} +{1\over \ep}+1+\cdots\,,\quad n>0\,,\cr
W^M_{n}&\sim {1\over \ep}+ 1 + {\ep}+\cdots\,,\quad  n\geq 0\,,\cr}
\qquad\eqalign{
L^L_{-n}&\sim 1+\ep+\ep^2+\cdots\,,\quad n\geq 0\,,\cr
L^L_{n}&\sim {1\over \ep}+1+\ep+\cdots\,,\quad n>0\,,\cr
W^L_{-n}&\sim {1\over \ep}+1+\ep+\cdots\,,\quad n\geq 0\,,\cr
W^L_{n}&\sim {1\over \ep^2}+{1\over\ep}+1+\cdots\,,n>0\,.\cr}}
In particular, the lowest degree components of generators
$L_{-n}^M$, $L_n^L$, $W^M_{-n}$, and $W^L_{n}$, $n>0$, map a given
basis vector onto another one, which is obtained by increasing the
power of the corresponding generator in \eqPBbm\ or \eqPBCa. Thus those
operators commute. The $\nu$-degrees of the ghost and antighost mode
operators are
\eqn\ghostdeg{
c_n^{[2]}\sim \ep\,,\quad b_n^{[2]}\sim {1\over\ep}\,,
\quad c_n^{[3]}\sim \ep^2\,,\quad b_n^{[3]}\sim {1\over \ep^2}\,,
\quad n\in\ZZ \,.}
\smallskip

By expanding the differential \brstdd\ in terms mode operators, and
then using \degrees\ and \ghostdeg, we find that $d$ is a sum of
operators with nonnegative degrees. Thus, we have a spectral sequence
$(\cE_r,d_r)$, $r\geq 0$, induced from the filtration defined by the
$\fdeg$-degree.\foot{See, \eg, [\BMPkar] for a more extensive discussion of
this spectral sequence and its applications in the context of
cohomology of the Virasoro and affine Lie algebras.}  Recall that the
first term in this sequence is given by $\cE_0=\cC$, while the
differential is $d_0=(d)_{\ep^0}$. Then $\cE_{r+1}=H(d_r,\cE_r)$ and
$d_{r+1}$ is induced from $\sum_{i=0}^{r+1}(d)_{\ep^i}$. In the present
case
\eqn\dzero{
d_0=\sum_{n>0} c_n^{[2]}(L^M_{-n})_{1/\ep}+c_{-n}^{[2]}
(L^L_{n})_{1/\ep}+c_n^{[3]}(W^M_{-n})_{1/\ep^2}+c_{-n}^{[3]}
(W^L_{n})_{1/\ep^2}\,,}
\ie, $(\cE_0,d_0)$ is the Koszul complex of the abelian algebra generated by
the leading terms of the $\cW_3$ algebra generators.  By the standard
argument (see, \eg, [\Knapp]) there is a contracting homotopy for the
differential $\dzero$ and therefore its cohomology is
concentrated on the states of the form
\eqn\suppofeone{
\eqalign{
v_0\,, \quad c_0^{[2]}v_0\,, \quad & c_0^{[3]}v_0\,, \quad
c_0^{[2]}c_0^{[3]} v_0\,,\cr &\vdots\cr v_{\vdeg-1}\,, \quad
c_0^{[2]}v_{\vdeg-1}\,, \quad & c_0^{[3]}v_{\vdeg-1}\,, \quad
c_0^{[2]}c_0^{[3]} v_{\vdeg-1}\,.\cr}} Those states span $\cE_1$, on
which the differential $d_1$ is explicitly given by
\eqn\donediff{
d_1=c_0^{[2]}(L_0^M+L_0^L)+c^{[3]}_0(\hbox{$1\over \sqrt{\be^M}$}\,
W_0^M- \hbox{$i\over \sqrt{\be^L}$}\,W_0^{L})\,.}  By evaluating this
operator on
\suppofeone\ we find that its cohomology is nonvanishing if and only if
\eqn\condonwgts{
h(\La^M,\al_0^M)+h(\La^L,\al^L_0)=0\,\quad {\rm and} \quad
w(\La^M,\al^M_0)-iw(\La^L,\al^L_0)=0\,.}  It follows from Lemma \thPBbd\
 that the most general solution to
those conditions is given by the weights $\La^M$ and $\La^L$
satisfying $-i(\La^L+\al^M_0\rh)=w(\La^M+\al_0^L\rh)$ for some $w\in
W$. The nonvanishing cohomology, \ie, the $\cE_2$ term, is then spanned
by the states
\eqn\secterm{
v_0\,,\quad c^{[2]}_0v_0\,,\quad c^{[3]}_0 v_{\vdeg-1}\,,\quad
c^{[2]}_0 c^{[3]}_0 v_{\vdeg-1}\,.}  Since those states are
annihilated by $d$, we also have $d_2=d_3=\ldots= 0$, so that the spectral
sequence collapses at this term, $\cE_2=\cE_3=\ldots=\cE_\infty$, and
 \secterm\ yields the entire cohomology
$H(\cW_3,M^{(\vdeg)}(\La^M,\al_0^M)\otimes\cgM(\La^L,\al_0^L))$. \Box
\smallskip

\noindent {\it Remark:} Theorem \redthm\ is a generalization of a
similiar result for the semi-infinite cohomology of the Virasoro and
affine Lie algebras [\Feigin,\FGZ] (see also [\BMPkar]).
\smallskip

The spectral sequence argument in the proof above relies
on the existence of a filtration with respect to which the degrees of
all generators were bounded from below by some power of
$\ep$. Replacing one of the modules by an arbitrary
module with a suitable filtration gives the following vanishing
theorem.

\thm\moreredthm
\proclaim Theorem \moreredthm. Let $F$ be a $\cW_3$ module and
$\fdeg:F\rightarrow \CC\,[\,\la^{-1},\la\,]$ a $\fdeg$-degree on
$F$, such that the $\fdeg$-degrees of all $\cW_3$ generators are bounded
from below,
\ie,
\eqn\bounding{L_{n}=\sum_{k\geq k_0} (L_{n})_{\la^k}\,,\quad
W_{n}=\sum_{k\geq k_0} (W_{n})_{\la^k}\,,\quad n\in\ZZ\,,}
for some $k_0\in\ZZ$. Then
\eqn\vancohm{
H^n(\cW_3,M^{(\vdeg)}(\La^M,\al_0^M)\otimes F)=0\,, \quad {\rm for}\quad n
\leq 2\,,} and
\eqn\vancohcm{
H^n(\cW_3,F\otimes \cgM(\La^L,\al_0^L))=0\,, \quad {\rm for}\quad n
\geq 6\,.}
\smallskip

\proof In the first case consider  the $\fdeg$-degree on
$M^{(\vdeg)}(\La^M,\al_0^M)\otimes F^{gh}$ defined as in the proof of
Theorem \redthm, but with $\ep=\la^{|k_0|+1}$. Then the $\fdeg$-degree
extended to $M^{(\vdeg)}(\La^M,\al_0^M)_N\otimes F\otimes F^{gh}$
yields a spectral sequence $(\cE_r,d_r)$ with
\eqn\newdzero{
d_0=\sum_{m>0} c_m^{[2]}(L^M_{-m})_{1/\ep}
+c_m^{[3]}(W^M_{-m})_{1/\ep^2}\,.}  As before the cohomology of $d_0$
simply picks up the highest weight vectors $v^M_i$ in the Verma module,
\ie,
\eqn\neweone{
\cE_1\cong \bigoplus_{i=0}^{\vdeg-1}\,\CC \,
v^M_i\otimes F\otimes F^{gh}_>\,,}
where $F^{gh}_{>}$ is generated  by $c_{-m}^{[2]}$ and $c_{-m}^{[3]}$,
$m\geq 0$,
acting on $|0\rangle_{gh}$. Thus all states in  $\cE_1$ have ghost
numbers $3+n$, with $n\geq 0$, which implies \vancohm.\smallskip

The second part of the theorem is proved similarly, except that $\cE_1
\cong \CC\,\overline v^L\otimes F\otimes F^{gh}_\leq\,,$
where $F^{gh}_\leq$ is generated by $c_{0}^{[2]}$,  $c_{0}^{[3]}$,
and  $b_{-n}^{[2]}$, $b_{-n}^{[3]}$, $n> 0$, from $|0\rangle_{gh}$.
\Box
\smallskip

\thm\fockfiltr
\proclaim Lemma \fockfiltr.
Define an $\fdeg$-degree on a  Fock space,
$F(\La,\al)$, by setting,
\eqn\defofffock{
\fdeg(f_{m_1\ldots m_M;n_1\ldots n_N})=\la^{M+N}\,.}
Then the action of $\cW_3$ on
$F(\La,\al)$ is bounded as in \bounding\ with $k_0=-3$.

\smallskip\proof
Note that $\al_{-n}^i$, $\al_{n}^i$, and $\al^i_0$, $i=1,2$, $n>0$,
act as $\la$, $1/\la$ and $1$, respectively.  The lemma follows
by examining the explicit formula \eqPBcdd\ for the generators. \Box

\def\ep{\epsilon}

\subsubsec{The $\bga$ symmetry of $H(\cW_3,\fC)$ }
\subsubseclab\SSslth

The vertex operator realization of $\hslth$, reviewed in Appendix \appPBC,
can be extended to act on the complex $\fC$, by the currents
\eqn\slthonfc{
H^i(z)=i\partial \ph^i\,,\quad i=1,2\,;\qquad E^\al(z)=\cV_{\al,0}\,,\quad
\al\in\De\,.}  From the explicit form of \optostate\ and \phcoc\
we find that this realization acts on the matter degrees of freedom
only, but for the phase factor that depends on the Liouville momentum.
The corresponding $\slth$ generators commute with $d$, and
thus their action descends to the cohomology.
\smallskip

Additional symmetry operators that commute with $d$ are the Liouville
momenta, $-ip^{L,i}$, with the corresponding currents $\partial
\ph^{L,i}(z)$, $i=1,2$. They obviously commute with the
$\slth$ algebra as well. The resulting  $\bga$  symmetry of
$H(\cW_3,\fC)$ will greatly simplify the following dicussion.
\smallskip

The levels, $h$, of operators in $\fC$ at a given ghost number are bounded
from below, as the only operators with  nonpositive dimension are
$\p^nc^{[2]}$, $n=0,1$, and $\p^nc^{[3]}$, $n=0,1,2$. If in addition
we require that a given  operator be annihilated by $L_0^{tot}$, we have
\eqn\energlev{
h=\half |-i\La^L+2\rh|^2- \half |\La^M|^2 -4\,,\quad (\La^M,\La^L)\in
L\,.}
Thus for a fixed Liouville momentum, $\La^L$, and a ghost
number, $n$, but arbitrary matter weight, $\La^M$, there is a finite
dimensional subspace of operators in $\fC$ whose level satisfies
\energlev. This subspace is clearly closed under the action of $\bga$,
which immediately yields the following result.

\thm\slthac
\proclaim Theorem \slthac. The cohomology $H(\cW_3,\fC)$ decomposes
into a direct sum of finite dimensional modules of $\bga$.

\subsubsec{A bilinear form on $\fC$ and $H(\cW_3,\fC)$}
\subsubseclab\SSbilinear

By combining the linear anti-involutions on
$\cA^M$, $\cA^L$ and $\fV^{gh}$ we obtain a linear
anti-involution $\om = \om_{\cA^M} \otimes \om_{\cA^L} \otimes
\om_{gh}$ on $\fC$.  A straightforward calculation show that
the differential $d$ behaves naturally under this anti-involution,
namely
\eqn\conjdiff{
\om(d) ~=~ d\,.
}
Similarly,
let $\cC(\La^M,\La^L)$ denote the
complex $ F(\La^M,0) \otimes F(\La^L,2i) \otimes F^{gh}$, then
by combining \conjdiff\ with the results of Theorems \thPBcb\ and \ghform,
we immediately have the following result
\thm\lemform
\proclaim Theorem \lemform.
\item{i.}
There exists a unique bilinear form $\vev{-|-}_{\cC}\,:\,
\cC(\La^M,w_0\cdot\La^L)\,\times \, \cC(\La^M,\La^L) ~\mapright{}~
\CC$, contravariant with respect to $\om$, and such that
$\vev{ \La^M,w_0\cdot \La^L | \La^M,\La^L}_{\cC} = 1$. This form is
non-degenerate on $\cC^{8-n}(\La^M,w_0\cdot\La^L)\,\times \,
\cC^n(\La^M,\La^L)$.
\item{ii.}
The differential $d$ is symmetric with respect to the form $\vev{-|-}_{\cC}$.
\item{iii.} The form $\vev{-|-}_{\cC}$ induces a non-degenerate
bilinear form on $H^{8-n}(\cW_3, F(\La^M,0) \otimes F(w_0\cdot \La^L,2i))
\,\times\, H^n(\cW_3, F(\La^M,0) \otimes F(\La^L,2i))$.\par

As an immediate important consequence, we have
\thm\cohduality
\proclaim Corollary \cohduality.  There is  an isomorphism
\eqn\dualcoh{
H^n(\cW_3,F(\La^M,0)\otimes F(\La^L,2i))~\cong~
H^{8-n}(\cW_3,F(\La^M,0)\otimes F(w_0\cdot\La^L,2i))\,,}
for all $(\La^M,\La^L)\in L$ and $n\in\ZZ$, which extends to an
isomorphism of $H(\cW_3,\fC)$ as a $\bga$ module.\par

\smallskip
\noindent
{\it Remark:} We will refer to \dualcoh\ as the ``duality'' of the
cohomology.

\subsec{The cohomology in the ``fundamental Weyl chamber''}
\subseclab\SSfundamental

In this section we determine $H(\cW_3,\fC)$ in the fundamental Weyl
chamber, \ie, for the Liouville weights satisfying $-i\La^L+2\rh\in
P_+$.  This computation relies on several results derived earlier --
the isomorphism $F(\La^L,2i)\cong\cgM(\La^L,2i) $, that holds for
$-i\La^L+2\rh$ in the fundamental Weyl chamber (Corollary
\thPBcda), the reduction theorem for the $\cW_3$ cohomology with
values in a tensor product of a (generalized) Verma and a
contragredient Verma modules (Theorem \redthm), and explicit
resolutions of the irreducible $\cW_3$ modules (Conjecture \thJMab)
together with the decomposition of Fock modules at $c=2$ (Theorem
\thPBda).
\smallskip

The cohomology $H(\cW_3,F(\La^M,0)\otimes F(\La^L,2i))$ is then
obtained as follows: First using the decomposition theorem for the
matter Fock space (Theorem \thPBda ), and the isomorphism in the
Liouville sector, it is sufficient to compute the cohomology
$H(\cW_3,L(\La,0)\otimes \cgM(\La^L,2i))$, where $\La=\La^M+\be$,
$\be\in Q_+$. The latter cohomology can be studied
through a spectral sequence associated with the resolution of the
irreducible module $L(\La,0)$ in terms of generalized Verma modules
obtained in Section \WVerres. Using the reduction theorem it is then easy to
show that this spectral sequence collapses at most at the second term,
and to compute its limit explicitly. The main result for the cohomology
is given in Theorems \fockcoh\  and \cohfweyl, and in Appendix \primeplots.

\subsubsec{$ H(\cW_3,L(\La,0)\otimes F(\La^L,2i))$
             with $-i\La^L+2\rh\in P_+$}
\subsubseclab\SSHfirst

In Section \WVerres\  we have argued that for a given irreducible $\cW_3$
module $L(\La,0)$, $\La\in P_+$, there exists a resolution $(\cM,\de)$
of $L(\La,0)$ in terms of $c=2$ (generalized) Verma modules
of highest weight $(h(\si\circ\La), w(\si\circ\La))$, \hfil\break
where
$\si\circ\La\equiv\La+\rh-\si\rh$ and $\si$ runs over the set
$W(\La)\subset \widetilde W$ given in \eqPBRe\ (see also Table
\tbBPa).  Replacing $L(\La,0)$ with this resolution allows us to
calculate $H(\cW_3,L(\La,0)\otimes F(\La^L,2i))$ via relatively
standard techniques applied to the resulting double complex.  A
cursory inspection of the resolutions displayed in Figures
\fiPBba--\fiPBbc\ shows that there are only a few ways in which
(generalized) Verma modules with the same highest weights arise; in
particular, they are distinguished by how they are joined by the arrows
representing the nontrivial homomorphisms comprising the differential
$\de$. As this structure is important in the calculation, we will
first discuss these different possibilities explicitly, using this
opportunity to introduce notation with which the result may be
conveniently stated.
\smallskip

The first case, Case I, is that a given space is isolated, \ie, it is
not joined by arrows to a space with the same highest weights.  This
occurs in all the resolutions of $L(\La,0),\, \La\in P_+$, but there
are actually two subcases: in Case Ia the isolated space is a Verma
module, $M(\si\circ\La,0)$, which appears for $\si\in W(\La)\cap W$ at
step $-\ell(\si)$; in Case Ib it is a generalized Verma module
$M^{(2)}(\si\circ\La,0)$, which only appears for $\si = \si_1$ at step
$-\ell(\si_1)=-1$ in the resolution of $L(0,0)$.  The next case, Case
II, has exactly two spaces with the same highest weights joined by an
arrow.  This only occurs in the following way,
\eqn\jctw{
\matrix{M(\si\circ\La,0)& \longrightarrow & M^{(2)}(\si\circ\La,0)\cr} \, ,}
and is present in all resolutions.  For $\si\in W(\La)\cap W$ the
``top space,'' $M^{(2)}(\si\circ\La,0)$, appears at step $-\ell(\si)$,
while for $\si = \si_1$ it appears at step $-1$ in the resolutions of
$L(\La,0)$ for $\La \in P_+\backslash P_{++}$.  The last case, Case
III, has exactly three spaces with the same highest weights joined by
an arrow.  This only occurs as
\eqn\jcth{
\matrix{M(\si\circ\La,0)\oplus M(\si\circ\La,0)& \longrightarrow &
M^{(2)}(\si\circ\La,0)\cr} \, ,}
for $\si = \si_1$ in the resolution of $L(\La,0)$ for $\La \in P_{++}$
(the top space occurring at step $-1$).\smallskip

We may now state the result.
\thm\fundcoh
\proclaim Theorem \fundcoh. Let $-i\La^L+2\rh\in P_+$. Then
\smallskip
\item{i.} $H(\cW_3,L(\La,0)\otimes F(\La^L,2i))\not=0$ if and only if
\eqn\condforweg{
-i\La^L+2\rh=\La+\rh-\si\rh ~=~ \si\circ\La\,,}
\item{} for some $\si\in W(\La)$.
\item{ii.} For a given $\La$, $\La^L$ and $\si$ satisfying
\condforweg,
\eqn\primest{
\dim\,H^m(\cW_3,L(\La,0)\otimes F(\La^L,2i))~=~
d(m,3-\ell(\si))\,,}
where
\eqn\dimnsns{d(m,n)= \de_{m,n}+2\,\de_{m,n+1}+\de_{m,n+2}\,,}
\item{} \ie, each  $\si\in W(\La)$
gives rise to an independent ``quartet'' of cohomology states at ghost
numbers $n$, $n+1$, $n+1$ and $n+2$, respectively, where $n=3-\ell(\si)$.
\smallskip

\proof Consider the double complex
$(\cM\otimes F(\La^L,2i)\otimes F^{gh},d,\de)$, obtained by
``replacing'' the irreducible module $L(\La,0)$ with the corresponding
resolution.%
\foot{Once more the technique employed here is quite
standard, and the reader can consult [\BMPkar] for an elementary
exposition in a similar context of the semi-infinite cohomology of the
Virasoro algebra.}  Since $H^n(\de,\cM)\cong \de^{n,0}\,L(\La,0)$, the
first spectral sequence associated with this double complex (see, \eg,
[\BT]) collapses at the first term to yield
\eqn\fstspsq{\eqalign{E_\infty^{p,q}&~\cong~
H^p(\cW_3,H^q(\de,\cM)\otimes F(\La^L,2i))\cr
&~\cong~ \de^{q,0} H^p(\cW_3,L(\La,0)\otimes F(\La^L,2i))\,.\cr}}
The $E_2'$-term of the second spectral sequence is given by
\eqn\othsctrm{
E'{}_2^{p,q}~\cong~ H^q(\de,H^p(\cWth, \cM\otimes F(\La^L,2i)))\,,}
and can be
computed  explicitly using the isomorphism $F(\La^L,2i)\cong
\cgM(\La^L,2i)$,  the
reduction theorem of Section \SSreduction, and the explicit form of the
resolutions (see Theorem \thJMab). Since
$H(\cW_3,M^{(\vdeg)}(\si\circ\La,0)\otimes\cgM(\La^L,2i))$
vanishes unless
$\La^L$ satisfies \condforweg, the first part of the theorem follows
immediately. Moreover, the reduction theorem implies that this
cohomology, when non-vanishing, arises only at
the highest weight of the given
$M(\si\circ\La,0)$.  Thus, depending on $\La$ and $\si$, we find
$(E'_r,\de_r)$, $r\geq 2$, to be given by one of the following three cases
(which correspond precisely to those introduced above).

\bigskip
\noindent
Case Ia.
$${\eqalign{&\matrix{E_1':\qquad&
\eqalign{0\cr\cr0\cr\cr0\cr}& \longrightarrow &\eqalign{
c^{[2]}_0 &c^{[3]}_0 v_0\cr\cr
c^{[2]}_0 v_0\,, &\quad  c^{[3]}_0 v_0\cr\cr
&v_0\cr}  & \longrightarrow & \eqalign{0\cr\cr0\cr\cr0\cr}\,, \cr}\cr
&\cr
&E_1' ~=~ E'_2 ~=~ \ldots~=~E'_\infty\,.\cr}
}$$
\bigskip
\bigskip

\noindent
Case Ib.
$${\eqalign{&\matrix{E_1':\qquad&
\eqalign{0\cr\cr0\cr\cr0\cr}& \longrightarrow &\eqalign{
c^{[2]}_0 &c^{[3]}_0 v_1\cr\cr
c^{[2]}_0 v_0\,, &\quad  c^{[3]}_0 v_1\cr\cr
&v_0\cr}  & \longrightarrow & \eqalign{0\cr\cr0\cr\cr0\cr}\,, \cr}\cr
&\cr
&E_1'~=~E'_2~=~\ldots~=~E'_\infty\,.\cr}
}$$
\bigskip
\bigskip

\noindent
Case II.
$$
\eqalign{
&\matrix{E_1':\qquad& \eqalign{0\cr\cr0\cr\cr0\cr}& \longrightarrow &\eqalign{
c^{[2]}_0 &c^{[3]}_0 v_0\cr\cr c^{[2]}_0 v_0\,, &\quad  c^{[3]}_0 v_0\cr\cr
&v_0\cr} & \mapright{\de_1} &\eqalign{ c^{[2]}_0 & c^{[3]}_0 v_1'\cr\cr
c^{[2]}_0 v_0'\,, &\quad c^{[3]}_0 v_1'\cr\cr &v_0'\cr} & \longrightarrow
&\eqalign{0\cr\cr0\cr\cr0\cr}\,, \cr
&&&&&&\cr&&&&&&\cr&&&&&&\cr
&&&&&&\cr&&&&&&\cr&&&&&&\cr
E_2':\qquad &\eqalign{0\cr\cr0\cr\cr0\cr}& \longrightarrow
&\eqalign{
c^{[2]}_0 &c^{[3]}_0 v_0\cr\cr
&\hskip -8pt c^{[3]}_0 v_0\cr\cr
&0\cr} & \mapright{\de_2} &\eqalign{
c^{[2]}_0 & c^{[3]}_0 v_1'\cr\cr
&\hskip -8pt c^{[3]}_0 v_1'\cr\cr
&0\cr}  & \longrightarrow &\eqalign{0\cr\cr0\cr\cr0\cr}\,, \cr}\cr
&\cr
&E_2'~=~E'_3~=~\ldots~=~E'_\infty\,.\cr}
$$
\bigskip
\bigskip

\noindent
Case III.
$$\eqalign{
&\matrix{E_1':\qquad& \eqalign{0\cr\cr0\cr\cr0\cr}&
\longrightarrow &\eqalign{
c^{[2]}_0 &c^{[3]}_0 v_0\cr\cr
c^{[2]}_0 v_0\,, &\quad c^{[3]}_0 v_0\cr\cr
&v_0\cr}
\,,\quad \eqalign{
c^{[2]}_0 &c^{[3]}_0 v_0'\cr\cr
c^{[2]}_0 v_0'\,, &\quad  c^{[3]}_0 v_0'\cr\cr
&v_0'\cr}
& \mapright{\de_1} &\eqalign{
c^{[2]}_0 & c^{[3]}_0 v_1''\cr\cr
c^{[2]}_0 v_0''\,,  &\quad c^{[3]}_0 v_1''\cr\cr
&v_0''\cr}  & \longrightarrow &\eqalign{0\cr\cr0\cr\cr0\cr}\,, \cr
&&&&&&\cr&&&&&&\cr&&&&&&\cr
&&&&&&\cr&&&&&&\cr&&&&&&\cr
E_2':\qquad& \eqalign{0\cr\cr0\cr\cr0\cr}& \longrightarrow &\eqalign{
c^{[2]}_0 &c^{[3]}_0 v_0^{(-)}\,,\quad c^{[2]}_0 c^{[3]}_0 v_0^{(+)}  \cr\cr
c^{[2]}_0 v_0^{(-)}\,, &
\quad c^{[3]}_0 v_0^{(-)}\,,\quad c^{[3]}_0 v_0^{(+)}\cr\cr
&\hskip 15pt v_0^{(-)}\cr}
&\mapright{\de_2} &\eqalign{
c^{[2]}_0 & c^{[3]}_0 v_1''\cr\cr
&\hskip -8pt c^{[3]}_0 v_1''\cr\cr
&0\cr}  & \longrightarrow &\eqalign{0\cr\cr0\cr\cr0\cr}\,, \cr}\cr
&\cr
&E_2'~=~E'_3~=~\ldots~=~E'_\infty\,.\cr}
$$
\bigskip
\vfill\eject

The notation for the states is  the same as in Theorem
\redthm, except that we have denoted the highest weight states from different
spaces by primes, and at the third term of the spectral resolution in
Case 3 we have introduced $v_0^{(\pm)} \equiv v_0 \pm v_0'$.  Each
diagram represents a double-graded complex $E'_r{}^{p,q}$, with the
ghost number, $p$, increasing in the vertical direction, starting with
$p=3$, which is the ghost number of the state $v_0$. The horizontal
grading, $q$, is induced from the resolutions.  In particular, each
quartet of states in $E_1'$ arises at the position, $q$, of a Verma
module $M(\si\circ\La,0)$ or a generalized Verma module
$M^{(2)}(\si\circ\La,0)$ in the resolution. The differential $\de_1:
E'_1{}^{p,q}\rightarrow E'_1{}^{p,q+1} $ is obtained from the
differential $\de$ in the resolutions, \ie, it maps, up to a sign, a
given state in the quartet onto the identical one in the quartet at
the next step (if such is present, it maps to zero otherwise).  For
example, in Case II we find $\de_1(v_0)=v_0'$,
$\de_1(c_0^{[2]}v_0')=c_0^{[2]}v_0'$, but $\de_1(c_0^{[3]}v_0)=0$,
etc. The resulting $E'_2$ terms are spanned by the elements listed in
the diagrams. Since $\de_2: E'_1{}^{p,q}\rightarrow E'_1{}^{p-1,q+2}
$, we find that in all cases $\de_2$ vanishes identically, and thus
the sequence collapses.  The second part of the theorem then follows
by comparing the limits of the two spectral sequences using
\eqn\jzz{
\bigoplus_{p+q=n}E{}^{p,q}_\infty~\cong~
\bigoplus_{p+q=n}E'{}^{p,q}_\infty \, ,
}
\ie, the so-called ``zig-zag procedure.''  \Box\hfil\smallskip

 In view of Theorem \fundcoh, it is rather natural to seek an explicit
description of the $H(\cW_3,L(\La,0)\otimes F(\La^L,2i))$ in terms of
quartets. A quartet with states of ghost number $n$, $n+1$, $n+1$ and
$n+2$ is parametrized by its lowest lying member, which will be called
a ``prime state,'' following the terminology introduced in [\Pope] for
similar states in the critical $\cW_3$ cohomology. We should stress,
however, that the decomposition of the cohomology into quartets is at
the level of vector spaces only, and that there is no intrinsic
characterization of prime states as specific cohomology classes. Let
$H_{pr}(\cW_3,L(\La,0)\otimes F(\La^L,2i))$ denote the space of prime
states. Then part (ii) of Theorem \fundcoh\ can be restated simply as
follows

\thm\restate
\proclaim Theorem \restate. Consider $\La$, $\La^L$ and $\si\in
W(\La)$ as in \condforweg.
Then
\eqn\primecoh{H^n_{pr}(\cW_3,L(\La,0)\otimes F(\La^L,2i))~\cong~
\cases{\CC & if $n=3-\ell(\si)\,,$\cr
0 & otherwise$\,.$\cr}}
and there is a (non-canonical) isomorphism (of vector spaces)
\eqn\fullvsprime{H^n~\cong~ H^n_{pr}\oplus H^{n-1}_{pr} \oplus
H^{n-1}_{pr} \oplus H^{n-2}_{pr}\,.}
\smallskip

We would like to conclude with a comment on a possible role of the
relative cohomology. One may be tempted to conjecture, by
extrapolating the known result for the Virasoro algebra
[\FGZ,\LZvir,\BMPvir], that the full cohomology is (non-canonically)
isomorphic to the direct sum of relative cohomologies ``shifted'' by
the ghosts zero modes, \ie, schematically, $H\cong H_{rel}\oplus
c_0^{[2]}H_{rel} \oplus c_0^{[3]}H_{rel} \oplus
c_0^{[2]}c_0^{[3]}H_{rel}$. If this was the case, it would be natural
to identify prime states with the relative cohomology states. It would
also explain the quartet structure of the cohomology. Unfortunately,
as discussed earlier, the relative cohomology, as well as its relation
to the full cohomology, is difficult to analyze, and we cannot give
any general arguments that would support such a conjecture. However,
one finds, at least in cases we have studied explicitly, that
representatives of prime states in cohomology can be chosen such that
they are annihilated by $b_0^{[2]}$, $b_0^{[3]}$ (and thus by $L_0^{tot}$
and $W_0^{tot}$).

\subsubsec{$H(\cW_3,F(\La^M,0)\otimes F(\La^L,2i))$ with
$-i\La^L+2\rh\in P_+$}
\subsubseclab\SHgen

The result for the cohomology $H(\cW_3,F(\La^M,0)\otimes F(\La^L,2i))$,
with $-i\La^L+2\rh\in P_+$,
follows immediately by applying Theorem \fundcoh\ to the decomposition of
$F(\La^M,0)$ into irreducible modules $L(\La,0)$ given in Theorem \thPBda.
The nontrivial contributions to the cohomology for a given Liouville
momentum come from $L(\La,0)$
in the decomposition such that $\La$ satisfies \condforweg.
Thus it is clearly convenient to collect such weights together.

\thm\setofshwe
\proclaim Definition \setofshwe. For $\La'+2\rh\in P_+$, define
$P(\La')$ as the set of all $\La\in P_+$ such that
$\La'+2\rh=\si\circ\La$, for some $\si\in W(\La)$.

\noindent
We have then proven the following result.

\thm\fockcoh
\proclaim Theorem \fockcoh. Let $-i\La^L+2\rh\in P_+$. Then
\eqn\dimfockcoh{
\dim\,H^n_{pr}(\cW_3,F(\La^M,0)\otimes F(\La^L,2i)) ~=~
\sum_{\La\in P(-i\La^L)} \sum_{\si\in W(\La)} \, \de_{n,3-\ell(\si)}\,
m^\La_{\La^M}\,,}
In particular, $H(\cW_3,F(\La^M,0)\otimes F(\La^L,2i))\not=0$ if and
only if $\La^L$ satisfies \condforweg\
for some $\La\in P_+$ and $\si\in W(\La)$ and  $m_{\La^M}^\La\not=0$.
\smallskip

The appearance of the multiplicities $m^\La_{\La^M}$ in \dimfockcoh\
is well understood in the light of Theorem \slthac.  Since
all states in $F(\La^M,0)\otimes F(\La^L,2i)$ have the
same weight, $(\La^M,-i\La^L)$, with respect to $\slth\oplus
(\uone)^2$, the multiplicities simply reflect the decomposition of
$H(\cW_3,\fC)$ into finite dimensional $\slth\oplus (\uone)^2$ modules.
\smallskip

  We may make this structure even more manifest as follows. Fix a pair
$(\La,\si)$, $\La\in P_+$ and $\si\in W(\La)$, and then determine the
Liouville weight $\La^L$ via \condforweg.  Now consider all matter Fock
spaces, $F(\La^M,0)$, that give rise to nonvanishing cohomology through
the irreducible module $L(\La,0)$ in their decomposition. From
\dimfockcoh\ we see that they fill up precisely one
$\slth\oplus(\uone)^2$ module $\cL(\La)\otimes\CC_{-i\La^L}$ in the
``prime cohomology''
-- or, more rigorously, a quartet of such modules in the full
cohomology.

\thm\triples \proclaim Lemma \triples. For $-i\La^L + 2\rho$ in the
fundamental Weyl chamber, the decomposition of $H(\cW_3,\fC)$  into
quartets of $\slth\oplus(\uone)^2$ irreducible modules is in one to one
correspondence with the space of pairs $(\La,\si)$, where $\La\in
P_+$ and $\si\in W(\La)$. Moreover, the space of such
pairs is a sum  of disjoint cones in the $(\La^M,-i\La^L)$ weight
space that are isomorphic with $P_+$.
\smallskip

\proof The first part of the Lemma is just a summary of the previous
discussion, so let us proceed to the cone decomposition.
{}From the definition of $W(\La)$ in \eqPBRe\
it is clear that if $(\La,\si)$ is a pair then so also
is $(\La+\la,\si)$, for all $\la\in P_+$.  Moreover, the set of
weights that give rise to a given $\si$ is determined by a
set of inequalities, each of the form $(\La,\al)\geq 0$, from which
the cone structure follows. \Box
\smallskip

This cone-like structure for the decomposition of the cohomology into
irreducible modules of $\slth\oplus(\uone)^2$, and its correspondence to
an extension of the Weyl group, will play a  key role in our extension
of the result to other Weyl chambers.  Let us therefore take a closer look
at how this correspondence arises in the fundamental chamber.   The
``tips'' of the cones in Lemma \triples\  can be determined explicitly
by examining the sets $W(\La)$.
Let $\cS^n$ be the set of cone tips at ghost number $n$.
The result for $\cS^n$ is given in Table \polytips\ below.
Moreover, we notice that the ``shift,'' $(-i\La^L+2\rh)-\La^M$, is
constant throughout each cone and equal to $\rh-\si\rh$,
$\si\in\widetilde W$.
Thus the set of cones, as parametrized by the shifts, say, are in
correspondence with the extension of the Weyl group, $\widetilde W$.
\smallskip

 To conclude this section, we summarize the result for the cohomology in
the fundamental Weyl chamber.

\thm\cohfweyl
\proclaim Theorem \cohfweyl. The cohomology $H(\cW_3,\fC_1)$
is isomorphic as an $\slth\oplus(\uone)^2$ module to the direct sum of
quartets of irreducible modules parametrized by  disjoint
cones $\{(\La,\La')+(\la,\la)\,|\,\la\in P_+\}$, \ie,
\eqn\decomfweyl{
H^n_{pr}(\cW_3,\fC_1)~\cong~ \bigoplus_{(\La,\La')\in
\cS^n}\bigoplus_{\la\in P_+} \, \cL(\La+\la)\otimes\CC_{\La'+\la}\,,}
where the sets $\cS^n$ (tips of the cones) are given in Table \polytips.

\bigskip
\tbl\polytips

\begintable
$\qquad n\qquad$ |  $\cS^n$ \cr
0 | $(0,0)$ \nr
1 | $(0,-2\La_1+\La_2)$, $(\La_1+\La_2,0)$,
$(0,\La_1-2\La_2)$\nr
2 | $\qquad (\La_1,-2\La_1)$, $(0,-\La_1-\La_2)$,
$(\La_2,-2\La_2)\qquad $\nr
3 | $(0,-2\La_1-2\La_2)$
\endtable
\bigskip
\centerline{Table \polytips. The sets $\cS^n$}

\bigskip\noindent
{\it Remarks:}
\item{i.} The nonvanishing cohomology in the fundamental Weyl chamber
arises in ghost numbers $0, \ldots, 5$, and exhibits a similar pattern
to that found in the non-critical cohomology of the
Virasoro algebra [\LZvir,\BMPvir].
\item{ii.} The pattern of the cohomology cones in Table \polytips\ can be
conveniently represented as a plot on the lattice of shifted Liouville
momenta, $-i\La^L+2\rh$, see Appendix \primeplots.
\item{iii.} The precise form of the resolutions, $(\cM,\de)$, required an
explicit computation of the embedding patterns of (generalized) Verma
modules. An independent partial confirmation of those results is
provided by a computation of cohomology spaces for low lying (shifted)
Liouville weights, \ie, an explicit verification  of \decomfweyl. This is
summarized in Appendix \APexplicit.
\item{iv.} Given the isomorphism
\eqn\isomorph{
H^n(\cW_3,F(\La^M,0)\otimes F(\La^L,2i))~\cong~
H^{8-n}(\cW_3,F(\La^M,0)\otimes F(w_0\cdot\La^L,2i))\,,} proved in Section
\SSbilinear, Theorem \cohfweyl\ also gives a complete result for the
cohomology $H(\cW_3,\fC_{w_0})$, \ie, for $-i\La^L+2\rh\in P_-$.
At the level of prime cohomology states
\isomorph\ reads
\eqn\isomorphpr{
H^n_{pr}(\cW_3,F(\La^M,0)\otimes F(w_0\cdot\La^L,2i))~\cong~
H^{6-n}_{pr}(\cW_3,F(\La^M,0)\otimes F(w_0\cdot\La^L,2i))\,.}
The reflection by the Weyl group element accompanied by a shift in the
ghost number in \isomorphpr\ suggests a generalization of
Theorem \cohfweyl\ to the other Weyl chambers.

\subsec{The  conjecture for $H(\cW_3,\fC)$}
\subseclab\SSgeneral

\subsubsec{Introduction}
\subsubseclab\SSgenintro

In this section we derive a conjecture for $H(\cW_3,\fC)$ by assuming
that there is a ``symmetry'' with respect to the action of the Weyl
group on the (shifted) Liouville momentum. In other words, if we
define $\fC_w$, $w\in W$, to be the subcomplex
of $\fC$ with $-i\La^L+2\rh\in
w^{-1}P_+$, then all cohomologies $H(\cW_3,\fC_w)$ should be related
in some sense.  For the case $w = w_0$, we saw at the end of the last
section that this relation is determined by duality.  Moreover, we
learned there that, loosely speaking, each Weyl group reflection of
the Liouville momentum should be accompanied by a shift in the ghost
number of the cohomology.  (This is also suggested by examining an
analogous problem in the cohomology of Lie algebras, as well as the
so-called generic regime of the $\cW_3$ cohomology (see, \eg,
[\BLNWnp,\BMPc]).)  Our aim is, therefore, to correlate the cone -- Weyl
correspondence with the Weyl reflection to other chambers,
incorporating the ghost number shift appropriately.  One can clearly
expect that there might be additional subtleties if $-i\La^L+2\rh$
lies close to the boundary of a Weyl chamber. Thus we develop an
ansatz for the cohomology with $-i\La^L+2\rh$ lying sufficiently deep
inside a Weyl chamber (referred to as the ``bulk region'') and then
use the results of explicit cohomology computations to extend it to a
complete conjecture.

\subsubsec{A vanishing theorem}
\subsubseclab\SSvanishh

Let us begin with an observation that $H^{n}(\cW_3,\fC)$ may be
nonzero only in a finite range of ghost numbers. The restriction is
given by the following vanishing theorem.

\thm\ghnumlim
\proclaim Theorem \ghnumlim. The cohomology
$H^{n}(\cW_3,\fC)$ is nonvanishing at most in ghost numbers
$n=0,\ldots,8$.
\smallskip

\proof Consider $H^n(\cW_3,F(\La^M,0)\otimes F(\La^L,2i))$.
Let $L(\La,0)$ be an irreducible module in the decomposition of
$F(\La^M,0)$. Then for all Verma modules $M^{(\vdeg)}(\si\circ\La,0)$,
$\vdeg=1$ or $2$, $\si\in W(\La)$, in the resolution of $L(\La,0)$,
the cohomology $H^n(\cW_3,M^{(\vdeg)}(\si\circ\La,0)\otimes F(\La^L,2i))$
vanishes for $n\leq 2$ (see Theorem \moreredthm). A straightforward
repetition of the double complex argument in the proof of Theorem
\fundcoh, would then give $3-4=-1$ as the lower bound for the ghost
number of the cohomology.  The ghost number $-1$ cohomology could only
arise from the Verma module, $M(w_0\circ \La,0)$, at level $-4$ in the
resolution -- in the language of the double complex we may be
more precise: from the ghost number 3 state in $E'{}_1^{3,-4}\cong
H^3(\cW_3,M(w_0\circ \La,0)\otimes F(\La^L,2i))$.  Thus in order to
increase the lower bound on the ghost number, we must show that
$\de_1': E'{}_1^{3,-4}\rightarrow E'{}_1^{3,-3} $ is an embedding.
\smallskip

Consider the factor in $\de_1'$ that arises from the
embedding $M(w_0\circ\La,0)\rightarrow M^{(2)}(w_0\circ\La,0)$.
{}From the isomorphism
$M^{(2)}(w_0\circ\La,0)/M(w_0\circ\La,0)\cong M(w_0\circ\La,0)$ we have a
short exact sequence
\eqn\lowestsect{
0\quad \longrightarrow\quad
M(w_0\circ\La,0) \quad \mapright{}\quad M^{(2)}(w_0\circ\La,0) \quad
\mapright{}\quad M(w_0\circ\La,0)\quad
\mapright{}\quad 0\,.}
By applying $H^3(\cW_3,\,-\,\otimes F(\La^L,2i))$ to \lowestsect, we
obtain a long exact sequence, from which the required embedding
is proved using
$H^{2}(\cW_3,M(w_0\circ\La,0)\otimes F(\La^L,2i))=0$. The upper bound
on the ghost number follows from \isomorph. \Box
\smallskip

Note that for $-i\La^L+2\rh\in P_+$ the ghost number of the
nonvanishing cohomology is between 0 and 5, which saturates the lower
bound imposed by Theorem \ghnumlim. As discussed in
Remark iv of the previous section, if the Liouville weight is
reflected by $w_0$, \ie, $-i\La^L+2\rh\in w_0P_+$, there is a
corresponding shift in the ghost numbers, which now range between $3$
and $8$, thus saturating the upper bound of the allowed values.  We
expect that the cohomology in the intermediate Weyl chambers
interpolates between those two extreme cases. If we require
consistency with duality (see Theorem \cohduality), and symmetry with
respect to interchange of the fundamental weights $\La_1$ and $\La_2$,
there is just one possibility left.

\thm\conjone
\proclaim Conjecture \conjone. The cohomology $H^n(\cW_3,\fC_w)$,
$w\in W$, is nonvanishing at most in ghost numbers
$n=\ell(w),\ldots,\ell(w)+5$.

\subsubsec{$H(\cW_3,F(\La^M,0)\otimes F(\La^L,2i))$}
\subsubseclab\SSgencoh
\tbl\gencones

For weights $-i\La^L+2\rh\in P_{+}$, Theorem \cohfweyl\ states that
the $\slth\oplus(\uone)^2$ content of prime cohomology is a direct sum
of the eight cones in Table \polytips.  We have seen that these cones,
parametrized by $\si\in\widetilde W$, arise at ghost number $3 -
\ell(\si)$: their tips are given in the table; their cone shift
$(-i\La^L+2\rh)-\La^M$, which is constant throughout the cone, is
easily calculated to be $\rh-\si\rh$.  Conversely, notice that if we
know that a cone with a shift $\de\la$ arises in cohomology, its tip
is determined by the ``lowest'' weight $\La\in P_+$ for which
$(\La^M,\La^L)=(\La,i(\La+\de\la-2\rh))$ gives rise to a nontrivial
cohomology state. Thus, given the cone structure and the set of all
possible cone shifts, we could determine the complete cohomology by
simply finding all such lowest weights $\La$ -- a finite computation
that can  be carried out.
\smallskip

 We will, indeed, assume that the cone structure generalizes to the
other Weyl chambers.  Hence we must formulate an ansatz for the cone
shifts.  The case $w = w_0$ suggests that the cones which arise for
$-i\La^L+2\rh\in w^{-1} P_+$, $w\in W$, should be related by a Weyl
reflection to the cones in the fundamental Weyl chamber.  More
precisely, let us introduce the notion of a $w$-twisted cone.

\thm\twistedco
\proclaim Definition \twistedco. We define a $w$-twisted
cone as a set of weights $\{(\La,\La')+(\la,w^{-1}\la)\,|\, \la\in P_+\}$,
where the tip $(\La,\La')$ has $\La\in P_+$. The shift
characterizing this cone is given by $w(\La'+2\rh)-\La$.

\noindent
Clearly the shift does not change when the cone is reflected from one
Weyl chamber to another.  The natural extension of our previous
results is the following conjecture for the decomposition of the
cohomology in the bulk region.

\thm\shiftsss
\proclaim Conjecture \shiftsss. For $-i\La^L+2\rh$ sufficiently inside
the Weyl chamber $w^{-1}P_+$ the cohomology $H(\cWth,\fC_w)$ is a
direct sum of $w$-twisted cones with the shifts $\rh-\si\rh$,
$\si\in\widetilde W$.
\smallskip

\nfig\twlengthbig{}
\nfig\twlengexs{}

It remains to find a proper ansatz for a shift in the ghost number
corresponding to a given reflection $w$.  Given $w\in W$ there is a
natural generalization of the length $\ell$, called the twisted length
[\FeFrtw,\BMPqg], which is defined by $\ell_w(w')=\ell(w^{-1}w') -
\ell(w^{-1})$, $w'\in W$.  This twisted length may again be extended
to $\widetilde W$ by using $w^{-1}\si_i=\si_i$, and a simple algorithm
for computing it is given in Figures \twlengthbig\ and \twlengexs.

\bigskip

\centerline{\vbox{\epsffile{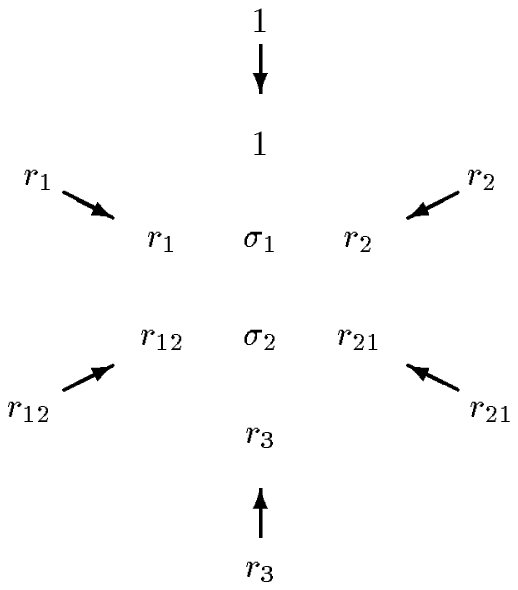}}}

\bigskip
\centerline{\vbox{\hsize 6in
\noindent
Figure \twlengthbig. Each twisted length $\ell_w(\si)$ increases in
the direction of the corresponding arrow ``$w\rightarrow$'' and is
constant along the transverse directions, as illustrated in Figure
\twlengexs\ below.}}
\bigskip

\bigskip

\line{\hfill  \vbox{\epsffile{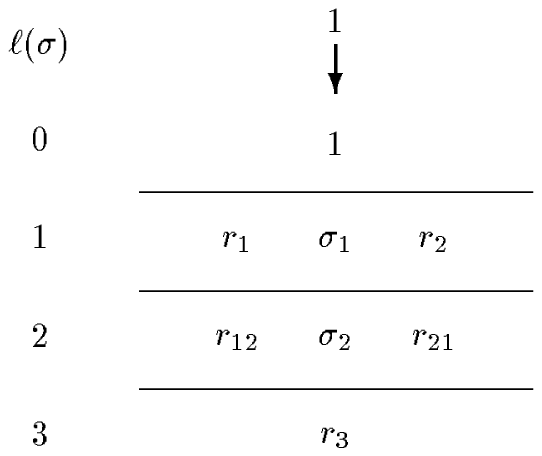}}\hskip 1.5cm
             \vbox{\epsffile{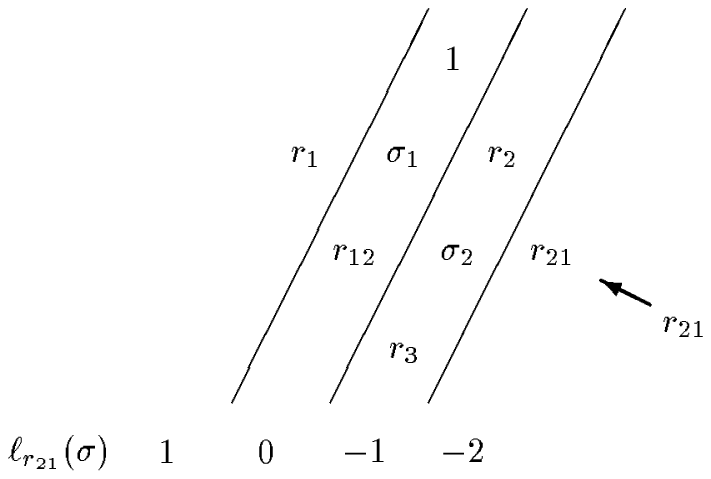}}\hfill }

\bigskip
\centerline{
\noindent
Figure \twlengexs. Examples of twisted lengths for $w=1$ and $w=r_{21}$. }
\bigskip

 Once again, a natural generalization of
our previous results follows.

\thm\twconess
\proclaim Conjecture \twconess. The ghost number  of the
prime cohomology state in a $w$-twisted cone with the shift
$\rh-\si\rh$ is equal to $3-\ell_w(\si)$, $w\in W$, $\si\in \widetilde W$.

\bigskip
\begintable
\quad $n$ \quad | \quad $w$ \quad | \quad $(\La,\La')_\si$ \quad \crthick
$0$ | $1$ |  $(0,0)_{r_3}$  \cr
$1$ | $1$ |  $(\La_2,\La_1-\La_2)_{r_{12}}$,
               $(\La_1+\La_2,0)_{\si_2}$,
             $(\La_1,-\La_1+\La_2)_{r_{21}}$ \nr
    | $r_1$ | $(0,-2\La_1+\La_2)_{r_{21}}$ \nr
    | $r_2$ | $(0,\La_1-2\La_2)_{r_{12}}$ \cr
$2$ | $1$ | $(2\La_2,-\La_2)_{r_1}$,
            $(0,-\La_1-\La_2)_{\si_1}$, $(2\La_1,-\La_1)_{r_2}$ \nr
    | $r_1$ | $(\La_1,-2\La_1)_{r_2}$, $(\La_2,-3\La_1+\La_2)_{\si_2}$,
              $(0,-4\La_1+2\La_2)_{r_3}$ \nr
    | $r_2$ | $(\La_2,-2\La_2)_{r_1}$, $(\La_1,\La_1-3\La_2)_{\si_2}$,
              $(0,2\La_1-4\La_2)_{r_3}$ \nr
    | $r_{12}$ | $(0,-3\La_2)_{r_2}$ \nr
    | $r_{21}$ | $(0,-3\La_1)_{r_1}$ \cr
$3$ | $1$ | $(\La_1+\La_2,-\La_1-\La_2)_1$ \nr
    | $r_1$ |  $(\La_2,-2\La_1-\La_2)_1$, $(\La_1,-4\La_1+\La_2)_{\si_1}$,
              $(\La_2,-5\La_1+2\La_2)_{r_{12}}$  \nr
    | $r_2$ | $(\La_1,-\La_1-2\La_2)_1$, $(\La_2,\La_1-4\La_2)_{\si_1}$,
              $(\La_1,2\La_1-5\La_2)_{r_{21}}$ \nr
    | $r_{12}$ | $(\La_2,-\La_1-3\La_2)_1$, $(0,\La_1-5\La_2)_{r_{21}}$,
                 $(\La_2,-5\La_2)_{\si_1}$ \nr
    | $r_{21}$ | $(\La_1,-3\La_1-\La_2)_1$, $(0,-5\La_1+\La_2)_{r_{12}}$,
                 $(\La_1,-5\La_1)_{\si_1}$ \nr
    | $r_3$ | $(0,-2\La_1-2\La_2)_1$ \cr
$4$ |  $r_1$ | $(0,-4\La_1-\La_2)_{r_1}$ \nr
    | $r_2$ | $(0,-\La_1-4\La_2)_{r_2}$ \nr
    | $r_{12}$ | $(\La_2,-2\La_1-4\La_2)_{r_1}$,
                 $(\La_1,-\La_1-5\La_2)_{\si_2}$,
                 $(0,-6\La_2)_{r_3}$ \nr
    | $r_{21}$ | $(\La_1,-4\La_1-2\La_2)_{r_2}$,
                 $(\La_2,-5\La_1-\La_2)_{\si_2}$,
                 $(0,-6\La_1)_{r_3}$ \nr
    | $r_3$ |  $(0,-3\La_1-3\La_2)_{\si_1}$, $(2\La_1,-4\La_1-3\La_2)_{r_2}$,
                $(2\La_2,-3\La_1-4\La_2)_{r_1}$ \cr
$5$
    | $r_{12}$ | $(0,-2\La_1-5\La_2)_{r_{12}}$ \nr
    | $r_{21}$ | $(0,-5\La_1-2\La_2)_{r_{21}}$ \nr
    | $r_3$ |  $(\La_1,-5\La_1-3\La_2)_{r_{21}}$,
$(\La_1+\La_2,-4\La_1-4\La_2)_{\si_2}$,
        $(\La_2,-3\La_1-5\La_2)_{r_{12}}$ \cr
$6$ | $r_3$ | $(0,-4\La_1-4\La_2)_{r_3}$ \endtable
\bigskip

\centerline{Table \gencones.\ The sets $\cS_w^n$ -- the weights
$(\La,\La')_\si$ satisfy $\La'+2\rh=w^{-1}(\La+\rh-\si\rh)$}
\medskip
\bigskip

\smallskip
\noindent
{\it Remarks:}
\item{i.} By construction, Conjectures \shiftsss\ and \twconess\ correctly
reproduce the cones and their ghost numbers in the  $w=1$ and $w=w_0$
Weyl chambers. In the other chambers they yield the correct range of
ghost numbers, in particular, those suggested by Conjecture \conjone.
\item{ii.} In the context of Lie algebras or affine Lie algebras, the
twisted length functions arise naturally in the  resolutions
of highest weight irreducible modules in terms of twisted Verma or
Wakimoto modules [\BMPkar]. Hypothetically, if analogue twisted resolutions
for positive energy $\cW_3$ modules exist, one would expect to
prove, following the steps in Section \SSfundamental, that
the  structure of the full cohomology is as conjectured above.
\item{iii.} One can also arrive at Conjectures  \shiftsss\ and
\twconess\ under seemingly weaker assumptions, by studying
the BV-algebra structure of $H(\cWth,\fC)$. This is discussed in
Section \Scomstr.
\smallskip

In the following we assume the validity of Conjectures \shiftsss\ and
\twconess,  and proceed to study their consequences.
As discussed above, to determine the full cohomology we need now only
calculate the cone tips.  We have carried out an exhaustive
computation of the dimensions of the cohomologies for low lying
weights, the results are summarized in Appendix \APexplicit.
{}From this we
determine the cone tips to be as listed in Table \gencones\
below, where $\cS^n_w$ denotes the set of $w$-twisted cone tips at
ghost number $n$. Finally, then, we have

\thm\fullcoh
\proclaim Theorem \fullcoh.  The cohomology
$H(\cWth,\fC)$ is isomorphic, as an $\bga$ module, to the
direct sum of quartets
of irreducible $\bga$ modules with the highest weights in a set of
disjoint cones $\{ (\La,\La') + (\la,w^{-1}\la)\,|\,\la\in P_+,\,
(\La,\La') \in \cS_w\}$, \ie,
\eqn\bigdecomp{
H_{\rm pr}^n(\cW_3,\fC)\ \cong\ \bigoplus_{w\in W}\, \bigoplus_{(\La,\La')\in
  \cS_w^n} \, \bigoplus_{\la\in P_+} \left( \cL(\La+\la) \otimes
  \CC_{\La' + w^{-1}\la} \right) \,,}
where the sets $\cS_w^n$ (tips of the cones) are given in Table \gencones.
\smallskip

\noindent
{\it Remark:}
Since a given cone and its Weyl reflection may overlap, the theorem
requires an explicit decomposition of $H(\cW_3,\fC)$  into disjoint cones
in the overlap region.  In all cases we resolve the ambiguity
by including the complete common region in only one of the cones.
In particular, this explains why some of the tips in the fundamental Weyl
chamber given in Table \polytips\ are shifted with respect to those
in Table \gencones\ with $w=1$.
\smallskip

\noindent
{\it Example.} To illustrate this ambiguity, let us consider as an
example all the cones in $H^1(\cW_3,\fC)$
characterized by the shift $3\La_1$.
We have already found such a cone in the untwisted sector, namely
$\{(0,\La_1 - 2\La_2)+(\la,\la)\,|\,\la\in P_+\}$, which appears in
Theorem \cohfweyl.  Examining the tables in Appendix \APexplicit, we
conclude from the appearance of the quartet with weights
$(\La_2,2\La_1 - 3\La_2)$ that there is an $r_2$-twisted cone with the
same shift, $3\La_1$.  The lowest weight state, \ie, the tip of this
$r_2$-twisted cone, would also be $(0,\La_1 - 2\La_2)$ -- in fact the
common boundary of the two cones is one dimensional, at $(n \La_1,
(n+1)\La_1 - 2\La_2),\, n>0$.
\smallskip

By retracing the steps which give Theorem \fockcoh\ from Theorem
\restate, we may derive from Theorem \fullcoh\ the result for
$H(\cW_3,L(\La,0) \otimes F(\La^L,2i))$ when $\La^L$ is arbitrary.

\thm\thBa
\proclaim Corollary \thBa.
Let $\La\in P_{+}$.
\item{(i)} The cohomology $H^n(\cWth,L(\La,0) \otimes F(\La^L,2i))$
is nontrivial only if there exist $w \in W,\ \si\in\widetilde W$ such
that
\eqn\eqBa{
-i\La^L + 2 \rh ~=~ w^{-1} (\La + \rh - \si \rh)\,.
}
\item{(ii)}
For $w,\si,\La$ and $\La^L$ as in \eqBa,
the cohomology $H^n_{pr}(\cWth,L(\La,0) \otimes F(\La^L,2i))$
is 1-dimensional if
\eqn\ennns{
n= 3-\ell_w(\si)= 3+ \ell(w^{-1}) - \ell(w^{-1}\si) \,,}
and
$$\matrix{
\si \in W\,,  &  \La\in P_+\,,&  w\in W\,, \cr
\si \in \{\si_1,\si_2\}\,,    &   \La\in P_{++}\,, &  w\in W\,, \cr
\si=\si_1\,,    &  (\La,\al_i) = 0\,,\La\neq0\,,  &
   w\in\, <\!r_i\!>\!\backslash W\,,\cr
\si=\si_2\,, & (\La,\al_i) = 0\,,\La\neq0\,, &
   w\in r_i(<\!r_i\!>\!\backslash W)\,, \cr}
$$
and vanishes otherwise.

\noindent {\sl In the case that certain weights $(\La,-i\La^L)$ and
certain ghost
number $n$ satisfy {\sl (i)} and {\sl (ii)} for more than
one choice of $(w, \si)$, the above should be understood in the
sense that the
corresponding cohomology is nevertheless 1-dimensional.}

\vfill\eject

%
%
\secno3
\newsec{BATALIN-VILKOVISKY ALGEBRAS}
\seclab\BValgebra

In this section we collect some general results on BV-algebras and
study a  class of examples that will be important for describing
explicitly the BV-algebra structure of $H(\cW_3,\fC)$. Most of this
section can be read independently from the rest of the paper.
\smallskip

The notion of BV-algebras first appeared in the work of the mathematician
J.\ Koszul [\Ko], where they were called (exact) coboundary G-algebras
(see also [\KosSch]). Independently, and at roughly the same time, the
physicists Batalin and Vilkovisky [\BV] constructed a particular
example of a BV-operator and applied it to the quantization of gauge
theories (see also [\Wia]).  Recently, Lian and Zuckerman [\LZbv],
Schwarz and Penkava [\MS] and Getzler [\Get] -- building upon earlier
work of Witten and Zwiebach [\Wi,\WiZw] -- recognized that BV-algebras
provide a proper framework for describing operator algebras in a large
class of topological field theories; in particular, in
two-dimensional string theory.

\subsec{G-algebras and BV-algebras}
\subseclab\SGBValg

\subsubsec{Definitions}
\subsubseclab\SSdefini

\thm\BVghal
\proclaim Definition \BVghal\ [\Ge]. A G-algebra (or Gerstenhaber algebra)
$(\fA,\,\cdot\,,[-,-])$
is a $\ZZ$-graded, supercommutative, associative algebra under
the ``dot'' product, $\cdot$, and a $\ZZ$-graded Lie superalgebra under the
bracket, $[-,-]$
(of degree $-1$), such that the (odd) bracket acts as a
superderivation of the algebra, \ie, $\fA=\bigoplus_{n\in\ZZ}\fA^n$
$$\eqalign{
\cdot \quad &:\quad \fA^m\times \fA^n \longrightarrow \fA^{m+n}\cr
[-,-] \quad &:\quad \fA^m\times \fA^n \longrightarrow
\fA^{m+n-1}\cr}$$
and for any homogeneous $a,b,c\in\fA$ (we define $|a|=n$ for $a\in\fA^n$)
\item{i.} $a\cdot b~=~(-1)^{|a||b|}b\cdot a$,
\item{ii.} $(a\cdot b)\cdot c~=~a\cdot (b\cdot c)$,
\item{iii.} $[a,b]~=~-(-1)^{(|a|-1)(|b|-1)}{[b,a]}$,
\item{iv.} $(-1)^{(|a|-1)(|c|-1)}[a,[b,c]] + {\rm cyclic} ~=~0$,
\item{v.} $[a, b\cdot c]~=~[a,b]\cdot c+(-1)^{(|a|-1)|b|}b\cdot [a,c]$.
\smallskip

Let $\fA$ be a $\ZZ$-graded supercommutative algebra.  We recall
that a first order superderivation of $\fA$ of degree $|K|$ is a
map $K\,:\, \fA^n \to \fA^{n+|K|}$ such that for all $a,b\in \fA$
\eqn\eqBPza{
K(a\cdot b) ~=~ (Ka)\cdot b + (-1)^{|K||a|} a\cdot (Kb)\,.
}
We will refer to \eqBPza\ as the (super) Leibniz rule.
Now, for all $a\in \fA$, we define the left multiplication
$l_a\,:\, \fA \to \fA$ by
\eqn\eqBPzb{
l_a(b) ~=~ a\cdot b\,,
}
then \eqBPza\ is equivalent to
\eqn\eqBPzc{
[K,l_a] - l_{Ka} ~=~ 0\,,
}
with an obvious definition of the (graded) bracket.
By induction, an $n$-th order superderivation of degree $|K|$ is a
map $K\,:\, \fA^n \to \fA^{n+|K|}$ such that
$[K,l_a] - l_{Ka}$ is an $(n-1)$-th order derivation for all $a\in \fA$.
For example, a second order derivation of degree $|K|$ satisfies
\eqn\BVscor{ \eqalign{
K(a\cdot b \cdot c) ~=~ & K(a\cdot b)\cdot c + (-1)^{|K||a|}
  a \cdot K(b\cdot c) + (-1)^{(|K|+|a|)|b|} b\cdot K(a\cdot c) \cr
  &  - (Ka)\cdot b \cdot c - (-1)^{|K||a|} a\cdot (Kb)\cdot c
  - (-1)^{|K|(|a|+|b|)} a\cdot b\cdot (Kc) \,,\cr}
}
for all $a,b,c\in \fA$.

\thm\BVbv
\proclaim Definition \BVbv\ [\Ko,\LZbv].  A BV-algebra
(or Batalin-Vilkovisky algebra)
$(\fA,\,\cdot\,,\De)$ is a $\ZZ$-graded, supercommutative, associative
algebra with a second order derivation $\De$ (BV-operator)
of degree $-1$ satisfying $\De^2=0$.\par

There is a close relation between the two classes of algebras; indeed, the
following lemma shows that any BV-algebra has the canonical structure of
a G-algebra.

\thm\thCd
\proclaim Lemma \thCd\ [\Wia,\LZbv,\MS]. For any BV-algebra
$(\fA,\,\cdot\,,\De)$,
the bracket
\eqn\eqCc{
[a,b] ~=~ (-1)^{|a|} \left( \De(a\cdot b) - (\De a)\cdot b -
  (-1)^{|a|} a\cdot(\De b) \right)\,,\qquad a,b\in \fA\,,}
introduces on $\fA$ the structure of a G-algebra. Moreover, the
BV-operator acts as a superderivation of the bracket
\eqn\eqCd{
\De [a,b] ~=~ [\De a,b] + (-1)^{|a|-1} [a,\De b]\,.}

By combining Definition \BVbv\ with Lemma \thCd, we obtain

\thm\triplec
\proclaim Theorem \triplec. Let $(\fA,\,\cdot\,,[-,-])$
be a $G$-algebra and $\De$ an operator of degree $-1$ satisfying
$\De^2=0$ and \eqCc. Then $(\fA,\,\cdot\,,\De)$ is a BV-algebra.
\smallskip

\proof One must only show that $\De$ satisfies identity \BVscor, which
follows directly by evaluating the left hand side in \BVscor\ as, \eg,
$\De(a\cdot(b\cdot c))$ using \eqCc, and then identity (v) for the
bracket. \SMu
\smallskip

It is clear from the definitions above that for any G-algebra the
subspace $\fA^0$ is an Abelian algebra with respect to the dot product.
Similarly, $\fA^1$ is a Lie algebra with respect to the bracket.
Moreover, by Definition \BVghal,  the map $\fA^1 \to {\rm Map}(\fA,\fA)$
defined by
\eqn\derofa{
a ~\mapsto~ K_a \quad {\rm with}\quad K_a(b) ~=~ [a,b]\,,\quad
a\in\fA^1\,,\, b \in \fA\,,
}
satisfies
\eqn\eqPBexa{ \eqalign{
K_a (b\cdot c) & ~=~ K_a(b) \cdot c + b\cdot K_a(c) \,,\cr
[K_a,K_b]    & ~=~ K_{[a,b]} \,,\cr}
}
\ie, $a \to K_a$ is a Lie algebra homomorphism from $\fA^1$ into the
Lie algebra  $\cD(\fA)$ of derivations of the ``dot algebra'' $\fA$.

Using \eqCd\ we also prove
\thm\LLaut
\proclaim Lemma \LLaut. Let $\fA$ be a BV-algebra and consider $a\in \fA^1$
which satisfies $[\De(a),b]=0$ for all $b\in \fA$, then
$K_a\,\De = \De \, K_a$.

To orient the reader, let us just note here a standard
class of examples of G-algebras: the algebra of polyvector fields on a
given manifold, with the operations of wedge product and Schouten
bracket (the bracket induced from the Lie bracket on vector fields),
is a G-algebra.  We next detail an abstraction of this example, before
returning to the simplest such: polyvectors on $\CC^N$.

\subsubsec{The G-algebra of polyderivations of an Abelian algebra}
\subsubseclab\SSpolyderaa

 We now summarize the canonical construction of the polyderivations of
an arbitrary Abelian algebra.  This is a standard example
of a G-algebra. For a complete discussion see, \eg,
[\Sch,\FNij,\Ko,\Kr,\KosSch]. \smallskip

First recall that if $(\cR,\,\cdot\,)$ is an Abelian algebra and $M$
an $\cR$-module then the space of derivations of $\cR$ with values in
$M$, $\cD(\cR,M)$ (or simply $\cD(\cR)$ if $M=\cR$), consists of those
$a\in\Hom(\cR,M)$ that satisfy the Leibniz rule, \ie,
\eqn\BVleib
{a(x\cdot y) = x\cdot a(y)+y\cdot a(x)\,,}
for any $x,y\in\cR$.

\thm\defpolvs
\proclaim Definition \defpolvs.
The  polyderivations  $\cP^n(\cR,M)$ of degree%
\foot{{\rm Note, that it is more conventional to use the term ``order''
in this context.}}
 $n$ are defined as
follows:
\smallskip
\item{i.} $\cP^0(\cR,M)~\cong~ M$, for $n=0$,
\item{ii.}  $\cP^1(\cR,M) ~\cong~ \cD(\cR,M)$, for $n=1$,
\item{iii.} $\cP^n(\cR,M)$, $n\geq 2$,  is the space of those
$a\in \cD(\cR,\cP^{n-1})$ that satisfy
\eqn\BVderv{
a(x,y)~=~-a(y,x)\,,\qquad x,y\in\cR\,,}
where $a(x,y)$ denotes the element $a(x)(y)\in\cP^{n-2}(\cR,M)$.
\smallskip

\noindent
Clearly, one may simply consider the $n$-th degree derivations
$\cP^n(\cR,M)$ as a subspace in $\Hom(\cR^{\otimes n},M)$.
Then we have

\thm\derivas
\proclaim Lemma \derivas. The polyderivations  $\cP^n(\cR,M)$ of
degree $n$ consist of those $a\in \Hom(\cR^{\otimes n},M)$ for which
$a(x_1,\ldots,x_n)=a(x_1)(x_2)\ldots(x_n)$ is completely antisymmetric
and satisfies the
Leibniz rule \BVleib\ in all of the arguments $x_1\,,\ldots\,,x_n$.
\smallskip

In the case where $M=\cR$, the space of polyderivations
$\cP(\cR,\cR)$ -- in the following denoted simply by $\cP(\cR)$ --
is itself a $\ZZ$-graded algebra, $\cP(\cR)=\bigoplus_{n\geq 0}
\cP^n(\cR)$,
with the product ``$\,\cdot\,$'' defined by induction\foot{For
simplicity we do not distinguish the degree 0 space in this induction
(here or later), and so we have extended the notation via $a(x) = 0$
for $a \in \cP^0(\cR,M)$.} using
\eqn\prodpol{
(a\cdot b)(x)~=~a\cdot b(x) + (-1)^{|b|} a(x)\cdot b\,.}
More explicitly,
\eqn\prodex{
(a\cdot b)(x_1\,,\ldots\,,x_{m+n})~=~(-1)^{mn}\sum_{\si\in S_{m+n}}
{\rm sgn}({\si}) \, a(x_{\si(1)},\ldots,x_{\si(m)})\cdot
   b(x_{\si(m+1)},\ldots,x_{\si(m+n)})\,,}
where $a\in\cP^m(\cR)$, $b\in\cP^n(\cR)$, and $x_i\in\cR$,
$i=1,\ldots,m+n$.  In addition, there is a natural bracket operation on
$\cP(\cR)$,

\thm\defsch
\proclaim Definition \defsch. The Schouten bracket on
$\cP(\cR)$ is the unique  bilinear map $[-,-]_S:\cP(\cR)\times
\cP(\cR) \rightarrow \cP(\cR)$ satisfying
\eqn\schbrack{[a,b]_S~=~\cases{0, & for $|a|=0\,, |b|=0$, \cr
          a(b), & for $|a|=1\,, |b|=0$,\cr
          -b(a), & for $|a|=0\,, |b|=1$,\cr}}
and
\eqn\schbrtwo{
[a,b]_S(x)~=~[a,b(x)]_S+(-1)^{|b|-1}[a(x),b]_S\,,\quad x\in\cR\,,}
          for all other $|a|,|b|$.

 In particular, it is straightforward to check that if $|a| >0$
and $x\in\cR$,
\eqn\jab{
[a,x]_S = a(x).
}

\thm\bvschal
\proclaim Theorem \bvschal. The algebra
$(\cP(\cR),\,\cdot\,,[-.-]_S)$ of polyderivations of an Abelian
algebra $\cR$, with the dot product \prodpol\ and the Schouten
bracket \schbrack\ is a G-algebra.

\subsubsec{Example: the BV-algebra of polyvectors on a free algebra, $\cC_N$}
\subsubseclab\SBVpolder

 To illustrate these ideas, we discuss in some detail a simple
example of a BV-algebra.  It can be considered as a model for more complicated
examples in the following sections.
\smallskip

Let $\cC_N\cong\CC\,[ x^1,\ldots, x^N]$ be a free Abelian
algebra.  It is straightforward to verify that the G-algebra of
polyderivations $\cP(\cC_N)$, constructed above, is nothing but the
algebra of polynomial polyvector fields on $\CC^N$, \ie,
\eqn\polasder{\cP(\cC_N)~\cong~ \bigoplus_{n=0}^N
\bigwedge{}^{\!n}\, \cD(\cC_N)\,.}
More explicitly, it is a free $\ZZ_2$-graded algebra with the
even generators $ x^1\,,\ldots, x^N\in\cC_N$ and the odd generators
$ x_1^*\,,\ldots, x^*_N\in \cD(\cC_N)$, where
\eqn\notforget{
x_i^*( x^j)~=~\de^j_i\,,\quad i,j=1,\cdots,N\,.}
Let
$\Ph=\Ph^{i_1\ldots i_m} x^*_{i_1}\cdots  x^*_{i_m} $. Then
\eqn\coeff{
\Ph^{i_1\ldots i_m}~=~{(-1)^{m(m-1)/2}}
\,\hbox{$1\over \LW{m\,!}$}\,\Ph( x^{i_1},\ldots, x^{i_m})\,.}
Given $\Psi=\Ps^{j_1\ldots j_n} x^*_{j_1}\cdots x^*_{j_n}$, the
product $\Ph\cdot\Psi$ is just the wedge product in the exterior
algebra \polasder,
\eqn\product{
\Ph\cdot\Psi~=~
(\Ph^{i_1\ldots i_m}\Ps^{j_1\ldots j_n}) x^*_{i_1}\ldots x^*_{i_m}
x^*_{j_1}\ldots x^*_{j_n}\,,} while the Schouten bracket is
the extension of the usual bracket on vector fields,
\eqn\bracket{\eqalign{
[\Ph,\Psi]_S ~=~
&\sum_{k=1}^m (-1)^{m+k}\Ph^{i_1\ldots i_m} x^*_{i_k}(\Psi^{j_1\ldots j_n})
 x_{i_1}^*\ldots\widehat x{}_{i_k}^*\ldots x_{i_m}^* x^*_{j_1}\ldots
 x_{j_n}^*\cr
&-(-1)^{(m-1)(n-1)}
\sum_{k=1}^n (-1)^{n+k}\Ps^{j_1\ldots j_n} x^*_{j_k}(\Ph^{i_1\ldots i_m})
 x^*_{j_1}\ldots\widehat x{}_{j_k}^*\ldots  x_{j_n}^*
 x^*_{i_1}\ldots x_{i_m}^*\,.\cr}}

\smallskip
Let $\imath(x)$, $x\in\cC_N$, be the usual evaluation operator,
$\imath(x):\cP^n(\cC_N)\rightarrow\cP^{n-1}(\cC_N)$, defined by
\eqn\evop{
\imath(x)\,\Ph(x_1,\ldots,x_{n-1}) ~=~ \Ph(x,x_1,\ldots,x_{n-1})\,,\quad
x_1,\ldots,x_{n-1}\in\cC_N\,.}
By the definition of the bracket we also have $\imath(x)=[\,-,x\,]_S$.
Note that these two forms
of the evaluation operator are always well-defined for any polyderivation
algebra $\cP(\cR)$.  In addition, for $x =  x^i$, we obtain
a representation specific to $\cP(\cC_N)$ -- by expansion in
the dual basis we find $\imath( x^i)={ \RSr\partial\over
\lower 1.5pt\hbox{$\scriptstyle \partial x^*_i$}}$,
\ie, it acts like  the right derivative with respect to
the Grassmann  variable $ x^*_i$.
\smallskip

Consider an operator
\eqn\BVoppl{
\De ~=~ -\hbox{${\RSr\partial\over\LWrr{\partial x^i}}{\RSr\partial\over
\LWrr{\partial x^*_i}}$}
\,.}
This operator, $\De$, is a second order derivation on $\cP(\cC_N)$,
and, by direct calculation, we also verify that in fact it turns
$\cP(\cC_N)$ into  a BV-algebra.

\thm\bvpolvs
\proclaim Theorem \bvpolvs\  [\Ko,\Wia].
The polyvector algebra $(\cP(\cC_N),\,\cdot\,,\De)$ is a
BV-algebra. The bracket induced by $\De$ is equal to the Schouten
bracket \bracket.
\smallskip

Finally, let us note that there is a canonical polyvector
of maximal degree, the volume form,
\eqn\eqJMvol{
\Om ~=~ \hbox{$1\over N\,!$}
  \ep^{i_1\ldots i_N} x^*_{i_1}\cdots x^*_{i_N}\,,
}
and that $(\cP(\cC_N)\,,\,\cdot\,,\De)$, as a BV-algebra,
is generated by $ x^1\,,\ldots, x^N$ and $\Om$.

\subsubsec{Algebra of polyderivations associated with a BV-algebra }
\subsubseclab\SSpolyder

For any BV-algebra $(\fA,\,\cdot\,,\De)$, the subspace $\fA^0$ is an
Abelian algebra with respect to the dot product. Thus, as we have just
seen, there is a naturally associated G-algebra; namely, $\cP(\fA^0)$,
the $G$-algebra of polyderivations with the Schouten bracket
$[-,-]_S$.  We now study the relation between $\cP(\fA^0)$ and the
$G$-algebra of $\fA$ with the $[-,-]$ induced by $\De$.

Assume that $\fA$ has components
of only nonnegative degree, \ie, $\fA=\bigoplus_{n\geq 0}\fA^n$. Then
there is a natural map, $\pi:\fA\rightarrow
\cP(\fA^0)$, which is defined by induction on the degree. For
$n=0$, $\pi$ is the identity map, \ie, $\pi(x)=x$ for $x\in\fA^0$. It
is then extended to $n>0$ using the condition
\eqn\defpi{
\pi(a)(x) ~=~ \pi([a,x])\,,}
for any $a\in\fA^n$ and $x\in\fA^0$.  It is easy to verify using the
properties of the bracket that $\pi(a)$ is indeed a polyderivation of
degree $n$ for all $a\in\fA^n$.

\thm\gmorphism
\proclaim Theorem \gmorphism\ [\LZbv].
Suppose $\fA^n =  0$ for $n<0$, then the map $\pi:\fA\rightarrow
\cP(\fA^0)$  is a G-algebra homomorphism between
$(\fA,\,\cdot\,,[-,-])$ and $(\cP(\fA^0),\,\cdot\,,[-,-]_S)$.
\smallskip

\proof One must show that $\pi$ is a homomorphism with respect to the
dot product and the bracket. Both follow easily by induction on the
sum of degrees of $a$ and $b$ in \BVleib\ and \schbrack-\schbrtwo.
Indeed, in the case of the product  we
have $\pi(a\cdot b)=\pi(a)\cdot\pi(b)$ for
$|a|=|b|=0$. For $|a|+|b|>0$, we find, using \defpi,
properties of the bracket,  and the
induction hypothesis, that
\eqn\indprod{
\eqalign{\pi(a\cdot b)(x)&~=~\pi([a\cdot b,x])\cr
&~=~\pi( a\cdot [b,x]) + (-1)^{|b|}\pi([a,x]\cdot b)\cr
&~=~\pi(a)\cdot \pi(b)(x)+(-1)^{|b|}\pi(a)(x)\cdot \pi(b)\cr
&~=~(\pi(a)\cdot\pi(b))(x)\,.\cr}}
In the case of the bracket, we first note that \defpi\ is equivalent
to
\eqn\fststep{
[\pi(a),\pi(b)]_S~=~\pi([a,b])\,\quad {\rm for} \quad |a|+|b|=1\,.}
Then the general step of the induction is shown similarly as above,
\eqn\genind{\eqalign{
\pi([a,b])(x)&~=~\pi([[a,b],x])\cr
&~=~\pi([a,[b,x]]) + (-1)^{|b|-1}\pi([a,x],b])\cr
&~=~[\pi(a),\pi(b)(x)]_S+(-1)^{|b|-1}[\pi(a)(x),\pi(b)]_S\cr
&~=~[\pi(a),\pi(b)]_S(x)\,,\cr}}
for all other $|a|,|b|$. \SMu

\smallskip

In the following we will consider BV-algebras for which $\pi$ is an
epimorphism, \ie, $\pi(\fA)=\cP(\fA^0)$, and, in addition, $\cP(\fA^0)$
is itself a BV-algebra with a BV-operator $\De_S$. The problem then is
to find a convenient criterium to determine whether $\pi$ is in fact a
homomorphism of BV-algebras.  This will often be answered by the
following result.

\thm\bvhomscr
\proclaim Theorem \bvhomscr. Suppose that the $G$-algebra
$\cP(\fA^0)$ admits a BV-structure $(\cP(\fA^0),\,\cdot\,,\De_S)$ and
that
\eqn\condforpi{\pi\De(a)=\De_S\pi(a)\,,}
 for all $ a\in\fA^1$.  Then $\pi$ is a homomorphism of BV-algebras.

\smallskip
\proof Recall that $\fA^n=0$ for $n<0$. Then
\condforpi\ is obviously satisfied for any $a\in
\fA^0$, as both sides must  vanish. The general case, with
$a\in\fA^n$, is proved by induction. Indeed, for any $x \in \fA^0$,
\eqn\gencasebv{\eqalign{
\pi(\De a)(x)&=\pi([\De a,x])\cr
&=\pi(\De[a,x])\cr
&=\De_S(\pi([a,x]))\cr
&=\De_S([\pi(a),x]_S)\cr
&=[\De_S\pi(a),x]_S\cr
&=\De_S\pi(a)(x)\,,\cr
}}
where we have used that $\De$ and $\De_S$ act as derivations on
$[-,-]$ and $[-,-]_S$, respectively. \SMu

\subsec{G-modules and BV-modules}
\subseclab\Smodules

The notion of a  G-module (BV-module) of a G-algebra (BV-algebra)
can be introduced by  generalizing the dot and the bracket
action (BV-operator) on the algebra itself.

\thm\gmodule
\proclaim Definition \gmodule. Let $(\fA,\,\cdot\,,[-,-])$ be a
G-algebra and $\fM=\bigoplus_{n\in\ZZ} \fM^n$ a graded module of
$\fA$. We call the action of the algebra $\fA$ on $\fM$ the
dot action, and thus call $\fM$ a dot algebra module of $\fA$.  Then
$\fM$ is a G-module of $\fA$ if there further exists a bracket map,
$$[-,-]_M:\fA^m\times \fM^n\longrightarrow \fM^{m+n-1}\,,$$
such that
\eqn\gmAA{
[\,a\cdot b,m\,]_M=a\cdot [\,b,m\,]_M+(-1)^{|a||b|}b\cdot [\,a,m\,]_M\,,}
\eqn\gmAB{
[\,a,b\cdot m\,]_M=[\,a,b\,]\cdot m +(-1)^{(|a|-1)|b|} b\cdot[\,a,m\,]_M\,,}
\eqn\gmAC{
[\,[\,a,b\,],m\,]_M=[\,a,[\,b,m\,]_M]_M
-(-1)^{(|a|-1)(|b|-1)} [\,b,[\,a,m\,]_M]_M\,.}
for all $a,b\in \fA$ and $m\in\fM$.
\smallskip

\noindent
{\it Remark:} Relations \gmAB\ and \gmAC\ may be interpreted
as the statement that the operators $[a,-]_M$,
$a\in\fA$, define a representation of the graded Lie algebra
$(\fA,[-,-])$, which acts as a graded derivation of the dot action of
$\fA$ on $\fM$.
\smallskip

\thm\defbvmod
\proclaim Definition \defbvmod. Let $(\fA,\,\cdot\,,\De)$ be a BV-algebra and
$\fM=\bigoplus_{n\in\ZZ} \fM^n$ a graded module of $\fA$ as a
dot algebra. Then $\fM$ is a BV-module of $\fA$ if there exists a map
$$\De_M:\fM^n\longrightarrow \fM^{n-1}\,,$$
such that $\De_M^2=0$ and for any $a,b\in \fA$ and $m\in\fM$,
\eqn\BVscormod{\eqalign{
\De_M(a \cdot b\cdot m)& = \De(a\cdot b)\cdot m +(-1)^{|a|}
a\cdot \De_M(b\cdot m)+(-1)^{(|a|-1)|b|}b\cdot \De_M(a\cdot m)\cr
&-(\De a)\cdot b\cdot m - (-1)^{|a|}a\cdot (\De b)\cdot m -
(-1)^{|a|+|b|}a\cdot b\cdot\De_M(m)\,.\cr}}
\smallskip

Clearly, a BV-module $\fM$ of a BV-algebra $\fA$ is also a G-module of
$\fA$ with the bracket defined by
\eqn\gmAE{
[a,m]_M=(-1)^{|a|} \big( \De_M(a\cdot m)-(\De a)\cdot m -(-1)^{|a|}
a\cdot(\De_M m)\big)\,,\quad  a\in \fA\,,\, m\in\fM\,,}
which measures to what extent $\De_M$ fails to be a derivation
of the dot action of $\fA$ on $\fM$.
\smallskip

Free modules on one generator, $\om$, provide simplest examples of
G-modules and/or BV-modules. They are spanned by expressions of
the form
\eqn\gmAF{
a_1\cdot[a_2,[\,\ldots\, [a_{n-1},a_n\cdot\om]_M\,\ldots\,]_M]_M\,,
\quad a_1,\ldots,a_n\in\fA\,,}
and
\eqn\gmAG{
a_1\cdot\De_M(a_2\cdot\De_M(\,\ldots \,
\De_M(a_n\cdot \om)\,\ldots\,)\,)\,,
\quad a_1,\ldots,a_n\in\fA\,,}
subject to the defining relations \gmAA-\gmAC\ and \BVscormod, respectively.
\smallskip

\subsubsec{Natural G-modules for the G-algebra $(\cP(\cR),\,\cdot\,,[-,-]_S)$}
\subsubseclab\SSpolydermod

For the G-algebra of polyderivations $(\cP(\cR),\,\cdot\,,[-,-]_S)$ of
an Abelian algebra $\cR$,  a natural class of G-modules  consists of
polyderivations $\cP(\cR,M)$, where $M$ is a suitable module of $\cR$.

\thm\gmodder{
\proclaim Theorem \gmodder. Suppose that $M$ is a module of $\cR$ on which
the Lie algebra $\cD(\cR)$  acts by derivations of the dot product
action of $\cR$, \ie, the representation $a\mapsto K_a$, $a\in \cD(\cR)$,
satisfies (see Section \SSdefini)
\eqn\gmBA{
K_a(K_b(m))-K_b(K_a(m))~=~K_{[a,b]_S}(m)\,, \quad
a,b\in \cD(\cR)\,, m\in M\,,}
\eqn\gmBB{
K_a(x\cdot m)~=~a(x)\cdot m + x\cdot K_a(m)\,,\quad
a\in\cD(\cR)\,, x\in \cR\,, m\in M\,.}
Then the space of polyderivations $\cP(\cR,M)$ naturally has the structure
of a G-module of  $(\cP(\cR),\,\cdot\,,[-,-]_S)$.
\smallskip

\proof The proof parallels that of Theorem \bvschal. The module
structure with respect to the dot product is defined by
\prodex, with $a\in \cP(\cR)$ and $b\in\cP(\cR,M)$. The bracket is
constructed by induction setting
\eqn\gmBR{\eqalign{
[a,m]_M & ~=~ 0\,,\quad a\in \cR\,,\,m\in M\,,\cr
[a,m]_M & ~=~ K_a(m)\,, \quad a\in \cP^1(\cR)\,,\,m\in M\,,\cr
[a,m]_M & ~=~ -m(a)\,, \quad a\in \cR\,, \, m\in \cP^1(\cR,M)\,,\cr}}
and then using
\eqn\gmBQ{
[a,m]_M(x) ~=~ [a,m(x)]_M+(-1)^{|m|-1}[a(x),m]_M\,,\quad  x\in\cR\,,}
for $|a|+|m|\geq 2$. \SMu
\smallskip

\noindent
{\it Remark:} We have assumed implicitly that the grading on
$\cP(\cR,M)$ as a $G$-module is the same as the degree of
polyderivations, \ie, $\fM^n=\cP^n(\cR,M)$. If we shift the grading of
the module by taking $\fM^n\rightarrow \fM^{n+k}$, an obvious
modification of the construction above equips $\cP(\cR,M)$ with
another $\cP(\cR)$ G-module structure. Clearly, all structures with
$k$ respectively even or odd are equivalent, so it makes sense to talk
about $\cP(\cR,M)$ and as an ``even'' or ``odd'' G-module of
$\cP(\cR)$.

\subsec{The BV-algebra of polyderivations of the ground ring algebra
$\cG_N$}
\subseclab\bvpolyderivations

In this section we construct explicitly the BV-algebra of
polyderivations of an Abelian algebra which is not free, but whose
generators satisfy a single quadratic relation.
\smallskip

\subsubsec{The ``ground ring'' algebra $\cG_N$}
\subsubseclab\SSgrrinal

Consider the Abelian algebra $\cG_N=\cC_{2N}/\cI$, where $\cI$ is the
ideal generated by the vanishing relation
\eqn\vanishideal{
h_{ij}\, x^i\cdot  x^j ~=~ 0\,,}
where the metric is
\eqn\metric{
(h_{ij}) ~=~ \left(\matrix{\bf 0 & \bf 1\cr \bf 1 & \bf 0\cr}
\right)\,,\quad i,j=1,\ldots, 2N \,.}
In the following this metric will be used to raise and lower
indices, \eg, we will write $ x_i=h_{ij} x^j$.
Denote by $p$ the projection $p:\cC_{2N}\rightarrow\cG_N$.
If no confusion can arise, we will write  $ x^i$ both for
a generator $ x^i$ in $\cC_{2N}$ as well as its
image $p( x^i)$ in $\cG_N$, and omit the dot in the products.
\smallskip

For $N=3$ the algebra $\cG_3$ is isomorphic with the ground ring
algebra in Section \SSground.  We will therefore refer to $\cG_N$ as a
ground ring algebra or simply a ground ring.
\smallskip

The free algebra $\cC_{2N}$ carries a natural action of the
Lie algebra of $\sotwon$ realized by the first order derivations
\eqn\sosixder{
\La_{ij} ~=~ x_i x^*_j- x_j x^*_i\,,\quad i,j=1,\ldots,2N\,.}
Clearly, $\La_{ij}(\cI)\subset \cI$, so the action of
$\sotwon$ descends to the ground ring $\cG_N$, with the generators
$ x^i$ transforming in the vector ($2N$-dimensional) representation.
Let us define a $\ZZ$-grading of $\cG_N$, the so-called $\cG$-degree,
by declaring an element of $\cG$-degree $m$ to be the sum of products of
precisely $m$ generators $x^i$, and let $\cG^m_N$ denote the
subspace of $\cG_N$ of elements of $\cG$-degree $m$.

Since the dot product in $\cG_N$ is commutative, and the constraint
\vanishideal\ merely amounts to subtracting the trace, we also
find the following result for the structure of the entire ground ring.
\thm\strgrr
\proclaim Theorem \strgrr.
Each $\cG^m_N$ has a basis consisting of elements of the form
\eqn\basisss{
P^{i_1\ldots i_m}~=~ x^{(i_1}\cdots x^{i_m)} \,,\quad m\geq 0\,.}
In other words $\cG^m_N$ is an irreducible finite dimensional module
of $\sotwon$, isomorphic with the space of completely symmetric
traceless $\sotwon$ tensors of rank $m$.  Thus,
the ground ring $\cG_N$ decomposes as a direct sum
of irreducible finite dimensional modules of $\sotwon$ as follows
\eqn\decgrr{
\cG_N ~=~ \bigoplus_{m=0}^\infty \cG^m_N\,,}
where each $\cG^m_N$ arises precisely once.
\smallskip

\subsubsec{A ``hidden symmetry'' of $\cG_N$}
\subsubseclab\hiddensymmetry

The ground ring $\cG_N$ acts on itself by left multiplication. Let
us denote by $ x^i$ both the generator and the corresponding
multiplication operator acting on the ground ring. The natural
problem is then to determine the Lie algebra of transformations of
$\cG_N$ which includes the multiplication operators together with the
$\sotwon$ symmetry generators  $\La_{ij}$.

\thm\sotwonpluss
\proclaim Theorem \sotwonpluss. The ground ring $\cG_N$ is  an
irreducible module of $\sotwontwo$. The explicit realization of the
$\sotwontwo$ generators is given by the following differential
operators on $\cG_N$:
\eqn\sotwonnn{
\eqalign{
M_i& ~=~ x_i\,,\cr
\La_{ij}&  ~=~ x_i\pxiu j - x_j \pxiu i\,,\cr
U_i& ~=~ (N-1)\pxiu i + x^j\pxiu j \pxiu i -\half  x_i
\pxiu j \pxid j\,,\cr
U& ~=~ x^i\pxiu i +(N-1)\,.\cr}}
\smallskip

\proof One verifies by straightforward algebra that the operators
in \sotwonnn\ preserve the constraint
\vanishideal, and thus are well defined  on $\cG_N$. Similarly, one
finds that they satisfy the commutation relations of the $\sotwontwo$
algebra, \eg, $[U_i,M_j]=-\La_{ij}+h_{ij}\,U$. \SMs
\smallskip

Theorem \sotwonpluss\ was first proved in [\BdFl] for $N=3$, and then
generalized in [\GeZl].

\subsubsec{Polyderivations of $\cG_N$}
\subsubseclab\polyderofcgn

A polyderivation $\Ph\in\cP^n(\cG_N)$ is completely determined by its
value on the ground ring generators, \ie, we have a natural injection
from $\Ph\in\cP^n(\cG_N)$ into the space  of multilinear alternating
maps $\End(\bigwedge{}^{\!n}\,\cG_N^1,\cG_N)$. The problem of determining
all polyderivations of $\cR_N$ thus amounts to identifying
which elements $\Ph$ in $\End(\bigwedge{}^{\!n}\,\cG_N^1,\cG_N\vps)$ are in
the image of this injection. This is resolved by the following
criterium.

\thm\poldercond
\proclaim Theorem \poldercond. An endomorphism
$\Ph\in\End(\,\bigwedge{}^{\!n}\,\cG_N^1,\cG_N\vps)$
determines a polyderivation of $\cG_N$  iff
\eqn\conscond{
\frde\,  \Ph ~=~ 0\,,}
where $\frde:\End(\,\bigwedge{}^{\!n}\,\cG_N^1,\cG_N\vps)\rightarrow
\End(\,\bigwedge{}^{\!n-1}\,\cG_N^1,\cG_N\vps)$ is the operator
$\frde= x_i  \imath( x^i)$.
\smallskip

We may also express \conscond\ more explicitly by expanding
$\Ph$ in the dual basis,
\eqn\formder{
\Ph ~=~ \Ph^{i_1\ldots i_n} x^*_{i_1}\ldots x^*_{i_n}\,,\quad
\Ph^{i_1\ldots i_n}\in\cR_N\,. }
Then we have

\proclaim Theorem \poldercond${}^\prime$. An endomorphism
$\Ph$ is a
polyderivation iff the coefficients of its expansion \formder\ satisfy
\eqn\realcond{
 x_i\cdot\Ph^{i\,i_1\ldots i_{n-1}} ~=~ 0\,,
\quad i_{1},\ldots,{i_{n-1}}=1,\ldots,2N\,.}
\smallskip

\proof If $\Ph$ is a polyderivation, the Leibniz rule yields
\eqn\prfcond{
\Ph( x_i\cdot x^i, x^{i_1},\ldots, x^{i_{n-1}}) ~=~
2 x_i\cdot\Ph( x^i, x^{i_1},\ldots, x^{i_{n-1}}) ~=~0\,.}

To extend an endomorphism $\Ph\in\End(\,\bigwedge
{}^n\cG_N^1,\cG_N\vps)$ to
the ground ring we may assume that it acts as a
derivation on the products of generators in $\cG_N$.  The conditions
\conscond, or equivalently
\realcond, guarantee then that if we evaluate $\Ph$ on arbitrary
elements of the ground ring, the final result does not depend on the
particular representation of those elements as linear combinations of
products of the generators. Also, essentially by construction,
the resulting $\Ph$ is a polyderivation, see Lemma \derivas. \SMs
\smallskip

The second part of the proof may also be rephrased as follows: First
lift $\Ph$ to a homomorphism $\widetilde \Ph$ in the ``covering
algebra'' $\cC_{2N}$, by choosing arbitrary elements
$\widetilde\Ph^{i_1\ldots i_n}$ that project onto $\Ph^{i_1\ldots
i_n}$ and set $\widetilde\Ph=\widetilde\Ph{}^{i_1\ldots i_n} x^*_{i_1}
\ldots  x^*_{i_n}$. Obviously $\widetilde \Ph$ is a
polyderivation of $\cC_{2N}$. The condition \realcond\ allows us
to first project $\widetilde\Ph$ to a polyderivation of $\cG_N$, and
then to show that the resulting polyderivation does not depend
on the choice of  lift $\Phi\rightarrow \widetilde\Ph$.
\smallskip

Note that if we consider $\End(\,\bigwedge{}^{\!*}\,\cG_N^1,\cG_N\vps)$
as a graded commutative algebra with the product induced
from the exterior product
in $\bigwedge{}^{\!*}\,\cG_N^1$ and the product in $\cG_N$, the subspace
${\rm ker}\,\frde$ is an ideal, which shows that the identification
of polyderivations $\cP(\cG_N)$ with ${\rm ker}\,\frde$ is in fact
an isomorphism of algebras.
\smallskip

Now we would like to solve the constraint \conscond\ and determine
explicitly all the polyderivations $\cP(\cG_N)$. Since
$ x^i\,\cG^m_N\subset \cG^{m+1}_N$, it is enough to consider
endomorphisms taking values in a subspace of fixed $\cR$-degree. Given
the basis \basisss\ in $\cG_N$, we may choose endomorphisms
\eqn\endbasis{
P_{i_1\ldots i_m}\, x^*_{j_1}\ldots x^*_{j_n}\,,
\quad i_1,\ldots, j_n=1,\ldots,2N\,,}
as the basis in $\cE^n_m=\End(\bigwedge{}^{\!n}\,\cG_N^1,\cG_N^m)$.
The action of $\sotwon$ on the ground ring in \sosixder\
extends to $\cE^n_m$, which, as an $\sotwon$ module,  is then
isomorphic with the tensor product of two $\sotwon$ modules;
the first one corresponds to completely symmetric traceless
tensors of rank $m$, and the second one to antisymmetric tensors
of rank $n$. Since  the operator $\frde:\cE^n_m\rightarrow
\cE^{n-1}_{m+1}$ commutes with the action of $\sotwon$, it can
only map between the same irreducible modules in the decomposition
of $\cE^n_m$ and $\cE^{n-1}_{m+1}$.
\smallskip

Recall that tensor representations of $\sotwon$ can be conveniently
enumerated using Young tableaux. We
will only need a small class of ``hook like'' tableaux, with $m$ boxes
in the first row and one box in the subsequent $n$ rows. Let us denote
the corresponding representation of $\sotwon$ by $[m;n]$, $m, n\geq
0$. In particular, $[0;0]$ is the identity representation,
$[1;0]\simeq[0;1]$ the vector representation, $[m;0]$ corresponds to
the completely symmetric traceless tensors of rank $m$, while $[1;n]$
to the completely antisymmetric tensors of rank $n+1$. Since an
antisymmetric tensor of rank $n$ is equivalent (dual) to a tensor
of rank $2N-n$, we also have $[1;n]\simeq [1;2N-n-2]$.  The final
subtlety is that the representation $[m;N-1]$ is a direct sum of two
irreducible ones.
\smallskip

The product of a traceless symmetric tensor of rank $m$  and an
antisymmetric tensor of rank $n$ decomposes as
\eqn\tenspr{
[m;0]\,\otimes\, [1;n-1] ~=~ [m+1;n-1]\,\oplus\,[m;n]
\,\oplus\,[m-1;n-1]\,\oplus\,[m;n-2]\,,\quad m,n\geq 2\,.}
This can be derived in two steps. In the first we use the usual rule
for multiplying Young tableaux of $\gltwon$, and obtain the first two
terms on the right hand side but with no traces subtracted. In the
second step we subtract the traces (in fact just a single trace) from
the first and the second term, which yields the third and the fourth
term, respectively. The decomposition \tenspr\ is valid for generic
values of $m\geq 2$ and $2N>n\geq 2$.  The following are the special
cases, where some terms on the right hand side in the decomposition
above are not present:

\eqn\speccs{\eqalign{{\sl i.}&\cr
{\sl ii.}&\cr {\sl iii.}&\cr {\sl iv.}&\cr {\sl v.}&\cr}
\qquad
\eqalign{
 [0;0]\,&\otimes\,[1,n-1] ~=~ [1,n-1]\,,\cr
[1;0]\,&\otimes\,[1,n-1]~=~ [2;n-1]\,\oplus\,[1;n]\,\oplus\,[1;n-2]\,,\cr
 [m;0]\,&\otimes\,[1;0]~=~[m+1;0]\,\oplus\,[m;1]\,\oplus\,[m-1,0]\,,\cr
[m;0]\,&\otimes\,[1;2N-2]~=~[m+1;2N-2]\,\oplus\,[m;2N-1]\,\oplus\,
[m;2N-3]\,,\cr
[m;0]\,&\otimes\,[1;2N-1]~=~[m+1;2N-1]\,.\cr}}
\smallskip

Now, let us go back to \conscond. To illustrate the method, we first
consider some of the exceptional cases.  It is clear from \realcond\
that there can be no polyderivation with $m=0$, so the simplest
nontrivial case is that of $m=1$ and $n=1$. The decomposition of
$\cE^1_1$ with respect to the $\sotwon$ action yields a direct sum of
three modules (see \speccs\ iii), spanned by $S_{i,j}$, $P_{i,j}$ and
$C$, $i,j=1\ldots,2N$, respectively, where%
\foot{Here and in the
following $\scriptstyle (\,\cdots\,)$ and $\scriptstyle [\,\cdots\,]$
denote the symmetrization and the antisymmetrization, respectively,
both normalized with strength one; \ie, for a completely symmetric
tensor $s_{(i_1\ldots i_n)}=s_{i_1\ldots i_n}$, and for a completely
antisymmetric tensor $a_{[i_1\ldots i_n]}=a_{i_1\ldots i_n}$.}
\eqn\basione{
S_{i,j} ~=~ x_{(i}\vps x^*_{j)} \,,\quad
  P_{i,j} ~=~ x_{[i}\vps x^*_{j]}\,,
  \quad C ~=~ x^i x^*_i\,.}
We find
\eqn\actionone{
\frde S_{i,j} ~=~ x_{(i} x_{j)}\,,\quad
  \frde P_{i,j} ~=~ 0\,, \quad
  \frde\, C ~=~ 0 \,,}
which shows that the space of polyderivations $\cP^1_1(\cG_N)$ is spanned
by $P_{i,j}$ and $C$. Note that $2P_{i,j}=\La_{ij}$, and thus we have
rederived the $\sotwon$ symmetry generators.
\smallskip

For $\frde : \cE^n_m \to \cE^{n-1}_{m+1}$ with
$m=1$ and $n\geq 2$, the following decompositions are relevant
\eqn\firstdec{
\eqalign{\cE^n_1\,&:\cr \cE^{n-1}_2\,&:\cr}
\qquad
\eqalign{& [2;n-1]\,\oplus\,[1;n]\,\oplus\,[1,n-2]\,,\cr
&
[3;n-2]\,\oplus\,[2;n-1]\,\oplus\,[2,n-3]\,\oplus\,[1;n-2]\,.\cr}}
By comparing the two decompositions, we conclude that the $[1;n]$ submodule
of $\cE^1_1$ must  lie in ${\rm ker}\,\frde$. Indeed,  $[1;n]$
is spanned by endomorphisms of the form
\eqn\genasymtns{
P_{i_1,j_1\ldots j_{n}} ~=~ x_{[i_1}\vps x^*_{j_1}\ldots x^*_{j_{n}]}\,,}
for which
\eqn\actonp{
\frde P_{i_1,j_1\ldots j_{n}} ~=~ \,\sum_{k=1}^n (-1)^{k+1}
 x_{[i_1}\vps x^*_{j_1}\ldots  x_{j_k}^* \ldots
 x^*_{j_{n}]} ~=~ 0\,.}
We also have
\eqn\genact{
\frde \, x_{i_1}\vps x^*_{j_1}\ldots x^*_{j_{n}} ~=~
  n\, x_{i_1}\vps x_{[j_1}\vps x^*_{j_2}\ldots x^*_{j_n]}\,.}
By decomposing both sides into traceless and trace components,
we see that $\frde$ has a nontrivial image in
both the $[2;n-1]$ and $[1;n-2]$ submodules of $\cE^{n-1}_2$,
and thus \genasymtns\ exhaust all polyderivations in this case.
\smallskip

The case $m\geq 2$ and $n=1$ is similar in that there is only one trace
in the decomposition of the tensor product. However, since
\eqn\gendec{
\eqalign{\cE^1_m\,&:\cr \cE^{0}_{m+1}\,&:\cr}
\qquad
\eqalign{& [m+1;0]\,\oplus\,[m;1]\,\oplus\,[m-1;0]\,,\cr
&
[m+1,0]\,,\cr}}
we find two $\sotwon$ representations in the decomposition
of $\cP^1_m$. The corresponding basis is given by
\eqn\basisextwo{
P_{i_1\ldots i_m,j} ~=~ x_{i_1}\vps\ldots  x_{i_{m-1}}\vps
 x_{[i_m}\vps x^*_{j]} + \cdots \,,}
and
\eqn\baseistwo{
C_{i_1\ldots i_{m-1}} ~=~ x_{i_1}\ldots  x_{i_{m-1}} C\,,}
where ``$\cdots$'' indicate explicit subtraction of trace
terms in $i_1,\ldots,i_m,j$.

In the generic case, for $m\geq 2$ and $2N-2\geq n\geq 2$, we have
\eqn\gendec{
\eqalign{\cE^n_m\,&:\cr \cE^{n-1}_{m+1}\,&:\cr}
\qquad
\eqalign{& [m+1;n-1]\,\oplus\,[m;n]\,\oplus\,[m-1;n-1]\,\oplus\,[m,n-2]\,,\cr
&
[m+2;n-2]\,\oplus\,[m+1;n-1]\,\oplus\,[m,n-2]\,\oplus\,[m+1;n-3]\,.\cr}}
The modules $[m;n]$ and $[m-1;n-1]$ lie in ${\rm ker}\,\frde$, with
a convenient basis given by
\eqn\genderbasone{
P_{i_1\ldots i_m,j_1\ldots j_n} ~=~ x_{i_1}\vps\ldots
  \, x_{i_{m-1}}\vps x_{[i_m}\vps x^*_{j_1}\ldots x^*_{j_{n}]}+ \cdots\,,}
and
\eqn\genderbastwo{
C_{i_1\ldots i_{m-1},j_1\ldots j_{n-1}} ~=~
  C\, P_{i_1\ldots i_{m-1},j_1\ldots j_{n-1}}\,,}
where all basis elements
\genderbasone\ and \genderbastwo\  are traceless in $i_1,\ldots,j_n$.
Since
\eqn\generalact{
\frde  x_{i_1}\vps\ldots  x_{i_m}\vps  x^*_{j_1}\ldots x^*_{j_{n}}
  ~=~ n\,  x_{i_1}\vps\ldots  x_{i_m}\vps  x_{[j_1}\vps x^*_{j_2}\ldots
  x^*_{j_{n}]}\,,}
we verify that \genderbastwo\ gives
all polyderivations $\cP^n_m(\cG_N)$. The explicit form
of the trace terms that must be subtracted on the right hand side in
\basisextwo\ and \genderbasone\ are given in Appendix \Apolder.
An equivalent, but more concise, expression will be also given in the
next section.
\smallskip

Although $[1;2N-2]\simeq[1,0]$, the  $n=2N-1$ case is quite different than
that with $n=1$.
For $m=1$ we find one solution, see \genasymtns,
\eqn\genasymtnsbig{
P_{i_1,j_1\ldots j_{2N-1}} ~=~ x_{[i_1}
  \vps x^*_{j_1}\ldots x^*_{j_{2N-1}]}
  ~=~\ep_{i_1j_1\ldots j_{2N-1}}  X\,.}
We will refer to $X$ as the ``volume element'' of $\cG_N$.  Explicitly,
\eqn\eqvolel{
X ~=~ \textstyle{ 1\over (2N)!} \ep^{i_1 i_2 \ldots i_{2N}}
  \, x_{i_1} x^*_{i_2} \ldots x^*_{i_{2N}}\,.}
For $m\geq 2$ we have
\eqn\gendec{
\eqalign{\cE^{2N-1}_m\,&:\cr \cE^{2N-2}_{m+1}\,&:\cr}
\qquad
\eqalign{& [m+1;2N-2]\,\oplus\,[m;2N-1]\,\oplus\,[m,2N-3]\,,\cr
&
[m+2;2N-3]\,\oplus\,[m+1;2N-2]\,\oplus\,[m,2N-3]\,\oplus\,[m+1;2N-4]\,.\cr}}
This leaves just one solution spanned by the elements, see  \genderbasone,
\eqn\bigcand{ x_{i_1}\vps
\ldots x_{i_{m-1}}\vps x_{[i_m}\vps x^*_{j_1}\ldots
 x^*_{j_{2N-1}]}\,.}
Using standard identities for $\sotwon$ tensors, we find
\eqn\schid{
 x_i\vps x_{[j_1}\vps x^*_{j_2}\ldots x^*_{j_{2N}]} ~=~
  -(2N-1)\,C\, h_{i[j_1} P_{j_2,j_3\ldots j_{2N}]}\,,}
which shows that only trace components are present in \bigcand.
Thus the basis in $\cP^{2N-1}_m$, $m\geq 2$, consists of elements
$C_{i_1\ldots i_{m-1},j_1\ldots j_{2N-2}}$ defined as
in \genderbastwo.
\smallskip

Finally, there is no solution for $n=2N$, which shows that the maximal
degree of a polyderivation of $\cG_N$ is equal to $2N-1$.
\smallskip

Let us extend the notation for the polyderivations in \genderbasone\
and \genderbastwo\ and set $P_{i_1\ldots i_m,j_1\ldots j_n}$ equal to
$P_{i_1\ldots i_m}$ for $n=0$, and  to $1$ for $m=n=0$. Similarly,
we set $C_{i_1\ldots i_m,j_1\ldots j_n}$ equal to
$C_{i_1\ldots i_m}$ for $n=0$, and  to $C$ for $m=n=0$.
We may now summarize the complete classification of the polyderivations
$\cP(\cG_N)$.

\tbl\tblpolyvec
\thm\polyvects
\proclaim Theorem \polyvects.
i. The space of polyderivations $\cP(\cG_N)$ is doubly graded,
\eqn\dblgrd{
\cP(\cG_N) ~=~ \bigoplus_{n=0}^{2N-1}\bigoplus_{m=0}^\infty
\cP_m^n(\cG_N)\,,}
\item{} by the
 degree $n$ of the derivation, $2N-1\geq  n\geq 0$, and the $\cG$-degree
$m$ of the coefficients in the ground ring, $m\geq 0$.  Depending on
$m$ and $n$ each of the subspaces $\cP_m^n(\cG_N)$ is a direct sum of
finite dimensional irreducible modules of $\sotwon$ which are listed
in Table \tblpolyvec.
\item{ii.} In  $\cP_m^n(\cG_N)$, $m,n\geq 0$,
the $[m;n]$ submodule is spanned by the
polyderivations $P_{i_1\ldots i_m,j_1\ldots j_n}$, while the
$[m-1;n-1]$ submodule -- by the polyderivations
$C_{i_1\ldots i_{m-1},j_1\ldots j_{n-1}}$. In cases where a given submodule
does not arise in the decomposition of $\cP_m^n(\cG_N)$, the corresponding
polyderivations $P_{i_1\ldots i_m,j_1\ldots j_n}$ and/or
$C_{i_1\ldots i_{m-1},j_1\ldots j_{n-1}}$ vanish.

\bigskip
\begintable
$\cP^n_m(\cG_N)$ |$n=0$ | $n=1$ | $2N-2\geq n\geq 2$ | $\quad n=2N-1\quad$ \cr
$m=0$ | $[0;0]$ | | | \cr
$m=1$ | $[1;0]$ | $[1;1]\,\oplus\,[0;0]$ |  $[1;n]$ | $[1;2N-1]$ \cr
$\quad m\geq 2\quad$ | $\quad[m;0]\quad$ |
$\quad[m;1]\,\oplus\,[m-1;0]\quad$ |
 $\quad[m;n]\,\oplus\,[m-1;n-1]\quad$ | $[m-1,2N-2]$
\endtable
\medskip
\centerline{Table \tblpolyvec. The decomposition of $\cP^n_m(\cG_N)$ into
$\sotwon$ modules.}
\bigskip

{}From the formulae for the basis in $\cP^n_m(\cG_N)$ we see that at
each degree $n$ there are polyderivations of $\cR$-degree $m=1$, which cannot
be obtained as products of polyderivations of lower degrees. The
question of how to  describe  explicitly $\cP(\cG_N)$ in terms of
generators and relations is then answered by the following theorem.

\thm\generat
\proclaim Theorem \generat. The graded, graded commutative
algebra $(\cP(\cG_N),\,\cdot\,)$ is  generated,
as a dot algebra, by $1$, the ground ring generators $ x^i$,
degree one derivation $C$, and degree $n-1$
polyderivations $P_{i_1\,,i_2\ldots i_n}$,  $n=2,\ldots, 2N$,
satisfying the relations:
\eqn\conone{  x_i x^i ~=~ 0\,,}
\eqn\contwo{ x_{[i}P_{i_1,i_2\ldots i_n]} ~=~ 0\,,}
\eqn\conthr{ x^i P_{i\,,j_1\ldots j_n}~=~ -\hbox{$n\over n+1$}CP_{j_1,j_2
\ldots j_n}\,,}
\eqn\confour{P_{i_1,i_2\ldots i_m}P_{j_1,j_2\ldots j_n} ~=~
(-1)^{m-1}\hbox{$m+n-1\over n$} x_{[i_1} P_{i_2,i_3\ldots i_m]j_1\ldots
j_n}\,,}
\eqn\confive{
C P_{i_1,i_2\ldots i_{2N}} ~=~ 0\,.}
\smallskip

\proof Clearly, the identities \conone-\confive\ are satisfied
in $\cP(\cG_N)$. This is easily verified using the explicit form of those
polyderivations in \basione\  and \genasymtns. On the other hand if we
consider the algebra generated by $ x^i$, $C$, and $P_{i_1,i_2\ldots
i_n}$, subject to relations \conone-\confour, all elements in this
algebra are linear  combinations of the products
\eqn\products{
 x_{i_1}\ldots x_{i_m}\,,\quad  x_{i_1}\ldots x_{i_m}C\,,\quad
 x_{i_1}\ldots x_{i_m}P_{j_1,j_2\ldots j_n}\,,\quad
 x_{i_1}\ldots x_{i_m}CP_{j_1,j_2\ldots j_n}\,.}
There is a natural action of the $\sotwon$ algebra on this space,
with respect to which the elements in \products\ transform as
$[m;0]$, $[m;0]$, $[m;0]\otimes[1;n-1]$ and $[m;0]\otimes[1;n-1]$,
respectively. Condition \contwo\ sets to zero the
$[m;n]$ and $[m-1;n-1]$ components in those tensor products, while
\conthr\ relates the trace component in the third product in
\products\ to the single nonvanishing component of the fourth
term in \products. This shows that the  elements of this space are
in one to one correspondence with the elements of $\cP(\cG_N)$, and,
in fact, establishes the required algebra isomorphism. \SMu

\subsubsec{The G-algebra structure of $\cP(\cG_N)$}
\subsubseclab\galgstr

The computation of the Schouten bracket, as defined in section
\SSpolyderaa, only involves evaluation of polyderivations on elements of
the algebra. Thus we may use similar arguments to those which led to
Theorem \poldercond\ to derive an explicit formula for the Schouten
bracket of two polyderivations.

\thm\expolbr
\proclaim Theorem \expolbr. Let
$\Ph=\Ph^{i_1\ldots i_m} x^*_{i_1}\cdots  x^*_{i_m} $ and
$\Psi=\Ps^{j_1\ldots j_n} x^*_{j_1}\cdots  x^*_{j_n}$,
$\Ph^{i_1\ldots i_m},\,\Psi^{j_1\ldots j_n}\in\cG_N$
 be two polyderivations. Then the Schouten bracket $[\Ph,\Psi]_S$ can be
computed explicitly as in \bracket, where we assume that $ x^*_{i_1},\ldots,
 x^*_{j_n}$ act as derivations on the products of ground ring generators.
\smallskip

\noindent
The following observation  is a simple consequence of the above
result.

\thm\homogen
\proclaim Theorem \homogen. The Schouten bracket on $\cP(\cG_N)$
is homogenous in both the degree and the $\cR$-degree,
namely
\eqn\brackmap{
[-,-]_S\,:\,\cP^{n_1}_{m_{1}}\times \cP^{n_2}_{m_{2}}
 ~\longrightarrow~ \cP^{n_1+n_2-1}_{m_{1}+m_2-1}\,.}
\smallskip

 We now explicitly calculate some fundamental brackets between certain
elements of the algebra, which will be required in the next section
where we determine the BV-operator underlying the Schouten
bracket. All of these results are obtained using Theorem \expolbr\ and the
explicit form of the polyderivations. First we have
\eqn\sophi{
[\La_{ij}, x_k]_S ~=~ h_{ik} x_j-h_{jk} x_i\,,}
which represents the  $\sotwon$ transformation of the ground ring
generator. More generally,
\eqn\sophigen{
[P_{i_1,i_2\ldots i_m}, x_i]_S ~=~ (-1)^{m-1}(n-1)h_{i[i_1}
P_{i_2,i_3\ldots i_m]}\,,}
as well as
\eqn\twopcom{
[P_{i_1,i_2\ldots i_m},P^{j_1,j_2\ldots j_n}]_S ~=~
(-1)^{m-1}(m+n-2) \de_{[i_1}{}^{[j_1} P_{i_2,i_3\ldots i_m]}
{}^{j_2\ldots j_n]}\,,\quad i,j\geq 1\,.}

\thm\commcphi
\proclaim Lemma \commcphi. For any $\Ph\in\cP^n_m(\cG_N)$,
\eqn\Cph{
[C,\Ph]_S ~=~ (m-n)\Ph\,,}
\smallskip

Using the Schouten bracket we can also write down explicitly the
decomposition of a product of two basis elements in $\cP(\cG_N)$
into its traceless and trace components.

\thm\genproduct
\proclaim Theorem \genproduct. For any $m,m'\geq 0$ and $n,n'\geq 1$,
\eqn\bigproduct{
\eqalign{
&P^{i_1\ldots i_mi_{m+1},i_{m+2}\ldots i_{m+n}}\, P_{j_1\ldots
  j_{m'}j_{m'+1},j_{m'+2}\ldots j_{m'+n'}} ~=~ \cr
&\qquad \hbox{$n+n'-1\over n'$}\,
  P^{i_1\ldots i_m}{}_{j_1\ldots j_{m'}}{}^{[i_{m+1}}{}_{j_{m'+1},}
  {}^{i_{m+2}\ldots i_{m+n}]}{}_{j_{m'+2}\ldots j_{m'+n'}}\cr
&\qquad\qquad +(-1)^n\,\hbox{$1\over 2N+m+m'-n-n'+2$}\,
  C\,[P^{i_1\ldots i_mi_{m+1},i_{m+2}\ldots i_{m+n}},
   P_{j_1\ldots j_{m'}j_{m'+1},j_{m'+2}\ldots j_{m'+n'}}]_S\,.\cr}}
Also, the bracket on the right hand side lies in the subspace spanned by
the $P$-type basis elements in $\cP_{m+m'-1}^{n+n'-1}$.

\smallskip

\proof The first term on the right hand side is determined so that the leading
terms on both sides agree, see \genderbasone. Then the second term
on the right
hand side must account for all the traces in the product, which
indeed is the case. This fact, as well as the second part of the
theorem, are shown by a straightforward (though somewhat lengthy)
calculation which has been outlined in Appendix \ASproof. \SMs

\subsubsec{The BV-algebra structure of $\cP(\cG_N)$}
\subsubseclab\bvpolyvectors

We will now construct a BV-operator $\De_S$ on $\cP(\cG_N)$,
whose bracket \eqCd\ coincides with the Schouten bracket.
Since the latter operation is both $\sotwon$ invariant as well
as homogenous with respect to both the degree and the $\cG$-degree,
we will seek a BV-operator which satisfies similar restrictions.

\thm\uniquebv
\proclaim Theorem \uniquebv. There exists at most one  BV-operator
$\De$ on $\cP(\cG_N)$ that is $\sotwon$ invariant,
homogenous of degree minus one, \ie,
$$\De\,:\,\cP_m^n(\cG_N) ~\longrightarrow~ \cP_{m-1}^{n-1}(\cG_N)\,,$$
and whose bracket $[-,-]$ coincides with the Schouten bracket
$[-,-]_S$.
\smallskip

\proof Let $\De_1$ and $\De_2$ be two such BV-operators. Then
their difference $D=\De_1-\De_2$ is an $\sotwon$ invariant first order
derivation on $\cP(\cG_N)$, and $D:\cP^n_m(\cG_N)
\rightarrow \cP_{m-1}^{n-1}(\cG_N)$. By examining the $\sotwon$
decomposition  of $\cP(\cG_N)$ given in Theorem \polyvects\ we
conclude that for any BV-operator $\De$ satisfying the assumptions
above
\eqn\firstvan{
\De  P_{i_1,i_2\ldots i_m}=0\,,\quad m\geq 1\,.}
Thus $D  P_{i_1,i_2\ldots i_m}=0$, and by
Theorem \generat\ and $D$ being a derivation, it follows that
$D$ is completely determined by its action on $C$.

Once more, since $\De x^i=0$ and $\De\La_{ij}=0$,  we find, using \sophi,
\eqn\firstone{
\De( x^i\La_{ij}) ~=~ [ x^i,\La_{ij}]_S ~=~ (2N-1) x_j\,.}
However,
\eqn\exxxc{
 x^i\La_{ij} ~=~ x^i( x_i\vps x^*_j- x_j\vps x^*_i)
  ~=~ - x_jC\,,}
so we also have
\eqn\seconex{
\De( x^i\La_{ij}) ~=~ -\De( x_jC) ~=~ -[ x_j,C]- x_j\De C\,.}
Comparing \firstone\ with \seconex, and using \Cph, we determine that
\eqn\resoneo{
\De C~=~-2(N-1) 1\,,\quad \De(C  x_i)~=~-(2N-1) x_i\,.}
This shows that $DC=0$, and concludes the proof of the theorem. \SMu

\thm\simplelemma
\proclaim Lemma \simplelemma. Let $\De$ be a BV-operator as in
Theorem \uniquebv, and $\Ph\in\cP^n_m(\cG_N)$ satisfies
$\De\Ph=0$. Then
\eqn\deloncph{
\De(C\Ph)~=~-(2N+m-n-2)\Ph\,.}
\smallskip

\proof Using \eqCc, \Cph\ and \resoneo, we obtain
\eqn\candph{
\De(C\Ph) ~=~ -[C,\Ph]+\De(C)\Ph ~=~-(m-n)\Ph -(2N-2)\Ph\,.}
\SMu
\smallskip

The main result of this section is the following explicit construction
of $\De_S$.

\thm\bvopmth
\proclaim Theorem \bvopmth.  There exists a  unique
BV-operator $\De_S$ on $\cP(\cG_N)$ that is $\sotwon$ invariant,
homogenous of degree minus one,
and whose bracket $[-,-]$ coincides with the Schouten bracket
$[-,-]_S$.  It is explicitly given by
\eqn\exformbv{
\De_S P_{i_1\ldots i_m,j_1\ldots j_n} ~=~ 0\,,\qquad
\De_S C_{i_1\ldots i_m,j_1\ldots j_n} ~=~ -(2N+m-n-2)
P_{i_1\ldots i_m,j_1\ldots j_n}\,.}
\smallskip

\proof First we want to
argue that a BV-operator  $\De_S$ satisfying the assumptions
of the theorem must be of the form \exformbv. Similarly as
in the proof of Theorem \uniquebv, the $\sotwon$ invariance
restricts $\De_S$ to
\eqn\morerestr{
\De_S P_{i_1\ldots i_m,j_1\ldots j_n}=0\,,\qquad
\De_S C_{i_1\ldots i_m,j_1\ldots j_n}= \la(m,n)
P_{i_1\ldots i_m,j_1\ldots j_n}\,,}
where $\la(m,n)$ are some arbitrary numbers to be determined.
However, since $C_{i_1\ldots i_m,j_1\ldots j_n}=C\,
P_{i_1\ldots i_m,j_1\ldots j_n}$ and $\De_S\,P_{i_1\ldots i_m,j_1\ldots
j_n}=0$, the second part of \exformbv\ follows then from Lemma
\simplelemma.

Clearly $\De_S^2=0$, so to complete the proof we must show that the
bracket of $\De_S$ coincides with the Schouten bracket, as the second
order derivation property of $\De_S$ will then follow from Theorem
\triplec. The equality between the bracket of $\De_S$ and the Schouten
bracket is demonstrated by explicit computation.  There are three
cases: the bracket of two $P$'s, of a $P$ and a $C$, and of two $C$'s.
For the first we may simply use the general formula for the product of
two $P$'s given in Theorem \genproduct. Indeed, by acting with $\De_S$
on both sides of \bigproduct\ we find
\eqn\firstcomm{\eqalign{
\De_S(P_{i_1\ldots i_mi_{m+1},i_{m+2}\ldots i_{m+n}}\,& P_{j_1\ldots
j_{m'}j_{m'+1},j_{m'+2}\ldots j_{m'+n'}}) \cr &=(-1)^{n-1}
[P_{i_1\ldots i_mi_{m+1},i_{m+2}\ldots i_{m+n}},
 P_{j_1\ldots j_{m'}j_{m'+1},j_{m'+2}\ldots j_{m'+n'}}]_S\,.\cr}}
The remaining two cases are proved in Appendix \ASproof. \SMs
\smallskip

We can now characterize $\cP(\cG_N)$ as a BV-algebra in terms of
generators and relations. In comparison with Theorem \generat,
the main simplification is that all generators $P_{i_1,i_2\ldots i_m}$
with $2N-1\geq m\geq 2$ are obtained from the volume element
$ X$, see \genasymtnsbig, and the ground ring generators $ x^i$.
Indeed, we may first rewrite \sophigen\ as
\eqn\sopigenn{\eqalign{
P_{i_1,i_2\ldots i_m}&=\hbox{$m+1\over m\,(2N-m)$}[ x^i,P_{i,i_1\ldots
i_m} ]_S\cr
&=\hbox{$m+1\over m\,(2N-m)$}\De_S( x^i P_{i,i_1\ldots
i_m})\,,\cr}}
where the last line follows from the relation between the bracket
and the BV-operator as well as
\eqn\trvl{
\De_S  x^i=0\,,\quad \De_S P_{i,i_1\ldots i_m}  =0\,.}
By iterating \sopigenn\ we obtain
\eqn\psxi{
P_{i_1,i_2\ldots i_{2N-k}}=(-1)^{k(k+1)/2} \hbox{$2N\over( 2N-k)\,k\,!$}
\ep_{i_1\ldots i_{2N-k}j_1\ldots j_k}\De_S( x^{j_1}\De_S(\ldots
\De_S( x^{j_k}  X)\ldots))\,,}
where $2N-1\geq k\geq 0$.
In particular, for $k=2N-1$, we find
\eqn\xidexi{
 x_i=(-1)^{N(2N-1)}\hbox{$2N\over (2N-1)\,!$}
\ep_{i\,j_1\ldots j_{2N-1}} \De_S( x^{j_1}\De_S(\ldots
\De_S( x^{j_{2N-1}}  X)\ldots))\,.}
Since $\De_S X=0$, we may also rewrite \psxi\ in terms of multiple
brackets.

\thm\bvgener
\proclaim Theorem \bvgener. The BV-algebra
$(\cP(\cG_N),\,\cdot\,,\De_S)$ is generated by $1$, the ground ring
generators $ x^i$, degree one derivation $C$, and the volume element
$ X$ of degree $2N-1$. The BV-operator and the `dot' product are
completely determined using
\eqn\fstcond{\De_S  x^i=0\,,\quad \De_S C=-2(N-1)\,1\,,\quad \De_S X=0\,,}
\eqn\scncond{\De_S ( x^i x^j)=0\,,\quad \De_S(C x^i)=-(2N-1) x^i\,,}
together with  \xidexi\ and the relations \conone-\confive\
expressed in terms of the right hand side  in \psxi.
\smallskip

\proof The proof is similar to that of Theorem \generat. We will
just outline the main steps, and leave the details for the reader.  In
the first step we show that the BV-algebra generated by $1$, $ x^i$,
$C$, and $ X$, satisfying all the relations above, is spanned by the
elements of the form
\eqn\newelem{
 x_{i_1}\ldots  x_{i_m}\De_S( x_{j_1}\ldots \De_S( x_{j_n} X)\ldots )
\,,\quad
 x_{i_1}\ldots  x_{i_m} C \De_S( x_{j_1}\ldots
\De_S( x_{j_n} X)\ldots )\,,}
where, in obvious notation, we set $m,n\geq 0$. Relations
\conone-\confive\ determine then the structure of the `dot' product
between those elements. The  next  step is to show
that the BV-operator $\De_S$ is completely determined  using \fstcond\ and
\scncond\ together with the defining relation \BVscor.  The only
nontrivial computation is to derive the second equation in \exformbv,
which, in the notation of  Theorem \bvgener, reads
\eqn\newdelc{
\De_S(C\De_S( x^{i_1}\ldots \De_S( x^{i_n} X)\ldots ))=
-n \De_S( x^{i_1}\ldots \De_S( x^{i_n} X)\,.}
For $n=1$, we find using \BVscor\ and \fstcond, \scncond, and \confive,
\eqn\trplid{\eqalign{
\De_S(C x^i X)&=\De_S(C x^i) X - C\De_S( x^i X) -(\De_S C) x^i X\cr
&= - x^i X-C\De_S( x^i X)\,.\cr}}
Since $\De_S^2=0$, acting with $\De_S$ on both sides of this equation
yields \newdelc\ for $n=1$. The general step of the induction is then
proved similarly. \SMs

\smallskip

Since $\De_S:\cP^n(\cG_N)\rightarrow \cP^{n-1}(\cG_N)$ satisfies
$\De_S^2=0$, it is natural to consider the homology of the complex
$(\cP(\cG_N), \De_S)$. This homology is easily computed using
Theorems \polyvects\ and \bvopmth.

\thm\cohofdel
\proclaim Theorem \cohofdel. The homology of $\De_S$ on $\cP(\cG_N)$
is  spanned by the volume element $ X$.
\smallskip

As we will see later in this paper, it is interesting to construct
extensions of the BV-algebra $(\cP(\cG_N),\,\cdot\,,\De)$ in which the
homology of $\De$ is trivial. In particular the BV-algebra of the
semi-infinite cohomology of the $\cW_3$ algebra is an extension of this
type.

\subsubsec{``Chiral'' subalgebras of $\cP(\cG_N)$}
\subsubseclab\chiral

\def\al{\si} \def\be{\rh}
There is a natural complex structure on $\cP(\cG_N)$ induced from the
decomposition of the ground ring generators into the ``holomorphic''
generators $ x_\al$ and the ``antiholomorphic'' generators $
x_{\dot\al}$, such that $( x_i)=( x_\al, x_{\dot\al})$,
$\al,\dot\al=1,\ldots,N$. With respect to this decomposition the only
nonvanishing components of the metric \metric\ are the $(1,1)$
components , $h_{\al\dot\al}=\de_{\al\dot\al}$, and the $\sotwon$
symmetry is broken to the $\slN$ subalgebra generated by the
derivations
\eqn\slnalg{
 D_{\al\dot\al}= x_\al\vps x^*_{\dot\al}- x_{\dot\al}\vps x^*_\al
-\hbox{$1\over N$} h_{\al\dot\al}
(x^{\dot\be} x^*_{\dot\be} -  x^\be x^*_\be)
\,,\quad \al,\dot\al=1,\ldots,N\,.}
\smallskip

Let us denote by $\cP_+(\cG_N)$ the BV-subalgebra of $\cP(\cG_N)$
generated by the holomorphic elements $ x_\al$, and
$P_{\al_1,\al_2\ldots \al_n}$, $\al,\al_1,\ldots \al_n
=1,\ldots,N$. Similarly let $\cP_-(\cG_N)$ be the BV-subalgebra
generated by the anti-homolomorphic elements. We will refer to
$\cP_+(\cG_N)$ and $\cP_-(\cG_N)$ as the chiral subalgebras of
$\cP(\cG_N)$.

\thm\holthm
\proclaim Theorem \holthm. The chiral subalgebra $\cP_+(\cG_N)$
(resp.\ $\cP_-(\cG_N)$) is spanned by the elements $P_{\al_1\ldots
\al_m,\be_1\ldots\be_n}$ (resp.\ $P_{\dot \al_1\ldots
\dot\al_m,\dot\be_1\ldots\dot\be_n}$). The BV-operator $\De_S$ restricted
to $\cP_+(\cG_N)$ (respectively\ $\cP_-(\cG_N)$) vanishes.
\smallskip

\proof The first part of the theorem follows from
Theorems \polyvects, \generat\ and \genproduct. In particular, since
$ x_\al^*( x_\be\vps)=0$,
\bigproduct\ implies that
\eqn\holommult{
 x_{\al_1}\ldots x_{\al_{m-1}}P_{\al_m,\be_1\ldots \be_n}=
P_{\al_1\ldots\al_m,\be_1\ldots \be_n}\,.}
The vanishing of the BV-operator follows then from  \exformbv. \SMu
\smallskip

Finally, let us note that the involution $\om_\cP$,
$\om_\cP^2=\om_\cP$, that exchanges the holomorphic and
antiholomorphic generators, \ie, $\om_\cP( x_\al)= x_{\dot\al}$,
$\om_\cP( x_{\dot \al}^*)= x^*_{\al}$ extends to all
polyderivations $\cP(\cG_N)$, such that $\om_\cP(\cP_+(\cG_N))\cong
\cP_-(\cG_N)$
\def\al{\alpha} \def\be{\beta}

\subsec{The $N=3$ case }
\subseclab\SSthree

The major motivation for explicitly constructing the BV-algebra $\cP(\cG_N)$
was to better understand the special case, $N=3$, which plays a
central role in Section \Sbvalgebra. We will now specialize the
results of Section \bvpolyderivations\ to this case, and then discuss
further a certain class of ``twisted'' G-modules of $\cP(\cR_3)$.
\smallskip

\subsubsec{The algebra $\cP(\cR_3)$}
\subsubseclab\SSpirthree

Consider the ground ring algebra $\cG_3$ as an $\slth$ module, where
$\slth\subset\sosi$ is the subalgebra defined in \slnalg.  If
$(s_1,s_2)$ denotes an $\slth$ irreducible module with the
Dynkin labels $s_1$ and $s_2$, respectively, then the branching
rule for an $\sosi$ module $[m;0]$ is given by
\eqn\branch{
[m;0]=\bigoplus_{s_1+s_2=m}(s_1,s_2)\,,}
and the following result is an immediate consequence of Theorem \strgrr.

\thm\modelspace
\proclaim Theorem \modelspace.  The ground ring algebra $\cG_3$
is a model space for the Lie algebra $\slth$, \ie, $\cG_3$ is a direct
sum of all finite dimensional irreducible modules of $\slth$, each
module present with multiplicity one.
\smallskip

 In the following, we will often write $\fP$ instead of $\cP(\cR_3)$
for the space of polyderivations of $\cR_3$.

 It is worth bringing out the simplicity of this result.  The ground
ring is generated by\foot{We recall that
$ x_\si=h_{\si\dot\si} x^{\dot\si}$,
$ x_{\dot \si}=h_{\dot\si\si} x^{\si}$.}
$x_\si$ and $x^\si$%
with the single relation,
$x_\si x^\si = 0$. Thus the elements of the ring are simply tensors
which are independently totally symmetric in their upper and lower
indices, and which vanish when an upper index is contracted with a
lower index -- this is precisely a tensorial presentation of the
irreducible representations of $\slth$.  The subspace of
$\cG_3$ spanned by monomials with $s_1$ factors of $x_\si$
and $s_2$ factors of $x_{\dot\si}$ makes up exactly one irreducible
$\slth$ representation $(s_1,s_2)$, \ie, that with highest weight
$\La=s_1\La_1+s_2\La_2$.   We will denote this subspace by $\cG_3(\La)$
in the following.  This further decomposition of $\cG_3$ may clearly
be considered as the decomposition under $\bga$, where the additional
$(\uone)^2$ generators just count the number of $x_\si$ and
$x_{\dot\si}$ in a given monomial.

\smallskip

To determine the decomposition of $\fP$ with respect to
$\slth$ we need the branching rules,
\eqn\bigbrone{\eqalign{
[m;1]&=\bigoplus_{{s_1+s_2=m-1}}
\left[\,(s_1,s_2)\oplus (s_1+1,s_2)\oplus (s_1,s_2+1)
\oplus (s_1+1,s_2+1)\,\right]\cr
[m;2]&=\bigoplus_{{s_1+s_2=m}}
\left[\, (s_1,s_2)\oplus (s_1-1,s_2-1)\,\right]
\oplus
\bigoplus_{{s_1+s_2=m-1}}\left[\,2\, (s_1,s_2)\oplus
(s_1+2,s_2)\oplus (s_1,s_2+2)\,\right]\,.\cr}} These
formulae are valid for $m\geq 1$. The summation runs over
$s_1,s_2\geq 0$, and terms with negative labels are to be
omitted. The branching rules for $[m;3]$, $[m;4]$ and $[m;5]$ are
obtained using isomorphisms $[m;k]\cong [m;5-k]$, $k=0,1,2$.
\smallskip

By comparing Table \tblpolyvec\ with the above branching rules, we
find that $\fP$ decomposes into a sum of disjoint ``cones'' of
$\slth$ modules, each cone being a direct sum of modules
$(s_1^0+s_1,s_2^0+s_2)$, $s_1,s_2\geq 0$.  In particular, for $n=1$
we find five cones with the tips $(s_1^0,s_2^0)$ equal to
$(0,0)$, $(0,0)$, $(0,1)$, $(1,0)$ and $(1,1)$, which correspond to
the derivations
\eqn\defofc{
C_+= x^{\dot\si} x^*_{\dot\si}\,,\quad
C_-= x^\si x^*_\si\,,}
\eqn\defofp{
P_{\si,\rh}\vps=\half(x_\si\vps x^*_\rh- x_\rh\vps x^*_\si)\,,
\quad
P_{\dot\si,\dot\rh}\vps=\half(x_{\dot\si}\vps x^*_{\dot\rh}
-x_{\dot\rh}\vps x^*_{\dot\si})\,.} and
\eqn\defoflam{
D_{\si\dot\si}=
x_\si\vps x^*_{\dot\si}-x_{\dot\si}\vps x^*_\si-\hbox{$1\over 3$}
h_{\si\dot\si} (x^{\dot\si} x^*_{\dot\si}-x^\si x^{*\si}) \,,} respectively.
In the following we will also find it convenient to define
\eqn\deffod{
D_\si=\ep_{\si\rh\pi}P^{\rh,\pi}\,,\quad
D_{\dot\si}=\ep_{\dot\si\dot\rh\dot\pi}P^{\dot\rh,\dot\pi}\,.}
\smallskip

Note that while the derivations $D_{\si\dot\si}$ generate the $\slth$
algebra, $C_+$ and $C_-$ yield the additional $(\uone)^2$ discussed
earlier. With this representation as derivations it is clear
that we have, in fact, a decomposition of all of $\fP$ into $\bga$
modules. The complete result is summarized in Theorem \polystable.
\smallskip

\noindent
{\it Remark:} The model space of $\slth$ can also be realized as
the space of polynomial functions on the algebraic variety
$A=N_+\backslash SL(3,\CC)$, where $N_+$ is the complex subgroup of
$SL(3,\CC)$ generated by the positive root generators [\BGG]. The
space $A$ is called the base affine space. In this realization of the
ground ring algebra $\cG_3$, the algebra of polyderivations
$\cP(\cG_3)$ is nothing but the algebra of polynomial polyvector
fields on $A$. This provides a beautiful geometric interpretation for
$\cP(\cG_3)$, and, in particular, gives a natural explanation of its
cone decomposition. We give a detailed discussion of this geometric
construction in Appendix \Saffine.

\subsubsec{The hidden symmetry structure}
\subsubseclab\hidsym

 As discussed in Section \hiddensymmetry\ , the ground ring is
a module for itself under left multiplication.  There is a
hidden symmetry algebra, which includes multiplication by ring
generators, for which the ground ring is an irreducible module.
In the case of $\cG_3$ this hidden symmetry algebra is $\soei$.
Under the chain of embeddings,
$\slth\subset\sosi\subset\soei$, the adjoint representations
of $\sosi$ and $\soei$ decompose with respect to $\slth$ as
\eqnn\decomada \eqnn\decomadb
$$\eqalignno{ {\rm ad}_{\sosi}&={\bf
8}\oplus ({\bf 3}\oplus {\bf \overline 3})\oplus {\bf 1}\,,&\decomada
\cr {\rm ad}_{\soei}&= {\bf 8}\oplus ({\bf 3}\oplus {\bf \overline
3})\oplus ({\bf 3}\oplus {\bf \overline 3})\oplus ({\bf 3}\oplus {\bf
\overline 3})\oplus {\bf 1}\oplus {\bf 1}\,.&\decomadb\cr}$$
The operators corresponding to the decomposition
\decomadb\ are
\eqn\extrathr{
D_{\si\dot\si}\,,\quad ( D_\si, D_{\dot\si})\,,\quad
(P_\si,U_{\dot\si})\,,\quad (U_\si,P_{\dot\si})\,,\quad C_+\,,\quad
C_-\,,} where, in addition to the first order derivations given in
\defofc-\defoflam, we also have zero and  second order
differential operators on $\cR_3$,
\eqn\twistfor{
P_\si = x_\si\,,\quad P_{\dot\si}=x_{\dot\si}\,,}
and
\eqn\twostforA{\eqalign{
U_\si &=\hbox{$2
{\RSr\partial\over\LWrr{\partial x^\si}}
+ x^\rh
{\RSr\partial\over\LWrr{\partial x^\rh}}
{\RSr\partial\over\LWrr{\partial x^\si}}
+ x_\rh
{\RSr\partial\over\LWrr{\partial x_\rh}}
{\RSr\partial\over\LWrr{\partial x^\si}}
- x_\si
{\RSr\partial\over\LWrr{\partial x_\rh}}
{\RSr\partial\over\LWrr{\partial x^\rh}} $}
\,,\cr
U_{\dot\si} &=\hbox{$2
{\RSr\partial\over\LWrr{\partial x^{\dot\si}}}
+ x^{\dot\rh}
{\RSr\partial\over\LWrr{\partial x^{\dot\rh}}}
{\RSr\partial\over\LWrr{\partial x^{\dot\si}}}
+ x_{\dot\rh}
{\RSr\partial\over\LWrr{\partial x_{\dot\rh}}}
{\RSr\partial\over\LWrr{\partial x^{\dot\si}}}
- x_{\dot\si}
{\RSr\partial\over\LWrr{\partial x_{\dot\rh}}}
{\RSr\partial\over\LWrr{\partial x^{\dot\rh}}} $}
\,.\cr}}
\smallskip
The derivations $D_{\si\dot\si}$, $D_\si$, $D_{\dot\si}$ and $C_+-C_-$
generate the $\sosi$ algebra of Section \bvpolyderivations.  As we
have seen in Section \polyderofcgn, this symmetry lifts from the
ground ring to the polyderivations. But there is no such lift for the
operators $U$, $U_\si$ and $U_{\dot\si}$, the reason being that they
do {\it not} act on $\cR_3$ as derivations.
\smallskip

 This structure arose out of the consideration of a particular
extension of $\slth$ to $\sosi\subset \soei$; namely, that for which
the pair $( D_\si, D_{\dot\si})$ corresponds to $(\bf 3\oplus\bf
\overline 3)$ in \decomada.
{}From \decomadb\ it is clear that this
extension may be done in three ways, utilising any of the pairs in
\extrathr.  In fact, the existence of three extensions is explained
by the triality of $\soei$, \ie, the three inequivalent representations
of $\soei$ of dimension eight.  It is interesting to understand the
extensions which involve the two remaining pairs in \extrathr\ since,
as explained in the next section, this leads to new modules of the
ground ring.  Indeed, since for a given choice of the extension to
$\sosi$ there are still two ways to assign a remaining $({\bf 3}\oplus
{\bf \overline 3})$ pair in \decomadb\ to the ring generators, this
will produce a total of six ground ring modules.

\subsubsec{``Twisted'' modules of $\fP$ }
\subsubseclab\twistedmodules

  The $\cG_3$ module discussed above  -- namely,
$\cG_3$ itself -- will be denoted by $M_1$.  We will now explicitly
construct the remaining ring modules alluded to there.  It will be
convenient in the following to denote by $M$ the vector space spanned
by monomials in $x_\si$ and $x_{\dot\si}$, modulo the constraint
$h^{\si\dot\si} x_\si x_{\dot\si}=0$. Clearly $M$ carries a
representation of $\slth$ as differential operators -- in fact,
precisely the $D_{\si\dot\si}$ in \defoflam\ -- and, for $\La = s_1
\La_1 + s_2 \La_2$, we may introduce (in analogy with $\cG_3(\La)$)
the subspaces $M(\La)$ spanned by monomials with $s_1$ factors of
$x_\si$ and $s_2$ factors of $x_{\dot\si}$.  The space $M$ may also
carry a realization of the ring $\cG_3$ by differential operators.
Indeed, $M_1$ is $M$ on which the generators act by the (zeroth-order)
differential operators $(P_\si,P_{\dot\si})$ in \twistfor.

\smallskip

\thm\twmodls
\proclaim Theorem \twmodls. The six pairs
of operators $( P_\si^w, P_{\dot\si}^w)$, $w \in W$, given by
\eqn\twpairs{
(P_\si,P_{\dot\si})\,,\quad ( D_\si,P_{\dot\si})\,,\quad
(P_\si, D_{\dot\si})\,,\quad (U_\si, D_{\dot\si})\,,\quad (
D_\si,U_{\dot\si})\,,\quad (U_\si,U_{\dot\si})\,,
}
for $w$ equal to $1$, $r_1$, $r_2$, $r_{12}$, $r_{21}$ and $r_3$,
respectively, define six inequivalent $\cG_3$ module structures as
differential operators acting on $M$.
\smallskip

\noindent
{\it Remark:} We will denote by $M_w$ the ground ring module defined
by the realization $( P_{\si}^w, P_{\dot\si}^w)$ on $M$.
\smallskip

\proof It is straightforward to verify that, for each
$w$, the differential operators $ P^w_\si$ and $ P^w_{\dot\si}$
commute and satisfy the constraint \vanishideal,
\ie, $h^{\si\dot\si}  P_\si^w P_{\dot\si}^w=0$.  Thus
the module structure is established.
Examining the explicit action of the differential operators
on monomials in $M(\La)$, one finds
immediately that the operators $ P_\si^w$ and $ P_{\dot\si}^w$
act as epimorphisms  between vector spaces, and
\eqn\wactring{\eqalign{
( P_{\si}^w)&: M(\La)\longrightarrow M(\La+w^{-1}\La_1)\,,\cr (
P_{\dot\si}^w)&: M(\La)\longrightarrow M(\La+w^{-1}\La_2)\,.\cr}}
Hence the module structures are inequivalent since they map
differently between irreducible $\slth$ modules.\SMu
\smallskip

\noindent
{\it Remark:} For all $w\in W$ the modules $M_w$ are
isomorphic to $\cG_3$ as $\slth$-modules.  The action \wactring\ is
the motivation for the labelling by Weyl group elements.  \smallskip

 Consistent with the comments at the end of the last subsection, the
existence of the six modules $M_w$ is equivalent to the existence of
six realizations of $\soei$ as differential operators on $M$.  We
display those realizations in Table \twistoprs\ below, one on each line
labelled by the corresponding $\slth$ Weyl group element,
$w \in W$.  The fact that the operators in each line of Table
\twistoprs\ independently generate $\soei$ is shown by explicit
computation.

\tbl\twistoprs
\bigskip\bigskip
\begintable
$\quad w\quad$ | $\quad P_\si^w\quad$ | $\quad P_{\dot\si}^w\quad$ |
$\quad D_\si^w\quad$ | $\quad D_{\dot\si}^w\quad$ |
$\quad D^w_{\si\dot\si}\quad$ |
$\quad U_\si^w\quad$ | $\quad U_{\dot\si}^w\quad$ |
$C^w_+$ | $C^w_-$ \crthick
$1$      | $P_\si$   | $P_{\dot\si}$ | $D_{\si}$ | $D_{\dot\si}$ |
$D_{\si\dot\si}$ | $U_\si$   | $U_{\dot\si}$ |
$C_+$          | $C_-$          \nr
$r_1$    | $D_{\si}$ | $P_{\dot\si}$ | $P_\si$   | $U_{\dot\si}$ |
$D_{\si\dot\si}$ | $U_\si$   | $D_{\dot\si}$ |
 $-C_+ -2$      | $C_+ +C_- +1$  \nr
$r_2$    | $P_\si$   | $D_{\dot\si}$ | $U_{\si}$ | $P_{\dot\si}$ |
$D_{\si\dot\si}$ | $D_\si$   | $U_{\dot\si}$ |
 $C_+ +C_- +1$  | $-C_- -2$      \nr
$r_{12}$ | $U_{\si}$ | $D_{\dot\si}$ | $P_\si$   | $U_{\dot\si}$ |
$D_{\si\dot\si}$ |  $D_\si$   | $P_{\dot\si}$ |
$-C_+ -C_- -3$ | $C_+$          \nr
$r_{21}$ | $D_{\si}$ | $U_{\dot\si}$ | $U_{\si}$ | $P_{\dot\si}$ |
$D_{\si\dot\si}$ |  $P_\si$   | $D_{\dot\si}$ |
$C_-$          | $-C_+ -C_- -3$ \nr
$r_3$    | $U_{\si}$ | $U_{\dot\si}$ | $D_{\si}$ | $D_{\dot\si}$ |
$D_{\si\dot\si}$ | $P_\si$   | $P_{\dot\si}$ |
 $-C_- -2$      | $-C_+ -2$
\endtable
\bigskip
\centerline{Table \twistoprs. Six realizations of the generators of $\soei$
                                                       on $M$.}

\bigskip

 The main result of this section then follows.

\thm\plvmodule
\proclaim Theorem \plvmodule. Let $\fP_w \equiv\cP(\cG_3,M_w)$ be the
algebra of polyderivations of $\cR_3$ with values in $M_w$.
Then $\fP_w$ is a G-module of $\fP$.

\proof In view of Theorem \gmodder\ it is sufficient to define
an action of $\fP^1$ on $M_w$ that satisfies \gmBA\ and \gmBB.  Given
that the ring generators are realized as differential operators
on $M_w$ as in Table \twistoprs, $P^w_i = (P^w_\si,P^w_{\dot\si})$,
a natural candidate for the generators of $\fP^1$ is the set
$P^w_{i,j} = (P^w_{\si,\rho},P^w_{\dot\si,\dot\rho},P^w_{\si,\dot\si},C^w)$,
constructed from the other generators in the table,
\eqn\suthgenp{
P^w_{\si,\rho} = \hbox{$1\over 2$} \epsilon_{\si\rho\pi}D^\pi \,,\quad
P^w_{\dot\si,\dot\rho} = \hbox{$1\over 2$} \epsilon_{\dot\si\dot\rho\dot\pi}
D^{\dot\pi} \,,\quad
P^w_{\si,\dot\si}=2\, D_{\si\dot\si}^w +
 \hbox{$2\over 3$}h_{\si\dot\si} (C^w_+-C^w_-) \,,\quad
C^w=C^w_++C^w_-\,.
}
It is straightforward to check that these operators satisfy the relations
\contwo\ and \conthr, where the dot product is realized as the ordinary
product of differential operators.  Thus, by Theorem \generat, the space
of differential operators spanned by monomials of $P^w_i$ multiplying
$P^w_{j,k}$ on the left gives a realization of $\fP^0 \oplus \fP^1$
as a dot algebra.  Moreover, we know that a given line of Table \twistoprs\
generates $\soei$ under commutation, independent of $w$.  So,
\gmBA\ and \gmBB\ hold for the generators realized as above.  But
then they hold for any element of the corresponding realization
of $\fP^0 \oplus \fP^1$.  Thus we have found, for each $w$,
a realization of $\fP^0 \oplus \fP^1$ as a G-algebra, where
the bracket operation is simply the commutator of differential
operators. \SMs

\smallskip
We will often refer to $M_w$ as the ($w$-) twisted module of $\cR_3$.
While $M_w$ is isomorphic as an $\slth$ module with $\cR_3$, it will
turn out convenient to twist the $(\uone)^2$ weights such that
$\La'\rightarrow w^{-1}(\La'+\rh)-\rh$.  This is precisely consistent
with Table \twistoprs\ if we use $C_\pm^w$ as the $(\uone)^2$
generators in the twisted module.  In particular the identity in $M$
has weights $(0,w^{-1}\rh-\rh)$ when considered as an element of $M_w$
and will be denoted by $\Om_w$. From now on we will denote the action
of the ring generators on $M_w$ simply by $x_\si$ or $x_{\dot\si}$.
Also we will use the
terminology ``twisted polyderivations'' or ``generalized polyvector
fields'' for $\fP_w$, $w\not=1$.
\smallskip

\subsubsec{A classification of twisted polyderivations}
\subsubseclab\SStwplcl

We now describe the spaces of twisted polyderivations, $\fP_w$,
especially for $w=r_1$ and $r_2$, in more detail. As in section
\polyderofcgn, the computation of those spaces may be posed as an
algebraic problem of finding all $\Ph\in
\End(\,\bigwedge^n\cR_3^1,M_w\vps)$, whose coefficients of expansion,
$\Ph=\Ph^{i_1\ldots i_n} x^*_{i_1}\ldots x^*_{i_n}$, $\Ph^{i_1\ldots
i_n}\in M_w$, satisfy the analogue of
\def\frgeneqn{\conscond} \frgeneqn, \ie,
\eqn\conforplvct{
x_i \cdot \Phi^{ii_1\ldots i_{n-1}}=0\,,\quad i_1,\ldots,
i_{n-1}=1,\ldots 6\,.}
\smallskip

The decomposition of $\fP_w$ with respect to $\bga$, whose action is
induced from that on $\cR_3$ and $M_w$, is crucial for solving
\conforplvct\ by reducing it to separate irreducible components.
However, since this symmetry is now smaller than
$\sosi$ in Section \polyderofcgn, the analysis is rather lengthy.%
\foot{Note that the ``standard'' $\sosi$ symmetry of the ground ring
yields a decomposition of $M_w$, $w\not=1,r_3$, into infinite
dimensional modules. Thus it seems simpler to work with a smaller
symmetry algebra, that yields a decomposition into finite-dimensional
modules only.}  For that reason, rather than discussing the general
case, let us illustrate the method with the simple example of
generalized vector fields with values in $M_{r_1}$, and then present
the complete solution for $\fP_{r_1}$ and $\fP_{r_2}$.
\smallskip

\tbl\gonevec
\exm\genvecone
\noindent
{\it Example \genvecone.}  An arbitrary generalized vector
field, $\Ph\in\fP_{r_1}^1$, that transforms in an $\slth$
representation with weight $\La$, is of the form
\eqn\exofgph{
\Ph=\sum_{\la\in P_1(\La)} \Ph^\si_{\la} x^*_\si
+\sum_{\la\in P_2(\La)} \Ph^{\dot\si}_{\la} x^*_{\dot\si}\,,}
where $\Ph_\la^\si,\Ph_\la^{\dot\si}\in M_{r_1}(\la)$, and $P_i(\La)$
consist of those weights $\la \in P_+$ for which $\cL(\La)$
arises in the tensor product $\cL(\la)\otimes\cL(\La_i)$, \ie,
\eqn\psets{
P_1(\La)=\{\La+\La_1,\La-\La_1+\La_2,\La-\La_2\}\cap P_+\,,\quad
P_2(\La)=\{\La+\La_2,\La+\La_1-\La_2,\La-\La_1\}\cap P_+\,.}

Since the operator $\frde$ in \frgeneqn\ is $\slth$-invariant,
we conclude that the component of
\conforplvct\ along the  representation with weight
$\La$ must vanish,
\eqn\forfg{
\big(\sum_{\la\in P_1(\La)} x_\si\cdot \Ph^\si_{\la_1}+
\sum_{\la\in P_2(\La)}
x_{\dot\si}\cdot\Ph^{\dot\si}_{\la_2} \,\big)_\La=0\,.}
Now, recall that that the action of the ground ring on $M_{r_1}$
merely amounts to shifting between the following representations, see
\wactring,
\eqn\grringact{
(x_\si):M_{r_1}(\La)\,\longrightarrow
\,M_{r_1}(\La-\La_1+\La_2)\,,\quad (x_{\dot\si}):
M_{r_1}(\La)\,\longrightarrow
\,M_{r_1}(\La+\La_2)\,.}  Therefore \forfg\ reduces to
\eqn\redcondfo{
x_\si\cdot \Ph^\si_{\La+\La_1-\La_2} + x_{\dot\si}\cdot
\Ph^{\dot\si}_{\La-\La_2} =0\,.}  Since the action of $(
x_{\dot\si})$ on $M_{r_1}$ has no zeros, we may always solve this equation
and express the components $\Phi^{\dot\si}_{\La-\La_2}$ in terms of
$\Phi^{\si}_{\La+\La_1-\La_2}$.

\bigskip\bigskip

\begintable
$\Ph$ | $\La'$ | $(\La,\La')$ \crthick
$\Ph_{\La+\La_2}^\si x^*_\si$ | $\quad r_1\La -2\La_1+\La_2\quad $ |
$\quad (0,-2\La_1+\La_2) \quad $ \nr
$\Ph_{\La+\La_1}^{\dot\si}x^*_{\dot\si}$ | $r_1\La-4\La_1+2\La_2$ |
$(0,-4\La_1+2\La_2)$ \nr
$\Ph_{\La-\La_1}^{\si}x^*_\si$ | $r_1\La-\La_1-\La_2$ |
$(\La_1,-2\La_1)$ \nr
$\Ph_{\La-\La_1+\La_2}^{\dot\si} x^*_{\dot\si} $ | $r_1\La-2\La_1+\La_2$ |
$(\La_1,-3\La_1+2\La_2)$ \nr
$\quad \Ph_{\La+\La_1-\La_2}^\si x^*_\si + \Ph_{\La-\La_2}^{\dot\si}
x^*_{\dot\si}\quad $ | $r_1\La-3\La_1$ | $(\La_2,-3\La_1+\La_2)$
\endtable
\smallskip

\centerline{Table \gonevec. The $r_1$-twisted
 cone decomposition of $\fP_{r_1}^1$}
\bigskip

 The weights $\La$, for which a solution arises in the sum \exofgph,
form a set of cones in $P_+$. This, in turn, translates into an
$r_1$-twisted cone decomposition of $\fP^1_{r_1}$.  We have summarized
the classification of $\fP_{r_1}^1$ in Table
\gonevec\ below, where we give a schematic form of each
vector field, its $(\uone)^2$ weight $\La'$, that follows immediately
from the coordinate expansion, and the tip of the corresponding cone,
$(\La,\La')$, determined by the lowest $\La$ for which a given
component exists.
\smallskip

An extension of this example to the higher degree polyderivations is
complicated by the fact that \conforplvct\ may have components in
several $\bga$ irreducible modules. Upon expanding the coordinates of
the polyvectors in a basis of $\bga$ invariant tensors we obtain a system
of linear equations. A simple counting of independent solutions
yields the following enumeration of $\fP_1$ and $\fP_2$.

\thm\pigenpol
\proclaim Theorem \pigenpol. The space of generalized polyderivations
$\fP_{r_i}$, $i=1,2$, decomposes into a direct sum of $r_i$-twisted
cones of $\bga$ modules,
\eqn\gpldecom{
\fP_{r_i}^n =\bigoplus_{(\La,\La')} \bigoplus_{\la\in P_+}
\cL(\La+\la) \otimes \CC_{\La'+r_i\la}\,,}
where, in the case of $\fP_{r_1}$, the tips of the cones,
$(\La,\La')$, satisfy $\La+2\rh=r_1(\La'+\rh-\si\rh)$, with $\La\in
P_+$ and $\si\in \widetilde W$ given in Table \pltwcnsi.%
\foot{{\rm See also Table  \qiuscones\ in the appendix.}}
 The result for
$\fP_{r_2}$ is obtained by interchanging the fundamental weights
$\La_1$ and $\La_2$, and letting $r_1\rightarrow r_2$, $\si\rightarrow
w_0\si w_0$.
\bigskip

\begintable
$\quad \si\,\backslash\,n\quad$ | $0$ | $1$ | $2$ | $3$ | $4$ | $5$
\crthick
$r_{21}$ | $\quad 0 \quad$ | $\quad 0$, $\La_1 \quad$ | $\La_1$ | | | \nr
$r_2$ | | $\La_1$ | $\quad \La_1$, $2\La_1 \quad$ | $2\La_1$ | | \nr
$r_3$ | | $0$ | $0$, $0$ | $0$ | | \nr
$\si_2$   | | $\La_2$ | $0$, $\La_2$ | $0$ | | \nr
$\si_1$   | | | $\La_1$ | $\La_1$, $\La_1$ | $\La_1$ | \nr
$1$   | | | $\La_2$ | $\quad 0$, $\La_1+\La_2 \quad$ | $\La_1$ | \nr
$r_{12}$ | | | $\La_2$ | $0$, $\La_2$ | $0$ | \nr
$r_1$ | | | | $\La_2$ | $\quad 0$, $\La_2 \quad$ | $\quad 0 \quad$
\endtable
\bigskip
\tbl\pltwcnsi
\centerline{Table \pltwcnsi. The weights $\La$ in the
$r_1$-twisted cone decomposition of   $\fP_{r_1}$}
\bigskip

A more explicit description of $\fP_{r_1}$ and $\fP_{r_2}$ is obtained
by studying the G-module action of $\fP$, and in particular of its
chiral subalgebras, $\fP_-$ and $\fP_+$, respectively.
Consider $\fP_{r_1}$. Since $x_\si\cdot\Om_{r_1}=0$, we find one
polyderivation of degree three,
\eqn\ofspom{
\Ga_1= -
\hbox{$1\over 432$} \ep^{\si\rh\pi}\,\Om_{r_1}x^*_\si x^*_\rh x^*_\pi\,,}
at weight $(0,-2\rh)$. Similarly, we have $\Ga_2\in\fP_{r_2}$.

\thm\freegen
\proclaim Theorem \freegen. Twisted polyderivations $\fP_{r_1}$
and $\fP_{r_2}$ are generated as free G-modules by $\fP_-$ and
$\fP_+$ acting on $\Ga_1$ and $\Ga_2$, respectively.
\smallskip

\proof See Appendix \freemodules. \SMd
\smallskip

By comparing the decomposition of $\fP_{r_i}$ with that of polyvectors
$\fP$, as given, \eg, in Theorem \polystable, we conclude that for
modules sufficiently deep inside the cones the two decompositions are
related by the Weyl reflection of $\La'+\rh-\si\rh$. This could have
been anticipated by looking at the action of the ground ring on the
twisted modules at the level of $\bga$ modules. The essential
difference between $M_1\simeq \cR_3$ and the twisted modules $M_w$,
$w\not=1$, is the presence of zeros in the action of some (for $w=r_1$
and $r_2$) or all (for $w=r_{12}$, $r_{21}$ and $r_3$) ground ring
generators, which explains why the the relation between different
$\fP_w$ holds only in the bulk, \ie, sufficiently deep inside
the cones.
\smallskip

In the remaining cases of polyderivations with values in $M_{r_{12}}$,
$M_{r_{21}}$ and $M_{r_3}$, the cone decomposition breaks down close
to the boundaries of the corresponding Weyl chambers. Once more this
is explained by the lines of states in the twisted modules that
are annihilated by {\it all} ground ring generators. The presence of
such states results in additional  solutions to
\conforplvct, beyond those  predicted by a naive counting
of equations and components.  However,  inside the
chambers, those special cases cannot arise, and once more we find
a similar result to the one in the fundamental Weyl chamber.

\tbl\quargvf
\thm\twplsumm
\proclaim Theorem \twplsumm. In the bulk the space of generalized polyvectors
$\fP_w$ is a direct sum of quartets of $\bga$ modules,
$\cL(\La)\otimes
\CC_{\,\La'}$ such that:
\item{i.}
Each quartet consists of modules of polyvectors of degree $n$, $n+1$,
$n+1$ and $n+2$, respectively,
\item{ii.}  The weights $\La\in P_+$ and
$\La'\in P$ satisfy $\La+2\rh=w^{-1}(\La'+\rh-\si\rh)$, where $\si\in
 \widetilde W$%
\foot{{\rm The assigment of $\si=\si_1$ and $\si=\si_2$  to particular
cones appears to be  arbitrary at this point, and our choice was
motivated by the results in Section \SSgenpol.}}
depends on $n$ and $w$ as given in Table \quargvf.

\bigskip\bigskip
\begintable
$\quad n\backslash w \quad$ | $1$ | $r_1$ | $r_2$ | $r_{12}$ |
$r_{21}$ | $r_3$ \cr
0 | $r_3$ | $r_{21}$ | $r_{12}$ | $r_2$ | $r_1$ | $1$ \nr
1 | $\quad r_{12}$, $\si_2$, $r_{21}\quad $ | $\quad r_3$, $\si_2$, $r_2
\quad$ | $\quad r_1$, $\si_2$, $r_3 \quad$ | $1$, $\si_1$, $r_{21}$ |
$r_{12}$, $\si_1$, $1$ | $r_2$, $\si_1$, $r_1$ \nr 2 | $r_1$, $\si_1$,
$r_2$ | $r_{12}$, $\si_1$, $1$ | $1$, $\si_1$, $r_{21}$ | $\quad r_1$,
$\si_2$, $r_3 \quad$ | $\quad r_3$, $\si_2$, $r_2 \quad$ | $\quad
r_{21}$, $\si_2$, $r_{12}\quad $ \nr 3 | $1$ | $r_1$ | $r_2$ |
$r_{12}$ | $r_{21}$ | $r_3$
\endtable
\bigskip
\centerline{Table \quargvf. The dependence of $\si$ on $n$ and $w$ in
the quartet decomposition of $\fP^n_w$ in the bulk}

\bigskip

While the G-module structure of $\fP_w$ over $\fP$ is rather obvious,
it is less clear whether it arises from some BV-module structure.
Although we cannot answer this question in general, we would like to
point out that in the case of $\fP_{r_i}$, $i=1,2$, there is a natural
candidate for a BV-operator. This operator, $\De$, is uniquely
determined by the condition
\eqn\defoftwdel{
\De_i\Ga_i=0\,,\quad i=1,2\,,}
together with the properties of the bracket. Using the explicit
parametrization of $\fP_{r_i}$ as free modules of $\fP_\mp$, given in
Appendix \freemodules, we then find
\eqn\twistbvop{
\De_i(\Ph_0\cdot[\,\Ph_1,[\,\ldots,[\,\Ph_n,\Ga_i]\,\ldots\,]\,])
=(-1)^{|\Ph_0|}[\,\Ph_0,[\Ph_1,[\,\ldots,[\,\Ph_n,\Ga_i]\,\ldots\,]\,]\,,}
where $\Ph_0\,,\ldots,\Ph_n\in \fP_\mp$.  Essentially by construction,
$\De_1$ and $\De_2$ turn $\fP_{r_1}$ and $\fP_{r_2}$ into a BV-module
of $\fP_-$ and $\fP_+$, respectively. We would like to conjecture that
in fact $\De_i$ defines on $\fP_{r_i}$ a BV-module structure with
respect to the full BV-algebra of polyderivations $\fP$. It appears
that a direct algebraic prove of this conjecture along the lines of
Section
\bvpolyvectors\ is rather cumbersome. In particular, it would require
a more explicit
enumeration of the bases in $\fP_{r_i}$ beyond the one given in
Appendix \freemodules.


\vfill\eject

%
%
\secno4
\newsec{THE BV-ALGEBRA OF THE $\cWth$ STRING}
\seclab\Sbvalgebra

\subsec{Introduction}
\subseclab\SBVintro

In this section we will study, with various degrees of rigour, the
algebraic structure of $H(\cW_3,\fC)$ that is induced by the VOA
structure of the underlying complex $\fC$.

\subsubsec{General results}
\subsubseclab\SBVgen

Let us denote by $\fH$ the cohomology space $H(\cW_3,\fC)$ considered
as an operator algebra. A straightforward application of the results
in [\WiZw,\WY,\LZbv,\MS,\Get] yields the following general theorem.

\thm\cftbv
\proclaim Theorem \cftbv. The cohomology space $\fH$ carries a
structure of a BV-algebra $(\fH,\,\cdot\,,b_0)$, where the product
``$\,\cdot\,$'' is induced from the normal ordered product in $\fC$,
while the BV-operator, $b_0\equiv b_0^{[2]}$, is the zero mode of the
Virasoro antighost field $b^{[2]}(z)$.
\smallskip

\proof Using \brstcurr\ and \brstdd, we find that $\fC$ carries an
action of the Virasoro algebra $T^{tot}(z)=[d,b^{[2]}(z)]$, with a
diagonalizable energy operator $L_0^{tot}$. Thus the complex $(\fC,d)$ is an
example of a topological chiral algebra, and the theorem
follows from the discussion in  [\LZbv], Section 3.9.4. \SMu
\smallskip

For completeness let us recall some explicit formulae (see
[\Wi,\LZbv]).  The dot product of two operators in $\fH$ is given by
\eqn\product{
(\cO \cdot \cO')(z) ~=~ {1\over {2\pi i}}\oint_{C_z}{dw\over {w-z}}
\cO(w)\, \cO'(z) \, .
}
It is graded commutative according to the ghost number of the
operators.  Since all non-trivial cohomology states are annihilated by
$L_0^{tot}$ (see Lemma \vanishw), all singular terms in the OPE on the
right hand side\ are trivial in cohomology.  Thus \product\ is also
equivalent, in cohomology, to
\eqn\prodaslim{
(\cO \cdot \cO')(z)~=~\lim_{w\rightarrow z}\,\cO(w) \cO'(z)\,.
}
The action of the BV-operator is
\eqn\bvopac{
(b_0\cO)(z)~=~\oint_{C_z} {dw \over {2\pi i}} \,(w-z)\,
b^{[2]}(w)\cO(z)\,.  } The associativity and graded commutativity of
the product at the level of cohomology, as well as the required
properties of the BV-operator (see Definition \BVbv) follow
immediately [\MS]. Moreover, one finds that the corresponding
bracket\foot{Note that in this notation the bracket does {\it not}
denote the commutator.} is simply obtained as
\eqn\barckop{
[\cO,\cO'](z) ~=~ (-1)^{gh(\cO)}\oint_{C_z}{dw\over {2 \pi i}}\,
(b^{[2]}_{-1}\cO)(w) \, \cO'(z) \, .}
\smallskip

The action of $\bga$ commutes with the BV-operator
$b_0$, and acts as a derivation of the dot product.  The latter
follows from the distributivity of the normal ordered product with
respect to the horizontal algebra defined by zero modes of spin one
currents. Thus we also have a refinement of Theorem \slthac.

\thm\slthaut
\proclaim Theorem \slthaut.
\item{i.} The symmetry algebra $\slth\oplus
(\uone)^2$ acts on $\fH$ by (infinitesimal) BV-algebra automorphisms.
\item{ii.} $\fH$ is a direct sum of irreducible finite dimensional modules
of $\bga$.
\smallskip

In determining the explicit structure of the BV-algebra
$(\fH,\,\cdot\,,b_0)$ we will distinguish between two types of
arguments.  The first type, referred to as ``kinematics,'' involves
arguments based on general properties of the cohomology, in particular
such ``kinematical'' characteristics as dimensions of  cohomology
at various matter and Liouville momenta, or decomposition with
respect to $\bga$. The second type of argument is based on
explicit computations. Those have been mostly carried out using the
algebraic manipulation program Mathematica$^{\rm TM}$
together with the CFT package OPEdefs [\Th].

\subsubsec{More notation}
\subsubseclab\SSmnotation

In the following we will need a convenient parametrization of the
operators in $\fH$. By examining the decomposition \bigdecomp, we find
that whenever $(\La^M,-i\La^L)$ satisfies \eqBa\ with $\si\in W$ (\ie, $
\si\not\in \{\si_1,\si_2\}$)
there is precisely one quartet of $\bga$ modules
with those weights.  Clearly the modules
at the ``bottom'' and the ``top'' of the quartet are unique, and we
will denote them by
\eqn\notforst{
\Psi_{\La^M,-i\La^L}^{(n)}\quad {\rm and} \quad
\Psi_{\La^M,-i\La^L}^{(n+2)}\,,}
respectively.  In cases where $\si \in \{\si_1,\si_2\}$
there is still only one module
with the lowest and the highest ghost number, but the difference
between those two ghost numbers is 3 rather than 2. To resolve
ambiguity in the remaining cases, where an $\bga$ module in cohomology
is not characterized uniquely by $(\La^M,-i\La^L)$ and $n$, we will
use either additional letters or labels.
\smallskip

We will often use the same notation, as in \notforst, for
the operator corresponding to the highest weight in a given module. Of
course in this case $\La^M$ and $\La^L$ are the momenta of the
operator.  Since $\bga$ acts by
automorphisms of the BV-algebra, it is convenient to express most of
the results at the level  of modules rather than operators. To
emphasize this we will  then write   ``$\,\sim\,$'' instead of the usual
equality sign.

\subsec{A preliminary survey of $\fH$}
\subseclab\Ssurvey

\subsubsec{The ground ring $\fH^0$}
\subsubseclab\SSground

We have seen in Section \SSgencoh\ that the entire cohomology at
ghost number 0 is concentrated in the fundamental Weyl chamber, and
consists of a single cone with tip $(0,0)$.  Let us examine the
lowest lying modules in $\fH^0$. First, there is the unit operator
\eqn\unoppsi{
{\bf 1}(z)~=~\Psi_{0,0}^{(0)}(z)\,.}
At Liouville weights $i\La_1$ and $i\La_2$ there is a triplet
$\Psi_{\La_1,\La_1}^{(0)}$ and an anti-triplet
$\Psi_{\La_2,\La_2}^{(0)}$. Let us denote their elements by $\widehat
x_\si$ and $\widehat x^{\si}$, $\si=1,\ldots,3$, respectively, with
\eqn\grringgenx{
\widehat x_1(z)~=~\Psi_{\La_1,\La_1}^{(0)}(z)\,,\quad
\widehat x{}^3(z)~=~\Psi_{\La_2,\La_2}^{(0)}(z)\,.}
Explicit expressions for those two operators were first computed in
[\BLNWnp]. They are given  in appendix \APexstates.

\thm\grconst
\proclaim Lemma \grconst\ [\BMPa]. In cohomology, the operators
$\widehat x_\si$ and $\widehat x^\si$, $\si=1,\ldots,3$, satisfy the
constraint
\eqn\grconstraint{
 \widehat x_\si\cdot \widehat x^\si ~=~ 0\,.}
\smallskip

\proof The left hand side\ in \grconstraint\ is a cohomology class of ghost
number zero, with Liouville momentum $-i\La^L=\La_1+\La_2$, and
transforming as a singlet under $\slth$. However, by
Theorem \cohfweyl, the only nontrivial cohomology at this Liouville
momentum  is in the adjoint representation of $\slth$, which implies
\grconstraint. \SMu

\smallskip
\noindent
{\it Remark:} It is rather straightforward to check by direct
computation of dimensions of the cohomology  that  there can
be no cohomology at weights $\La^M=0$, $-i\La^L=\La_1+\La_2$ and ghost
number 0 (see Appendix \APexplicit). However, verifying
\grconstraint\ directly, using the explicit representatives of the ground
ring generators given in Appendix \APexstates, is clearly a formidable
computation (which we have not attempted to perform). Recently,
another  verification of \grconstraint\ has been given in [\CJZhu].
\smallskip

\thm\cohringgen
\proclaim Theorem \cohringgen. The associative abelian algebra
generated by $1$, $\widehat x_\si$ and $\widehat x^\si$,
$\si=1,\ldots,3$, is isomorphic with $\cR_3$, \ie,
\eqn\bigprod{
(\Psi_{\La_1,\La_1}^{(0)})^m\cdot(\Psi_{\La_2,\La_2}^{(0)})^n ~\sim~
\Psi_{m\La_1+n\La_2,m\La_1+n\La_2}^{(0)}\,.}
\smallskip

\proof In view of Lemma \grconst, and the discussion in Section \SSgrrinal,
we must only show that the product of the highest weight operators in
\bigprod\  does not vanish in cohomology. This can be done by
examining explicitly the representatives \exone\ and \exdotthree\
together with the BRST current \brstact.
\smallskip

Define $\phi^{\pm,i} = \phi^{M,i} \pm i \phi^{L,i}$, $i=1,2$. In terms
of those fields the highest weight operators have the form
\eqn\hwopsex{
\widehat x_1(z) ~=~ P_{x_1}\, e^{i \La_1 \cdot \phi^+}\,,\quad
\widehat x^3(z)~=~ P_{x^3}\,
e^{i \La_2 \cdot \phi^+}\,,} where the operator-level 2 prefactors $P_{x_1}$
and $P_{x^3}$ are
\eqn\pisdep{
\eqalign{
P_{x_1} &~=~ - \hbox{$\sqrt{3}\over 4$}\, (\,2\sqrt{3} \pr\phi^{-,1}
\pr\phi^{-,1} + 5 \pr\phi^{-,1} \pr\phi^{-,2} + \pr\phi^{-,2}
\pr\phi^{-,1}\,) + \ldots\,, \cr
P_{x^3} &~=~ -\hbox{$3\over 4$}
(\pr\phi^{-,1} \pr\phi^{-,1} - 3\pr\phi^{-,2} \pr\phi^{-,2}\,) +
\ldots \,.\cr}}
The dots in
\pisdep\ stand for terms with $c^{[j]}$, $b^{[j]}$, $\pr\phi^{+,i}$,
and their derivatives, as well as the derivatives $\pr^{n}\phi^{-,i}$, $n>1$;
the terms that have been written explicitly are the only ones
that depend solely on $\pr\phi^{-,i}$

\smallskip
We will refer to a polynomial in $\pr\phi^{-,i}$ as a ``leading
term'' of an operator.  Remarkably, the leading terms in \pisdep\ do
{\it not} depend on the choice of a representative in the cohomology
class of $\widehat x_1$ and $\widehat x^3$.  Indeed, the only ghost
number $-1$ operators at the same weights and operator-level 2 are
$b^{[2]}e^{i\La_i\cdot \ph^{+}}$, and the assertion follows by
examining the residuum of the first order pole in the OPE with the
BRST current.
\smallskip

Consider the normal ordered product
\eqn\normalhigh{\eqalign{
(\widehat x_1)^m (\widehat x^3)^n &~\equiv~
(\widehat x_1\, \ldots\,(\widehat x_1(\widehat x^3\,\ldots\,(
\widehat x^3\widehat x^3)\ldots))\ldots)\cr
&~=~ P_{mn} e^{i (m \La_1 + n \La_2) \cdot \phi^+} \, . \cr}}
The prefactor $P_{mn}$ is a level $2m+2n$, ghost number zero
operator. The proof of the theorem now reduces to showing that
\smallskip
\item{i.} The leading term in $P_{mn}$ is the product of
the leading terms of all factors in \normalhigh.
\item{ii.} This term does not depend on the choice of operator in the
cohomology class of the product.
\smallskip

Recall that the normal ordered product of two operators
is just the first nonsingular term in their OPE.  Since
\eqn\phiopes{
\phi^{\pm,i} (z) \phi^{\pm,j} (w) ~\sim~ {\rm regular} \, , \quad
\phi^{+,i} (z) \phi^{-,j} (w) ~\sim~ - 2\de^{ij}\,\log(z-w) +
{\rm regular} \,,} there are no contractions between the exponentials
in computing \normalhigh, as all of them depend only on $\ph^{+,i}$.
Contributions to the normal ordered product that would yield
additional leading terms beyond the product of those in \pisdep\ can
arise only after the Taylor expansion cancels pole terms arising from
contractions between the prefactors and the exponentials, and between
the prefactors. However,  a moment's thought reveals that all such terms
in which  $\pr\ph^{-,i}$ is present must also have as  factors either
other fields, or derivatives of $\pr\ph^{-,i}$. This proves claim (i).
\smallskip

To show (ii), we must examine contributions to the residuum of the
first order pole in the OPE of the BRST current with an arbitrary
operator that has ghost number $-1$, and the same momenta and level
as $(\widehat x_1)^m (\widehat x^3)^n$. A similar argument to the one
above shows that none of the terms arising via Taylor expansion can
yield a polynomial in $\pr\ph^{-,i}$. Thus the only terms we need to
be concerned with are obtained through a single contraction of
$b^{[i]}$ with $c^{[i]}$, as otherwise we would have either higher
order poles or uncancelled ghost operators. In fact the only
possibility is that the BRST current has a term of the form
$c^{[i]}\times P[\pr\ph^{-,i}]$ which upon contraction with a term of
the form $b^{[i]} \times \widetilde P[\pr\ph^{-,i}] $ would contribute
to the leading term.  The result (ii) then follows as a simple
consequence of the fact that, as read from \brstcurr, the BRST current
has no such term when expanded in $\phi^{\pm,i}$. \SMu
\smallskip

As a direct consequence of Theorems \fockcoh\  and \cohringgen\ we
obtain the following isomorphism.

\noindent
\thm\groundring
\proclaim Theorem \groundring.   The ground ring $\fH^0$,  of the
BV-algebra $\fH$, is isomorphic to $\cR_3$.
The isomorphism $\pi:\fH^0\rightarrow\cR_3$ is explicitly given by
\eqn\defofpi{
\pi(\widehat x_\si)~=~x_\si\,,\quad \pi(\widehat x_{\dot\si})
  ~=~ x_{\dot\si}\,,\quad \si,\dot\si~=~1,2,3\,.}
\smallskip

Since $\fH^n\cong0$ for $n<0$ (see Theorem \ghnumlim), we can extend
$\pi$ to a $G$-algebra homomorphism $\pi:\fH \rightarrow \fP\equiv
\cP(\cR_3)$, as discussed in Section \SSpolyder.  Our immediate goal
is to use $\pi$ to establish a precise relation between $\fH$ and
$\fP$ as BV-algebras.  First, however, we need to study in some detail
further explicit cohomology states at higher ghost numbers.

\subsubsec{$\fH^1$: the $\bga$ symmetry of $\fH$ revisited}
\subsubseclab\SSautom

Since $\fH$ is a BV-algebra with BV-operator $b_0$, there is a Lie
algebra action of $\fH^1$ on $\fH$ defined by $\Ph\mapsto[\Psi,\Ph]$,
$\Psi\in\fH^1$, $\Ph\in\fH$ (see Section \SSpolyder). Moreover, by
Lemma \LLaut, the derivation $[\Psi,\,-\,]$ commutes with
the BV-operator if
$b_0\Psi$ has vanishing bracket with $\fH$.  We will now show
that the algebra $\bga$, as introduced in Section \SSslth\ ,
does in fact arise as a subalgebra of $\fH^1$ in this way.
\smallskip

Consider $\fH^1$ at the Liouville weight $\La^L=0$. From
Theorem \cohfweyl, or simply Table \zerozero, we find that it consists of
three $\slth$ modules, the adjoint and two singlets.
\smallskip

The highest weight operator in the adjoint representation is
\eqn\slthreeop{
\Psi^{(1)}_{\La_1+\La_2,0}(z)
 ~=~ (\,-c^{[2]} - \hbox{$\sqrt{3\over 2}$}i\pr\ph^{M,1}c^{[3]} +
\hbox{$1\over\sqrt{2}$} i\pr\ph^{M,2}c^{[3]} + b^{[2]}\pr
c^{[3]}c^{[3]}\,) \cV_{\La_1+\La_2,0}\,.}
It  satisfies
\eqn\bonslth{
(b^{[2]}_{-1} \Psi^{(1)}_{\La_1+\La_2,0})(z) ~=~
-\cV_{\La_1+\La_2,0}(z)\,,}
\ie, its action on $\fH$ defined by the bracket \barckop, is the same as that
of the $\slth$ automorphism.  Clearly, the same holds for the
remaining operators in the octet. Let us denote them by $\widehat
D_{\si,\dot\si}(z)$, where
\eqn\higslth{
\widehat D_{1,\dot 3}(z) ~=~ \Psi^{(1)}_{\La_1+\La_2,0}(z)\,.}

\thm\slthalgebra
\proclaim Theorem \slthalgebra. The operators
$\widehat D_{\si,\dot\si}$ close under the bracket onto
the $\slth$ algebra.
\smallskip

\proof Recall that
\eqn\tebnspr{
{\bf 8}\otimes _a\bf 8 ~=~ 8\oplus 10\oplus \overline{10}\,.}
Since at  $\La^L=0$ and
ghost number $n=1$ there is no cohomology in either the
${\bf 10}$ or the ${\bf \overline{10}}$ of $\slth$, the $\widehat
D_{\si,\dot\si}$'s span a subspace in $\fH^1$ which is closed under the
bracket. The theorem now follows by noting that the action of this
algebra on the ground ring coincides with $\slth$.   \SMu
\smallskip

In fact we also have
\eqn\pionpiss{
\pi(\widehat D_{\si,\dot\si}) ~=~
D_{\si,\dot\si}\,,\quad\si,\dot\si~=~1,2,3\,.}
\smallskip

The two singlets can be understood as a part of the quartet associated
with the identity operator. This quartet consists of ${\bf 1}(z)$,
$C^{[2]}(z)$, $C^{[3]}(z)$ and $C^{[23]}(z)$, where
$C^{[23]}(z)=(C^{[2]}\cdot C^{[3]})(z)$. The ghost number one
operators $C^{[2]}(z)$ and $C^{[3]}(z)$ are given in
\stctwo\ and \stcthree, respectively.  Note that neither of them
depends on the matter fields, $\ph^{M,i}$, so they
indeed transform as singlets under $\slth$.
\smallskip

{}From the explicit formulae we find
\eqn\expciipieces{
\eqalign{
C^{[2]}\ ~=~ &-\ 4\partial c^{[2]}- (\al_1+\al_2)\cdot\partial\ph^L
c^{[2]}+ \ldots\,,\cr
C^{[3]}\ ~=~ &\ -(\al_1-\al_2)\cdot\partial\ph^L c^{[2]}+ \ldots\,,\cr}}
where the dots stand for terms without $c^{[2]}$ or its derivatives.
Thus
\eqn\biioncc{
(b^{[2]}_{-1}C^{[2]})(z)~=~-(\al_1+\al_2)\cdot \partial \ph^L(z)\,,\quad
(b^{[2]}_{-1}C^{[3]})(z)~=~-(\al_1-\al_2)\cdot \partial \ph^L(z)\,,}
which shows that $C^{[2]}$ and $C^{[3]}$ are the $(\uone)^2$ generators
we are looking for. Moreover, if we set
\eqn\ohhatccc{
\widehat C(z) ~=~ C^{[2]}(z)\,,\quad
\widehat C_{\pm}(z)~=~\half(C^{[2]}(z)\pm C^{[3]}(z))\,,}
then, \cf, \defofc,
\eqn\pioncccc{
\pi(\widehat C) ~=~ C\,,\quad
\pi(\widehat C_{\pm})~=~C_{\pm}\,.}
\smallskip
As another straightforward consequence of \expciipieces, we will
obtain explicit formulae for the action of the BV-operator $b_0$.

\thm\techlemmacp
\proclaim Theorem  \techlemmacp. Let $\Ph\in \fH$ be an arbitrary
operator  with the Liouville momentum $-i\La^L=t_1\La_1+t_2\La_2$
satisfying $b_0\Ph=0$. Then
\eqn\ctwothph{
b_0(C^{[2]}\cdot\Ph)~=~-(4+t_1+t_2)\Ph\,,\quad
b_0(C^{[3]}\cdot\Ph)~=~-(t_1-t_2)\Ph\,,}
\eqn\ctwoththph{
b_0(C^{[23]}\cdot\Ph)~=~-(t_2-t_1)C^{[2]}\cdot\Ph-(4+t_1+t_2)C^{[3]}\Ph\,.}
\smallskip

\proof For $\Ph$ of the form \genformofpsi, the action of $b_0$ in
\bvopac\ simply amounts to setting to zero all the terms in the
polynomial prefactor $P$ that do not contain $\partial c^{[2]}$ as a
factor, and removing $\partial c^{[2]}$ from all the terms in which it
is present. Thus, $b_0\Ph=0$ implies that $\partial c^{[2]}$ is absent
from all the terms in $P$.  But then $(\partial c^{[2]} \Ph)(z)\not=0$,
and from \expciipieces\ we have
\eqn\ctwothphp{
(C^{[2]}\cdot\Ph) ~=~-(4+t_1+t_2)(\partial c^{[2]} \Ph)
+\Ph'\,,\quad
(C^{[3]}\cdot\Ph)~=~-(t_1-t_2)(\partial c^{[2]} \Ph)+\Ph''\,,}
where $b_0\Ph'=b_0\Ph''=0$.
These equations imply \ctwothph, and \ctwoththph\
is then obtained using the second order derivation property of $b_0$,
see \BVscor. \SMu

\thm\booncis
\proclaim Corollary \booncis.
\eqn\boonciss{
\hbox{$1\over 4$} (b_0 C^{[2]})(z)~=~ -
  \,{\bf 1}(z)\,,\quad (b_0 C^{[3]})(z)~=~0\,.}
\smallskip

\proof Take $\Ph={\bf 1}$ in \ctwothph. \SMu

Note that by kinematics we must have $b_0\widehat
D_{\si,\dot\si}=0$, which,
together with Corollary \booncis, proves directly  that the algebra
generated by $\widehat D_{\si,\dot\si}$ and $\widehat C_\pm$
commutes with the BV-operator on $\fH$.
\smallskip

\thm\intertwns
\proclaim Theorem \intertwns. The G-algebra homomorphism $\pi$
is equivariant with respect to the action of $\bga$  on $\fH$ and $\fP$.
\smallskip

\proof  Since $\pi$ is a G-algebra homomorphism,
the equivariance of $\pi$ with respect to $\bga$ follows from
\pionpiss\ and \pioncccc. Indeed,  for
any $\Psi\in\fH$ we have
\eqn\equivproof{
\pi([\,\widehat D,\Psi\,])~=~ [\,\pi(\widehat D),\pi(\Psi)\,]\,,}
where $\widehat D=\widehat D_{\si,\dot\si}$, $\widehat C{}_+$ or
$\widehat C{}_-$. \SMu

\subsubsec{More $\fH^1$}
\subsubseclab\SSmorehone

The operators at the tips of the two remaining $w=1$ cones in Table
\gencones\ have Liouville momenta
$-i\La^L=-\La_1+\La_2$ and $\La_1-\La_2$, and transform under $\slth$
as the triplet, $\bf 3$, and the anti-triplet, $\bf
\overline 3$, respectively. We denote them
by antisymmetric tensors $\widehat P_{\dot\rh,\dot\si}$ and $\widehat
P_{\rh,\si}$.  The highest weight operators are
\eqn\hgsklth{
\widehat P_{\dot 2,\dot 3}(z)~=~\Psi^{(1)}_{\La_1,-\La_1+\La_2}(z)\,,\quad
\widehat P_{1,2}(z)~=~\Psi^{(1)}_{\La_2,\La_1-\La_2}(z)\,,}
and their explicit expressions can be found in Appendix \APexstates.
\smallskip

Now consider the action of $\widehat P_{\rh,\si}$ on the ground ring.
First we must have $[\widehat P_{\mu,\rh} , \widehat x^{\dot\si}]=0$,
because at the total Liouville momentum of this operator, given by
$-i\La^L=2\La_1-\La_2$, there is no cohomology with ghost number zero.
Similarly, the other bracket, $[\widehat P_{\mu,\rh} , \widehat
x^{\si}]$, must be a linear combination of the generators in the
triplet of $\slth$.  By an explicit computation we verify that in fact
\eqn\ohthosepi{
[\widehat P_{\mu,\rh} , \widehat x^{\si}]~=~-\half(\de^\si_\mu\widehat
x_\rh-\de^\si_\rh\widehat x_\mu)\,.
}
A similar result also holds for
$\widehat P_{\dot\si,\dot\rh}$, and we conclude that
\eqn\goodproj{
\pi(\widehat P_{\rh,\si})~=~ P_{\rh,\si}\,,\quad
\pi(\widehat P_{\dot\rh,\dot\si})~=~ P_{\dot\rh,\dot\si}\,.}
\smallskip

At this point we have considered all cones in $\fH^1$, except for the two
twisted cones with $w=r_1$ and $w=r_2$ in Table \gencones.  By
comparing the weights of operators in the twisted cones with those of the
polyvectors, in Table \hmsandplvc, we conclude that all these
operators must act trivially on the ground ring.  In other words the
bracket between those operators and the elements of the ground ring
must vanish.
\smallskip

The  operators at the tips of the two twisted cones will be denoted by
\eqn\ohomegas{
\widehat\Om_{r_1}(z) ~=~ \Psi^{(1)}_{0,-2\La_1+\La_2}(z)\,,\quad
\widehat\Om_{r_2}(z) ~=~ \Psi^{(1)}_{0,\La_1-2\La_2}(z)\,.}
They are given in \omegarone\ and \omegartwo, respectively.

\subsubsec{An extension of $\sosi$}
\subsubseclab\SSextens

Now, we wish to  examine whether the $\sosi$ symmetry of the ground ring is
realized by  operators in $\fH^1$. First let us set
\eqn\fullpiss{
\widehat P_{\si,\dot\si} ~=~ -\widehat P_{\dot\si,\si}
 ~=~ \half \widehat D_{\si\dot\si}+\hbox{$1\over 6$} h_{\si\dot\si}
  \,(\widehat C_+-\widehat C_-)\,,}
and combine them with $\widehat P_{\rh,\si}$ and
$\widehat  P_{\dot\rh,\dot\si}$ to $\widehat P_{i,j}$,
$i,j=1,\ldots,6$. Then, from
\pionpiss, \pioncccc\ and
\goodproj, we find that $\pi(\widehat P_{i,j})= P_{i,j}$, or, simply,
that the $\widehat \La_{ij}=2\widehat P_{i,j}$ act on the ground ring
as the $\sosi$ algebra.
\smallskip

However, when we consider the bracket between the operators $\widehat
\La_{ij}$ in $\fH^1$, we find that they do not  form a Lie subalgebra
isomorphic with $\sosi$, but rather generate an (infinite dimensional)
extension of $\sosi$. In particular we find

\thm\noslfour
\proclaim Lemma \noslfour.
\eqn\twopistoom{
[\widehat P_{\dot\mu,\dot\nu}, \widehat P_{\dot\rh,\dot\si}]~=~
-\hbox{${1\over 48}$} \, (\ep_{\dot\mu\dot\nu\dot\rh}\widehat x_{\dot\si}-
     \ep_{\dot\mu\dot\nu\dot\si}\widehat x_{\dot\rh})\cdot
 \widehat{\Om}_{r_1}\,,
\quad
[\widehat P_{\mu,\nu}, \widehat P_{\rh,\si}] ~=~
 -\hbox{${1\over 48}$} \, (\ep_{\mu\nu\rh}\widehat x_{\si}-
     \ep_{\mu\nu\si}\widehat x_{\rh})\cdot\widehat{\Om}_{r_2}\,.}

\proof  By kinematics, the general
form of the first  bracket at the level of modules is
\eqn\kinofslf{
[\Psi^{(1)}_{\La_1,-\La_1+\La_2},\Psi^{(1)}_{\La_1,-\La_1+\La_2}]
  ~\sim~ n \Psi^{(1)}_{\La_2,-2\La_1+2\La_2}\,,}
where $n=0$ or $1$. The operator on the right hand side turns out to be
a product
\eqn\prodofomr{
\Psi^{(1)}_{\La_1,-2\La_1+2\La_2}~\sim~\Psi^{(0)}_{\La_2,\La_2}\cdot
\Psi^{(1)}_{0,-2\La_1+\La_2}\,.}
The general form in \twopistoom\ follows then by the $\bga$ covariance, and
the overall normalization factor is fixed by explicitly evaluating the
bracket between a single pair of operators. \SMu
\smallskip

Let us state, without further detail, that all other brackets between
the operators $\widehat P_{i,j}$ close as expected, thus the $\sosi$
commutation rules are violated only by \twopistoom.

\subsubsec{A summary for $\fH^1$}
\subsubseclab\SSsummary

We may summarize the structure of $\fH^1$ as follows.

\thm\summofone
\proclaim Theorem \summofone. Let us set, according to the cone
decomposition in Table \gencones,
\eqn\decomofhone{
\fH^1~=~\fH^1_1\oplus\fH^1_{r_1}\oplus\fH^1_{r_2} \,.}
Then $ \fH^1_1\cong\pi(\fH^1_1)\cong\fP^1$ and $\ker\,\pi\cong
\fH^1_{r_1}\oplus\fH^1_{r_2}$.
\smallskip

\proof We have shown that all the generators of $\fP^1$ as a module over the
ground ring, are obtained as the image under $\pi$ of the tips of
cones in $\fH^1$. Thus $\pi(\fH^1)=\fP^1$. By comparing Table
\gencones\ with Table \hmsandplvc\ we see that $\pi$ must be an
isomorphism between $\fH^1_1$ and $\fP^1$, and vanish on $\fH^1_{r_1}$
and $\fH^1_{r_2}$. \Box
\smallskip

\thm\genofhone
\proclaim Corollary \genofhone. The subspace
$\fH^1_1$ is generated as an $\fH^0$ dot module by $\widehat C=\widehat
C_++ \widehat C_-$ and $\widehat P_{i,j}$.

\subsec{The relation between $\fH$ and $\fP$  }
\subseclab\Spolybv

\subsubsec{The main theorem}
\subsubseclab\SSmain

\thm\maintheorem
\proclaim Theorem \maintheorem.
\item{i.} The map
$\pi:\fH\rightarrow \fP$ is a BV-algebra homomorphism between
$(\fH,\,\cdot\,,b_0)$ and $(\fP,\,\cdot \,,\De_S)$.
\item{ii.} Let $\frak{I} \equiv {\rm Ker\,}\pi$ be a BV-ideal of $\fH$.
We have an exact sequence of BV-algebras
\eqn\exactseq{
\matrix{
0&\mapright{}&\frak{I}&\mapright{}&\fH&\mapright{\pi}&\fP&\mapright{}&
0\,.\cr}}
There exists a dot
algebra homomorphism $\imath:\fP\rightarrow \fH$, such that
$\pi\circ\imath={\rm id}$, \ie, the sequence splits as a sequence of
$\imath(\fP)$ dot modules.
\smallskip

In the following subsections we will give a complete proof of this theorem.

\subsubsec{$\pi(\fH)=\fP$}
\subsubseclab\SSpi

Consider the unique ghost number 5 singlet with the Liouville momentum
$-i\La^L=-2\La_1-2\La_2$,
\eqn\volumeop{
\widehat X(z)~=~\Psi_{0,-2\La_1-2\La_2}^{(5)}\,,}
given in \volumeom. Using the $\sosi$ notation,  define
\eqn\thepisss{
\widehat P_{i_1,i_2\dots i_m} ~=~ \hbox{${1\over {(6-m)!}} {6\over {m}}$}\,
 \epsilon_{j_1\dots j_{6-m} i_1 \dots i_m} [\,\widehat
x^{j_{6-m}},\dots, [\,\widehat x^{j_1},\widehat X\,]\dots\, ] \,, }
where $2\leq m\leq 5$. It is easy to verify by an explicit computation
that \thepisss\ agrees with the previous definition of the operators
$\widehat P_{i,j}$.
\smallskip

Again, by explicit computation we find that, for $m=1$, \thepisss\ extends to
\eqn\comtox{
\widehat P_i~=~\widehat x_i\,,}
which implies
\eqn\projofX{
\pi(\widehat X)~=~X\,.}
In fact,  \comtox\ implies a stronger result.

\thm\pisonpis
\proclaim Lemma  \pisonpis. For all $1\leq m\leq 5$,
\eqn\pissonpiss{
\pi(\widehat P_{i_1,i_2\ldots i_m})~=~P_{i_1,i_2\ldots i_m}\,.}
\par

Therefore, we have
\thm\piofhonp
\proclaim Theorem  \piofhonp.
\eqn\piofhonP{\pi(\fH)~=~\fP\,.}

\proof  Lemma \pisonpis, together with \projofX, shows that all generators
of the dot algebra $\fP$, given in Theorem \generat, are in the image of
$\pi$. \Box
\smallskip

In appendix \APexstates\ we have listed the complete set of operators
$\widehat P_{i_1,i_2\ldots i_m}$ corresponding to the highest weights
of all $\slth$ modules.

\subsubsec{$\pi$ is a BV-algebra homomorphism}
\subsubseclab\SSpihom

Using Theorem \bvhomscr, the first part of Theorem \maintheorem\ is
proved by the following lemma.

\thm\piisbv
\proclaim Lemma \piisbv. For  $\Psi\in\fH^1$,
\eqn\piandbv{
\pi(b_0\Psi)~=~\De_S\pi(\Psi)\,.}
\smallskip

\proof By kinematics we have $b_0\Psi=0$, for all $\Psi\in\fH^1_{r_1}
\oplus \fH^1_{r_2}\cong\ker\,\pi$, so \piandbv\ holds there.

\noindent
In $\fH^1_1$ we can use Corollary \genofhone\ to conclude that, since
$b_0$ and $\De_S$ are second order derivations and $\pi$ is a
dot algebra homomorphism, it is sufficient to verify \piandbv\ on
$\widehat C$ and $\widehat P_{i,j}$, and on their products with a
single ground ring generator.
Note that by  kinematics, together with \fullpiss\ and \boonciss, we
must have  $b_0\widehat P_{i,j}=0=\De_S P_{i,j}$. Similarly,
\fstcond\ and \boonciss\ show that \piandbv\ holds for $\widehat C$
and $C$. Then
\eqn\ohxprodpi{
\pi (\, b_0(\widehat x_i\cdot \widehat P_{i,j})\,)
  ~=~\pi(\,[\, \widehat x_i, \widehat
   P_{j,k}\,]\,) ~=~
[\,x_i,P_{j,k}\,] ~=~ \De_S(x_i\cdot P_{j,k})\,.}
The last case, $\widehat C\cdot \widehat x_i$, is proved using
\scncond\ and Theorem \techlemmacp. \SMu

\subsubsec{An embedding $\imath:\fP\rightarrow \fH$}
\subsubseclab\SSembed

We have seen in Section \Ssurvey, and in particular Section
\SSsummary, that a simple kinematical analysis yields a unique
embedding of $\fP^0\oplus \fP^1$ into $\fH$. However, this is not the
case at higher ghost numbers, where at some momenta there are more
states in the cohomology than in the corresponding polyderivations.
\smallskip

The simplest example is at ghost number two along the boundaries of
the fundamental Weyl chamber. Indeed, by comparing Table
\hmsandplvc\ with Table \polytips, or Table \gencones, we find that
the $\bga$ modules with the highest weights
$(\La_1+n\La_2,-\La_1+(n+1)\La_2)$ and
$(n\La_1+\La_2,(n+1)\La_1-\La_2)$, $n\geq 0$, are doubly degenerate in
$\fH^2$, but nondegenerate in $\fP^2$. The same phenomenon is present
at higher ghost numbers.
\smallskip

The problem then is to find an embedding $\imath :\fP\rightarrow
\fH$ which preserves as much of the BV-algebra structure of
$\fP$ as possible. We have already seen
(Section \SSextens) that at ghost number one the
image of $\fP^1$ in $\fH^1$ is not closed under the bracket. Thus the
most one can expect is to embed $\fP$ as dot algebra. In that case,
although $\fP$ is generated as a BV-algebra by the ground ring, $C$
and the ``volume element'' $X$, all of which embed uniquely into
$\fH$, it is necessary to define the embedding of the remaining
generators $P_{i_1,i_2\ldots i_m}$, $m=2,\ldots,5$.

\thm\imathdef
\proclaim Theorem \imathdef. Let us define
\eqn\defiotaaone{
\imath(C)~=~\widehat C\,,\quad \imath(X)~=~\widehat X\,,}
\eqn\defiotaatwo{
\imath(P_{i_1,i_2\ldots i_m})~=~\widehat P_{i_1,i_2\ldots i_m}\,,\quad
1\leq m\leq 5\,.}
Then $\imath$ extends uniquely to a dot algebra embedding  of $\fP$
into $\fH$.
\smallskip

\proof Clearly, it is sufficent to prove that the elements $\widehat
C$, $\widehat X$, and $\widehat P_{i_1,i_2\ldots i_m}$,
$m=1,\ldots,5$, satisfy \conone-\confive\ in Theorem \generat.  The
last relation,
\eqn\ctimesx{\widehat C\cdot \widehat X~=~0\,,}
is
verified easily using \pissenonec, \stctwo, \stcthree\ and
\omegartwo. It also follows by kinematics, as there is no cohomology
with $-i\La^L=-2\La_1-2\La_2$ and $n=6$ (see Table \zerozero). Thus we
must show, beyond our previous result \grconstraint, that
\eqn\JJeon{
\widehat x_{[i}\cdot\widehat  P_{i_1,i_2\dots i_m]} ~=~ 0\,, }
\eqn\JJetw{
\widehat x^i \cdot\widehat  P_{i,j_1\dots j_m} ~=~
 -\hbox{$m\over {m+1}$}\, \widehat C \cdot\widehat P_{j_1,j_2\dots j_m}\,,}
\eqn\JJeth{
\widehat P_{i_1,i_2\dots i_m} \cdot\widehat  P_{j_1,j_2\dots j_n}
  ~=~ (-1)^{m-1} \hbox{${m+n-1\over n}$}\,
\widehat x_{[i_1}\cdot\widehat P_{i_2,i_3\dots i_m]j_1\dots j_n} \, ,}
where $m,n=1,\ldots,5$ and $\widehat P_{i_1,i_2\ldots i_6}=\ep_{i_1i_2\ldots
i_6}\widehat X$.
\smallskip

Using the complete antisymmetry of the multiple bracket, which
follows immediately from (iv) in Definition \BVghal, we may invert
\thepisss\ as
\eqn\invthepiss{
[\,\widehat x{}^{j_{6-m}},[\,\widehat
x{}^{j_{5-m}}\,\ldots\,,[\,\widehat x{}^{j_1},\widehat X\,]\,\ldots
\,]\,]~=~\hbox{$1\over 6\,(m-1)\,!$}\, \ep^{j_1\ldots j_{6-m} i_1\ldots
i_m}\,\widehat P_{i_1,i_2\ldots i_m}\,,\quad m=1,\ldots,5\,.}
This implies (see \identone)
\eqn\nicecomm{
[\,\widehat x{}^i,\widehat P_{i_1,i_2\ldots
i_m}\,]~=~(m-1)\,\de^i{}_{[i_1}\, \widehat P_{i_2,i_3\ldots i_m]}\,\,.}
\smallskip

Now, for arbitrary $\Psi\in\fH$ we have
\eqn\comwithzero{
\widehat x_i\cdot[\,\widehat x^i,\Psi\,] ~=~ \half \,[\,\widehat
x_i\cdot \widehat x^i,\Psi\,] ~=~ 0\,,}
so \JJeon\ follows from \nicecomm\ after multiplication by $\widehat x_i$.
\smallskip

\def\hbbb#1{\hbox{$#1$}}
The second relation, \JJetw\ is proved by induction on $m$. First we
have, using \ctimesx, \pioncccc\ and \Cph,
\eqn\firstpartt{
0 ~=~ [\,\widehat x,\widehat C\cdot\widehat X\,] ~=~ -\widehat
C\cdot[\,x^i,\widehat X\,]+x^i\cdot\widehat X\,,}
which by
\thepisss\ is equivalent to \JJetw\ for $m=5$.  Now suppose that
\JJetw\ is true for $m = 5,\, 4, \, \ldots,\, n+1$, with $1<n<5$.
Then\foot{We use a shorthand notation for index structure that is
either obvious or irrelevant, and write $\{m-1\}$ for $m$ indices, \eg,
$\widehat P_{\{m-1\}}$ for the ghost number $m-1$ operator $\widehat
P_{i_1,i_2\ldots i_m}$. }
\eqn\JJind{
\eqalign{
\widehat C \cdot\widehat P_{\{n-1\}} &~=~\hbbb {n+1\over {n (6-n)}}\, \widehat
C \cdot [ x^i,P_{i,\{n-1\}}]
\cr
&~=~ \hbbb{n+1\over {m (6-n)}}\,
\big(- [\,\widehat x^i,\widehat C \cdot\widehat P_{i,\{n-1\}}\,] + [\,\widehat
x^i,\widehat C\,] \cdot\widehat P_{i,\{m-1\}}\,\big)
\cr
&~=~\hbbb {n+2\over {n (6-n)}}\,[\,\widehat x^i,\widehat x^j\cdot \widehat
P_{j,i \{n-1\}}\,] -
\hbbb{n+1\over {n (6-n)}}\,\widehat x^i \cdot\widehat P_{i,\{n-1\}} \cr
&~=~\hbbb {n+1\over n} (-\hbbb {6-n-1 \over {6-n}} -\hbbb {1\over {6-n}})
\,\widehat x^i \cdot\widehat P_{i,\{n-1\}} \cr
& ~=~ - \hbbb{n+1\over n}\,\widehat x^i \cdot\widehat P_{i,\{n-1\}} \, . \cr
}}
\smallskip

Finally, let us consider the last relation \JJeth\ .  We have
\eqn\toohigh{
\widehat P_{\{m_1\}}\cdot \widehat P_{\{m_2\}} ~=~0 \quad {\rm if}\quad
m_1+m_2\geq 6\,.}
This can be proved by noting that for all but two pairs
$(m_1,m_2)$ there are simply no operators in the complex $\fC$ with
the Liouville momentum and the ghost number of the product on the
left hand side\ in \toohigh. The two exceptions, $(2,4)$ and $(3,3)$, can be
reduced to the other cases using
\eqn\exceptionone{
[\,\widehat x^i,\widehat P_{i,\{4\}}\cdot \widehat P_{\{2\}}\,]~=~
\hbbb{5\over 6}\,\widehat P_{\{4\}}\cdot \widehat P_{\{2\}} +
\widehat P_{i,\{4\}}\cdot [\,\widehat x^i,\widehat P_{\{2\}}\,]\,,}
\eqn\exceptionone{
[\,\widehat x^i,\widehat P_{i,\{3\}}\cdot \widehat P_{\{3\}}\,] ~=~
\hbbb{8\over 5}\,\widehat P_{\{3\}}\cdot \widehat P_{\{3\}} +
\widehat P_{i,\{3\}}\cdot [\,\widehat x^i,\widehat P_{\{3\}}\,]\,,}
which follow from the  distributivity of the bracket and \nicecomm.
\smallskip

On the one hand \toohigh\ proves \JJeth\ for $m+n\geq 8$. On the
other hand \JJeth\ clearly is true if $m=1$ and $n$ arbitrary. Then
the complete proof of \JJeth\ is obtained by induction on $m+n$ and
$m$, using \nicecomm\ and \thepisss. This completes the proof of
Theorem \imathdef\ and thus also of Theorem \maintheorem. \SMu

\subsec{The bulk structure of $\fH$}
\subseclab\Scomstr

We have seen in the previous section that the action of the BV-algebra
$\fH$ on its ground ring $\fH^0$ leads to a projection $\pi$ from
$\fH$ onto polyvector fields, $\fP$.  For a given ghost number $n$
cohomology class, the components of the projection are simply the
ring elements isomorphic to its $n$-times iterated bracket with
the ground ring generators.  For elements in the kernel of $\pi$,
there is clearly some point at which this iteration of brackets vanishes,
though in general there will be a non-trivial result after some
number of iterations less than $n$.  Identifying this last nontrivial
stage will allow us to refine our study of the kernel of $\pi$.
In fact, this construction yields a homomorphism from
$\fH^n$ into polyderivations $\cP(\cR_3,\fH^{n-k})$, $k\leq n$, the
homomorphism $\pi$ corresponding to the maximal case $k=n$.  Our main
observation is that in the bulk, \ie, for the Liouville momenta
sufficiently deep inside Weyl chambers, the cohomology
$H(\cW_3,\fC)$ admits a description in terms of ``generalized
polyvector fields'' associated with twisted modules of the ground ring
of the type introduced in Section \twistedmodules. In particular, this
result gives then a partial proof of Conjecture
\shiftsss, in the sense that it establishes the lower bound on the
cohomology.
\smallskip

Most of the results below are obtained by a combination of kinematical
arguments and explicit computations.  While a more rigorous treatment
along the lines of the discussion in Section \polyderofcgn\ or the
proof of Theorem \cohringgen\ could be given, the details of such
proofs are rather cumbersome, at least in comparison with their
counterparts in the fundamental Weyl chamber. We will thus mainly
limit our discussion to a general summary of the results.

\subsubsec{Twisted modules of $\fH^0$}
\subsubseclab\SStwmod

An examination of the pattern of cohomology states (see Table
\gencones\ or, more conveniently, the figures in Appendix \primeplots)
reveals that in each Weyl chamber the cohomology with the lowest ghost
number forms precisely one (twisted) cone, $\widehat M_w$, of $\bga$
modules with the highest weights $(\La,w^{-1}(\La+\rh)-\rh)$, $\La\in
P_+$, $w\in W$. The operators at the tips of those cones,
\eqn\tipoftwc{
\widehat\Om_w(z)~=~\Psi_{0,w^{-1}\rh-\rh}^{\left(\ell(w)\right)}\,,}
can be found in Appendix \SStwgen. Note that for $w=1$ we have
the identity operator, while for $w=r_1$ and $r_2$ these are exactly
the two ghost number one operators which already appeared in Section
\SSextens.
\smallskip

Now we would like to understand the dot action of the ground ring on
each of the cones $\widehat M_w$.  The simple fact that there is only
one $\slth$ module at each Liouville momentum in $\widehat M_w$ allows
us to determine most of ground ring action by a purely kinematical
analysis.

\thm\genmodcoh
\proclaim Theorem \genmodcoh. The twisted cones, $\widehat M_w$, $w\in
W$, are closed under the dot product action of the ground ring, \ie,
$\fH^0 \,\cdot \widehat M_w\subset \widehat M_w$, and as $\fH^0$
modules they are isomorphic to the corresponding twisted $\cR_3$ modules,
$M_w$, introduced in Section \twistedmodules.
\smallskip

\proof   Clearly, the decomposition of each cone into $\slth$ modules
is that of a model space, and thus identical with that of the ground
ring, $\fH^0$. In fact, for $w=1$, $\widehat M_1\cong\fH^0$.  More
interesting are $w=r_1$ and $r_2$, where we observe that,
as $\slth$ modules, $\widehat M_{r_1}\cong
\fH^1_{r_1}$ and $\widehat
M_{r_2}\cong \fH^1_{r_2}$, respectively. We will now
outline the main steps of the proof for those two cases.
\smallskip

Consider $\widehat M_{r_1}$ first.  By acting with the ground
generators on the tip of this cone we obtain
\eqn\grgenonom{
\Psi^{(0)}_{\La_1,\La_1}\cdot \Psi^{(1)}_{0,-2\La_1+\La_2}~\sim~
0\,,\quad \Psi^{(0)}_{\La_2,\La_2}\cdot
\Psi^{(1)}_{0,-2\La_1+\La_2}~\sim~ \Psi^{(1)}_{\La_2,-2\La_1+2\La_2}\,.}
In fact, it is not too difficult to verify, by examining the leading
terms in the first product, that subsequent action of the anti-triplet of
ground ring generators always yields a nonvanishing result. This
proves that the operators along the boundary,
$(n\La_2,-2\La_1+(n+1)\La_2)$, of $\widehat M_{r_1}$ are
\eqn\onebdry{
\widehat x_{\dot\si_1}\cdot\ldots\cdot \widehat
x_{\dot\si_n}\cdot\widehat\Om_{r_1}\,, \qquad
\dot\si_1,\ldots,\dot\si_n=1,\ldots,3\,,\quad n\geq 0\,.}
\smallskip

To obtain the remaining operators in the cone we must study the
bracket action of $\fH^1$ on $\widehat M_{r_1}$, in particular those
of
\eqn\thdees{
\widehat \cD_\si~=~\ep_{\si\mu\rh}[\,\widehat P^{\mu,\rh},-\,]\,,\quad
\widehat \cD_{\dot\si}~=~\ep_{\dot\si\dot\mu\dot\rh}[\,\widehat
P^{\dot\mu,\dot\rh},-\,]\,.}
Note that, when acting on the ground
ring, $\widehat \cD_\si$ and $\widehat \cD_{\dot\si}$ are the first order
differential operators $D_\si^{(1)}$ and $D_{\dot\si}^{(1)}$ given in
\polydisba.  Once more we verify explicitly that
\eqn\brackonom{
[\,\Psi_{\La_1,-\La_1+\La_2}^{(1)},\Psi^{(0)}_{0,-2\La_1+\La_2}\,]~\sim~
\Psi^{(1)}_{\La_1,-3\La_1+2\La_2}\,,\quad
[\,\Psi_{\La_2,\La_1-\La_2}^{(1)},\Psi^{(0)}_{0,-2\La_1+\La_2}\,]~\sim~
0 \,,}
which suggests that the other boundary of the cone, $(n
\La_1,-(n+2)\La_1+(n+1)\La_2)$, is realized by
\eqn\othbry{
\widehat \cD_{\si_1}\ldots \widehat \cD_{\si_n}\widehat\Om_{r_1}\,,\quad
\si_1,\ldots,\si_n=1,\ldots,3\,,\quad n\geq 0\,.}
Since
\eqn\comofom{
[\widehat\Om_{r_i},\widehat\Om_{r_j}]~=~0\,,\quad i,j=1,2\,,}
we find, by repeatedly using
\twopistoom, \brackonom\ and the Jacobi identity for the bracket, that
the $\slth$ tensor in \othbry\ is completely symmetric in
$\si_1\,\ldots,\si_n$. The ``leading term'' type argument
shows that those operators span the required $\slth$ module.
\smallskip

Combining \onebdry\ with \othbry, and using the fact that the actions
of $\widehat \cD_{\si}$ and $\widehat x_{\dot\si}$ commute, we find that an
explicit basis in $\widehat M_{r_1}$ consists of elements $\widehat
x_{\dot \si_1}\cdot\ldots\cdot \widehat x_{\dot\si_m}\cdot \widehat
\cD_{\si_1}\ldots \widehat D_{\si_n}\widehat\Om_{r_1}$, $m,n\geq 0$.
Moreover, since $\widehat x^{\si} \widehat \cD_{\si}=0$, this basis
also gives an explicit isomorphism $\pi_{r_1}:\widehat
M_{r_1}\rightarrow M_{r_1}$ of $\bga$ modules,
\eqn\basisinrone{
\pi_{r_1}(\widehat x_{\dot \si_1}\cdot\ldots\cdot \widehat
x_{\dot\si_m}\cdot \widehat \cD_{\si_1}\ldots \widehat
D_{\si_n}\widehat\Om_{r_1})~=~
x_{\dot \si_1}\cdot\ldots\cdot
x_{\dot\si_m}\cdot  \cD_{\si_1}\ldots
D_{\si_n}\Om_{r_1}
\,,\quad m,n\geq 0\,.}
Using \grgenonom, or equivalently, $\widehat
x_{\si}\cdot\widehat\Om_{r_1}=0$, it is straightforward to evaluate
the action of the triplet of the ground ring generators on the basis
elements \basisinrone, with the result precisely that given in
\wactring. Thus $\pi_{r_1}$ is also an isomorphism of $\widehat M_{r_1}$
and $M_{r_1}$ as ground ring modules.
\smallskip

The proof in the case of $\widehat M_{r_2}$ is similar. In the
remaining three cones, $w=r_1r_2$, $r_2r_1$ and $r_3$, one cannot
construct explicit bases of $\widehat M_w$ in terms of polyvectors
acting on the corresponding operators at the tips of the
cones. (However, it is easy to verify that the elements of the form
$\widehat \cD_{\si_1}\ldots
\widehat \cD_{\si_n}\widehat\Om_{r_2r_1}$ and, similarly,
$\widehat \cD_{\dot\si_1}\ldots
\widehat  \cD_{\dot\si_n}\widehat\Om_{r_1r_2}$ span
one of the boundaries in the
respective cones.)  In those cases our claim is based on first noting
that by kinematics the action of the ground ring, if nontrivial, must
be of the twisted type as stated in the theorem, and then verifying it
by evaluating the products of the ground ring generators with the
operators lying close to the tips of the cones. \SMu

\thm\honeoneact
\proclaim Theorem \honeoneact. The ground ring modules isomorphisms
$\pi_w:\widehat M_w\,\rightarrow M_w$ are equivariant with respect to
the Lie algebra action of $\fH_1^1\cong\imath(\fP^1)$ and $\fP^1$ on
$\widehat M_w$ and $M_w$, respectively.
\smallskip

\proof The proof of this theorem is similar to the one above. \SMd

\subsubsec{Interpretation of $\fH$ in terms of twisted polyderivations}
\subsubseclab\SSgenpol

We have found that the lowest ghost number subspaces of $\fH$ in each
of the Weyl chambers may be identified with the twisted modules of the
ground ring. The problem is then to extend the isomorphism $\pi_w$ to
a map between the higher ghost number cohomology and twisted
polyderivations $\fP_w\equiv\fP(\cR_3,M_w)$ of the ground ring. The
result may be summarized as follows.

\thm\genproj
\proclaim Theorem \genproj. There is a natural map, $\pi_w$, that
identifies $\widehat M_w$ and $M_w$,  and maps
$\Phi\in\fH^{\ell(w)+n}$, with $-i\La^L+2\rh$ sufficiently deep inside
$w^{-1}P_+$, onto a generalized polyderivation $\pi_w(\Phi)\in\fP^n_w$,
given by
\eqn\generpolof{
\pi_w(\Ph)( x_{i_1},\ldots, x_{i_n}) ~=~ \pi_w(
[\,\ldots\, [\,[\,\Ph,\widehat x_{i_1}\,],\ldots\,],\widehat
x_{i_n}\,]) \,.}
\smallskip

\proof Clearly $\pi_1$ is just the homomorphism $\pi$.
In the other cases, although the right hand side in \generpolof\ is well
defined for any $\Ph$, the restriction on the Liouville weight is
imposed to ensure that the multiple bracket lies in $\widehat
M_w$. For such $\Ph$, the proof that $\pi_w(\Ph)$ is a twisted polyderivation
requires that we check the conditions in Lemma \derivas, which in fact
follow immediately using elementary properties of the dot product and
the bracket. \SMu
\smallskip

A more interesting question is to what extent $\pi_w$ is an
isomorphism between generalized polyvectors, $\fP_w$, and a subspace
of $\fH$.  In this respect a comparison of Theorem \pigenpol, which
gives an enumeration of all twisted polyderivations in the bulk, with
Theorem \fullcoh\ for the cohomology, leads to the conclusion that
in the bulk,
\eqn\asympisom{
\fH^n ~\cong~
H^n(\cW_3,\fC)~\approx~ \bigoplus_{w\in W}\cP^{n-\ell(w)}(\cR_3,M_w)\,.}
In fact, we should  interpret this equality  as a lower
bound for the cohomology, and thus a partial proof of Conjecture
\shiftsss.
\smallskip

The description of the cohomology in terms of twisted polyderivations
in Theorem \genproj\ breaks down close to the origin of the
lattice of shifted
Liouville momenta, because of the presence of operators that have
vanishing brackets with some or even all ground ring generators, and
therefore cannot be ``detected'' by \generpolof.  A particularly
interesting example is the ``special operator''
\eqn\specst{
\Psi^{(2)}_{0,-\La_1-\La_2}(z)~=~c^{[2]}c^{[3]}\,\cV_{0,-\La_1-\La_2}\,.}
By explicit evaluation of all products and all brackets of this
operator with the generators of $\imath(\fP)$ we find

\thm\spstdb
\proclaim Lemma \spstdb. The doublet of operators
$(\Psi^{(2)}_{0,-\La_1-\La_2} ,\widehat C\cdot
\Psi^{(2)}_{0,-\La_1-\La_2} )$ is invariant under dot product and
bracket with the elements of $\imath(\fP)$.
\smallskip

In particular, Lemma \spstdb\ implies that the dot products and the
brackets of the special operator with all ground ring generators
vanish.

\subsec{Towards the complete structure of $\fH$}
\subseclab\Scomplstr

It remains an open problem to understand how the bulk regions of
cohomology, parametrized in terms of twisted polyderivations of the
ground ring, are ``glued'' together. We will suggest here a possible
answer to this question that is essentially based on explicit
computations of products and brackets between the low lying operators
in $\fH$.
\smallskip

As in the analogous problem for the Virasoro cohomology, which has
been exhaustively discussed in [\LZbv] and is summarized here in
Appendix \AVir, the starting point is to understand the action of the
BV-operator $b_0$ on $\fH$.  The next step will be to unravel the
structure of $\fH$ as a module of $\fP$.

\subsubsec{The BV-operator $b_0$}
\subsubseclab\SScomplstrint

As a simple application of the results in Section \SSautom\ we have

\thm\cohofbo
\proclaim Theorem \cohofbo. The cohomology of $b_0$ on $\fH$ is
trivial.

\proof Suppose that $b_0\Psi=0$, where $\Psi\in\fH$ has the Liouville
momentum $-i\La^L=t_1\La_1+t_2\La_2$. From \ctwothph\ we find that
unless $t_1=t_2=-2$, either $\widehat C_+$ or $\widehat C_-$ yield a
contracting homotopy for $b_0$. In the exceptional case we find that
there is simply a quartet of operators in the complex, $\fC$, all of
which are nontrivial in cohomology (see Table
\hhmfour). Those are
$$
T_{0,-2\La_1-2\La_2}~=~c^{[2]}\p
c^{[3]}c^{[3]}\cV_{0,-2\La_1-2\La_2}\,,
$$
\eqn\tachquar{
T_{0,-2\La_1-2\La_2}^{[2]}~=~\p c^{[2]} c^{[2]}\p
  c^{[3]}c^{[3]}\cV_{0,-2\La_1-2\La_2}\,,\qquad
T_{0,-2\La_1-2\La_2}^{[3]}~=~c^{[2]}\p^2 c^{[3]} \p
  c^{[3]}c^{[3]}\cV_{0,-2\La_1-2\La_2}\,,}
$$
T_{0,-2\La_1-2\La_2}^{[23]}
 ~=~\p c^{[2]} c^{[2]}\p^2 c^{[3]} \p
  c^{[3]}c^{[3]}\cV_{0,-2\La_1-2\La_2}\,.
$$
They  form two doublets under $b_0$,
\eqn\tachdbl{
b_0\,T_{0,-2\La_1-2\La_2}^{[2]} ~=~ T_{0,-2\La_1-2\La_2}\,,\quad
b_0\,T_{0,-2\La_1-2\La_2}^{[23]} ~=~ T_{0,-2\La_1-2\La_2}^{[3]}\,,}
which shows that indeed the cohomology of $b_0$ is trivial. \SMu
\smallskip

\noindent
{\it Remark:} Note that $T_{0,-2\La_1-2\La_2}$ is the tachyon
operator, proportional to $\widehat\Om_{w_0}$, while
$T_{0,-2\La_1-2\La_2}^{[23]}=1728\sqrt{3} \widehat X$.  More
generally, the tachyon operators arise at momenta $(\La,w\La-2\rh)$,
$\La\in P_+$, $w\in W$ [\BLNW]. The quartet of cohomology operators
associated with each tachyon is then given by \tachquar,
but with $\cV_{\La,w\La-2\rh}$. It decomposes into two doublets under
the action of $b_0$, as in \tachdbl.
\smallskip

An immediate consequence of Theorem \cohofbo\ is that all cohomology
states are paired into doublets. This does not yet explain the quartet
structure, which one might want to associate with the  presence of
another BV-type operator. A naive candidate for such an operator is
$b^{[3]}_0$. It turns out, however, that the latter is not a well
defined operator on $\fH$, as is easily seen in the following example:
\eqn\crsubsp{
b_0^{[3]}C^{[2]}(z)~=~ 8( b^{[2]}c^{[3]})(z)\,.}
The operator on the right hand side is not annihilated by $d$.
\smallskip

Another consequence of Theorem \cohofbo\ is that the image of
polyderivations, $\imath(\fP)$, in $\fH$ is not closed under $b_0$.
Indeed, given \piandbv, this would contradict Theorem \cohofdel. An
obvious example of an operator that is mapped by $b_0$ outside
$\imath(\fP)$ is $\widehat X$, the image of the cohomology class, $X$,
of $\De$.  Let us denote $\widehat \Ga=b_0 \widehat X$.  It follows
from \tachdbl\ that this operator is nonzero.
\smallskip

The non-closure of $\imath(\fP)$ under $b_0$ also implies non-closure
under the bracket, and we have seen an example to that
effect in Section \SSextens.


\subsubsec{The dual decomposition of  $\fH$}
\subsubseclab\SSdualdec

The description of $\fH$ in terms of
polyderivations $\fP_w$, for $w=r_1$ and $r_2$, may be generalized
so as to also include the states at the boundaries of the regions.
Together with the duality of $\fH$, this will allow an explicit
description of the dot module structure of $\fH$ over $\fP$.
\smallskip

Consider $\fI=\ker\,\pi$. By Theorem \maintheorem, $\fI$ is a
BV-ideal. Thus it is also a BV-module of $\fH$, provided we set
$\De_M=\De\big|_\fI$. Moreover, we have $\fI^n=0$ for $n<1$, and
$\fI^1\cong \widehat M_{r_1}\oplus \widehat M_{r_2}$. Consider $\fI^1$
as a Lie algebra.

\thm\bronid
\proclaim Lemma \bronid. $\fI^1$ is an Abelian Lie algebra, \ie, the
bracket $[\,-\,,\,-\,]$ vanishes when restricted to $\fI^1$.
\smallskip

\proof The vanishing of the bracket on $\widehat M_{r_1}$ and
$\widehat M_{r_2}$ follows by kinematics. Indeed, for  $\Ph\in \widehat
M_{r_1}(\La)$ and $\Ph'\in\widehat
M_{r_1}(\La')$
the bracket, $[\,\Ph,\Ph'\,]$, has the Liouville weight
\eqn\brmmms{
r_1(\La+\La'+2\rh)-2\rh ~=~
r_1\left((\La+\La'+\al_1)+\rh\right)-\rh\,,} and thus must vanish because
the irreducible representation
with the highest weight $\La+\La'+\al_1$ cannot arise in the
tensor product $\La\otimes \La'$.  As for the bracket between
$\widehat M_{r_1}$ and $\widehat M_{r_2}$, we start with \comofom\ and
then proceed by induction also using the explicit bases in $\widehat
M_{r_i}$ constructed in Section \SStwmod, the vanishing
relations
\eqn\vanishomeg{ \widehat x_\si\cdot\widehat \Om_{r_1}
{}~=~ 0\,,\quad \widehat \cD_{\dot\si}\widehat \Om_{r_1} ~=~ 0\,,\quad
\widehat x_{\dot\si}\cdot\widehat\Om_{r_2} ~=~ 0\,,\quad
\widehat \cD_{\si}\widehat \Om_{r_2} ~=~ 0\,,}
and the properties of the bracket. \SMu

\smallskip
\noindent
{\it Remark:} Similar arguments show that, given $\widehat
\Omega_{r_i}\cdot \widehat \Omega_{r_j}=0$, $i,j=1,2$, we must have
$\widehat M_{r_i}\cdot \widehat M_{r_j}=0$ as well.
\smallskip

Since $\fI^1$ is a ground ring module, as well as  the lowest
ghost number subspace in $\fI$, it is natural to repeat the
construction of Section \Spolybv. Namely, consider the map
\eqn\mappip{
\pi'\equiv\pi_{r_1}\oplus \pi_{r_2}\,:\, \fI^n\,\longrightarrow
\cP^{n-1}(\cR_3,\fI^1)\cong \fP^{n-1}_{r_1}\oplus\fP^{n-1}_{r_2}\,,}
which is equal to the identity on $\fI^1$, while for $n\geq 2$ it is
given by the multiple brackets \generpolof.  (Since $\fI$ is a
BV-ideal, all brackets \generpolof\ lie in $\fI$ for all
$\Ph\in \fI$, the map $\pi'$ is well-defined on $\fI$, which of course
includes the bulk region in Theorem \genmodcoh.)  It is
straightforward to verify by induction on the ghost number $n$, \cf,
Section \SSpolyder, that
\eqn\dothomie{
\pi'(\Ph\cdot \Psi) ~=~ \pi(\Ph)\cdot \pi'(\Psi)\,,\qquad
\Ph\in\fH\,,\quad \Psi\in\fI\,,}
where the product on the right hand side corresponds to the dot action
of $\fP$ on $\fP_{r_1}\oplus \fP_{r_2}$. In fact, the latter space is a
G-module of $\fP$ (see Theorem \plvmodule), and  a
stronger result holds.

\thm\gmodofi
\proclaim Theorem \gmodofi. The map $\pi'$ is a G-morphism between  the
G-module $\fI$ of $\fH$ and the G-module $\fP_{r_1}\oplus \fP_{r_2}$
of $\fP$; \ie, in addition to \dothomie\ we also have
\eqn\brahomie{
\pi'([\,\Ph,\Psi\,]) ~=~ [\,\pi(\Ph),\pi'(\Psi)\,]_M\,, \qquad
\Ph\in\fH\,,\quad \Psi\in\fI\,.}
\smallskip

\proof Let $\Ph\in \fH^m$ and $\Psi\in \fI^n$. Once more the proof
follows by induction on $m+n$. In particular, for $m=0$ and $n=1$ both
sides of \brahomie\ vanish -- the left one because $\fI^0\cong 0$,
while the right one by the definition of the bracket action of $\fP$
on $\fP_{r_1}\oplus \fP_{r_2}$. Next take $m=1$ and $n=1$.  Using
decomposition \decomofhone\ and Lemma \bronid, the only case in which
both sides do not vanish automatically is for  $\Ph\in
\fH^1_1\cong\imath(\fP)$. Then the equality follows from the isomorphism of
$\widehat M_{r_i}$ and $M_{r_i}$ as G-modules. The general step of the
induction is now completed similarly as in the proof of Theorem
\maintheorem, using the definition of the bracket action of
$\fP$. \SMu

\thm\conjfori
\proclaim Conjecture \conjfori. Consider $\fP_{r_1}\oplus\fP_{r_2}$ as a
BV-module of $\fP$, with the (conjectured) BV-operator
$\De=\De_1\oplus \De_2$ defined in \defoftwdel\ and \twistbvop.  Then
$\pi'$ is a BV-morphism between BV-modules.
\smallskip

The ideal $\fI$ at weights $(0,-2\La_1-2\La_2)$ is spanned by the
operators $T_{0,-2\La_1-2\La_2}$, $T^{[2]}_{0,-2\La_1-2\La_2}$ and
$T^{[3]}_{0,-2\La_1-2\La_2} \approx \widehat \Ga$. By an explicit
computation we verify that while
\eqn\projofga{
\pi'(\widehat \Ga)=\Ga_1+\Ga_2\,,}
the other two operators are mapped to zero. This shows that $\pi'$ has
a nontrivial cokernel. In fact, by examining the $\bga$ decomposition of
$\fI$ and $\fP_{r_1}\oplus\fP_{r_2}$, as well as a number of explicit
checks, we conclude that $\pi'$ is onto except at the weight
$(0,-2\La_1-2\La_2)$. Let us denote the image $\pi'(\fI)=\fP'$.
\smallskip

Let $\fI'=\ker\,\pi'$. Using \dothomie\ and \brahomie\ we show that
$\fI'$ is a G-ideal (conjecturally, a BV-ideal) of $\fH$. It will turn
out convenient to factor out from $\fI'$ the doublet, $\fD_{sp}$, of
special states introduced in Lemma \spstdb, and write
\eqn\decofdblt{
\fI'\cong \fH(-)\oplus \fD_{sp}\,.}
Consider the quotient $\fH(+)\cong\fH/\fH(-)$. Note that as a vector
space $\fH(+)$ is
isomorphic with $\fP\oplus\fP'\oplus\fD_{sp}$. By examining the $\bga$
decomposition of $\fH(-)$ and $\fH(+)$ we concluded that each
comprises precisely ``one half of the cohomology'' in the following
sense.

\thm\bvminus
\proclaim Conjecture \bvminus. Let  $\langle\,-,-\,\rangle_\fH$ be
the nondegenerate
bilinear form on $\fH$, introduced in Section \SSbilinear. Then
\smallskip
\item{i.} The form $\langle\,
-,-\,\rangle_\fH$ vanishes identically on $\fH(-)$.
\item{ii.}  As a vector space, $\fH(+)$ is isomorphic with the
dual subspace to
$\fH(-)$ in $\fH$ with respect to this form.
\smallskip

Most of this conjecture follows from the $\bga$ decomposition. Only
the cases where at the same Liouville weight there are states both in
$\fH(+)$ and $\fH(-)$ require a more detailed analysis. We have checked
explicitly some of those cases for low lying weights. The extension to the
general case is then consistent with  the expected module structure
of both spaces with respect to the dot action of polyderivations to
be discussed shortly.
\smallskip

The question now is whether one can construct $\fH(+)$ as a (natural)
subspace in $\fH$. In other words we would like to find an extension
of the embedding $\imath : \fP\rightarrow \fH$ to $\fP'$ and
$\fD_{sp}$. The embedding of the special doublet is unambiguous.
However, simple kinematics shows that such an embedding on $\fP'$ is
ambiguous in the overlap regions with $\imath(\fP)$ and $\fH(-)$. To
resolve this problem we propose to proceed as in Section \SSmain, and
use the explicit parametrization of $\fP'$ in terms of free modules of
the chiral subalgebras given in Appendix \freemodules. Thus we set
\eqn\imofga{
\imath(\Ga_i)=\widehat \Ga\,,}
and then require that $\imath(\fP_{r_i})$ is freely generated from
$\widehat \Ga$ in $\fH$ as a G-submodule of the respective
holomorphic subalgebra $\imath(\fP_-)$ or $\imath(\fP_+)$. More
explicitly, this construction yields
\eqn\expliii{
\imath(\Ph_0\cdot[\,\Ph_1,[\,\ldots,[\,\Ph_n,\Ga_i]\,\ldots\,]\,])
=\imath(\Ph_0)\cdot[\imath(\Ph_1),[\,\ldots,[\,\imath(\Ph_n),\widehat
\Ga\,]\,\ldots\,]\,]\,,\quad \Ph_0\,,\ldots,\Ph_n\in\fP_\mp\,.}
{}From now on we will identify $\fH(+)$ with the image $\imath(\fP\oplus
\fP'\oplus \fD_{sp})\subset \fH$.
\smallskip

To summarize, we have constructed an explicit decomposition
\eqn\decomofh{\fH\cong\fH(-)\oplus \fH(+)\,,}
where $\fH(\pm)$ is completely isotropic with respect to the bilinear
form on $\fH$, and $\fH(+)$ is dual to $\fH(-)$.
\smallskip

The duality between $\fH(-)$ and $\fH(+)$, due to  the ``hermiticity'' of
the ground ring generators with respect to the  bilinear form (which can
be proved using explicit expressions in Appendix  \SSgrring), holds as
the duality of  ground ring modules.
\smallskip

It follows from Conjecture \conjfori\ that $\fH(-)$ should be a
BV-ideal in $\fH$.  A combination of kinematics and explicit checks
suggest that $\fH(+)\subset \fH$ is a submodule with respect to the
dot action of the subalgebra $\imath(\fP)\subset \fH$.
\smallskip

Finally,  let us compare the result above with the one
for the BV-algebra associated with the Virasoro string.\foot{see, the
summary in Appendix \AVir} The decomposition \decomofh\ of the algebra
as a ground ring module is an analogue of the similar decomposition of
$H(\cW_2,\fC)$ [\LZbv]. However, unlike in the Virasoro case, now
$\fH(+)$ is much larger than the algebra of polyderivations of the
ground ring.

\subsubsec{Concluding remarks and open problems}
\subsubseclab\SSconject

The above analysis of the BV-algebra $(\fH,\,\cdot\,,b_0)$
is completely consistent with the result for the cohomology in
the fundamental Weyl chamber and its -- partially conjectural -- extension
to the remaining Weyl chambers. We have also exhibited a
geometric structure underlying this algebra; it turns out to be modeled on
the algebra of (generalized) polyderivations on the base affine space
of $SL(3,\CC)$.  Still, the problem of properly understanding the global
structure of this algebra remains open.
\smallskip

To conclude this paper we would like to review briefly two
problems that we could not solve completely at this stage. Quite
likely some of them will require a qualitatively new insight and not
merely a refinement of the approach advocated above.
\smallskip

\noindent
{\it 1. The proof of the cohomology for $-i\La^L+2\rh \not\in P_+\cup
w_0P_+$}.

It is rather surprising that, unlike in the case of the Virasoro
algebra, the cohomology of the $\cW_3$ algebra with values in the
tensor product of two Fock modules cannot be computed directly, \ie,
without resorting to an indirect procedure. While the formal
origin of this difficulty is clear -- the presence of quartic terms
precludes any simple spectral sequence argument -- one might
hope that by a suitable field redefinition the problem could  become
tractable.
\smallskip

More along the lines of the proof in the fundamental Weyl chamber one
could try to construct a new class of highest weight modules of the
$\cW_3$ algebra that are ``dual'' to the $c^L = 98$ Fock modules
$F(\La^L,2i)$, $-i\La^L+2\rh \not\in P_+\cup w_0P_+$, in the sense of
the reduction theorem (Theorem \redthm), just as contragredient Verma
modules are ``dual'' to Fock modules when $-i\La^L+2\rh \in P_+$.
(This cohomological ``construction'' of modules is largely motivated
by the counterpart problem in the representation theory of affine Lie
algebras, where the corresponding cohomology is that with respect to
the (twisted) nilpotent subalgebra -- see [\FeFrtw], and also
[\BMPkar], for further details.)  By constructing resolutions of
$c^M=2$ irreducible modules in terms of those new modules one could
compute the cohomology in a straightforward manner.
\smallskip

\noindent
{\it 2. Is $\fH$ generated by $\imath(\fP)$ as a BV-algebra?}

We have seen that $\imath(\fP)$ is not closed under $b_0$. In fact the
subspace generated by the bracket and dot action of $\imath(\fP)$ on
itself contains at least the subspace $\fH(+)$.  This follows from the
discussion in Section \SSdualdec, and the observation that the special
doublet, $\fD_{sp}$, lies in the subspace spanned by elements of the
form $[\,\widehat x_\si,[\,\widehat x_{\dot\si},b_0\widehat X]$ and
$\widehat C\cdot [\,\widehat x_\si,[\,\widehat x_{\dot\si}, b_0\widehat
X]$.
\smallskip

While $\imath(\fP)$ is closed under the dot product, this is not the
case for $\fH(+)$; dot products of some elements in $\fH(+)
\cap \fI$ lie in $\fH(-)$.\foot{Note that in the Virasoro case the dot
product on $\fI$ is always zero.} For instance, the ``square'' of the
special state yields
\eqn\sqspst{
\Psi_{0,-\La_1-\La_2}^{[2]}\cdot \Psi_{0,-\La_1-\La_2}^{[2]}
\sim T^{[2]}_{0,-2\La_1-2\La_2}\,.}
Similarly, further products between $\fH(+)\cap
\fI$ and $\fI'$ are nonvanishing. A good example is given by the
products of the
tips of twisted  cones $\widehat\Om_w$, which lie in $\fI'$ for
$w=r_{12}$, $r_{21}$ and $r_3$. We find
\eqn\profomeg{
\widehat \Om_{r_1}\cdot \widehat \Om_{r_{12}}\sim \widehat
\Om_{r_3}\,, \quad \widehat \Om_{r_2}\cdot \widehat \Om_{r_{21}}
\sim \widehat \Om_{r_3}\,.}

To gain some  insight into  the full structure of $\fH$ we have
studied the BV-subalgebra $\fH_{\rm singl}$, consisting of all
elements of $\fH$ transforming as singlets under $\slth$. This algebra is
finite-dimensional and is spanned by 19 quartets, that are easily read
off from Table \gencones. The elements of  $\imath(\fP)$ form a quartet at
the Liouville weight $0$, three doublets (with respect to $b_0$) at
$\La_1-2\La_2$, $-2\La_1+\La_2$ and $-\La_1-\La_2$, and a single element,
$\widehat X$, at $-2\La_1-2\La_2$. It appears that those elements
generate the entire $\fH_{\rm singl}$ as a BV-algebra.
\smallskip

At this point it is tempting to conjecture that also $\fH$ is
generated from $\imath(\fP)$. Unfortunately, we were not able to calculate
any nontrivial example, beyond the singlet subalgebra, that would
support such conjecture. If this conjecture turned out false, one
would have to understand what is the significance of the (proper)
subalgebra generated by $\imath(\fP)$ inside $\fH$.

\vfill\eject

%
%
%
\noindent
{\bf APPENDICES}
\bigskip

\appendix{A}{Verma modules at $c=2$}
\applab\appPBA

\appsubsec{Primitive vectors}
\appsubseclab\appPBAa
\bigskip

%
%
\tbl\tbPBaa
\begintable
$\,\,h\quad$| $M[0,0]$ | $M[1,1]$ | $M[3,0]$ | $M[0,3]$ |
$M[2,2]$ | $M[4,1]$ | $M[1,4]$ | $M[3,3]$ | $\cdots$ \cr
0 | $u_{00}$   ||||||||\nr
1 | $(v_{11},u_{11})$ | $u_{11}$ |||||||\nr
3 | $u_{30}$ | $u_{30}$ | $u_{30}$ | |||||\nr
  | $u_{03}$ | $u_{03}$ | | $u_{03}$ | | | | | \nr
4 | $(w_{22}, v_{22}, u_{22})$ | $(v_{22}, u_{22})$ | $u_{22}$ | $u_{22}$
  | $u_{22}$ ||||\nr
7 | $\vdots$ |  $(v_{41}, u_{41}) $ |
    $(v_{41}, u_{41}) $  |  $ u_{41} $ | $u_{41}$
  | $u_{41}$ ||| \nr
  | $\vdots$ | $(v_{14} ,u_{14})$ | $u_{14}$ | $(v_{14} ,u_{14})$
  | $u_{14}$ || $u_{14}$ ||\nr
9 | $\vdots$ | $(w_{33}, v_{33}, u_{33})$ | $(v_{33}, u_{33})$ |
    $(v_{33}, u_{33})$ | $(v_{33}, u_{33})$ | $u_{33}$ | $u_{33}$ | $u_{33}$
|\nr
  $\vdots$ | $\vdots$ |$\vdots$ |$\vdots$ |$\vdots$ |$\vdots$ |$\vdots$
  |$\vdots$ |$\vdots$ |$\vdots$
\endtable
\bigskip

\centerline{Table \tbPBaa. Primitive vectors in $M[s_1,s_2]$ (triality $0$)}
\bigskip\bigskip

%
\tbl\tbPBab
\begintable
$\,\,h\quad$|   $M^{(2)}[1,1]$ | $M^{(2)}[2,2]$ | $M^{(2)}[4,1]$ |
$M^{(2)}[1,4]$ | $M^{(2)}[3,3]$ |  $\cdots$ \cr
1 | $(v_{11},u_{11})$ ||||| \nr
3 | $u_{30},\,u_{30}'$ |  ||||\nr
  | $u_{03},\,u_{03}'$ |  | | | | \nr
4 | $(w_{22}, v_{22},u_{22}),\,u'_{22}$ | $(v_{22},u_{22})$ ||||\nr
7 | $\vdots$ |  $(v_{41},u_{41})$
  | $(v_{41},u_{41})$ ||| \nr
  | $\vdots$ |  $(v_{14},u_{14})$ || $(v_{14},u_{14})$ ||\nr
9 | $\vdots$ | $(w_{33}, v_{33},u_{33}),\,u_{33}'$ | $(v_{33},u_{33})$ |
    $(v_{33},u_{33})$ | $(v_{33},u_{33})$ |\nr
     $\vdots$ |$\vdots$ |$\vdots$ |$\vdots$
  |$\vdots$ |$\vdots$ |$\vdots$
\endtable
\bigskip

\centerline{Table \tbPBab. Primitive vectors in $M^{(2)}[s_1,s_2]$
(triality $0$)}
\vfil\eject

%
\tbl\tbPBac
\begintable
$\,\,h\quad$| $M[1,0]$| $M[0,2]$ | $M[2,1]$ | $M[1,3]$ | $M[4,0]$ |
$M[3,2]$ | $M[0,5]$ | $M[2,4]$ | $\cdots$ \cr
\txt{1\over3} | $u_{10}$ ||||||||\nr
\txt{4\over3} | $u_{02}$ | $u_{02}$ |||||||\nr
\txt{7\over3} | $(v_{21},u_{21})$ | $u_{21}$ | $u_{21}$ ||||||\nr
\txt{13\over3}| $(v_{13},u_{13})$ | $(v_{13},u_{13})$ | $u_{13}$ |
$u_{13}$ ||||| \nr
\txt{16\over3}| $u_{40}$ |$u_{40}$ | $u_{40}$ |  | $u_{40}$ ||||\nr
\txt{19\over3}| $(w_{32}, v_{32},u_{32})$ | $(v_{32},u_{32})$ |
$(v_{32},u_{32})$ | $u_{32}$
	       | $u_{32}$ | $u_{32}$ ||| \nr
\txt{25\over3}| $\vdots$ |$u_{05}$ | $u_{05}$ |  $u_{05}$ | | | $u_{05}$ ||\nr
\txt{28\over3}| $\vdots$ | $(w_{24}, v_{24}, u_{24})$ | $(v_{24},u_{24})$ |
$(v_{24},u_{24})$ |
		 $u_{24}$ |  $ u_{24}$ | $ u_{24}$ |$ u_{24}$ |\nr
$\vdots$ |$\vdots$ |$\vdots$ |$\vdots$ | $\vdots$ |$\vdots$ |$\vdots$
|$\vdots$ |$\vdots$|$\vdots$
\endtable
\bigskip

\centerline{Table \tbPBac. Primitive vectors in $M[s_1,s_2]$ (triality $1$)}
\bigskip\bigskip

%
\tbl\tbPBad
\begintable
$\,\,h\quad$| $M^{(2)}[2,1]$ |
  $M^{(2)}[1,3]$ | $M^{(2)}[3,2]$ | $M^{(2)}[2,4]$ | $\cdots$ \cr
\txt{7\over 3} | $(v_{21},u_{21})$ ||||\nr
\txt{13\over 3}| $(v_{13},u_{13})$ | $(v_{13},u_{13})$ ||| \nr
\txt{16\over 3}| $u_{40},\,u_{40}'$ ||||\nr
\txt{19\over 3}| $(w_{32}, v_{32}, u_{32}),\,u_{32}'$ |
		 $(v_{32},u_{32})$ | $(v_{32},u_{32})$  || \nr
\txt{25\over 3}| $\vdots$ |$u_{05},\,u_{05}'$ ||| \nr
\txt{28\over 3}| $\vdots$ |$(w_{24}, v_{24},u_{24}),\,u_{24}'$ |
$(v_{24},u_{24})$ |
		 $(v_{24},u_{24})$ | \nr
$\vdots$ |$\vdots$ |$\vdots$ |$\vdots$ |$\vdots$ | $\vdots$
\endtable
\bigskip

\centerline{Table \tbPBad. Primitive vectors in $M^{(2)}[s_1,s_2]$
(triality $1$)}
\vfil\eject

\appsubsec{Irreducible modules}
\appsubseclab\appPBAb
\bigskip

%
\tbl\tbPBae
\begintable
$ h \backslash [s_1,s_2]$ | $[0,0]$ | $[1,1]$ | $[3,0]$ | $[2,2]$ | $[4,1]$ |
$[3,3]$ \cr
0  | 1  |    |    |    |    |    \nr
1  | 0  | 1  |    |    |    |    \nr
2  | 1  | 2  |    |    |    |    \nr
3  | 2  | 3  | 1  |    |    |    \nr
4  | 3  | 6  | 1  | 1  |    |    \nr
5  | 4  | 10 | 3  | 2  |    |    \nr
6  | 8  | 16 | 5  | 5  |    |    \nr
7  | 10 | 27 | 9  | 8  | 1  |    \nr
8  | 17 | 42 | 14 | 16 | 2  |    \nr
9  | 24 | 64 | 25 | 26 | 4  | 1  \nr
10 | 36 | 98 | 37 | 45 | 8  | 2
\endtable
\bigskip

\centerline{Table \tbPBae.  ${\rm dim\,}L[s_1,s_2]_{(h)}$ (triality 0)}
\bigskip\bigskip

%
\tbl\tbPBaf
\begintable
$ h \backslash [s_1,s_2]$ | $[1,0]$ | $[0,2]$ | $[2,1]$ | $[1,3]$ | $[4,0]$ |
$[3,2]$ | $[0,5]$ | $[2,4]$ \cr
\txt{1\over3}  | 1  |    |    |    |    |    |    |    \nr
\txt{4\over3}  | 1  | 1  |    |    |    |    |    |    \nr
\txt{7\over3}  | 2  | 1  | 1  |    |    |    |    |    \nr
\txt{10\over3} | 3  | 3  | 2  |    |    |    |    |    \nr
\txt{13\over3} | 6  | 4  | 4  | 1  |    |    |    |    \nr
\txt{16\over3} | 9  | 8  | 7  | 2  | 1  |    |    |    \nr
\txt{19\over3} | 15 | 12 | 13 | 4  | 1  | 1  |    |    \nr
\txt{22\over3} | 22 | 21 | 21 | 8  | 3  | 2  |    |    \nr
\txt{25\over3} | 35 | 31 | 35 | 14 | 5  | 5  | 1  |    \nr
\txt{28\over3} | 51 | 50 | 55 | 24 | 10 | 9  | 1  | 1  \nr
\txt{31\over3} | 77 | 73 | 87 | 40 | 15 | 17 | 3  | 2
\endtable
\bigskip

\centerline{Table \tbPBaf. ${\rm dim\,}L[s_1,s_2]_{(h)}$ (triality 1)}
\vfil\eject

\appsubsec{Verma modules}
\appsubseclab\appPBAc
\bigskip

%
\tbl\tbPBag
\begintable
$ h \backslash  S$ | $\{u_{00}\}$ | $\{v_{11}\}$ |
$\{u_{11}, w_{22}\}$ | $\{u_{11}\}$ | $\{u_{30}, u_{03}, v_{22}\}$
| $\{u_{30}, u_{03}\}$ |
$\{u_{30}, v_{22}\}$ | $\{u_{30}\}$ | $\{v_{22}\}$ | $\{ u_{22} \}$ \cr
0 | 1  |    |    |    |    |    |    |    |    |    \nr
1 | 2  | 2  | 1  | 1  |    |    |    |    |    |    \nr
2 | 5  | 4  | 2  | 2  |    |    |    |    |    |    \nr
3 | 10 | 8  | 5  | 5  | 2  | 2  | 1  | 1  |    |    \nr
4 | 20 | 17 | 11 | 10 | 4  | 3  | 3  | 2  | 2  | 1  \nr
5 | 36 | 32 | 22 | 20 | 10 | 8  | 7  | 5  | 4  | 2  \nr
6 | 65 | 57 | 41 | 36 | 20 | 15 | 15 | 10 | 10 | 5
\endtable
\bigskip

\centerline{Table \tbPBag. ${\rm dim\,}M(S)_{(h)}$ for $S \subset M[0,0]$}
\bigskip\bigskip

%
\tbl\tbPBah
\begintable
$ h \backslash  S$ | $\{u_{10}\}$ | $\{v_{21}\}$ |
$\{v_{21}, u_{20}\}$  \cr
\txt{1\over3}  | 1  |    |    \nr
\txt{4\over3}  | 2  |    | 1  \nr
\txt{7\over3}  | 5  | 2  | 3  \nr
\txt{10\over3} | 10 | 4  | 7  \nr
\txt{13\over3} | 20 | 10 | 14 \nr
\txt{16\over3} | 36 | 19 | 27 \nr
\txt{19\over3} | 65 | 38 | 50 \nr
\txt{22\over3} | 110| $*$| $*$
\endtable
\bigskip

\centerline{Table \tbPBah. ${\rm dim\,}M(S)_{(h)}$ for $S \subset M[1,0]$}
\vfil\eject

%
\tbl\tbPBai
\begintable
$ h \backslash  S$ | $\{ u_{11} \}$ |  $\{ u_{30}, u_{03}, v_{22} \}$ |
$\{ u_{30}, u_{03}\}$ | $\{ u_{30}, u_{03}, w_{33} \}$ |
$\{ u_{03}, w_{33}, v_{41} \}$ | $\{u_{30}\}$ | $\{v_{22}\}$ \cr
1 | 1  |    |    |    |    |    |    \nr
2 | 2  |    |    |    |    |    |    \nr
3 | 5  | 2  | 2  | 2  | 1  | 1  |    \nr
4 | 10 | 4  | 3  | 3  | 2  | 2  | 2  \nr
5 | 20 | 10 | 8  | 8  | 5  | 5  | 4  \nr
6 | 36 | 20 | 15 | 15 | 10 | 10 | 10 \nr
7 | 65 | 38 | 30 | 30 | 21 | 20 | 20 \nr
8 | 110| 68 | 52 | 52 | $*$| 36 | 40 \nr
9 | 185| 121| 94 | 95 | $*$| 65 | 71 \nr
10| 300| 202| 155| 157| $*$| 110| 128
\endtable
\bigskip

\centerline{Table \tbPBai. ${\rm dim\,}M(S)_{(h)}$ for $S
\subset M[1,1]$}\bigskip
\vfil\eject


\appendix{B}{Vertex Operator Algebras associated to root lattices}
\applab\appPBC

 In this appendix we explicitly construct a VOA, in the chiral algebra
$\fV$ of two free scalar fields with momenta lying on the root lattice
of $\slth$, which includes the currents of the affine Lie algebra
$\hslth$.  As discussed under \eqPBcaa, the principal condition which
we must account for is the ``statistics'' of the VOA -- that under
interchange of order the OPEs of any two fields in the VOA are related
by analytic continuation.

 Let $Q$ be the root lattice of a simple simply-laced
Lie algebra $\bfg$, and let $c_\al$ be a set
of (momentum dependent) operators on $Q$.
The VOA associated to
the lattice $Q$ involves, in particular, the assignment of a vertex operator
$\cV_\al(z) = V_\al(z) c_\al$ to each $\al\in Q$, where
$V_\al(z) = e^{i\al\cdot \ph(z)}$ and $c_\al$ is chosen such that
$\hat{c}_\al \equiv e^{iq\cdot \al} c_\al$ satisfies
\eqn\eqCoa{ \eqalign{
\hat{c}_\al \hat{c}_\be & = e^{i\pi (\al,\be)} \hat{c}_\be \hat{c}_\al\,,\cr
\hat{c}_0 & = 1 \,,\cr}
}
for all $\al,\be\in Q$.
Note that \eqCoa\ is precisely required to implement the statistics condition,
since for any two exponential operators, $e^{i\la\cdot \ph(z)}$
and $e^{i\la'\cdot \ph(z)}$,
\eqn\eqJMphase{
e^{i\la\cdot \ph(z)} e^{i\la'\cdot \ph(w)} ~=~ (z-w)^{\la\cdot\la'}
e^{i\la\cdot \ph(z) + i\la'\cdot \ph(w)} \, .
}
The extension of the statistics condition to the exponential operators
corresponding to the rest of the root lattice is discussed later.  It
is enough to just consider the purely exponential operators in $\fV$
since contributions to the OPE from the polynomial field prefactors
are clearly meromorphic and satisfy the condition automatically.

  We may interpret \eqCoa\ as the statement that $\hat{c}_\al$ defines
a central extension of $Q$ by the group $\ZZ/2\ZZ \cong \{\pm1\}$
[\FK].  Such central extensions are uniquely specified by a 2-cocycle
$\ep\,:\,Q\times Q \mapright{} \{\pm1\}$, satisfying
\eqn\eqCoc{
\ep(\al,\be) \ep(\al+\be,\ga) = \ep(\al,\be+\ga) \ep(\be,\ga) \,,
}
\eqn\eqCod{
\ep(\al,\be) = e^{i\pi (\al,\be)} \ep(\be,\al)\,,
}
\eqn\eqCoe{
\ep(\al,0) = 1 \,,
}
for all $\al, \be, \ga \in Q$, through%
\foot{Given a 2-cocycle $\ep(\al,\be)$, the construction of $\hat{c}_\al$
is outlined in Section 5 of [\GNOS].  Despite the slight abuse of
language we will call the $c_\al$ ``phase-cocycles.'' }
\eqn\eqCof{
\hat{c}_\al \hat{c}_\be = \ep(\al,\be) \hat{c}_{\al + \be}\,.
}
Clearly, the consistency of \eqCof\ implies the 2-cocycle
condition \eqCoc, \ie, the fact that $\ep\in H^2(Q,\ZZ/2\ZZ)$,
while \eqCod\ and \eqCoe\ follow from \eqCoa.

A 2-cocycle $\ep$, satisfying \eqCoc-\eqCoe, is easily
constructed as follows [\FK].  In addition to \eqCoc-\eqCoe\ we
may impose a bilinearity condition
\eqn\eqCog{ \eqalign{
\ep(\al + \be,\ga) & = \ep(\al,\ga) \ep(\be,\ga) \,,\cr
\ep(\al,\be+\ga) & = \ep(\al,\be)\ep(\al,\ga) \,.\cr}
}
Then $\ep$ is completely specified by its values $\ep(\al_i,\al_j)$,
where $1\leq i \leq j \leq \ell$ and $\{\al_i\}_{i=1}^{\ell}$ is
a basis of $Q$ (\ie, a simple root system).

In our case, where $\bfg \cong \slth$, we may simply choose
\eqn\eqCoh{
\ep(\al_1,\al_1) = \ep(\al_2,\al_2) = \ep(\al_1,\al_2) = 1\,,}
from which it follows, \eg
\eqn\eqCoi{
\ep(\al_1,-\al_1) = \ep(\al_2,-\al_2) = 1 \,,
}
while
\eqn\eqCoj{
\ep(\al_2,\al_1) = \ep(\al_3,-\al_3) = -1 \,.
}
In fact, for arbitrary $\al, \be \in Q$ we then have
\eqn\eqCok{
\ep(\al,\be) = e^{i\pi (\La_2,\al)(\La_1,\be)} \,.
}
A $c_\al$, satisfying \eqCoa\ and \eqCof\ for the 2-cocycle \eqCok\ is
explicitly given by
\eqn\eqCol{
c_\al(p) = e^{i\pi p\cdot \xi(\al)} \,,
}
where
\eqn\eqCom{
\xi(\al) = (\La_2,\al)\La_1\,.
}

Now, if we restrict ourselves to $\al\in \De$ the modes of the vertex
operators $\cV_\al(z)$ will provide  a realization of $\hslth$
on $ \bigoplus_{\al\in Q} \, F(\al,0)$ ismorphic to $L(\La_0)$ (\ie
the so-called basic representation), albeit not in the ``conventional''
form.  In particular we would like to have $\ep(\al_3,-\al_3) =1$.
Clearly, a 2-cocycle $\ep$ satisfying
\eqCoc-\eqCoe\ is not unique, but can be modified
by a coboundary $\de(\ep)$, $\ep\,:\,Q \mapright{} \{ \pm1\}$, \ie
\eqn\eqCon{
\ep'(\al,\be) = \ep(\al,\be) \et(\al) \et(\be) \et(\al+\be)\,.
}
This corresponds to a change
\eqn\eqCoo{
\hat{c}'_\al = \et(\al) \hat{c}_\al\,.
}
(We need to take $\et(0)=1$ to preserve \eqCoa\ or, equivalently,
\eqCoe.)
We can use this `gauge freedom' to choose $\ep(\al,\be)$ such that
\eqn\eqCop{
\ep(\al,-\al) = 1 \,,
}
for all $\al\in Q$ or, equivalently,
\eqn\eqCoq{
\hat{c}_\al \hat{c}_{-\al} = 1\,.
}
For example we can take
\eqn\eqCor{
\et(\al) = \cases{ 1 & for $(\La_1,\al) \geq 0$ \cr
                   e^{i\pi (\La_1,\al)(\La_2,\al)} & for $(\La_1,\al)
                   < 0 \,.$ \cr}
}
Note that with this choice of cocycle we automatically have
\eqn\eqCos{ \eqalign{
\overline{\om}_{\cA} ( \hat{c}'_\al ) & =
  \overline{\om}_{\cA} ( \et(\al) e^{iq\cdot \al} e^{i\pi \al\cdot
  \xi(\al)} ) \cr
& = \et(\al) \et(-\al) e^{i\pi \al\cdot \xi(\al)} \hat{c}'_{-\al} \cr
& = \hat{c}'_{-\al} \cr}
}
such that
\eqn\eqCot{
\overline{\om}_{\cA} ( \cV_\al(z) ) = \left(
  {1\over z} \right)^{\al^2}  \cV_{-\al} ({1\over z})\,,
}
\ie, this choice makes the realization of $\hslth$ unitary with respect
to the Hermitean form defined by $\overline{\om}_{\cA}$.

\vfill\eject

\appendix{C}{Tables for resolutions of $c=2$ irreducible modules}
\applab\appPBB

\bigskip\bigskip

%
\tbl\tbPBba
\begintable
$h$ | $L[0,0]$ |$\crM{0}$ | $\crI{0}$ |$\crM{-1}$ |$\crI{-1}$ |$\crM{-2}$
|$\crI{-2}$ |$\crM{-3}$ |$\crI{-3}$ | $\crM{-4}$ \cr
0 | 1  | 1  |    |    |    |    |    |    |    | \nr
1 | 0  | 2  | 2  | 2  |    |    |    |    |    | \nr
2 | 1  | 5  | 4  | 4  |    |    |    |    |    | \nr
3 | 2  | 10 | 8  | 10 | 2  | 2  |    |    |    | \nr
4 | 3  | 20 | 17 | 20 | 3  | 4  | 1  | 2  | 1  | 1  \nr
5 | 4  | 36 | 32 | 40 | 8  | 10 | 2  | 4  | 2  | 2  \nr
6 | 8  | 65 | 57 | 72 | 15 | 20 | 5  | 10 | 5  | 5  \nr
7 | 10 | 110| 100| 130| 30 | 40 | 10 | 20 | 10 | 10
\endtable
\bigskip

\centerline{Table \tbPBba.  Dimensions for $L[0,0]$ resolution}
\bigskip\bigskip

%
\tbl\tbPBbb
\begintable
$h$ | $L[1,0]$ |$\crM{0}$ | $\crI{0}$ |$\crM{-1}$ |$\crI{-1}$ |$\crM{-2}$
|$\crI{-2}$ |$\crM{-3}$ |$\crI{-3}$ | $\crM{-4}$ \cr
\txt{1\over3}  | 1  | 1  |    |    |    |    |    |    || \nr
\txt{4\over3}  | 1  | 2  | 1  | 1  |    |    |    |    || \nr
\txt{7\over3}  | 2  | 5  | 3  | 4  | 1  | 1  |    |    || \nr
\txt{10\over3} | 3  | 10 | 7  | 9  | 2  | 2  |    |    || \nr
\txt{13\over3} | 6  | 20 | 14 | 20 | 6  | 7  | 1  | 1  || \nr
\txt{16\over3} | 9  | 36 | 27 | 40 | 13 | 15 | 2  | 2  || \nr
\txt{19\over3} | 15 | 65 | 50 | 76 | 26 | 32 | 6  | 7  | 1  | 1  \nr
\txt{22\over3} | 22 | 110| 88 | 137| 49 | 61 | 12 | 14 | 2  | 2
\endtable
\bigskip

\centerline{Table \tbPBbb.  Dimensions for $L[1,0]$ resolution}
\bigskip
\bigskip

\vfil\eject

%
\tbl\tbPBbc
\begintable
$h$ | $L[1,1]$ | $\crM{0}$ | $\crI{0}$ |$\crM{-1}$ |$\crI{-1}$ |$\crM{-2}$
|$\crI{-2}$ |$\crM{-3}$ |$\crI{-3}$ | $\crM{-4}$ \cr
1  | 1  | 1  |    |    |    |    |    |    || \nr
2  | 2  | 2  |    |    |    |    |    |    || \nr
3  | 3  | 5  | 2  | 2  |    |    |    |    || \nr
4  | 6  | 10 | 4  | 6  | 2  | 2  |    |    || \nr
5  | 10 | 20 | 10 | 14 | 4  | 4  |    |    || \nr
6  | 16 | 36 | 20 | 30 | 10 | 10 |    |    || \nr
7  | 27 | 65 | 38 | 60 | 22 | 24 | 2  | 2  || \nr
8  | 42 | 110| 68 | 112| 44 | 48 | 4  | 4  || \nr
9  | 64 | 185| 121| 202| 81 | 92 | 11 | 12 | 1  | 1  \nr
10 | 98 | 300| 202| 350| 148| 170| 22 | 24 | 2  | 2
\endtable
\bigskip

\centerline{Table \tbPBbc.  Dimensions for $L[1,1]$ resolution}
\bigskip
\bigskip

%
\tbl\tbJMbd
\begintable
$h$ | $L(2,0)$ |$\crM{0}$ | $\crI{0}$ |$\crM{-1}$ |$\crI{-1}$ |$\crM{-2}$
|$\crI{-2}$ |$\crM{-3}$ |$\crI{-3}$ | $\crM{-4}$ \cr
\txt{4\over3}  | 1  | 1  |    |    |    |    |    |    || \nr
\txt{7\over3}  | 1  | 2  | 1  | 1  |    |    |    |    || \nr
\txt{10\over3}  | 3  | 5  | 2  | 2  |   |   |    |    || \nr
\txt{13\over3} | 4  | 10 | 6  | 7  | 1  | 1  |    |    || \nr
\txt{16\over3} | 8  | 20 | 12 | 14 | 2  | 2  | 1  | 1  || \nr
\txt{19\over3} | 12  | 36 | 24 | 30 | 6 | 7 | 1  | 1  || \nr
\txt{22\over3} | 21 | 65 | 44 | 56 | 12 | 14 | 2  | 2  |   |   \nr
\txt{25\over3} | 31 | 110| 79   | 105| 26 | 31 | 5 | 5 |   |   \nr
\txt{28\over3} | 50 | 185| 135 | 182| 47 | 58 | 11 | 12 | 1  | 1
\endtable
\bigskip

\centerline{Table \tbJMbd.  Dimensions for $L(2,0)$ resolution}
\vfil\eject

\appendix{D}{Summary of explicit computations}
\applab\APexplicit
\appsubsec{Introduction}
\appsubseclab\SSapecintro

In this appendix we summarize the results of explicit computations of
the cohomologies \hfill\break $H(\cW_3,F(\La^M,0)\otimes F(\La^L,2i))$
that are required to determine tips of all cones in the proof of
Theorem \fullcoh. Given a Liouville weight, $\La^L$, the corresponding
matter weight, $\La^M$, is chosen to be the lowest lying positive
weight such that $(\La^M,\La^L)\in L$. This assures that the
cohomology will include states from all irreducible $\bga$ modules
$\cL(\La)\otimes \CC_{-i\La^L}$ that may arise at this particular
Liouville momentum. The number of states in each irreducible module is
given by the multiplicity $m^\La_{\La^M}$.
\smallskip

For a given $(\La^M,\La^L)$ the cohomology may arise only in the
finite dimensional subcomplex that is annihilated by $L_0^{tot}$.  All
operators in this subcomplex are of the form
$P\cV_{\La^M,-i\La^L}(z)$, where the prefactor $P$ is a polynomial
in all the fields (see Sections \Swcohprob\ and \SSslth), whose
dimension is equal to
\eqn\levelll{
h=\half|-i\La^L+2\rh|^2-\half|\La^M|^2-4\,.}
Thus the number, $d(h,n)$, of linearly independent
operators at ghost number $n$ is given by expanding the
partition function
\eqn\partfunct{
q^{-4} \,\prod_{m=0}^\infty
(1+tq^m)^2(1+t^{-1}q^{m+1})^2(1-q^n)^{-4}=
\sum_{h,n\in\ZZ}d(h,n)q^ht^n\,.}
\smallskip

To compute the action of the differential on the complex it is
necessary to determine the OPE of the BRST current with all operators
in a basis. Because of the algebraic complexity of this computation,
we have used the algebraic manipulations program Mathematica$^{\rm TM}$
together with the CFT package OPEdefs [\Th].  As a result we obtain at each
ghost number, $n$, a $d(h,n)\times d(h,n+1)$ complex matrix,
$(d_{n,n+1})$, of the differential.  The dimension of the kernel of
$d$ is then found as the number of zero eigenvalues of the
$d(h,n)\times d(h,n)$ hermitian matrix $(d_{n,n+1})^\dagger
(d_{n,n+1})$. Here the product of matrices is computed exactly, but
the eigenvalues in most cases are found using a numerical routine.
\smallskip

Because of symmetry that  exchanges the fundmental
weights $\La_1$ and $\La_2$, it is sufficient to compute only ``half''
of the cases.  As a consistency check, we have included, however, the
results that could be deduced using duality \dualcoh.
\smallskip

The results are summarized in Section \Appthetables.  The tables are
arranged according to the value of the level, $h$, defined in
\levelll.  Given $h$, we first determine which ghost numbers, $n$, may arise,
and what are the dimensions, $\dim\,C^n=d(h,n)$, of the corresponding
subspaces in the complex. Then for various choices of
$(\La^M,-i\La^L)$ we list dimensions, $\dim\,K^n$, of the kernels and
dimensions, $\dim\,H^n$, of cohomologies. The latter are computed
using
\eqn\cohdimmm{
\dim\,H^n=\dim\,K^n-(\dim\,C^{n-1}-\dim\,K^{n-1})\,.}
\smallskip

The cones can be identified by matching dimensions of the cohomologies
with the multiplicities of the modules that could be present.
Starting with the low (shifted) Liouville weights, this gives a
systematic way of determining the boundaries of all the cones.
\smallskip

As an illustration, let us verify Theorem \fullcoh\ at  weights $(0,0)$.
The representations and the corresponding multiplicities are:
\hfill\break

\begintable
$\La$ | $ \quad 0 \quad $ | $ \quad \La_1+\La_2 \quad $ | $ \quad 3\La_1$ |
$ \quad 3\La_2 $ | $ \quad 2\La_1+2\La_2 \quad$ | $ \quad \cdots  \quad $
\cr
$ \quad m^\La_0 \quad$ | 0 | 2 | 1 | 1 | 3 | $\cdots$
\endtable
\bigskip

The contribution from each cone to the cohomology is read off from
Table \gencones. This yields the result in the table below, which agrees
with an explicit computation.
\tbl\zerozero

\bigskip\bigskip
\begintable
$\La\,\backslash\, n$ |  \quad 0  \quad |  \quad 1  \quad |  \quad 2 \quad |
 \quad 3 \quad
|  \quad 4 \quad |  \quad 5 \quad  |  \quad 6 \quad
|  \quad 7  \quad |  \quad 8  \quad \cr
$0$ | 1 | 2 | 1 | | | | | | \nr
$\La_1+\La_2 $ | | 2 | 4 | 2 | | | | | \nr
$3\La_1 $ | | | 1 | 2 | 1 | | | | \nr
$ 3\La_2$ | | | 1 | 2 | 1 | | | | \nr
$\La_1+\La_2 $ | | | 2 | 4 | 2 | | | | \nr
$ \quad 2\La_1+2\La_2 \quad $ | | | | 3 | 6 | 3 | | | \cr
$\dim\,H^n$ | 1  | 4  | 9  | 13  | 10 | 3  | 0 | 0 | 0
\endtable
\medskip
\centerline{ Table \zerozero: $\,H(\cW_3,F(0,0)\otimes F(0,2i))$}
\bigskip

\noindent
All other cases are analyzed similarly.

\bigskip

\appsubsec{The tables}
\appsubseclab\Appthetables

\bigskip
\begintable
\raise-6pt\hbox{$(\La^M,-i\La^L)$} | $n$ |  \quad 3\quad  | \quad 4\quad
| \quad 5\quad   \nr
 | $\dim\,C^n$      | 1 | 2 | 1 \crthick
$(0,-2\La_1-2\La_2)$  | \quad $\dim\,K^n$  \quad  | 1 | 2  | 1 \nr
 | $\dim\,H^n$   | 1 | 2  | 1 \cr
$(\La_1,-\La_1-2\La_2)$ |
 \quad $\dim\,K^n$  \quad  | 1 | 2  | 1 \nr
 | $\dim\,H^n$   | 1 | 2  | 1  \cr
$(\La_2,-\La_1-3\La_2)$ | \quad $\dim\,K^n$  \quad  | 1 | 2  | 1 \nr
 | $\dim\,H^n$   | 1 | 2  | 1 \cr
$(0,-\La_1-4\La_2)$ | $\dim\, K^n$ | 1 | 2 | 1 \nr
 | $\dim \,H^n$ | 1 | 2 | 1 \cr
$\quad(\La_1,-2\La_1-3\La_2)\quad$ | \quad $\dim\,K^n$  \quad
 | 1 | 2  | 1 \nr
 | $\dim\,H^n$   | 1 | 2  | 1
\endtable
\medskip
\tbl\hhmfour
\centerline{Table \hhmfour: $h=-4$}
\bigskip

\vfill\eject

\begintable
\raise-6pt\hbox{$(\La^M,-i\La^L)$}
 | $n$ | \quad 2\quad | \quad 3\quad  | \quad 4\quad
| \quad 5\quad  | \quad 6\quad  \nr
 | $\dim\,C^n$     | 2 | 8 | 12 | 8 | 2 \crthick
$(0,-\La_1-\La_2)$ |  \quad $\dim\,K^n$  \quad | 1 | 5 | 8  | 6 | 2 \nr
 | $\dim\,H^n$ | 1 | 4 | 5  | 2 |  0 \cr
$(\La_2,-2\La_2)$ | \quad $\dim\,K^n$  \quad | 1 | 4 | 7  | 6 | 2 \nr
 | $\dim\,H^n$ | 1 | 3 | 3  | 1 |  0 \cr
$(0,-3\La_2)$ |  \quad $\dim\,K^n$  \quad | 1 | 5 | 8  | 6 | 2 \nr
 | $\dim\,H^n$ | 1 | 4 | 5  | 2 | 0 \cr
$(\La_1,-4\La_2)$ | \quad $\dim\,K^n$  \quad | 1 | 4 | 8  | 7 | 2 \nr
 | $\dim\,H^n$ | 1 | 3 | 4  | 3 | 1 \cr
$(0,-\La_1-4\La_2)$ |  \quad $\dim\,K^n$  \quad | 0 | 4 | 9  | 7 | 2 \nr
 | $\dim\,H^n$ | 0 | 2 | 5  | 4 | 1 \cr
$(\La_2,-2\La_1-4\La_2)$ |  \quad $\dim\,K^n$  \quad | 0 | 3 | 8  | 7 | 2 \nr
 | $\dim\,H^n$ | 0 | 1 | 3  | 3 | 1 \cr
$\quad(0,-3\La_1-3\La_2)\quad$ | \quad
   $\dim\,K^n$  \quad | 0 | 4 | 9  | 7 | 2 \nr
 | $\dim\,H^n$ | 0 | 2 | 5  | 4 | 1
\endtable
\medskip
\tbl\hhmthree
\centerline{Table \hhmthree: $h=-3$}

\bigskip
\bigskip
\begintable
\raise-6pt\hbox{$(\La^M,-i\La^L)$}
 | $n$ | \quad 1 \quad | \quad 2\quad | \quad 3\quad
| \quad 4\quad | \quad 5\quad  | \quad 6\quad
| \quad 7 \quad \nr
   | $\dim\,C^n$ | 1 | 12 | 39 | 56 | 39 | 12 | 1 \crthick
$(\La_1,-\La_2)$ | \quad $\dim\,K^n$ \quad | 0
   | 3  | 15 | 30 | 28 | 11 | 1  \nr
 | $\dim\,H^n$ | 0 | 2  | 6 | 6 | 2  | 0  | 0 \cr
$(\La_1,\La_1-3\La_2)$ |\quad $\dim\,K^n$  \quad |
      0 | 3 | 15 | 30 | 28 | 11 | 1 \nr
 | $\dim\,H^n$ | 0 | 2 | 6 | 6 | 2| 0 | 0 \cr
$(\La_2,\La_1-4\La_2)$ |  \quad $\dim\,K^n$  \quad |
      0 | 2 | 15 | 32 | 29 | 11 | 1 \nr
 | $\dim\,H^n$ | 0 | 1 | 5 | 8 | 5 | 1 | 0 \cr
$(\La_2,-5\La_2)$ |
 \quad $\dim\,K^n$  \quad | 0 | 2 | 15 | 32 | 29 | 11 | 1 \nr
 | $\dim\,H^n$ | 0 | 1 | 5 | 8 | 5 | 1 | 0 \cr
$(\La_1,-\La_1-5\La_2)$ |
 \quad $\dim\,K^n$  \quad | 0 | 1 | 13 | 32 | 30 | 11 | 1 \nr
 | $\dim\,H^n$ | 0 | 0 | 2 | 6 | 6| 2 | 0 \cr
$\quad (\La_1,-3\La_1-4\La_2)\quad $ |
 \quad $\dim\,K^n$  \quad | 0 | 1 | 13 | 32 | 30 | 11 | 1 \nr
 | $\dim\,H^n$ | 0 | 0 | 2 | 6 | 6| 2 | 0
\endtable
\medskip
\tbl\hhmtwo
\centerline{Table \hhmtwo: $h=-2$}

\vfill\eject

\begintable
 \raise-6pt\hbox{$(\La^M,-i\La^L)$}
  |  $n$ | \quad 1 \quad | \quad 2\quad | \quad 3\quad
| \quad 4\quad | \quad 5\quad  | \quad 6\quad
| \quad 7 \quad \nr
 | $\dim\,C^n$     | 8 | 56 | 152 | 208 | 152 | 56 |
8 \crthick
$(0,\La_1-2\La_2)$ | \quad $\dim\,K^n$  \quad |
    1 | 11 | 51 | 105 | 104 | 48 | 8 \nr
| $\dim\,H^n$ | 1  | 4 | 6  | 4 | 1 | 0 | 0 \cr
$(0,\La_1-5\La_2)$ |
 \quad $\dim\,K^n$  \quad | 0 | 10 | 52 | 108 | 106 | 48 | 8 \nr
 | $\dim\,H^n$ | 0 | 2 | 6 | 8 | 6 | 2 | 0 \cr
$\quad (0,-2\La_1-5\La_2)\quad $ |
 \quad $\dim\,K^n$  \quad | 0 | 8 | 49 | 107 | 107 | 49 | 8 \nr
  | $\dim\,H^n$ | 0 | 0 | 1 | 4 | 6 | 4 | 1
\endtable
\medskip
\tbl\hhmone
\centerline{Table \hhmone: $h=-1$}

\bigskip\bigskip

\begintable
\raise-6pt\hbox{$(\La^M,-i\La^L)$}  | $n$ |  \quad 0  \quad |  \quad 1
\quad |  \quad 2 \quad |
 \quad 3 \quad
|  \quad 4 \quad |  \quad 5 \quad  |  \quad 6 \quad
|  \quad 7  \quad |  \quad 8  \quad \nr  |
$\dim\,C^n$     | 2  | 39 | 208 |  513 | 684 | 513 | 208 | 39 | 2 \crthick
$(0 ,0 )$ |
 \quad $\dim\,K^n$  \quad | 1 | 6 | 43 | 178 | 345 | 342 | 171 | 37 | 2 \nr
 | $\dim\,H^n$ | 1  | 4 | 9 | 13 | 10 | 3 | 0 | 0 | 0 \cr
$(\La_1 , -\La_2 )$ |
 \quad $\dim\,K^n$  \quad | 0 | 3 | 41 | 176 | 344 | 342 | 171 | 37 | 2 \nr
 | $\dim\,H^n$ | 0  | 1 | 5 | 9 | 7 | 2 | 0 | 0 | 0 \cr
$(\La_2 , 2\La_1-3\La_2 )$ |
 \quad $\dim\,K^n$  \quad | 0 | 3 | 41 | 176 | 344 | 342 | 171 | 37 | 2 \nr
 | $\dim\,H^n$ | 0 | 1 | 5  | 9 | 7 | 2 | 0 | 0 | 0 \cr
$(0,2\La_1-4\La_2)$ |
 \quad $\dim\,K^n$  \quad | 0 | 2 | 41 | 180 | 348 | 343 | 171 | 37  | 2 \nr
 | $\dim\,H^n$ | 0 | 0 | 4 | 13 | 15 | 7 | 1 | 0 | 0 \cr
$(\La_1,2\La_1-5\La_2)$ |
 \quad $\dim\,K^n$  \quad | 0 | 2 | 38 | 176 | 348 | 344 | 171 | 37 | 2 \nr
 | $\dim\,H^n$ | 0 | 0 | 1 | 6 | 11 | 8 | 2 | 0 | 0 \cr
$(\La_1 , \La_1-6\La_2)$ |
 \quad $\dim\,K^n$  \quad | 0 |  2 |  39 | 177 | 347 | 343 | 171 | 37 | 2 \nr
 | $\dim\,H^n$ | 0 | 0 | 2 | 8 | 11 | 6 | 1 | 0 | 0 \cr
$( 0 , -6\La_2 )$ |
 \quad $\dim\,K^n$  \quad | 0  | 2  | 38 |  177 | 351 | 346 | 171 | 37 | 2 \nr
 | $\dim\,H^n$ | 0  | 0  |  1 | 7 | 15 | 13 | 4 | 0 | 0 \cr
$(\La_2 , -\La_1-6\La_2 )$ |
 \quad $\dim\,K^n$  \quad | 0  | 2  | 37  | 173 | 347 | 346 | 172 | 37
| 2 \nr
 | $\dim\,H^n$ | 0  | 0  | 0 | 2 | 7 | 9 | 5 | 1 | 0 \cr
$(\La_2,-3\La_1-5\La_2)$ |
 \quad $\dim\,K^n$  \quad | 0 | 2 | 37 | 173 | 347 | 346 | 172 | 37  | 2 \nr
 | $\dim\,H^n$ | 0 | 0 | 0 | 2 | 7 | 9 | 5 | 1 | 0 \cr
$\quad (0 , -4\La_1-4\La_2 ) \quad$ |
 \quad $\dim\,K^n$  \quad | 0 | 2  | 37 | 174 | 349 | 348 | 174 | 38 |
                       2 \nr
 | $\dim\,H^n$ | 0  | 0  | 0  | 3 | 10 | 13 | 9 | 4 | 1
\endtable
\medskip
\tbl\hhmzero
\centerline{Table \hhmzero: $h=0$}

\vfill\eject
\appendix{E}{A graphical representation of  $H_{pr}(\cW_3,\fC)$}
\applab\primeplots
\vfill
\epsffile{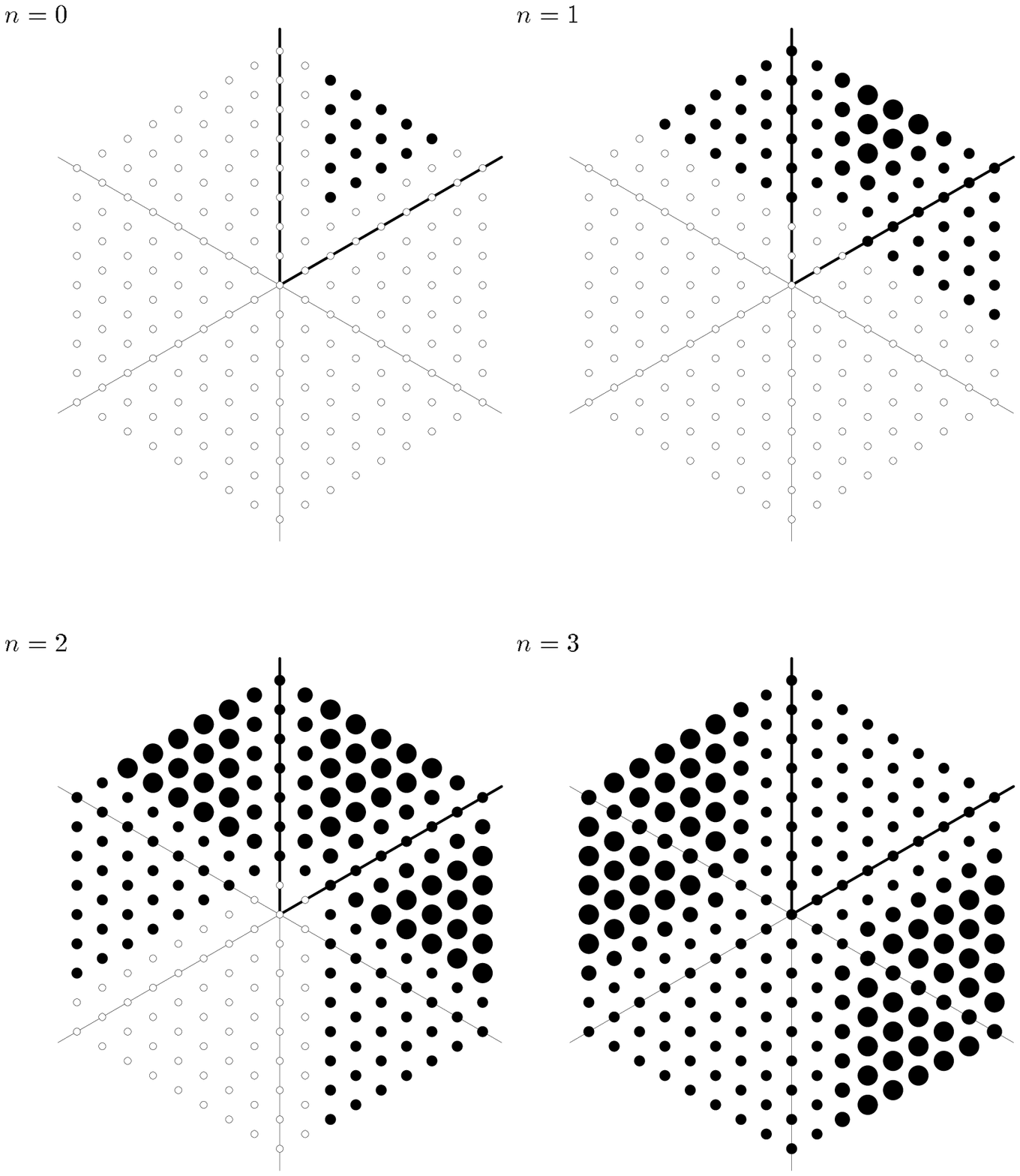}
\vfill\eject
\epsffile{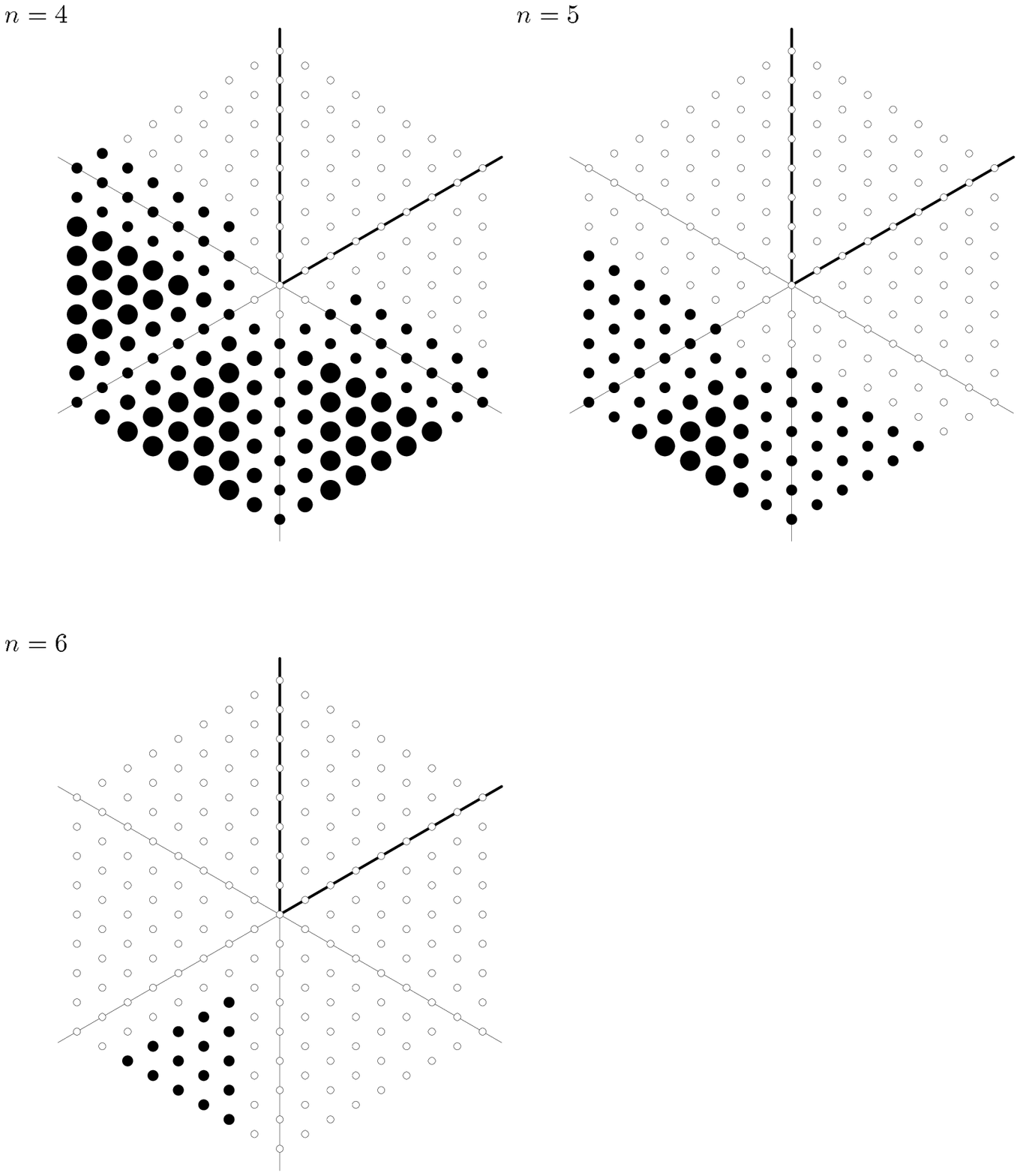}
\bigskip

\nfig\plotttsss{}
\noindent Figure \plotttsss. A schematic representation of
$H^n_{pr}({\cal W}_{3}, {\fC})$ (\cf, Table \gencones). The
points on the lattice correspond to shifted Liouville momenta,
$-i\Lambda^L +2\rho$, and the dots of increasing size indicate 0, 1, 2
and 3 irreducible $\slth$ modules of prime states. The
boundary of the fundamental Weyl chamber is outlined by thick lines.

\vfill\eject

\appendix{F}{Polyderivations $\cP(\cG_N)$}
\applab\Apolder

In this appendix we derive additional results on the polyderivations
$\cP(\cG_N)$ and, in particular, complete the proofs of Theorems
\genproduct\ and \bvopmth.

\appsubsec{Preliminary results}
\appsubseclab\Apprelim

Let
\eqn\ttensor{
T_{i_1\ldots i_m,j_1\ldots j_n}= x_{i_1}\vps\ldots x_{i_m}\vps
 x^*_{j_1}\ldots  x^*_{j_n}\,.}
As an $\sotwon$ tensor, $T_{i_1\ldots i_m,j_1\ldots j_m}$ is symmetric
and traceless in $i_1,\ldots, i_m$, and antisymmetric in
$j_1,\ldots,j_n$.

\thm\masterlemma\
\proclaim Lemma \masterlemma. Define the trace
\eqn\tracee{
\widehat T_{i_1\ldots i_m,j_1,\ldots j_n}=g^{ij}\,
T_{i\,i_1\ldots i_m,j\,j_1,\ldots j_n}\,.}
Then
\eqn\tracefactors{\eqalign{
T{}^{\,i_1\ldots i_m}{}_{,j_1\ldots j_n}&=\widetilde T{}^{\,i_1\ldots
i_m}{}_{,j_1,\ldots j_n}
+ a(m,n)\, \de^{(i_1}{}_{[j_1}\widehat T^{\,i_2\ldots i_m)}{}_{,j_2\ldots
j_n]}\cr & + b(m,n)\, \de^{(i_1}{}_{[j_1}\widehat
T{}^{\,i_2\ldots i_{m-1}}{}_{j_2,}{}^{i_m)}{}_{j_3\ldots j_n]}
+c(m,n)\,g^{(i_1i_2}\widehat
T{}^{\,i_3\ldots i_m)}{}_{[j_1,j_2\ldots j_n]}\,,\cr}}
where
\eqn\coefficients{\eqalign{
a(m,n)&=d(m,n)\,\left[(2N+2m-2)(2N+m-n-1)-2N\right]\,,\cr
b(m,n)&=-2 d(m,n)\,(m-1)(n-1)\,,\cr
c(m,n)&=-d(m,n)\,(2N+m-n)(m-1)\,,\cr}}
and
\eqn\dfactor{
d(m,n)={mn\over(2N+m-n)(2N+m-n-2)(2N+2m-2)}\,,} is the decomposition of
$T$ into its traceless and  trace components $\widetilde
T$ and $\widehat T$, respectively.
\smallskip

\proof One verifies by explicit algebra that $\widetilde T$ defined by
\tracefactors\ is indeed traceless in all pairs of indices
$i_1,\ldots, j_n$. \SMu

\thm\piciex
\proclaim Lemma \piciex. In the above notation we have
\eqn\pisandcis{
P_{i_1\ldots i_m,j_1\ldots j_n}=\widetilde T_{i_1\ldots
i_{m-1}[i_m,j_1\ldots j_n]}\,,\quad
C_{i_1\ldots i_m,j_1\ldots j_n}=\widehat
T_{i_1\ldots [i_m,j_1\ldots j_n]}\,.}

\proof See, the definitions in Section \polyderofcgn. \SMu
\smallskip

To avoid confusion, let us denote
\eqn\notone{
E^{i_1\ldots i_m}= x^{i_1}\ldots  x^{i_m}\,.}
{}From the decomposition \tracefactors\ and  Lemma \piciex\ we find

\thm\multlemma
\proclaim Lemma \multlemma.
\eqn\identthree{
E^{i_1\ldots i_{m}}P_{j_1,j_2\ldots j_n}=
P^{i_1\ldots i_m}{}_{j_1,j_2\ldots j_n}-
\hbox{$m(n-1)\over 2N+m-n$}\,
\de^{(i_1}{}_{[j_1}\vps C^{i_2\ldots i_m)}{}_{j_2,j_3\ldots
j_n]}\vps\,,}
\eqn\identmulti{
E^{i\,i_1\ldots i_{m}}P_{i,j_1\ldots j_n}=
-\hbox{$n\over n+1$}\,C^{i_1\ldots i_m}{}_{,j_1\ldots j_n}\,.}

\smallskip

Clearly these relation allow a convenient construction of the entire
basis in $\cP(\cG_N)$ in terms of products of the ``generating
elements'' $E_{i_1\ldots i_m}$, $P_{j_1,j_2\ldots j_n}$, and $C$.

\appsubsec{Proof of Theorem \genproduct}
\appsubseclab\ASproof

We may now proceed with the proof of Theorem \genproduct. The startegy
is to first consider the products of generating elements, and then
extend the result to arbitrary basis elements.
\smallskip

\noindent
{\it Case 1.} Since
\eqn\prodofes{
E_{i_1\ldots i_m}E_{j_1\ldots j_m}=E_{i_1\ldots j_n}\,,}
equation \bigproduct\ is clearly satisfied in this case.
\smallskip

\noindent
{\it Case 2.} Consider the Schouten bracket
\eqn\commone{
[E^{i_1\ldots i_m},P_{j_1,j_2\ldots j_n}]_S=
m(n-1)\,\de^{(i_1}{}_{[j_1}\vps x^{i_2}\ldots  x^{i_m)}
 x_{j_2}\vps  x^*_{j_3}\ldots x^*_{j_n]}\,.}
Substituting the decomposition \tracefactors\ in the rhs, we find that
all the trace terms vanish upon symmetrization in $i_1,\ldots,i_m$ and
antisymmetrization in $j_1,\ldots,j_n$, and we obtain
\eqn\identone{
[E^{i_1\ldots i_m},P_{j_1,j_2\ldots j_n}]_S=
m(n-1)\,\de^{(i_1}{}_{[j_1}\vps\,P^{i_2\ldots i_m)}{}_{j_2,j_3\ldots
j_n]}\vps\,.}
Hence, \identthree\ can be rewritten as
\eqn\newproduct{
E_{i_1\ldots i_{m}}P_{j_1,j_2\ldots j_n}= P_{i_1\ldots
i_mj_1,j_2\ldots j_n}-
\hbox{$1\over 2N+m-n$}\,C\,[E_{i_1\ldots i_m},P_{j_1,j_2\ldots
j_n}]_S\,,}
which proves Theorem \genproduct\ in this case.
\smallskip

\noindent
{\it Case 3.} The last special case follows easily from
identities in Section \galgstr. There we find
\eqn\productpis{
P_{i_1,i_2\ldots i_m} P_{j_1,j_2\ldots j_n}=
(-1)^{m-1}\,\hbox{$ m+n-1\over n$}\, x_{[i_1}\,P_{i_2,i_3\ldots
i_m  ] j_1\ldots j_n}\,,}
which, combined with \identthree\ and \sophigen, yields
\eqn\productpispr{
P_{i_1,i_2\ldots i_m} P^{j_1,j_2\ldots j_n}= (-1)^{m-1}\,\hbox{$
m+n-1\over n$}\,P_{[i_1i_2,i_3\ldots i_m]}{}^{j_1\ldots j_n} -
\hbox{$m+n-2\over 2N-m-n+2$}\de_{[i_1}{}^{[j_1}C_{i_2,i_3\ldots
i_m]}{}^{j_2\ldots j_n]}\,.}
However, we also have, see \twopcom,
\eqn\piscomm{
[P_{i_1,i_2\ldots i_m},P^{j_1,j_2\ldots j_n}]_S ~=~
(-1)^{m-1}\, (m+n-2)\de_{[i_1}\vps{}^{[j_1}\,P_{i_2,i_3\ldots
i_m]}\vps{}^{j_2\ldots j_n]}\,.}
This shows that \productpispr\ is indeed equivalent to \bigproduct.
\smallskip

Before we discuss the general case, let us simplify the notation, and
write $E_{(m)}$ for $E_{i_1\ldots i_m}$, $P_{(m,n)}$ for $P_{i_1\ldots
i_mi_{m+1} , i_{m+2}\ldots i_{m+n}}$, and  $C_{(m,n)}$ for
$C_{i_1\ldots i_mi_{m+1} , i_{m+2}\ldots i_{m+n}}$. Also, for any
$\sotwon$ tensor $T$, let
$(\!( T )\!)$ denotes its  traceless component.  Finally, let
\eqn\geeee{g(m,n)=\hbox{$1\over 2N+m-n$}\,.}
In this notation we may rewrite \newproduct\ as
\eqn\newprrr{
E_{(m)}P_{(0,n)}=(\!(E_{(m)}P_{(0,n)})\!)- g(m,n)
C[E_{(m)},P_{(0,n)}]_S\,,} where $(\!(E_{(m)}P_{(0,n)})\!)=P_{(m,n)}$,
and
\productpispr\ as
\eqn\twopispr{P_{(0,m)}P_{(0,n)}=(\!(P_{(0,m)}P_{(0,n)})\!)+ (-1)^m g(2,m+n)
C\,[P_{(0,m)},P_{(0,n)}]_S\,.}

\thm\auxlemma
\proclaim Lemma \auxlemma.
\eqn\auxident{
E_{(m)}C\,[E_{(m')},P_{(0,n')}]_S=\hbox{$1\over 1+mg(m',n')$}\,
\left(C\,[E_{(m+m')},P_{(0,n')}]_S-C\,[E_{(m)},P_{(m',n')}]\right)\,.}

\proof Using \newprrr, \prodofes, \Cph, and the Leibnitz rule for the
bracket, we obtain
\eqn\letsgo{
\eqalign{
E_{(m)}&C\,[E_{(m')},P_{(0,n')}]_S = C\,[E_{(m+m')},P_{(0,n')}]_S
-C\,[E_{(m)},E_{(m')}P_{(0,n')}]_S\cr
&=C\,[E_{(m+m')},P_{(0,n')}]_S-C\,[E_{(m)},P_{(m',n')}]
-mg(m',n')C\,E_{(m)} [E_{(m')},P_{(0,n')}]_S\,,
\cr}}
which implies \auxident. \SMu
\smallskip

Now, using Lemma \auxlemma\ and the identities above, we find
\eqn\etimesp{\eqalign{
E_{(m)}P_{(m',n')}&=E_{(m)}\left(E_{(m')}P_{(0,n')}+g(m',n')C\,
[E_{(m')},P_{(0,n')}]_S\right)\cr
&=P_{(m+m',n')}-g(m+m',n')C[E_{(m+m')},P_{(0,n')}]_S +
g(m,n)C\,E_{(m)}[E_{(m')},P_{(0,n')}]_S\cr
&=P_{(m+m',n')}-g(m+m',n')C\,[E_{(m)},P_{(m',n')}]_S\,,
\cr}}
which agrees with \bigproduct.
\smallskip

Finally, in the general case, we find using \newprrr, \twopispr,
\etimesp\ and \Cph,
\eqn\gencase{\eqalign{
P_{(m,n)}P_{(m',n')}=(\!(P_{(m,n)}P_{(m',n')})\!)+(-1)^n
g(m+m'+1,n'-1)\, C \, [P_{(m,n)},P_{(m',n')}]_S\,.}}
We omit the details of this somewhat lengthy, but otherwise completely
straightforward algebra. This proves the first part of Theorem
\genproduct.
\smallskip

To show that the bracket on the left hand side in \gencase\ is a
linear combination of traceless elements we proceed
similarly. Let us only illustrate the method on a simpler case of the
bracket in
\etimesp.  Using the same identities that led to \etimesp, as well
Theorem \genproduct\ in Case 2 above, and the Jacobi identity for the
bracket, we find
\eqn\commalg{\eqalign{
[E_{(m)},&P_{(m'.n')}]_S=E_{(m')}[E_{(m)},P_{(0,n')}]_S-mg(m',n')E_{(m)}
[E_{(m')},P_{(0,n')}]_S\cr
&\quad-g(m',n')C\,[E_{(m)},[E_{(m')},P_{(0,n')}]_S\cr
&=(\!(E_{(m')}[E_{(m)},P_{(0,n')}]_S)\!)-mg(m',n')(\!(E_{(m)}
[E_{(m')},P_{(0,n')}]_S)\!)\cr
&\quad +
\left[\,g(m+m'-1,n'-1)\left(1+mg(m',n')\right)-g(m',n')\,\right]
C\,[E_{(m)},[E_{(m')},P_{(0,n')}]_S\,.\cr}} Since the sum of the two
terms inside the square bracket vanishes, we find that the bracket on
the left hand side is indeed traceless. In the general case one
reduces the bracket to a manifestly traceless expression using the
same identities and, in addition, the already proven result in Case
3. This concludes the proof of Theorem \genproduct. \SMu

\appsubsec{Proof of Theorem \bvopmth}
\appsubseclab\Appproof

The remaining two cases are, schematically, $\De_S(CP)$ and $\De_S(CC)$.
In the first one,  on the one hand  we have
\eqn\cpdelta{\eqalign{
&\De_S(C_{(m,n)}P_{(m',n')})-\De_S(C_{(m,n)})P_{(m',n')}-(-1)^nC_{(m,n)}
\De_S(P_{(m',n')})\cr
&\,\,=-(2N+m+m'-n-n'+2)(\!(P_{(m,n)}P_{(m',n')})\!)+(2N+m-n)P_{(m,n)}
P_{(m',n')}\cr
&\,\,=-(m'-n'+2)(\!(P_{(m,n)}P_{(m',n')})\!)\cr & \qquad +(-1)^n
(2N+m-n)g(m+m'+1,n+n'-1)\, C \, [P_{(m,n)},P_{(m',n')}]_S\cr
}}
On the other hand,
\eqn\realbigstuff{\eqalign{
[C_{(m,n)},&P_{(m',n')}]_S=C[P_{(m,n)},P_{(m',n')}]_S+(-1)^{n'(n-1)}
[C,P_{(m',n')}]_SP_{(m,n)}\cr
&=-(-1)^{n}(m'-n'+2)(\!(P_{(m,n)}P_{(m',n')})\!)\cr &\qquad
+\left(1-(m'-n'+2)g(m+m'+1,n+n'-1)\right)\, C \,
[P_{(m,n)},P_{(m',n')}]_S\,.
\cr}}
Since
\eqn\oneextra{
(2N+m-n)g(m+m'+1,n+n'-1)=1-(m'-n'+2)g(m+m'+1,n+n'-1)\,,}
the relation between $\De_S$ and $[-,-]_S$ also holds in this case.
The last case is proved similarly. \SMu

\vfil\eject

\appendix{G}{BV-algebra of polyvectors on the base affine space $A(G)$}
\applab\Saffine

In this appendix we describe the geometric construction of BV-algebras
associated with model spaces of classical Lie algebras. Specialization
to $\slth$ gives a geometric counterpart of the algebraic construction
in Sections \bvpolyderivations\ and \SSthree.

\appsubsec{The base affine space $A(G)$ }
\appsubseclab\SSaffine

Let $G$ be a finite-dimensional classical Lie group, and let $\Ep(G)$
be the space of regular functions on $G$ (\ie, polynomial functions
in the matrix elements of $g\in G$). The space $\Ep(G)$ carries a
left and the right regular representations of $G$. Explicitly, we have
for $f\in \Ep(G)$ and $g, g'\in G$
\eqn\eqAa{
L(g) \cdot f(g') = f(g^{-1}g') \,,\qquad
R(g) \cdot f(g') = f(g'g)\,.}
\smallskip

The generators  $X_A$ of the Lie algebra $\bfg$ of $G$ satisfy
commutation relations
\eqn\commutrel{
[X_A,X_B]=f_{AB}{}^C\,X_C\,.}  We will denote by $L_A=L(X_A)$ and
$R_A=R(X_A)$ the vector fields on $G$ corresponding to the
representations in \eqAa. They span two commuting algebras $\bfg_L$
and $\bfg_R$, respectively, both isomorphic with $\bfg$. A classical
result in representation theory is the decomposition of $\Ep(G)$ into
finite dimensional irreducible modules of $\bfg_R\oplus \bfg_L$.

\thm\thAb
\proclaim Theorem \thAb\ [Peter-Weyl].
For any finite-dimensional simple Lie group $G$, the
decomposition of $\Ep(G)$ under the action of $\bfg_R \oplus \bfg_L$
is given by
\eqn\peterweyl{
\Ep(G) ~\cong~ \bigoplus_{\La\in P_+}(  \cL(\La)\otimes \cL(\La^*)  )\,.}
Here $\cL(\La)$ and $\cL(\La^*)$ are finite dimensional
irreducible modules of $\bfg$
with the highest weights $\La$ and $\La^*$, respectively, and
$\La^*=-w_0\La$.
\smallskip

Let $\bfg=\bfnp\oplus\bfh\oplus\bfnm$ be the Cartan decomposition of
$\bfg$, and $N_+$, $H$, and $N_-$ the complex subgroups of $G$ generated by
those subalgebras.  Following [\BGG], we define the base affine space
of $G$ as the quotient $A=N_+\backslash G$. The space of regular
functions $\Ep(A)$ on $A$ consists of those functions in $\Ep(G)$ that
are invariant under $N_+^L$, and carries a representation of
$\bfg_R\oplus\bfh_L$.  So, from Theorem \thAb, we
immediately conclude that

\thm\thBa
\proclaim Theorem \thBa\ [\BGG]. Under the action of
$ \bfg_R\oplus\bfh_L $ we have
\eqn\modeldec{
\Ep(A) ~\cong~ \bigoplus_{\La\in P_+} ( \cL(\La)\otimes \CC_{\,\La^*} )\,,}
where $\CC_{\,\La^*}$ denotes the 1-dimensional representation of
$\bfh_L$ with weight $\La^*$. In other words, $\Ep(A)$ is a model
space for $\bfg$.
\smallskip

\noindent
{\it Example:} $G=SL(n,\CC)$

First let us construct explicitly the decomposition of $\Ep(G)$
corresponding to \peterweyl. In terms of matrix elements
$(g_{\dot\si\si})$, $g_{\dot\si\si}=\de_{\dot\si\rh}g^{\rh}{}_\si$,
the generators of $\bfg_R$ and $\bfg_L$ are%
\foot{Strictly speaking these formulae
define the action of $\bfg_R$ and $\bfg_L$ on the pull-back of
$\Ep(G)$ to the functions on $GL(n,\CC)$.}
\eqn\genvectf{\eqalign{
R_{\si\dot\si}&=g_{\dot\rh\si}{\partial\over\partial
g_{\dot\rh}{}^{\dot\si}}-\hbox{$1\over 3$} \de_{\si\dot\si}
g_{\dot\rh}{}^{\dot \kappa}{\partial\over\partial
g_{\dot\rh}{}^{\dot\kappa} }\,,\cr
L_{\si\dot\si}&=-g_{\dot\si\rh}{\partial\over\partial
g^\si{}_\rh}+\hbox{$1\over 3$} \de_{\si\dot\si}
g^{\kappa}{}_{\rh}{\partial\over\partial
g^\kappa{}_\rh}
\,,\cr}}
with  $\si,\dot\si=1,2,3$.
We introduce the following elements $\De_k\in \Ep(G)$
by means of minors of $g\in G$,
\eqn\eqAd{
\De_k = \left| \matrix{ g_{n-k+1 1} & \ldots & g_{n-k+1 k} \cr
\vdots & & \vdots\cr g_{n 1} & \ldots & g_{nk}
\cr} \right|\,,\qquad k=1,\ldots,n\,. }

Now, for $\si>\dot\si$, acting with $L_{\si\dot\si}$ replaces the
$\si$-th row by the $\dot\si$-th row, and $R_{\si\dot\si}$ replaces
the $\dot\si$-th column by the $\si$-th column, so it is clear that all
$\De_k$ are annihilated by $\bfnp^L \oplus \bfnp^R$.  (Note that
$\De_{n-1}=1$.)  The action of the Cartan subalgebra generators $h_i^L =
L_{ii} - L_{i+1 i+1}$, $h_i^R = R_{ii} - R_{i+1 i+1}$, $i=1,\ldots,n-1$, is
given by
\eqn\eqAf{
h_i^L \De_k = \de_{i,n-k} \De_k\,,\qquad h_i^R \De_k = \de_{i,k}
\De_k\,. }
Thus we find that the highest weight vector corresponding to the
weight $\La=\sum_i s_i\La_i$ in the decomposition \peterweyl\
is realized by the function $\prod_i \De_i^{s_i}$. By
invariance with respect to $N_+^L$ this is also the highest weight
vector corresponding to $\La$ in the decompostion of $\Ep(A)$ in
\modeldec.
\smallskip

For $\slth$ the elements in $\Ep(A)$ corresponding to  the fundamental
representations are explicitly given by
\eqn\xisfrompw{
 x_\si=g_{3\si}\,,\quad
 x^\si=\ep^{\si\rh\kappa}g_{2\rh}g_{3\kappa}\,,\qquad \si=1,2,3\,,}
with $ x_1=\De_1$ and $ x^3=\De_2$. Evidently, these functions
satisfy the constraint
\eqn\constraint{
 x^\si x_\si=0\,.}
Moreover, $\Ep(A)$ is spanned by the polynomials of those functions,
and thus we have shown that $\Ep(A)$,  for $A = SL(3,\CC)$,
provides an explicit realization
of the ground ring algebra $\cG_3$. Of course, the latter is also an
immediate  consequence of Theorems \modelspace\ and \thBa.

\appsubsec{Polyvectors on $A(G)$}
\appsubseclab\SSaffpol

Geometric objects on $A$ can be studied effectively using standard
techniques of induced representations. A good example of that is the
description of $\Ep(A)$ given by Theorem $\thBa$. In a similar spirit,
we will therefore define polyvector fields on $A$ as regular sections
of homogenous vector bundles over $A$,%
\foot{This description of polyvectors was suggested to us by
Gregg Zuckerman.}  rather than, as would be more natural if we worked
in the smooth category, through differential operators acting on
$\Ep(A)$. In the case of vector fields, the equivalence of the two
approaches follows immediately from the explicit construction of all
differential operators on $A$ in [\GKa,\GKb,\BGG]. We will
discuss this briefly below.
\smallskip

Since $A=N_+\backslash G$, the tangent space to $A$ at
the origin is isomorphic with $\bfnp\backslash \bfg$. Let $\pi$ denote
the  representation of $\bfnp$ on $\bfnp\backslash \bfg$
arising from the right action of $N_+$ on $G$,
as well as its extension to $\bigwedge{}^*(\bfnp\backslash \bfg)$.

\thm\polyvectors
\proclaim Definition \polyvectors. The  space $\cP^n(A)$ of polyvectors
of order $n$ on $A$ is the space of regular sections of the
vector bundle $G\times{}_{N_+} \bigwedge{}^n(\bfnp\backslash \bfg)$.
\smallskip

We recall that the total space of the bundle
$G\times{}_{N_+}\bigwedge{}^n (\bfnp\backslash \bfg)$ consists of
pairs $(g,t)$, $g\in G$, $t\in\bigwedge{}^n(\bfnp\backslash \bfg)$,
subject to an equivalence relation $(g,t)\sim(ng,\pi(n)t)$, $n\in
N_+$. Thus an $n$-vector field on $A$, defined as a section
of this bundle, is given by a function $\Ph:G\rightarrow
\bigwedge{}^n(\bfnp\backslash \bfg)$ such that
\eqn\polyvecdef{
\Ph(ng)=\pi(n)\,\Ph(g)\,,\qquad n\in N_+\,,\quad g\in G\,.}
or, in an infinitesimal form,
\eqn\polyinf{
L(x)\Ph(g)=-\pi(x)\Phi(g)\,,\qquad x\in\bfnp\,,\quad g\in G\,.}
\smallskip

Corresponding to the Cartan decomposition of $\bfg$ into $\bfnp$,
$\bfh$ and $\bfnm$, let us split the generators $X_A$ of $\bfg$ into
$X_\al$, $X_i$, and $X_{-\al}$, where $\al\in\De_+$ and $i=1,\ldots
\ell$, and denote the generators of $\bfh\oplus\bfnm$ by $X_a$.
Now, consider the Clifford algebra of the vector space $\bfg\oplus
\bfg'$, where $\bfg'$ is the dual of $\bfg$.  In physicists'
language this algebra is realized by the ghost operators $c(x)$,
$x\in\bfg$, and the antighost operators $b(x')$, $x'\in\bfg'$. Let us
set $c_A=c(X_A)$ and $b^A=b(X^A)$, where $X_A$ and $X^A$ are the dual
bases of $\bfg$ and $\bfg'$, respectively. Then the anticommutation
relations between the ghost and the antighost operators are
\eqn\ghostsntt{
[c_A,c_B]=0\,,\quad [b^A,b^B]=0\,,\quad
[c_A,b^B]=\de_A{}^B\,,\quad A,B=1,\ldots,\dim\,\bfg\,.}

We define a ghost Fock space, $F$, as the module of the
Clifford algebra of ghosts with highest vector $\om$, called the ghost
vacuum, satisfying
\eqn\fdghvac{
b^{-\al}\,\om =0\,,\quad c_{\al}\,
\om =0\,,\quad b^i\, \om =0\,,\qquad \al\in\De_+\,,\quad
i=1,\ldots,\ell\,.}  The Fock space $F$ is a
highest weight module of $\bfg$, with the action of $\bfg$ explicitly
given by,
\eqn\pirepr{
\pi_A=f_{AB}{}^C\,c_Cb^B\,,}
where $\pi_A=\pi(X_A)$. Since
$\sum_{\al\in\De_+}f_{i\al}{}^\al=(\al_i^\vee,2\rh)$, we also find
that the weight of the ghost vacuum $\om$ is equal to $2\rh$. The
Fock space $F=\bigoplus_n F^n$ is graded by the ghost number, with
the ghost number of the vacuum set to zero. With respect to this grading
$c_A$ and $b^A$ have ghost number $+1$ and $-1$, respectively.

\thm\smallspace%
\proclaim Lemma \smallspace.
Define
\eqn\tttt{
T=\{t\in F\,|\, c_\al t=0\,, \al\in\De_+\}\,.}  Then $T=\bigoplus_n
T^n$ is an $\bfnp$-submodule of $F$, and $T^n$ is isomorphic with
$\bigwedge {}^n (\bfnp\backslash \bfg)$.
\smallskip

\proof For any $\al\in\De_+$ we have
\eqn\nplusrep{\eqalign{\pi_\al  ~=~ &
 \str \al{-\be}{-\ga}c_{-\ga}b^{-\be}+
 \str \al{-\be}{i}c_{i}b^{-\be} \cr
&+
 \str \al{\be}{ \ga}c_{\ga}b^{ \be}+
 \str \al{i}{ \ga}c_{ \ga}b^{i} +
 \str \al{-\be}{ \ga}c_{ \ga}b^{-\be}\,,\cr}}
where the (implicit) summations run over $\be,\ga\in\De_+$ and
$i=1,\ldots,\ell$.  Only the first two terms in \nplusrep\ act
nontrivially on $T$. Clearly, $T$ is closed under this action. Since
\eqn\comofx{
[X_\al,X_i]=0\,,\quad [X_\al,X_{-\be}]=\str \al{-\be}{-\ga}X_{-\ga} +
\str \al{-\be}{i}X_i\,,}
the second part of the lemma follows by identifying the basis
$c_a\om$ in $T^1$ with the basis $X_a$ in
$\bfnp\backslash\bfg$. \SMu

In the following, we will use interchangeably both notations $T^n$ and
$\hbox{$\bigwedge$}{}^n (\bfnp \backslash \bfg)$.

\smallskip
Let $\Ep(G,F) = \Ep(G) \otimes_{\CC} F $
denote the space of regular functions on $G$ with
values in $F$.  Then
\eqn\bigrepr{
\Pi(x)\equiv L(x)+\pi(x)\,,\qquad x\in\bfg\,,}
defines a representation of $\bfg$ on $\Ep(G,F)$. Using Lemma
$\smallspace$ and the defining relation \polyinf, we find the
following characterization of $\cP(A)$  inside $\Ep(G,F)$.

\thm\inside%
\proclaim Lemma \inside. The space of polyvectors $\cP(A)$ is the
subspace of those $\Ph\in\Ep(G,F)$ satisfying
\eqn\polyes{
c(x)\,\Ph=0\,,\quad \Pi(x)\,\Ph=0\,, \qquad {\rm for\ all\ }x\in\bfnp\,.}
\smallskip

Now, let us turn to the $\bfg$-module structure of the polyvectors and
a generalization of Theorem \thBa. The (right) action of $G$ on $A$
lifts to $\cP(A)$ as $R(g')\cdot\Ph(g) =\Ph(gg')$, $g,g'\in G$. When
necessary we will write $\bfg_R$ as above when talking about the
corresponding right action of $\bfg$. Since
$[\,\bfh,\bfnp]\subset\bfnp$, the constraint \polyes\ is invariant under
the modified left regular action of $\bfh$ (call it $\bfh_L$) defined
by \bigrepr, shifted so that the weight of the constant function is
equal to 0, \ie, we set $\Pi'(x)=\Pi(x)-2\rh(x)$, $x\in\bfh$. Moreover,
$\bfg_R$ and $\bfh_L$ commute.

\thm\frobenius
\proclaim Theorem \frobenius. The space of $n$-vectors, $\cP^n(A)$, is a
completely reducible module of $\bfh_L\oplus\bfg_R$, with the
decomposition given by
\eqn\frobdec{\eqalign{
\cP^n(A)& ~\cong~ \bigoplus_{\La\in P_+} {\rm Hom}_\bfg (\cL(\La),\cP^n(A))
\otimes \cL(\La)\cr
&~\cong~ \bigoplus_{\La\in P_+} \homhom \otimes
\cL(\La)\,,\cr}} where $\bfh$ and $\bfg$ act on the first and the second
factor in the tensor product, respectively.
\smallskip

\proof The complete reducibility is  a consequence of
Theorem \thBa. Then \frobdec\ is just the  Frobenius reciprocity.
Let us briefly  recall its proof.  Suppose that under the
right action of $\bfg$ a set of polyvectors $\Ph_I$ spans an
irreducible module $\cL(\La)$, \ie,
\eqn\condonph{\Ph_I(gg') = \Ph_J(g) D(g')^J{}_I\,,\quad
g,g'\in G\,.} It is convenient to think about this set as defining a
function ${\bf \Ph}:G\rightarrow \End\,(\cL(\La),\hbox{$\bigwedge$}{}^n
(\bfnp \backslash \bfg))$. Setting
$g=e$, where $e$ is the identity, and $g'=n\in N_+$, we find, using
\polyvecdef\ and \condonph, that
\eqn\crucial{
\pi(n){\bf \Ph}(e) ~=~ {\bf \Ph}(n) ~=~ {\bf \Ph}(e)D(n)\,,}
\ie, ${\bf \Ph}(e)$ is an element of  $\homhom$. \SMu
\smallskip

{}From  \condonph\ it also follows that $\bPh$ satisfies
\eqn\diffph{L(x)\bPh(g)=-\bPh(g)D(g^{-1}x g)\,, \quad
x\in\bfg,\,\,g\in G\,,}
or simply
\eqn\infdiff{
L(x)\bPh(e)=-\bPh(e)D(x)\,,\quad x\in\bfg\,.}
\smallskip

The space $\homhomt$ is easy to characterize. Since $\cL(\La)$ is cyclic
over $\bfnp$, a homomorphism $\bf \Ph$ of $\cL(\La)$ into an arbitrary
$\bfnp$-module is completely determined by the image of the lowest
weight vector $v_{w_0\La}$.
Clearly, $\Ph(v_{w_0\La})$ is restricted by
$\pi(n){\bf \Ph}v_{w_0\La}=0$ for all $n\in\bfnp$ such
that $nv_{w_0\La}=0$.

\thm\homscond
\proclaim Lemma \homscond. A pair $(v_{w_0\La},t)$, where $v_{w_0\La}$
is the lowest weight vector of $\cL(\La)$ and $t\in T$, defines a
homomorphism of $\cL(\La)$ into $T$ provided $t$ is a solution to the
equations
\eqn\condforim{
e_i^{(\La^*+\rh,\al_i)}\,t=0\,, \qquad i=1,\ldots,\ell\,,}
with $\La^*=-w_0\La$.
\smallskip

\proof Any $\bfnp$-homomorphism can be extended to a unique
$\cU(\bfnp)$-homomorphism, where $\cU(\bfnp)$ is the enveloping
algebra of $\bfnp$. The elements of $\cU(\bfnp)$, that annihilate the
lowest weight vector $v_{w_0\La}$ of $\cL(\La)$, form a left ideal
$\cI(\La)\subset\cU(\bfnp)$ generated by the powers
$e_{i}^{(\La^*+\rh,\al_i)}$ of simple root generators (see, \eg
[\Dixmier]), which implies \condforim. \SMu
\smallskip

The $\bfh_L$-weight of an element $t\in T$ is of the form
$\sum_{\al\in\De_+} n_\al \al$, where $n_\al=0,-1$.  Thus, in
particular, $\la+2\rh\in P_+$. By \infdiff, the resulting
$\bfh_L$-weight of the homomorphism defined by $(v_{w_0\La},t)$ is
equal to $\la-w_0\La=\la+\La^*$.
\smallskip

Now consider homomorphisms into the trivial $\bfnp$-module
$\bigwedge^0(\bfnp\backslash \bfg) \cong \CC$.  In this case
\condforim\ is satisfied for any weight $\La\in P_+$.  Obviously
 ${\rm Hom\,}_{\bfnp} (\cL(\La),\CC)
\cong \CC_{\La^*}$ as an $\bfh_L$-module, and thus \frobdec\ reduces to
\modeldec\ in agreement with Theorem \thBa.
\smallskip

For higher order polyvectors, with $n\geq 1$, we find that for a
weight $\La$ sufficiently deep inside the fundamental Weyl chamber
$P_+$, there are no restrictions on $t$ due to \condforim, and
$\homhomtn\cong T^n$. However, if $\La$ lies close to the boundary of
$P_+$, the constraint \condforim\ becomes nontrivial. The following
observation is  quite useful in the computation of all polyvectors.

\thm\coness
\proclaim Lemma \coness. Given $t\in T$, consider the
set $\cC(t)$ of all weights $(\La,\La')\in P_+\otimes (P_+-2\rh)$
for which there exists a homomorphism that maps the lowest weight
vector in $\cL(\La)$ to $t$, and has the $\bfh_L$-weight equal to
$-w_0\La'$.  Then $\cC(t)$ is a cone
\eqn\connes{\cC(t)=\{S(t) + (\la,\la)\,|\, \la\in P_+\}\,,}
with the ``tip'' $S(t)$.
\smallskip

\proof  From the computation of the $\bfh_L$ weight of a homomorphism
it is clear that, for $t$ fixed, $\La'$ is determined by $\La$, as
well as that a shift $\La\rightarrow \La+\la$ induces the same shift
$\La'\rightarrow \La'+\la$.  So it is sufficient to show that the
set of $\La$'s is a cone in $P_+$. Now, for any $\La,\la\in P_+$ we
have $\cI(\La+\la)
\subset \cI(\La)$.  Thus if $\La$ corresponds to a
nontrivial homomorphism, then so does $\La+\la$. The lowest lying
$\La$ weight with the required property is then determined by
\condforim\ and the minimal powers of simple root generators that
annihilate $t$. \SMu
\smallskip

Thus the problem of computing all  homomorphisms is reduced
to that of determining a finite set of cones. Later we will give the
complete solution in the case of $\slth$.
\smallskip

To conclude, let us summarize the main steps of this construction of
polyvector fields, and comment on the relation between $\cP(A)$ and
the polyderivations $\cP(\Ep(A))$.  Given a weight $\La\in P_+$ and an
element $t\in T$ of ghost number $n$ satisfying \condforim, we first
construct a homomorphism $\bPh(e)$ by setting $\bPh(e)
v_{w_0\La}=t$. Using $\condonph$ we extend $\bPh(e)$ to a function on
$G$ with values in $\End(\cL(\La),T^n)$.  The components $\Ph_I$ of
$\bPh$, with respect to some basis in $\cL(\La)$, lie in $\Ep(G,T^n)$,
satisfy \polyes, and have an expansion of the form
\eqn\compexp{
\Ph_I(g)=\Ph^{a_1\ldots a_n}_I(g)\,c_{a_1}\ldots
c_{a_n}\om \,, \quad I=1,\ldots,\dim\,\cL(\La).}
\smallskip

Let $B_-=HN_-$ be the Borel subgroup in $G$, which, using the Gauss
decomposition, may locally be identified  with $A$.  The vector
fields $L_a$ on $B_-\subset G$ give then a local trivialization of the
tangent bundle of $A$. In the language of Definition \polyvectors,
they correspond to sections $(nb,\pi(n)X_a)$, where $b\in B_-$, $n\in
N_+$ and $X_a\in \bfnp\backslash\bfg$. Thus we identify $c_{a_1}\ldots
c_{a_n}\om$ with the exterior product $L_{a_1}\wedge
\ldots\wedge L_{a_n}$ of vector fields. One should remember
that in general the vector fields $L_a$  cannot be extended to the
entire base affine space. However, if we identify
a polyvector $\Ph=\Ph_I$ in \compexp\ with
\eqn\ghostvecexp{
\Ph=\Ph^{a_1\ldots a_n}L_{a_1}\wedge \ldots
\wedge L_{a_n}\,,}
then  \polyvecdef\ together with \polyes\ imply that the this polyvector
field is globally well defined on $A$, and thus defines a
polyderivation of $\Ep(A)$.
\smallskip

\appsubsec{Example of $\slth$}
\appsubseclab\suthreepv

As an illustration for the above discussion, let us now determine
polyvectors on the base affine space of $\slth$ and show that indeed
they reproduce all  polyderivations of the ground ring algebra
$\cG_3\cong\Ep(A)$.
\smallskip

Consider $\cP^1(A)$. The basis of  $T^1$ consists of states obtained
by acting with the ghost operators
$c_1$, $c_2$, $c_{-\al_1}$, $c_{-\al_2}$, and $c_{-\al_3}$ on
the vacuum. The $\bfnp$-module structure of $T^1$ is summarized by the
diagram in Figure \nfig\nplusmodone{}\nplusmodone, in which each arrow
corresponds to a nontrivial action of a given generator $e_\al$.
This must  be compared with \condforim, which for a weight $\La=
s_1\La_1+s_2\La_2$ reads
\eqn\condforimth{
e_1^{s_2+1}\,t=0\,,\quad e_2^{s_1+1}\,t=0\,.}
Clearly, depending on $t\in T^1$, the constraint
\condforim\ is satisfied provided
$s_1\geq 1$ for $c_{-\al_2}$, $s_2\geq 1$ for $c_{-\al_1}$, and
$s_1,s_2\geq 1$ for $c_{-\al_3}$. There is no restriction on $\La$ for
$t$ equal to $c_1$ or $c_2$. These five cases correspond to five cones
of vector fields on $A$, which we will now compute.

\bigskip
\vbox{\epsffile{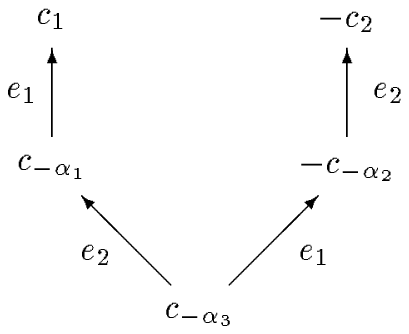}}
\medskip
\centerline{Figure \nplusmodone. The $\bfnp$-module structure of
$T^1\simeq\bfnp\backslash\bfg$.}
\bigskip\smallskip

Using  \xisfrompw, we find the following identities
\eqn\derofg{\eqalign{
{\partial\over \partial g_{2\si}}&=
\ep^{\si\rh\kappa}g_{3\rh}{\partial
\over \partial  x^\kappa}\,,\cr
{\partial \over\partial g_{3\si}}&={\partial\over\partial x_\si}
-\ep^{\si\rh\kappa}g_{2\rh}{\partial \over \partial
 x^\kappa}\,.\cr}}
Substituting those identities in \genvectf, it is straightforward,
though somewhat laborious, to obtain an explicit formula for a
polyvector if its ``coordinates'' $\Ph^{a_1\ldots a_n}$ in
\ghostvecexp\ are known. One should note that although intermediate
steps of this calculation may involve explicitly group elements (as in
\derofg), the final result for a polyvector field can always
be expressed in terms of polynomials in the ground ring generators
$ x^\si$ and $ x_\si$, and the derivatives,
$\partial\over\partial x^\si$ and $\partial\over\partial x_\si$.
\smallskip

In the simplest nontrivial example we take $\La=0$ and $t=c_1$
or $t=c_2$. The corresponding vector fields are $h_1=L_{11}-L_{22}$
and $h_2=L_{22}-L_{33}$, respectively. Using \genvectf\ and \derofg,
we find
\eqn\liuvvec{
L_{11}-L_{22}= x^\si\hbox{${\partial\over\partial  x^\si}$}\,,\quad
L_{22}-L_{33}= x_\si\hbox{${\partial\over\partial  x_\si}$}\,,}
which, as expected, coincide with $C_-$ and $C_+$ in \defofc.
Similarly, the
vector fields at the tips of the other three cones, corresponding to
the pairs $(\La_1,c_{-\al_2})$, $(\La_2,c_{-\al_1})$ and
$(\La_1+\La_2,c_{-\al_3})$, reproduce the   derivations
$P_{\si,\rh}$, $P_{\dot\si,\dot\rh}$ and $\La_{\si\dot\si}$ in \defofp.
\smallskip

For the higher ghost numbers the action of $\bfnp$ on $T^n$ follows
from the diagram in Figure \nplusmodone. The computation of the cones
of polyvectors and representatives of the tips is essentially the
same as above. The complete result may be summarized as follows:

\tbl\hmsandplvc
\thm\polystable
\proclaim Theorem \polystable. The space $\cP(A)$ is isomorphic  as a
$\slth\oplus (\uone)^2$ module to the direct sum of irreducible
modules $\cL(\La)\otimes\CC_{\La'}$ with weights $(\La,\La')\in
P_+\oplus (P_+-2\rh)$ lying in the set of disjoint cones
$\{(\La(t),\La'(t))+(\la,\la)\, | \,t\in T\,, \la\in P_+\}$, \ie
\eqn\poldecom{
\cP(A)\cong \bigoplus_{t\in T}\bigoplus_{\la\in P_+}
\cL(\La(t)+\la)\otimes\CC_{\La'(t)+\la}\,,}
where the tips of the cones, $(\La(t),\La'(t))$, $t\in T$, satisfy
$\La'(t)+2\rh=\La(t)+\rh-\si\rh$, $\si\in\widetilde W$. They are
listed in Table \hmsandplvc\ together with the corresponding
polyvectors $D^{(n)}{}_{\mu_1\ldots\mu_n}$ and $\widetilde
D^{(n)}{}_{\mu_1\ldots\mu_n}$ that are linear
combinations of the  generators in $\cP(\cR_3)$
explicitly given by%
\foot{{\rm Here, and in Table \hmsandplvc, $\wedge$ is the
exterior product of vector fields, while $\wedge'$ implies in addition
subtraction of all $\slth$ invariant  traces.}}
\eqn\polydisaa{
D^{(0)}=1\,,}
\eqn\polydisba{
D^{(1)}=x_{\si}\hbox{$\partial\over\partial
x_{\si}$}\,,\quad D^{(1)}{}_{\si}=\ep_{\si\mu\rh}\,
x^{\mu}\hbox{$\partial\over\partial x_{\rh}$}\,,\quad
D^{(1)}{}_{\dot\si}=\om_\cP(D^{(1)}{}_{\si})\,,\quad
\widetilde D^{(1)}=\om_\cP(D^{(1)})\,,}
\eqn\polydisbb{
D^{(1)}{}_{\si\dot\si}=x_\si\hbox{$\partial\over\partial x^{\dot\si}$}
- x_{\dot\si}\hbox{$\partial\over\partial x^{\si}$} - \, {\rm
trace}\,, }
\eqn\polydisca{
D^{(2)}=\ep_{\si\mu\rh}\,
            x^{\si}\hbox{$\partial\over\partial
            x_{\mu}$}\wedge\hbox{$\partial\over\partial
            x_{\rh}$}\,,\quad
\widetilde D^{(2)}=\om_\cP(D^{(2)})\,,}
\eqn\polydiscb{
D^{(2)}{}_{\si}=x_{\si}
              \hbox{$\partial\over\partial x_{\rh}$}\wedge
              \hbox{$\partial\over\partial x^{\rh}$}-
              x_{\rh}
              \hbox{$\partial\over\partial x_{\rh}$}\wedge
              \hbox{$\partial\over\partial x^{\si}$}+
              x^{\rh}
              \hbox{$\partial\over\partial x^{\rh}$}\wedge
              \hbox{$\partial\over\partial x^{\si}$}\,,\quad
D^{(2)}{}_{\dot\si}=\om_\cP(D^{(2)}{}_{\si})\,,}
\eqn\polydiscc{
D^{(2)}{}_{\si\rh}=\ep_{\si\mu\nu}(
              x^\mu
              \hbox{$\partial\over\partial x^{\rh}$}
              -\half
              x_\rh
              \hbox{$\partial\over\partial x_\mu$})\wedge
              \hbox{$\partial\over\partial
              x_\nu$}+(\si\leftrightarrow\rh)
              \,,\quad
D^{(2)}{}_{\dot\si\dot\rh}=\om_\cP(D^{(2)}{}_{\si\rh})\,,}
\eqn\polydisda{
D^{(3)}=x_\si\hbox{$\partial\over\partial x_{\si}$}\wedge
	\hbox{$\partial\over\partial x_{\rh}$} \wedge
	\hbox{$\partial\over\partial x^{\rh}$}
-x^\si\hbox{$\partial\over\partial x_{\rh}$}\wedge
	\hbox{$\partial\over\partial x^{\rh}$}\wedge
	\hbox{$\partial\over\partial x^{\si}$}\,,}
\eqn\polydisdb{D^{(3)}{}_{\si}=\ep_{\mu\nu\rh}
	(x^\rh \hbox{$\partial\over\partial
	x^{\si}$}-\hbox{$1\over 3$} x_\si \hbox{$\partial\over\partial
	x_{\rh}$})\wedge\hbox{$\partial\over\partial x_{\mu}$}\wedge
	\hbox{$\partial\over\partial x_{\nu}$}\,,\quad
D^{(3)}{}_{\si}=\om_\cP(D^{(3)}{}_{\si})\,,}
\eqn\polydisdc{
D^{(3)}{}_{\si\dot\si}=(x_\si\hbox{$\partial\over\partial x^{\rh}$}
	-x_{\rh}\hbox{$\partial\over\partial x^\si$})\wedge
        \hbox{$\partial\over\partial x^{\dot \si}$}\wedge
	\hbox{$\partial\over\partial x_{\rh}$}
+ (x_{\dot\si}\hbox{$\partial\over\partial x^{\dot\rh}$}
	-x_{\dot\rh}\hbox{$\partial\over\partial x^{\dot\si}$})\wedge
        \hbox{$\partial\over\partial x^{ \si}$}\wedge
	\hbox{$\partial\over\partial x_{\dot\rh}$}-{\rm trace} \,,}
\eqn\polydisea{
D^{(4)}{}_{\si}=\ep_{\si\rh\pi}\ep^{\kappa\mu\nu}(
x_\kappa\hbox{$\partial\over\partial x_{\rh}$}-\hbox{$1\over 3$}
x^\rh\hbox{$\partial\over\partial x^\kappa$})\wedge
\hbox{$\partial\over\partial x_\pi$}\wedge
\hbox{$\partial\over\partial x^\mu$}\wedge
\hbox{$\partial\over\partial x^\nu$}\,,\quad
D^{(4)}{}_{\dot\si}=\om_\cP(D^{(4)}{}_{\si})\,,}
\eqn\polydisea{
D^{(5)}=\ep_{\si\rh\pi}\ep^{\kappa\mu\nu}
(x^\pi\hbox{$\partial\over\partial x^\nu$}-
x_\nu\hbox{$\partial\over\partial x_\pi$})\wedge
\hbox{$\partial\over\partial x_\si$}\wedge
\hbox{$\partial\over\partial x_\rh$}\wedge
\hbox{$\partial\over\partial x^\kappa$}\wedge
\hbox{$\partial\over\partial x^\mu$}\,.}

\smallskip

At this point it is interesting to ask whether the two spaces
$\cP(\cG_3)$ and $\cP(A)$ coincide. By decomposing the polyderivations
$\cP(\cG_3)$ (see Theorem \polyvects) with respect to
$\slth\subset\sosi$ and comparing with Table \hmsandplvc\  we conclude
that indeed $\cP(\cG_3)\cong\cP(A)\,.$

\vfill\eject

\begintable
$t$ | $(\La(t),\La'(t))_\si$ | $D$ \crthick 1 | $(0,0)_{r_3}$ | $D^{(0)}$ \cr
$c_1$ | $(0,0)_{r_3}$  | $\widetilde D^{(1)}$ \nr
$c_2$ | $(0,0)_{r_3}$  | $D^{(1)}$ \nr
$c_{-\al_1}$ | $(\La_2,\La_1-\La_2)_{r_{12}}$ | $D^{(1){}}{}_{\dot\si}$ \nr
$c_{-\al_2}$ | $(\La_1,-\La_1+\La_2)_{r_{21}}$ | $D^{(1)}{}_{\si}$ \nr
$c_{-\al_3}$ | $(\La_1+\La_2,0)_{\si_2}$ | $D^{(1)}{}_{\si\dot\si}$ \cr
$c_1c_2$ | $(0,0)_{r_3}$ | $\widetilde D^{(1)}\wedge  D^{(1)}$  \nr
$c_1c_{-\al_1}$ | $(0,\La_1-2\La_2)_{r_{12}}$ | $\widetilde D^{(2)}$ \nr
$c_2c_{-\al_2}$ | $(0,-2\La_1+\La_2)_{r_{21}}$ | $D^{(2)}$ \nr
$c_2c_{-\al_1}$ | $(\La_2,\La_1-\La_2)_{r_{12}}$ | $D^{(1)}\wedge
                                         D^{(1)}{}_{\dot\si}$ \nr
$c_1c_{-\al_2}$ | $(\La_1,-\La_1+\La_2)_{r_{21}}$ | $\widetilde D^{(1)}\wedge
                                         D^{(1)}{}_{\si}$  \nr
$c_{-\al_1}c_{-\al_2}+c_2c_{-\al_3}$ | $(\La_2,-\La_1)_{\si_2}$ |
       $D^{(2)}{}_{\dot\si}$  \nr
$c_{-\al_1}c_{-\al_2}+ c_1c_{-\al_3}$ | $(\La_1,-\La_2)_{\si_2}$ |
       $D^{(2)}{}_{\si}$ \nr
$c_{-\al_1}c_{-\al_2}$ | $(\La_1+\La_2,0)_{\si_1}$  |
$\quad D^{(1)}{}_{\si}\wedge'
        D^{(1)}{}_{\dot\si}\quad $ \nr
$c_{-\al_1}c_{-\al_3}$ | $(2\La_2,-\La_2)_{r_1}$  |
$D^{(2)}{}_{\dot\si\dot\rh}$ \nr
$c_{-\al_2}c_{-\al_3}$ | $(2\La_1,-\La_1)_{r_2}$ |
           $D^{(2)}{}_{\si\rh}$ \cr
$c_1c_2c_{-\al_1}$ | $(0,\La_1-2\La_2)_{r_{12}}$ | $D^{(1)}\wedge
                                       \widetilde D^{(2)}$ \nr
$c_1c_2c_{-\al_2}$ | $(0,-2\La_1+\La_2)_{r_{21}}$ | $\widetilde D^{(1)}\wedge
                                            D^{(2)}$ \nr
$\quad c_1c_{-\al_1}c_{-\al_2} + c_1c_2c_{-\al_3}
-c_2c_{-\al_1}c_{-\al_2}\quad $ |
                   $(0,-\La_1-\La_2)_{\si_2}$ | $D^{(3)}$ \nr
$c_1c_{-\al_1}c_{-\al_2}+c_1c_2c_{-\al_3}$| $(\La_2,-\La_1)_{\si_1}$ |
              $\widetilde D^{(1)}\wedge D^{(2)}{}_{\dot\si}$ \nr
$c_2c_{-\al_1}c_{-\al_2}-c_1c_2c_{-\al_3} $ |$(\La_1,-\La_2)_{\si_1}$ |
                                     $D^{(1)}\wedge
                                         D^{(2)}{}_{\si}$  \nr
$c_1c_{-\al_1}c_{-\al_3}$ | $(\La_2,-2\La_2)_{r_1}$  |
$D^{(3)}{}_{\dot\si}$ \nr
$c_2c_{-\al_2}c_{-\al_3}$ | $(\La_1,-2\La_1)_{r_2}$ | $D^{(3)}{}_{\si}$ \nr
$c_2c_{-\al_1}c_{-\al_3}$ | $(2\La_2, -\La_2)_{r_1}$  |$
                   D^{(1)}\wedge D^{(2)}{}_{\dot\si\dot\rh}$ \nr
$c_1c_{-\al_2}c_{-\al_3}$ |$(2\La_1, -\La_1)_{r_2}$ |$\widetilde D^{(1)}
    \wedge
                      D^{(2)}{}_{\si\rh}$ \nr
$c_{-\al_1}c_{-\al_2}c_{-\al_3}$ |
     $\quad (\La_1+\La_2, -\La_1-\La_2)_1\quad$ | $ D^{(3)}{}_{\si\dot\si}$ \cr
$c_1c_2c_{-\al_1}c_{-\al_2}$ | $(0,-\La_1-\La_2)_{\si_1}$ |
                          $\widetilde D^{(2)}\wedge D^{(2)}$  \nr
$c_1c_2c_{-\al_1}c_{-\al_3}$ | $(\La_2,-2\La_2)_{r_1}$ |
                 $ D^{(1)}\wedge D^{(3)}{}_{\dot\si}$  \nr
$c_1c_2c_{-\al_2}c_{-\al_3}$ | $(\La_1,-2\La_1)_{r_2}$ | $\widetilde
                D^{(1)}\wedge D^{(3)}{}_{\si}$   \nr
$c_1c_{-\al_1}c_{-\al_2}c_{-\al_3}$| $(\La_1, -\La_1 - 2\La_2)_1$ |
               $D^{(4)}{}_{\si}$ \nr
$c_2c_{-\al_1}c_{-\al_2}c_{-\al_3}$ | $(\La_2, -2\La_1 - \La_2)_1$ |
               $D^{(4)}{}_{\dot\si}$ \cr
$c_1c_2c_{-\al_1}c_{-\al_2}c_{-\al_3}$ | $(0,-2\La_1-2\La_2)_1$ | $D^{(5)}$
      \endtable

\bigskip

\centerline{Table \hmsandplvc. The decomposition of $\fP$ into cones
of $\slth\oplus(\uone)^2$ modules. }
\vfill\eject

\appsubsec{The BV-algebra structure of $\cP(A)$ }
\appsubseclab\SSbvaff

The space $\Ep(G,F)$ has a natural structure of the dot algebra, with
respect to which $\cP(A)\subset \Ep(G,F)$ is a subalgebra. In particular
given two polyvector fields $\Ph=\Ph^{a_1\ldots a_m} c_{a_1}\ldots
c_{a_m}\om$ and $\Psi=\Psi^{b_1\ldots b_n} c_{b_1}\ldots c_{b_n}\om$,
their product is
\eqn\prodinghst{
\Ph\cdot\Ps=\Ph^{a_1\ldots a_m}\Psi^{b_1\ldots b_n}  c_{a_1}\ldots
c_{a_m} c_{b_1}\ldots c_{b_n}\om\,.} Define
$\de:\Ep(G,F^n)\rightarrow \Ep(G,F^{n-1})$  by
\eqn\bvghost{
\de=b^A L_A -\half f_{AB}{}^C b^A b^B c_C\,.}

\thm\bvprop
\proclaim Theorem \bvprop. {The operator $\de$  satisfies
\smallskip
\item{i.} $\de^2=0$.
\item{ii.} $[\de,c_A]=\Pi_A$, $A=1,\ldots,\dim\bfg$.
\item{iii.} $\de(\cP(A))\subset \cP(A)$.
\item{iv.} On $\cP(A)$, $\de=-\De'$, where
\eqn\bvplgh{
\De'=-b^a L_a + \half \str abc  b^ab^bc_c\,.}}

\proof The first property is equivalent to the Jacobi identity in $\bfg$.
Similarly (ii) follows by simple algebra using
\ghostsntt.  Lemma \inside\ and (ii) imply (iii). Finally,
by expanding $\de$, we find
\eqn\expanofd{\eqalign{\de & =  - \De' \cr &+ b^\al( L_\al -
\str\al{-\be}i b^{-\be}c_i + \str\al{-\be}{-\ga} b^{-\be}c_{-\ga} )
\cr &- (\half \str\al\be\ga b^\al b^\be + \str\al i \ga b^\al b^i +
\str\al{-\be}\ga b^\al b^{-\be})c_\ga\,.\cr}}
Note that the second line is equal to $b^\al\Pi_\al$, so it vanishes
on $\cP(A)$. Similarly, the last line being proportional to $c_{\ga}$
vanishes on $\cP(A)$ as well.  \SMu

\thm\bvoponpa
\proclaim Theorem \bvoponpa. The operator $\De'$ defined in \bvplgh\ is a
BV-operator on $\cP(A)$.
\smallskip

\proof Consider  $F$ as a polynomial algebra generated
by odd Grassmann elements $c_a$ and $b^{-\al}$. Then $b^a$ acts on $F$
by the commutator $[b^a,-]$, and thus is a first order derivation of
$F$. Similarly, $L_a$ is a first order derivation on $\Ep(G)$. Now,
consider $\De'$ on $\Ep(G,F)\cong\Ep(G)\otimes F$ rather than on
$\cP(A)$. Then the first term in $\De'$ is a tensor product of two
first order derivations, each acting on a different factor in the
tensor product of the two algebras, $\Ep(G)$ and $F$. It is
straightforward to verify that such a tensor product of first order
derivations is always a second order derivation. The second term in
$\De'$ acts only on $F$, and, by commuting $c_a$ to the left using
\ghostsntt, it becomes  a sum of terms with the second and first order
derivatives in the Grassmann variables. Thus it is also a second
order derivation on $\Ep(G,F)$. This shows that $\De'$ is a
second order derivation on $\Ep(G,F)$. Since $\cP(A)\subset\Ep(G,F)$ is
invariant under $\De'$, this proves the theorem. \SMu
\smallskip

\noindent
{\it Remarks:}
\item{i.}. The differential $\de$ defines an equivariant version
of the twisted homology introduced in  [\FeFrtw] as an analogue
of the  semi-infinite homology in the category of  finite-dimensional Lie
algebras. Here, that semi-infinite character is determined by the
choice of the ghost Fock space $F$.
\item{ii.} Since $b^{\al}$ acts like a multiplication, rather than a
derivation, on $F$, the full operator $\de$ is not a second order
derivation on $\Ep(G,F)$.

\smallskip

Let us compute the bracket between two vector fields induced by $\de$.
For $\Ph=\Ph^ac_a$, $\Ps=\Ps^ac_a$ we find (\cf,  \eqCc),
\eqn\commofvec{\eqalign{
[\Ph,\Ps]&=-\De'(\Ph\Psi)+(\De' \Ph)\Ps -\Ph\De'(\Ps)\cr
&= (\Ph^aL_a\Psi^c- \Ps^aL_a\Ph^c  + \str abc
\Ph^a\Psi^b)\,c_c\,,\cr}}
which, using \comofx\ and \ghostvecexp, is just the usual commutator
between vector fields. The generalization to higher order polyvectors
is essentially the same, and we omit the details.

\thm\brackofde
\proclaim Lemma \brackofde. The bracket induced by $\De'$  on $\cP(A)$
is the usual commutator between polyvector fields on $A$.
\smallskip

Clearly, $\De'$ commutes with the action of $\bfg$ on $\cP(A)$. It also
commutes with the action of $\bfh_L$, which is generated, up to a
constant, by the operators $\Pi_i=[c_i,\de]=[c_i,\De']$. To evaluate
$\De'$ explicitly on an irreducible $\bfg$-module of polyvectors
corresponding to a homomorphism $\bPh(e)$ it is
sufficient to determine the vector $\De'\bPh(e)v_{w_0\La}$. Using
\infdiff, \bvplgh\ and that $xv_{w_0\La}=0$ for all $x\in\bfnm$, we
obtain
\eqn\bvonhoms{\eqalign{
\De'\bPh(e)v_{w_0\La}&=-b^a\bPh(e)D_av_{w_0\La}+\half\str abc
b^ab^bc_c\bPh(e)v_{w_0\La}\cr
&=-b^i\bPh(e)D_iv_{w_0\La}+\half\str abc
b^ab^bc_c\bPh(e)v_{w_0\La}\cr
&=(\al_i^\vee,\La^*)b^i\bPh(e)v_{w_0\La}+
(\half\str{-\al}{-\be}{-\al-\be}b^{-\al}b^{-\be}c_{-\al-\be}+
\str i{-\al}{-\al}b^ib^{-\al}c_{-\al})\,\bPh(e)v_{w_0\La}\,.\cr}}
In particular, on vector fields, this reduces to
\eqn\donvectf{
\De'\bPh(e)v_{w_0\La}=(\al_i^\vee,\La^*+2\rh)b^i\bPh(e)v_{w_0\La}\,,
\quad \bPh\in\cP(A)\,.}
\smallskip

Finally, we have an analogue of Theorem \cohofdel.
\thm\cohofdelpr
\proclaim Theorem \cohofdelpr. The homology of $\De'$ on the polyvector
fields $\cP(A)$ is given by
\eqn\cohofdel{
H^n(\De',\cP(A))\cong \De^{n,D}\,\CC\,,} where $D=\dim(A)$. The
representative of the nontrivial homology is the polyvector
corresponding to the homomorphism defined by $\La=0$ and
$t=\prod_{i=1}^\ell c_i\,\prod_{\al\in\De_+}c_{-\al}\om$.
\smallskip

\proof Note that $c_i$, $i=1,\ldots,\ell$, are well-defined operators
on $\cP(A)$. Then, similarly as above, we find that for any
$\bPh\in\cP(A)$
\eqn\comofdandc{\eqalign{
[\De',c_i]\,\bPh(e)&=-\bPh(e)D_i+(\al_i^\vee,2\rh)\bPh(e)+
\sum_{\al\in\De_+}
(\al_i^\vee,\al)c_{-\al}b^{-\al}\bPh(e)\cr
&=(\al_i^\vee,\La^*+2\rh+\la)\bPh(e)\,,\cr}} where $\la$ is the weight
of ghosts in $t=\bPh(e)v_{w_0\La}$.  In particular for $C=\rh^ic_i$,
where $\rh=\rh^i\al_i$, we get
\eqn\homotop{
[\De',C]\bPh(e)=(\rh^\vee,\La^*+2\rh+\la)\bPh(e)\,.}  Since
$\la+2\rh\in P_+$, we find that the homology of $\De'$ is concentrated
on polyvectors with $\La=0$ and $\la=-2\rh$. The space of those
polyvectors is isomorphic with $\bigwedge{}^*\bfh$, spanned by the
products of $c_i$'s. From \bvonhoms\ we find that on this subspace
$\De$ reduces to $2 \sum_ib^i$, and the theorem follows by an elementary
evaluation of the homology in the reduced case. \SMu
\smallskip

\noindent
{\it Example:} $G=SL(3,\CC)$

The vector fields on $A$ that extend $\slth$ symmetry to the $\sosi$
symmetry in Section \bvpolyderivations, are $P_{\si,\rh}$ and
$P_{\dot\si,\dot\rh}$, corresponding to  the homomorphisms
$(\La_1,c_{-\al_2})$ and  $(\La_2,c_{-\al_1})$, respectively, and
$C_+-C_-$ corresponding to $(0,c_2-c_1)$. From \donvectf\ it
follows easily that all are annihilated by $\De'$. This, combined
with invariance with respect to $\slth$ and \eqCd\ yields

\thm\twodeleq
\proclaim Theorem \twodeleq. The BV operators $\De_S$ and $\De'$ satisfy
$\De'=-\De_S$, \ie,  $\cP(\cG_3)\cong\cP(A)$, as  BV-algebras.
\smallskip

\proof Given the $\sosi$ invariance, we must only verify the overall
normalizaton of both operators. Since $C=C_++C_-$ we find
using  \donvectf,
$\De'C=4$, which thus agrees with \fstcond.

\vfill\eject

\appendix{H}{Free modules of $\fP_\pm$}
\applab\freemodules

In this appendix we outline an explicit construction of a free
G-module on one generator of the chiral subalgebras, $\fP_+$ and
$\fP_-$.  An immediate application of this result is
to prove Theorem \freegen.
\smallskip

\tbl\gengr
Consider the holomorphic subalgebra $\fP_+$. As a dot-algebra it is
generated by
\eqn\holgen{
x_\si\,,\quad D^\si=\ep^{\si\rh\pi}P_{\rh,\pi}\,,\quad
P_{\si,\rh\pi}=\ep_{\si\rh\pi} P\,,}
subject to relations, see \contwo\ and  \confour,
\eqn\relone{
x_\si\cdot
D^\si=0\,,}
\eqn\reltwo{
D^\si\cdot
D^\rh=-\hbox{$3\over 2$}
\ep^{\si\rh\pi}x_\pi\cdot P\,,\quad
D^\al\cdot P=0\,,\quad P\cdot P=0\,.}  Since the bracket (and the
BV-operator) on $\fP$ vanishes when restricted to $\fP_+$, the free
G-module, $\fM_\Ga$,  is spanned by elements of
the form\foot{We will omit here the subscripts on the bracket and the
BV-operator.}
\eqn\arelfree{
\Ph_0\cdot[\,\Ph_1,[\,\ldots,[\,\Ph_n,\Ga\,]\,\ldots\,]\,]\,, \quad
\Ph_0\,,\ldots\,,\Ph_n\in\fP_+\,.}
In fact, given \gmAB, it is sufficient that the $\Ph_i$, $i\geq 1$, run over
the set of generators \holgen. Denote by
\eqn\broper{
\p_\si=[\,x_\si,-\,]\,,\quad
\cD^\si=[\,D^\si,-\,]\,,\quad
\cP=[\,P,-\,]\,,}
the generators of the bracket action of $\fP_+$ on $\fM_\Ga$.
{}From \gmAB\ and \gmAC, and the vanishing of the bracket on
$\fP_+$, it follows that -- together with the operators
$1$, $x_\si$, $D^\si$ and $P$,
corresponding to the dot action of $\fP_+$ on $\fM_\Ga$ -- they generate
a graded commutative  algebra,
$\fQ_+=\bigoplus_{n\in\ZZ}\fQ^n_+$.

\bigskip\bigskip
\begintable
$n$ | $-1$ | $0$ | $1$ | $2$ \cr
$\quad\Ph_I\quad$  | $\quad
\p_\si\quad$ | $\quad x_\si\,, \cD^\si\quad$ |
$\quad D^\si\,,\cP\quad$ | $\quad P\quad$
\endtable
\medskip
\centerline{Table \gengr. The generators of $\fQ_+$}
\bigskip

In addition to \relone\ and \reltwo,
\eqn\relthree{
x_\si\cD^\si+D^\si\p_\si=0\,,}
\eqn\relfour{
D^\si\cD^\rh - D^\rh\cD^\si=-\hbox{$3\over 2$}
\ep^{\si\rh\pi}\,(x_\pi\cP + P\p_\pi)\,,}
\eqn\relfive{D^\si\cP+P\cD^\si=0\,,\quad
P\cP=0\,,}
exhaust the defining relations between the generators of $\fQ_+$.
\smallskip

Since $\fM_\Ga$ is freely generated by $\fQ_+$ acting on $\Ga$, the
problem of determining the free module is equivalent to that of
computing $\fQ_+$, \ie, the quotient of a free graded commutative
algebra by the ideal generated by the relations \relone, \reltwo\ and
\relthree-\relfive.
\smallskip

Consider the subalgebra $\fQ_+'$  generated by the operators
$D^\si$, $P$,  $\p_\si$ and $\cP$. Clearly it is
finite dimensional and nonvanishing for $n=-3,\ldots,2$. The quotient
algebra $\fQ_+/(\fQ_+\fQ_+')$, is spanned by monomials
\eqn\firstmonom{
x_{\si_{1}}\ldots x_{\si_{\scriptstyle s_1}}\cD^{\rh_1}\ldots\,
\cD^{\rh_{\scriptstyle s_2}}\,,
\quad s_1,s_2\geq 0\,,}
of order zero, so that the $\fQ^n_+$ are nonzero for the same range of
$n$ as $\fQ_+'$.  Moreover, all monomials with the same $s_1$ and $s_2$
comprise a single $\bga$ module with the highest weight
$(\La,r_2\La)$, where $\La=s_1\La_1+s_2\La_2$, the reason being that
by \relthree\ all modules spanned by the trace components in
\firstmonom\ vanish in the quotient.  This heuristically shows
that $\fQ_+$ can be decomposed into a direct sum of disjoint
$r_2$-twisted cones of $\bga$ modules, although to determine the set
of those cones we must study  products of the monomials
\firstmonom\ with $\fQ_+'$. Given the finite number of cases to be
considered, this can be done explicitly.
\smallskip

In the case  $n=2$ we must consider all expressions of the form
\eqn\typeone{x_{\si_{1}}\ldots
x_{\si_{\scriptstyle s_1}}\cD^{\rh_1}\ldots
\,\cD^{\rh_{\scriptstyle s_2}}P\,,\quad s_1,s_2\geq 0\,,}
whose trace components still
vanish because of \reltwo\ and \relthree. This yields a single cone of
$\bga$ modules with the tip, $(0,\La_1-2\La_2)$, at the weights of
$P$.
\smallskip

For $n=1$ all terms can be reduced, using \relone, \reltwo, \relfour\
and \relfive, to just three types:
\eqn\typethree{%
x_{\si_{1}}\ldots x_{\si_{\scriptstyle s_1}}\cD^{\rh_1}\ldots
\,\cD^{\rh_{\scriptstyle s_2}}\,P
\,\p_{\pi}\,,\quad s_1,s_2\geq 0\,,}
\eqn\typetwo{%
x_{\si_{1}}\ldots x_{\si_{\scriptstyle s_1}}\cD^{\rh_1}\ldots
\,\cD^{\rh_{\scriptstyle s_2}}\cP\,,\quad
s_1,s_2\geq 0\,,}
\eqn\typefour{%
x_{\si_{1}}\ldots x_{\si_{\scriptstyle s_1}}\cD^{\rh_1}\ldots
\,\cD^{\rh_{\scriptstyle s_2}}\,D^{\rh_{\scriptstyle s_2+1}}
\,,\quad s_1,s_2\geq 0\,.}
As for $n=2$, we have no trace   in $(\si,\rh)$ in \typethree.
Thus the decomposition into $\slth$ modules is that of the
tensor product $(s_1,s_2)\otimes (1,0)$, giving rise to three
cones, $(\La_1,2\La_1-2\La_2)$, $(\La_2,3\La_1-2\La_2)$ and $(0,
3\La_1-3\La_2)$.
The traceless component in \typetwo\ yields a cone at $(0,
\La_1-2\La_2)$, while the trace component, using \relthree\ and
\relfive, is equivalent the last cone in \typethree. Finally,
 we may assume complete symmetry in
$\rh_1,\ldots,\rh_{s_2+1}$ of
\typefour, as all terms with mixed symmetry can be
expressed in terms of \typethree\ and \typetwo. Also, the trace terms
either vanish due to \relone\ or are reduced to \typethree\ and
\typetwo\ using \relthree\ and \reltwo. This leaves just a single cone
$(\La_2,\La_1-\La_2)$.
\smallskip

\tbl\qiuscones
All the remaining cases can be analyzed similarly.  The complete
decomposition of $\fQ_+$ into $r_2$-twisted cones of $\bga$ modules is
given in Table \qiuscones. We list there both the weights,
$(\La,\La')$, of the tips of the cones as well as elements of $\fQ_+'$
that give rise to them upon multiplication with monomials \firstmonom.
\smallskip

To obtain the $\bga$ decomposition of a free module $\fM_\Ga$, where
an $\slth$ singlet $\Ga$ has order $m$ and  the $(\uone)^2$
weight $\La_\Ga$, one must merely shift $n\,\rightarrow\, n+m$ and
$(\La,\La')\,\rightarrow (\La,\La'+\La_\Ga)$. The corresponding result
for $\fP_-$ modules is obtained by interchanging the fundamental
weights $\La_1$ and $\La_2$.
\smallskip

\noindent
{\it Proof of Theorem \freegen:} There is a unique G-homomorphism
$\iota: \fM_{\Ga_1} \rightarrow
\fP_{r_1}$ of $\fP_-$ modules defined by
$\iota(\Ga_1)=\Ga_1$.  Since the twisted cone decompositions of both
modules are identical, we first verify that the images of all
tips of the cones do not vanish. To conclude the proof we must show
that the entire $\fP_{r_1}$ is generated from those tips by the action
of $x_{\dot\si}$ and $[\,D^{\dot\si},\,-\,]$. Since those
operators generate the underlying $\cR_3$ module  $M_{r_1}$, the
required extension to higher order polyderivations seems rather
obvious. \SMd

\vfill\eject

\begintable
$\quad n\quad$ | $\fQ_+$ | $(\La,\La')$ \crthick
2 | $P$ | $(0,\La_1-2\La_2)$ \cr
1 |    $D^\rh$ | $(\La_2,\La_1-\La_2)$ \nr
  | $\cP$ | $(0,\La_1-2\La_2)$ \nr
  | $P\,\p_\si$ |
$(\La_1,2\La_1-2\La_2)$, $(\La_2,3\La_1-2\La_2)$, $(0,3\La_1-3\La_2)$
\cr
0
  | $1$ | $(0,0)$ \nr
  | $D^\mu\p_\si$ | $(\La_1+\La_2,2\La_1-\La_2)$,
                   $(2\La_2,3\La_1-\La_2)$, $(0,2\La_1-\La_2)$ \nr
  | $\p_\si\,\cP$ | $\quad (\La_1, 2\La_1-2\La_2)$,
                                       $(0,3\La_1-3\La_2)$,
                                       $(\La_2,3\La_1-2\La_2)\quad $ \nr
  | $P\,\p_\si\p_\rh$ | $(0,4\La_1-2\La_2)$,
                                     $(\La_1,4\La_1-3\La_2)$,
                                     $(\La_2,3\La_1-2\La_2)$ \cr
$-1$ | $\p_\si$ | $(\La_1,\La_1)$, $(\La_2,2\La_1)$,
                            $(0,2\La_1-\La_2)$ \nr
     | $D^\mu\p_\si\p_\pi$ |
                             $(\La_1,3\La_1-\La_2)$,
                             $(\La_2,4\La_1-\La_2)$,
                             $(2\La_2,3\La_1-\La_2)$ \nr
     | $\p_\si\p_\rh\cP$ |
          $(0,4\La_1-2\La_2)$, $(\La_1,4\La_1-3\La_2)$,
          $(\La_2,3\La_1-2\La_2)$ \nr
     | $\p_\si\p_\rh\p_\pi P$ |
           $(0,4\La_1-2\La_2)$ \cr
$-2$ | $\p_\si\p_\rh$ |
       $(0,3\La_1)$, $(\La_1,3\La_1-\La_2)$, $(\La_2,2\La_1)$ \nr
     |  $\quad D^\mu \p_\si\p_\rh\p_\pi\quad $  |
       $(\La_2,4\La_1-\La_2)$ \nr
   | $\p_\si\p_\rh\p_\pi\cP$ |
     $(0,4\La_1-2\La_2)$ \cr
$-3$ |  $\p_\si\p_\rh\p_\pi$ |
     $(0,3\La_1)$
\endtable
\medskip
\centerline{Table \qiuscones. The $r_2$-twisted cone decomposition
            of $\fQ_+$}

\vfill\eject

\appendix{I}{Some explicit cohomology states}
\applab\APexstates

In this appendix we give a complete list of explicit representatives
of the cohomology classes that are required for the calculations
discussed in Section \Sbvalgebra. We have listed only operators
corresponding to the highest weights in $\bga$ modules. Other
operators in those modules can be obtained through the action of the
$\slth$ currents \slthonfc.  The normalization has been chosen to
simplify formulae in section \Sbvalgebra.

\appsubsec{The ground ring generators}
\appsubseclab\SSgrring
\eqn\exone{
\eqalign{
\Psi^{(0)}_{\La_1,\La_1}\ =&\  \big(-4 b^{[2]} b^{[3]} c^{[2]} c^{[3]} -
  \hbox{${10\over {\sqrt{3}}}$} b^{[2]} b^{[3]} \pr c^{[3]} c^{[3]} +
\hbox{$\sqrt{3}$} b^{[2]} \pr^2 c^{[3]} -
  \hbox{$2 \sqrt{3}$} b^{[3]} c^{[2]} +
  4 b^{[3]} \pr c^{[3]}
 \cr   &
- \hbox{$2 i \sqrt{6}$} \pr\phi^{M,1} b^{[2]} \pr c^{[3]} -
  \hbox{$3 i \sqrt{2}$} \pr\phi^{M,1} b^{[3]} c^{[3]} -
  4 \pr\phi^{M,1} \pr\phi^{M,2} b^{[2]} c^{[3]} -
  \hbox{$2 \sqrt{3}$} \pr\phi^{M,1} \pr\phi^{M,2}
\cr  &
- \hbox{$i\sqrt{3}$} \pr\phi^{M,1} \pr\phi^{L,1} b^{[2]} c^{[3]} +
  \hbox{$3 i$}\pr\phi^{M,1} \pr\phi^{L,1} -
  i\pr\phi^{M,1} \pr\phi^{L,2} b^{[2]} c^{[3]} +
  \hbox{$i\sqrt{3}$} \pr\phi^{M,1} \pr\phi^{L,2}
\cr  &
- \hbox{$i\sqrt{6}$} \pr\phi^{M,1} \pr b^{[2]} c^{[3]} -
  \hbox{$2 i\sqrt{2}$} \pr\phi^{M,2} b^{[2]} \pr c^{[3]} -
  \hbox{$i\sqrt{6}$} \pr\phi^{M,2} b^{[3]} c^{[3]} +
  \hbox{${4\over {\sqrt{3}}}$} \pr\phi^{M,2} \pr\phi^{M,2} b^{[2]} c^{[3]}
\cr   &
+2 \pr\phi^{M,2} \pr\phi^{M,2} -
  i\pr\phi^{M,2} \pr\phi^{L,1} b^{[2]} c^{[3]} +
  \hbox{$i\sqrt{3}$} \pr\phi^{M,2} \pr\phi^{L,1} -
  \hbox{${i\over {\sqrt{3}}}$}\pr\phi^{M,2} \pr\phi^{L,2} b^{[2]} c^{[3]}
\cr   &
+ i\pr\phi^{M,2} \pr\phi^{L,2} -
  \hbox{$i\sqrt{2}$} \pr\phi^{M,2} \pr b^{[2]} c^{[3]} -
  \hbox{$3 \sqrt{2}$} \pr\phi^{L,1} b^{[2]} c^{[2]} -
  \hbox{$2 \sqrt{6}$} \pr\phi^{L,1} b^{[2]} \pr c^{[3]}
\cr  &
- \hbox{$3 \sqrt{2}$} \pr\phi^{L,1} b^{[3]} c^{[3]} -
  \hbox{$\sqrt{3}$} \pr\phi^{L,1} \pr\phi^{L,1} b^{[2]} c^{[3]} +
  3 \pr\phi^{L,1} \pr\phi^{L,1} -
  2 \pr\phi^{L,1} \pr\phi^{L,2} b^{[2]} c^{[3]}
\cr   &
+ \hbox{$2 \sqrt{3}$} \pr\phi^{L,1} \pr\phi^{L,2} -
  \hbox{$\sqrt{6}$} \pr\phi^{L,2} b^{[2]} c^{[2]} -
  \hbox{$2 \sqrt{2}$} \pr\phi^{L,2} b^{[2]} \pr c^{[3]} -
  \hbox{$\sqrt{6}$} \pr\phi^{L,2} b^{[3]} c^{[3]}
\cr   &
-\hbox{${1\over {\sqrt{3}}}$}\pr\phi^{L,2} \pr\phi^{L,2} b^{[2]} c^{[3]} +
  \pr\phi^{L,2} \pr\phi^{L,2} +
  \hbox{${2\over {\sqrt{3}}}$} \pr b^{[2]} b^{[2]} c^{[2]} c^{[3]} +
  \hbox{${20\over 3}$} \pr b^{[2]} b^{[2]} \pr c^{[3]} c^{[3]}
\cr   &
+ 2 \pr b^{[2]} c^{[2]} +
  \hbox{${8\over {\sqrt{3}}}$} \pr b^{[2]} \pr c^{[3]} +
  6 \pr b^{[3]} c^{[3]} +
  \hbox{$\sqrt{6}$} \pr^2\phi^{L,1} b^{[2]} c^{[3]} -
  \hbox{$3 \sqrt{2}$} \pr^2\phi^{L,1}
\cr   &
+\hbox{$\sqrt{2}$} \pr^2\phi^{L,2} b^{[2]} c^{[3]} -
  \hbox{$\sqrt{6}$} \pr^2\phi^{L,2} +
  \hbox{$\sqrt{3}$} \pr^2 b^{[2]} c^{[3]}\,\big)\,\cV_{\La_1,\La_1}\,, \cr
}}
\eqn\exdotthree{
\eqalign{
\Psi^{(0)}_{\La_2,\La_2}\  =&\  \big(
4 b^{[2]} b^{[3]} c^{[2]} c^{[3]} -
   \hbox{${10\over {\sqrt{3}}}$} b^{[2]} b^{[3]} \pr c^{[3]} c^{[3]}
   + \sqrt{3} b^{[2]} \pr^2 c^{[3]} -
  \hbox{$2 \sqrt{3}$} b^{[3]} c^{[2]}
 \cr   &
- 4 b^{[3]} \pr c^{[3]} +
  \hbox{$2 \sqrt{3}$} \pr\phi^{M,1} \pr\phi^{M,1} b^{[2]} c^{[3]} -
  3 \pr\phi^{M,1} \pr\phi^{M,1} -
  \hbox{$4 i \sqrt{2}$} \pr\phi^{M,2} b^{[2]} \pr c^{[3]}
 \cr   &
+ \hbox{$2 i \sqrt{6}$} \pr\phi^{M,2} b^{[3]} c^{[3]} -
  \hbox{${2\over {\sqrt{3}}}$} \pr\phi^{M,2} \pr\phi^{M,2} b^{[2]} c^{[3]} +
  \pr\phi^{M,2} \pr\phi^{M,2} -
\hbox{${4 i\over {\sqrt{3}}}$} \pr\phi^{M,2} \pr\phi^{L,2} b^{[2]} c^{[3]}
  \cr   &
 - 4 i \pr\phi^{M,2} \pr\phi^{L,2} -
  \hbox{$2 i \sqrt{2}$} \pr\phi^{M,2} \pr b^{[2]} c^{[3]} +
  \hbox{$2 \sqrt{6}$} \pr\phi^{L,2} b^{[2]} c^{[2]} -
  \hbox{$4 \sqrt{2}$} \pr\phi^{L,2} b^{[2]} \pr c^{[3]}
 \cr   &
+ \hbox{$2 \sqrt{6}$} \pr\phi^{L,2} b^{[3]} c^{[3]} -
\hbox{${4\over {\sqrt{3}}}$} \pr\phi^{L,2} \pr\phi^{L,2} b^{[2]} c^{[3]} -
  4 \pr\phi^{L,2} \pr\phi^{L,2} +
\hbox{${2\over {\sqrt{3}}}$} \pr b^{[2]} b^{[2]} c^{[2]} c^{[3]}
 \cr   &
- \hbox{${20\over 3}$} \pr b^{[2]} b^{[2]} \pr c^{[3]} c^{[3]} -
  2 \pr b^{[2]} c^{[2]} +
\hbox{${8\over {\sqrt{3}}}$} \pr b^{[2]} \pr c^{[3]}
 \cr   &- 6 \pr b^{[3]} c^{[3]} +
  \hbox{$2 \sqrt{2}$} \pr^2\phi^{L,2} b^{[2]} c^{[3]} +
  \hbox{$2 \sqrt{6}$} \pr^2\phi^{L,2} +
  \hbox{$\sqrt{3}$} \pr^2 b^{[2]} c^{[3]}\,\big)\,\cV_{\La_2,\La_2}\,. \cr
}}

\appsubsec{The identity quartet}
\appsubseclab\SSident

\eqn\identqu{
{\bf 1}(z)\,,}
\eqn\stctwo{ \eqalign{
C^{[2]}(z)\  =\  &-4(\pr c^{[2]}+\pr b^{[2]}\pr c^{[3]} c^{[3]}+
b^{[2]} \pr^2 c^{[3]} c^{[3]})-
\hbox{$1\over\sqrt{2}$}(\pr\ph^{L,1}+\hbox{$\sqrt{3}$}
\pr\ph^{L,2}) (c^{[2]}+b^{[2]}\pr c^{[3]}c^{3]})\cr &-
\hbox{$1\over\sqrt{2}$}(\sqrt{3}\pr\ph^{L,1}-
\pr\ph^{L,2})\pr c^{[3]}
+\sqrt{2}(\sqrt{3}\pr^2\ph^{L,1}+\pr^2\ph^{L,2})c^{[3]}
-\pr\ph^{L,1}\pr\ph^{L,2}c^{[3]}\cr &-
\hbox{$\sqrt{3}\over 2$}(\pr\ph^{L,1}\pr\ph^{L,1}-
\pr\ph^{L,2}\pr\ph^{L,2} )c^{[3]}\,,
\cr}}
\eqn\stcthree{ \eqalign{
C^{[3]}(z)\ =\ & -4\sqrt{3}\pr^2c^{[3]}-
\hbox{$\sqrt{3\over 2}$}(\sqrt{3}\pr\ph^{L,1}-\pr\ph^{L,2} )
(c^{[2]}+b^{[2]}\pr c^{[3]}c^{[3]})-
\sqrt{3}(\sqrt{2}\pr^2\ph^{L,1}+3\pr^2\ph^{L,2})c^{[3]}\cr &-
3\hbox{$\sqrt{3\over 2}$}(\pr\ph^{L,1}+
\sqrt{3}\pr\ph^{L,2})\pr c^{[3]} -
\hbox{$\sqrt{3}\over 2$}
(\pr\ph^{L,1}\pr\ph^{L,1}+
2\sqrt{3}\pr\ph^{L,1}\pr\ph^{L,2}+ \pr\ph^{L,2}\pr\ph^{L,2} )c^{[3]}\,, \cr}}
\eqn\identctwoth{
C^{[23]}(z)=(C^{[2]}\cdot C^{[3]})(z)\,.}


\appsubsec{Generators of $\fH_1^n$, $ n\geq 1$}
\appsubseclab\SShonegen
The following is a complete list of operators generating $\fH^n_1$,
$n\geq 1$, under the dot product action of $\fH^0$ and the bracket
action of $\slth$.  They are obtained by explicitly evaluating the
multiple commutators \thepisss. The normalization of $\widehat X$ in
\volumeom\ is chosen such that \comtox\ holds, with the ground ring
generators normalized as in \exone\ and \exdotthree.
\smallskip

\noindent
$n=1$:

\eqn\pissenone{
\widehat P_{\dot 2,\dot 3}=\Psi^{(1)}_{\La_1,-\La_1+\La_2}\,,\quad
\widehat P_{1,\dot 3}=\half \Psi^{(1)}_{\La_1+\La_2,0}\,,\quad
\widehat P_{1,2}=\Psi^{(1)}_{\La_2,\La_1-\La_2}\,,}
\eqn\pissenonec{
\widehat P_{\si,}{}^\si=\widehat C_+-\widehat C_-\,,\quad
\widehat C=\widehat C_++\widehat C_-\,.}

\eqn\ponemonetwo{
\eqalign{
\Psi^{(1)}_{\La_1,-\La_1+\La_2}\  =&\  \hbox{${1\over 36}$}
\big(12 \hbox{$\sqrt{3}$} b^{[2]} c^{[2]} \pr c^{[3]} +
    15 b^{[2]} \pr^2 c^{[3]} c^{[3]} -
    18 b^{[3]} c^{[2]} c^{[3]} -
    24 \hbox{$\sqrt{3}$} b^{[3]} \pr c^{[3]} c^{[3]}
\cr    &
 -3 i \hbox{$\sqrt{6}$} \pr\phi^{M,1} b^{[2]} c^{[2]} c^{[3]} -
    21 i \hbox{$\sqrt{2}$} \pr\phi^{M,1} b^{[2]} \pr c^{[3]} c^{[3]} -
    9 i \hbox{$\sqrt{2}$} \pr\phi^{M,1} c^{[2]} -
    18 i \pr\phi^{M,1} \pr\phi^{L,2} c^{[3]}
 \cr    &
-12 i \hbox{$\sqrt{6}$} \pr\phi^{M,1} \pr c^{[3]} -
    3 i \hbox{$\sqrt{2}$} \pr\phi^{M,2} b^{[2]} c^{[2]} c^{[3]} -
    7 i \hbox{$\sqrt{6}$} \pr\phi^{M,2} b^{[2]} \pr c^{[3]} c^{[3]} -
    3 i \hbox{$\sqrt{6}$} \pr\phi^{M,2} c^{[2]}
\cr      &
-6 i \hbox{$\sqrt{3}$} \pr\phi^{M,2} \pr\phi^{L,2} c^{[3]} -
    12 i \hbox{$\sqrt{2}$} \pr\phi^{M,2} \pr c^{[3]} +
    12 \hbox{$\sqrt{2}$} \pr\phi^{L,2} b^{[2]} c^{[2]} c^{[3]} -
    2 \hbox{$\sqrt{6}$} \pr\phi^{L,2} b^{[2]} \pr c^{[3]} c^{[3]}
 \cr     &
- 6 \hbox{$\sqrt{6}$} \pr\phi^{L,2} c^{[2]} -
    12 \hbox{$\sqrt{3}$} \pr\phi^{L,2} \pr\phi^{L,2} c^{[3]} -
    24 \hbox{$\sqrt{2}$} \pr\phi^{L,2} \pr c^{[3]} +
    36 \pr b^{[2]} \pr c^{[3]} c^{[3]}
\cr    &+ 18 \hbox{$\sqrt{2}$} \pr^2\phi^{L,2} c^{[3]} +
6 \hbox{$\sqrt{3}$} \pr^2 c^{[3]}\big)\cV_{\La_1,-\La_1+\La_2}\,,
\cr}}

\eqn\slthreeopp{
\Psi^{(1)}_{\La_1+\La_2,0}(z)
= (\,c^{[2]} + \hbox{$\sqrt{3\over 2}$}i\pr\ph^{M,1}c^{[3]} -
\hbox{$1\over\sqrt{2}$} i\pr\ph^{M,2}c^{[3]} - b^{[2]}\pr
c^{[3]}c^{[3]}\,) \cV_{\La_1+\La_2,0}\,.}

\eqn\ptwoonemtwo{
\eqalign{
\Psi^{(1)}_{\La_2,\La_1-\La_2}\ =\ &\hbox{${1\over 36}$}
\big(12 \hbox{$\sqrt{3}$} b^{[2]} c^{[2]} \pr c^{[3]} -
    15 b^{[2]} \pr^2 c^{[3]} c^{[3]} +
    18 b^{[3]} c^{[2]} c^{[3]} -
    24 \hbox{$\sqrt{3}$} b^{[3]} \pr c^{[3]} c^{[3]} \cr
    & - 6 i \hbox{$\sqrt{2}$} \pr\phi^{M,2} b^{[2]} c^{[2]} c^{[3]} +
    14 i \hbox{$\sqrt{6}$} \pr\phi^{M,2} b^{[2]} \pr c^{[3]} c^{[3]} +
    6 i \hbox{$\sqrt{6}$} \pr\phi^{M,2} c^{[2]} -
    18 i \pr\phi^{M,2} \pr\phi^{L,1} c^{[3]} \cr
    & - 6 i \hbox{$\sqrt{3}$} \pr\phi^{M,2} \pr\phi^{L,2} c^{[3]} -
    24 i \hbox{$\sqrt{2}$} \pr\phi^{M,2} \pr c^{[3]} +
    6 \hbox{$\sqrt{6}$} \pr\phi^{L,1} b^{[2]} c^{[2]} c^{[3]} +
    3 \hbox{$\sqrt{2}$} \pr\phi^{L,1} b^{[2]} \pr c^{[3]} c^{[3]} \cr
    & + 9 \hbox{$\sqrt{2}$} \pr\phi^{L,1} c^{[2]} -
    9 \hbox{$\sqrt{3}$} \pr\phi^{L,1} \pr\phi^{L,1} c^{[3]} -
    18 \pr\phi^{L,1} \pr\phi^{L,2} c^{[3]} -
    12 \hbox{$\sqrt{6}$} \pr\phi^{L,1} \pr c^{[3]} \cr
    & + 6 \hbox{$\sqrt{2}$} \pr\phi^{L,2} b^{[2]} c^{[2]} c^{[3]} +
    \hbox{$\sqrt{6}$} \pr\phi^{L,2} b^{[2]} \pr c^{[3]} c^{[3]} +
    3 \hbox{$\sqrt{6}$} \pr\phi^{L,2} c^{[2]} \cr
    & - 3 \hbox{$\sqrt{3}$} \pr\phi^{L,2} \pr\phi^{L,2} c^{[3]} -
    12 \hbox{$\sqrt{2}$} \pr\phi^{L,2} \pr c^{[3]} -
    36 \pr b^{[2]} \pr c^{[3]} c^{[3]} +
    9 \hbox{$\sqrt{6}$} \pr^2\phi^{L,1} c^{[3]} \cr
    & + 9 \hbox{$\sqrt{2}$} \pr^2\phi^{L,2} c^{[3]} +
    6 \hbox{$\sqrt{3}$} \pr^2 c^{[3]}\big)\, \cV_{\La_2,\La_1-\La_2}\,, \cr}}

\noindent
$n=2$:

\eqn\pisstwo{
\widehat P_{1,23}=\Psi^{(2)}_{0,-2\La_1+\La_2}\,,\quad
\widehat P_{1,\si}{}^\si=\Psi^{(2)}_{\La_1,-\La_2}\,,\quad
\widehat P_{1,2\dot 3}=\Psi^{(2)}_{2\La_2,-\La_2}\,, }
\eqn\pisstwotwo{
\widehat P{}_{\dot 1,\dot 2\dot 3}=\Psi^{(2)}_{0,\La_1-2\La_2}\,,\quad
\widehat P{}_{\dot 3,\dot\si}{}^{\dot\si}=\Psi^{(2)}_{\La_2,-\La_1}\,,\quad
\widehat P{}_{{\dot 3,1 2}}=\Psi^{(2)}_{2\La_1,-\La_1}\,. }

\eqn\FGaa{\eqalign{
\Psi^{(2)}_{0,-2\La_1+\La_2}=&\
\hbox{${1\over 108}$}\big(
 6 b^{[2]} c^{[2]} \pr^2 c^{[3]} c^{[3]} -
    6 b^{[2]} \pr c^{[2]} \pr c^{[3]} c^{[3]} +
    12 \hbox{$\sqrt{3}$} b^{[2]} \pr^2 c^{[3]} \pr c^{[3]} c^{[3]} +
    6 \hbox{$\sqrt{3}$} c^{[2]} \pr^2 c^{[3]} \cr
    & +  3 \hbox{$\sqrt{6}$} \pr\phi^{L,1} c^{[2]} \pr c^{[3]} +
    9 \pr\phi^{L,1} \pr\phi^{L,1} \pr c^{[3]} c^{[3]} +
    6 \hbox{$\sqrt{3}$} \pr\phi^{L,1} \pr\phi^{L,2} \pr c^{[3]} c^{[3]} -
    3 \hbox{$\sqrt{6}$} \pr\phi^{L,1} \pr c^{[2]} c^{[3]} \cr
    & +  9 \hbox{$\sqrt{2}$} \pr\phi^{L,1} \pr^2 c^{[3]} c^{[3]} +
    3 \hbox{$\sqrt{2}$} \pr\phi^{L,2} c^{[2]} \pr c^{[3]} +
    3 \pr\phi^{L,2} \pr\phi^{L,2} \pr c^{[3]} c^{[3]} -
    3 \hbox{$\sqrt{2}$} \pr\phi^{L,2} \pr c^{[2]} c^{[3]} \cr
    & +  3 \hbox{$\sqrt{6}$} \pr\phi^{L,2} \pr^2 c^{[3]} c^{[3]} +
    6 \pr b^{[2]} c^{[2]} \pr c^{[3]} c^{[3]} +
    6 \pr c^{[2]} c^{[2]} -
    6 \hbox{$\sqrt{3}$} \pr c^{[2]} \pr c^{[3]} \cr
    & +  3 \hbox{$\sqrt{6}$} \pr^2\phi^{L,1} c^{[2]} c^{[3]} -
    9 \hbox{$\sqrt{2}$} \pr^2\phi^{L,1} \pr c^{[3]} c^{[3]} +
    3 \hbox{$\sqrt{2}$} \pr^2\phi^{L,2} c^{[2]} c^{[3]} -
    3 \hbox{$\sqrt{6}$} \pr^2\phi^{L,2} \pr c^{[3]} c^{[3]} \cr
    & +  21 \pr^2 c^{[3]} \pr c^{[3]} +
    \pr^3 c^{[3]} c^{[3]}\big)\,\cV_{0,-2\La_1-\La_2}\,, \cr}}

\eqn\FGab{\eqalign{
\Psi^{(2)}_{0,\La_1-2\La_2}=&\ \hbox{${1\over 108}$}
\big(-6 b^{[2]} c^{[2]} \pr^2 c^{[3]} c^{[3]} +
    6 b^{[2]} \pr c^{[2]} \pr c^{[3]} c^{[3]} +
    12 \hbox{$\sqrt{3}$} b^{[2]} \pr^2 c^{[3]} \pr c^{[3]} c^{[3]} +
    6 \hbox{$\sqrt{3}$} c^{[2]} \pr^2 c^{[3]} \cr
    & + 6 \hbox{$\sqrt{2}$} \pr\phi^{L,2} c^{[2]} \pr c^{[3]} -
    12 \pr\phi^{L,2} \pr\phi^{L,2} \pr c^{[3]} c^{[3]} -
    6 \hbox{$\sqrt{2}$} \pr\phi^{L,2} \pr c^{[2]} c^{[3]} -
    6 \hbox{$\sqrt{6}$} \pr\phi^{L,2} \pr^2 c^{[3]} c^{[3]} \cr
    & - 6 \pr b^{[2]} c^{[2]} \pr c^{[3]} c^{[3]} -
    6 \pr c^{[2]} c^{[2]} -
    6 \hbox{$\sqrt{3}$} \pr c^{[2]} \pr c^{[3]} +
    6 \hbox{$\sqrt{2}$} \pr^2\phi^{L,2} c^{[2]} c^{[3]} \cr
    & + 6 \hbox{$\sqrt{6}$} \pr^2\phi^{L,2} \pr c^{[3]} c^{[3]} -
    21 \pr^2 c^{[3]} \pr c^{[3]} -
    \pr^3 c^{[3]} c^{[3]}\big)\,\cV_{0,-\La_1-2\La_2}\,, \cr}}

\eqn\FGad{\eqalign{
\Psi^{(2)}_{\La_1,-\La_2}=&\
\hbox{${1\over 108}$}\big(12 b^{[2]} c^{[2]} \pr c^{[3]} c^{[3]} +
    12 \hbox{$\sqrt{3}$} c^{[2]} \pr c^{[3]} +
    3 i \hbox{$\sqrt{6}$} \pr\phi^{M,1} c^{[2]} c^{[3]} -
    6 i \hbox{$\sqrt{2}$} \pr\phi^{M,1} \pr c^{[3]} c^{[3]} \cr
    & +  3 i \hbox{$\sqrt{2}$} \pr\phi^{M,2} c^{[2]} c^{[3]} -
    2 i \hbox{$\sqrt{6}$} \pr\phi^{M,2} \pr c^{[3]} c^{[3]} +
    6 \hbox{$\sqrt{6}$} \pr\phi^{L,1} c^{[2]} c^{[3]} +
    6 \hbox{$\sqrt{2}$} \pr\phi^{L,1} \pr c^{[3]} c^{[3]} \cr
    & + 6 \hbox{$\sqrt{2}$} \pr\phi^{L,2} c^{[2]} c^{[3]} +
    2 \hbox{$\sqrt{6}$} \pr\phi^{L,2} \pr c^{[3]} c^{[3]} +
    12 \pr^2 c^{[3]} c^{[3]}\big)\,\cV_{\La_1,-\La_2}\,, \cr}}

\eqn\FGac{\eqalign{
\Psi^{(2)}_{\La_2,-\La_1}=&\
\hbox{${1\over 54}$}\big(-6 b^{[2]} c^{[2]} \pr c^{[3]} c^{[3]} +
    6 \hbox{$\sqrt{3}$} c^{[2]} \pr c^{[3]} +
    3 i \hbox{$\sqrt{2}$} \pr\phi^{M,2} c^{[2]} c^{[3]} +
    2 i \hbox{$\sqrt{6}$} \pr\phi^{M,2} \pr c^{[3]} c^{[3]} \cr
    & +  6 \hbox{$\sqrt{2}$} \pr\phi^{L,2} c^{[2]} c^{[3]} -
    2 \hbox{$\sqrt{6}$} \pr\phi^{L,2} \pr c^{[3]} c^{[3]} -
    6 \pr^2 c^{[3]} c^{[3]}\big)\,\cV_{\La_2,-\La_1}\,, \cr}}

\eqn\FGae{
\Psi^{(2)}_{2\La_2,-\La_2} =
\hbox{${1\over {18\sqrt{3}}}$}\big (3 c^{[2]} c^{[3]} -
   \hbox{$2 \sqrt{3}$}\pr c^{[3]}
c^{[3]}\big)\,\cV_{2\La_2,-\La_2}\,,}

\eqn\FGaf{
\Psi^{(2)}_{2\La_1,-\La_1} =
-\hbox{${1\over {18\sqrt{3}}}$} \big(3 c^{[2]} c^{[3]} +
   \hbox{$2 \sqrt{3}$}\pr c^{[3]} c^{[3]}\big)\,\cV_{2\La_1,-\La_1}\,,}

\noindent
$n=3$:
\eqn\pissthreeoh{
\widehat P{}_{1,23\dot 3}=\Psi^{(3)}_{\La_2,-2\La_2}\,,\quad
\widehat P{}_{\dot 1,\dot 2\dot 3
1}=\Psi^{(3)}_{\La_1,-2\La_1}\,,\quad
\widehat P{}_{1,2\dot 2\dot
3}=\Psi^{(3)}_{\La_1+\La_2,-\La_1-\La_2}\,,\quad
\widehat P{}_{\si,}{}^\si{}_\rh{}^\rh=\Psi^{(3)}_{0,0}\,.}

\eqn\FGag{\eqalign{
\Psi^{(3)}_{\La_2,-2\La_2}=&\
-\hbox{${1\over 864}$}
\big(12 c^{[2]} \pr^2 c^{[3]} c^{[3]} +
    9 \hbox{$\sqrt{2}$} \pr\phi^{L,1} c^{[2]} \pr c^{[3]} c^{[3]} +
    3 \hbox{$\sqrt{6}$} \pr\phi^{L,2} c^{[2]} \pr c^{[3]} c^{[3]} \cr
    & +6 \hbox{$\sqrt{3}$} \pr c^{[2]} c^{[2]} c^{[3]} -
    6 \pr c^{[2]} \pr c^{[3]} c^{[3]} +
    7 \hbox{$\sqrt{3}$} \pr^2 c^{[3]} \pr c^{[3]}
c^{[3]}\big)\,\cV_{\La_2,-2\La_2} \,,  \cr}}

\eqn\FGah{\eqalign{
\Psi^{(3)}_{\La_1,-2\La_1}=&\
\hbox{${1\over 864}$}
\big(-12 c^{[2]} \pr^2 c^{[3]} c^{[3]} -
    6 \hbox{$\sqrt{6}$} \pr\phi^{L,2} c^{[2]} \pr c^{[3]} c^{[3]} +
    6 \hbox{$\sqrt{3}$} \pr c^{[2]} c^{[2]} c^{[3]} + \cr
    &6 \pr c^{[2]} \pr c^{[3]} c^{[3]} +
    7 \hbox{$\sqrt{3}$} \pr^2 c^{[3]} \pr c^{[3]}
c^{[3]}\big)\,\cV_{\La_1,-2\La_1}\,,  \cr}}

\eqn\FGak{\Psi^{(3)}_{\La_1+\La_2,-\La_1-\La_2}=\hbox{${1\over 48}$}
c^{[2]} \pr c^{[3]} c^{[3]} \,\cV_{\La_1+\La_2,-\La_1-\La_2}\,,}

\eqn\FGal{\Psi^{(3)}_{0,0}=\hbox{${1\over {32\sqrt{3}}}$}
\big(2 \pr c^{[2]} c^{[2]} c^{[3]} -
    \pr^2 c^{[3]} \pr c^{[3]} c^{[3]}\big)\,\cV_{0,0}\,.}

\noindent
$n=4$:
\eqn\ohhiiijj{
\widehat P{}_{1,23\dot 2\dot3}=\Psi^{(4)}_{\La_1,-\La_1-2\La_2}\,, \quad
\widehat P{}_{\dot 1,\dot 2\dot 3 12}=\Psi^{(4)}_{\La_2,-2\La_1-\La_2}\,.}

\eqn\FGam{
\Psi^{(4)}_{\La_1,-\La_1-2\La_2}=
-\hbox{${1\over {360\sqrt{3}}}$}
\big(c^{[2]} \pr^2 c^{[3]} \pr c^{[3]} c^{[3]} +
   \hbox{$\sqrt{3}$} \pr c^{[2]} c^{[2]} \pr c^{[3]} c^{[3]}\big)
\,\cV_{\La_1,-\La_1-2\La_2}\,,}

\eqn\FGan{
\Psi^{(4)}_{\La_2,-2\La_1-\La_2} =
-\hbox{${1\over {360\sqrt{3}}}$}\big(c^{[2]} \pr^2 c^{[3]}
\pr c^{[3]} c^{[3]} -
   \hbox{$\sqrt{3}$} \pr c^{[2]} c^{[2]} \pr c^{[3]} c^{[3]}\big)\,
\cV_{\La_2,-2\La_1-\La_2}\,.}

\noindent
$n=5$:
\eqn\ohfive{
\widehat X=\Psi_{0,-2\La_1-2\La_2}^{(5)}\,.}
\eqn\volumeom{\Psi_{0,-2\La_1-2\La_2}^{(5)}
=\hbox{$1\over 1728\sqrt{3}$}\p c^{[2]}c^{[2]}\p^2 c^{[3]}
\p c^{[3]} c^{[3]}\,\cV_{0,-2\La_1-2\La_2}\,.}


\appsubsec{Twisted modules of the ground ring}
\appsubseclab\SStwgen

\eqn\omegaone{
\widehat \Omega_1(z) ~=~ {\bf 1}(z)\,,}
\eqn\omegarone{
\widehat \Omega_{r_1}(z) ~=~ 2
 \big(b^{[2]} \pr c^{[3]} c^{[3]}  +
c^{[2]} + \hbox{$\sqrt{2}$} \pr\phi^{L,2} c^{[3]} +
\hbox{$\sqrt{3}$} \pr c^{[3]}\big)\,\cV_{0,-2\La_1+\La_2}\,,
}
\eqn\omegartwo{
\widehat \Omega_{r_2}(z) ~=~
\big(- 2 b^{[2]} \pr c^{[3]} c^{[3]} - 2 c^{[2]} +
\hbox{$\sqrt{6}$} \pr \phi^{L,1} c^{[3]}  +
\hbox{$\sqrt{2}$} \pr\phi^{L,2}c^{[3]} + 2 \hbox{$\sqrt{3}$} \pr
c^{[3]} \big)\,\cV_{0,\La_1-2\La_2}\,,
}
\eqn\omegaonetwo{
\widehat \Omega_{r_{12}}(z) ~=~ c^{[2]}c^{[3]}\,\cV_{0,-3\La_2}
\,,}
\eqn\omegatwoone{
\widehat \Omega_{r_{21}}(z) ~=~ c^{[2]}c^{[3]}\,\cV_{0,-3\La_1}
\,,}
\eqn\omegathree{
\widehat \Omega_{r_{3}}(z) ~=~ c^{[2]} \pr c^{[3]} c^{[3]}
 \,\cV_{0,-2\La_1-2\La_2} \,.}

\vfill\eject

\appendix{J}{The BV-algebra of $2D$ $\cW_2$ string}
\applab\AVir

 In this appendix we summarize the main features of the BV-algebra of
$2d$ gravity coupled to $c=1$ matter, the so-called $2D$ $\cW_2$
(Virasoro) string.  The reader should consult
[\LZvir,\BMPvir,\Wi,\WiZw,\WY,\MMMO] and especially [\LZbv] for additional
discussion and detailed proofs. We follow here the notation introduced
earlier, in the context of the $\cW_3$ string, to emphasize both the
similarities and the differences between the two cases.  In this
appendix, and only here, $P$ and $Q$ denote the $\sltw$ weight and the
root lattices, respectively, while $F^{gh}$ is the Fock space of the
Virasoro ghosts, \ie, the $j=2$ $(bc)$-system.

\appsubsec{The cohomology problem}
\appsubseclab\AVcoh

Recall that a Feigin-Fuchs module, $F(\La,\al_0)$, of the Virasoro
algebra is  parametrized by the momentum, $\La$, of the underlying
Fock space of a single scalar field and a background charge $\al_0$.
If we  interpret $\La$ as an $\sltw$ weight, then the highest weight
of the Virasoro module is $h=\half(\La,\La+2\al_0\rh)$. The central
charge is given by $c=1-6\al^2_0$.
\smallskip

An important problem in the study of $2D$ Virasoro string is to
compute the operator algebra, $\fH$, obtained as the semi-infinite
(BRST) cohomology of the VOA associated with tensor products,
$F(\La^M,0)\otimes F(\La^L,2i)$, of Feigin-Fuchs modules with $c=1$
and $c=25$, respectively. More precisely, we consider the $\cW_2$
analogue of the complex $(\fC,d)$ in Section \Swcohprob, defined by
$L\subset P\times iP$.
\smallskip

The vertex operator realization of $\widehat\sltw$ on the $c=1$ Fock
spaces, together with the Liouville momentum operator, $-ip^L$,
give rise to a $\sltw\oplus \uone$ symmetry on $\fC$, that commutes
with the BRST operator and yields a decomposition of the cohomology,
$H(\cW_2,\fC)$, into a direct sum of finite dimensional
modules.  The complete description of the cohomology space is then
given by the following theorem due to [\LZvir,\BMPvir,\Wi].

\tbl\vircones\nfig\virreltbl{}
\thm\sltcoh
\proclaim Theorem \sltcoh. The relative cohomology
$H_{rel}(\cW_2,\fC)$ is isomorphic, as an $\sltw\oplus \uone$ module, to
the direct sum of finite dimensional irreducible modules with
highest weights in a set of disjoint cones
$\{(\La,\La')+(\la,w\la)\,|\,\la\in P_+\}$,
\ie
\eqn\vircohdec{H^n_{rel}(\cW_2,\fC)\ \cong\ \bigoplus_{w\in
W}\,\,\bigoplus_{(\La,\La')\in S_w^n}\,\, \bigoplus_{\la\in P_+}
\cL(\La+\la)\otimes \CC_{\La'+w\la}\,,}
where the sets $S_w^n$ are given in Table \vircones.

\bigskip
\begintable
$\quad n \quad$ | $\quad w\quad$ | $S_w^n$ \crthick
0 | $1$ | $(0,0)_{-1}$ \cr
1 | $1$ | $(\rh,-\rh)_1$ \nr
  | $-1$ | $\quad (0,-2\rh)_1 \quad$ \cr
2 | $-1$ | $(0,-4\rh)_{-1}$ \endtable
\medskip
\centerline{Table \vircones. The tips $S_w^n$ in the decomposition
of $H_{rel}(\cW_2,\fC)$}.
\medskip

\noindent
{\it Remark:} The required cohomology, $H(\cW_2,F(\La^M,0)\otimes
F(\La^L,2i))$, may be determined either directly, see, \eg, [\BMPvir],
or by decomposing $F(\La^M,0)$ into irreducible modules, and then
computing $H_{rel}(\cW_2,L(\La,0)\otimes F(\La^L,2i))$, see [\LZvir].
The latter cohomology is nonvanishing if and only if the weights $\La$
and $\La^L$ satisfy $-i\La^L+2\rh=w(\La+\rh-\si\rh)$ for some
$w,\si\in W$.  This is different than in the $\cW_3$ case, summarized
in Corollary \thBa, where $\si$ is required to run over an extension
of the Weyl group of $\slth$.  Thus, the $\cW_3$ cohomology displays
qualitatively new features, which, in this particular case, can be
explained by the more complex embedding pattern of primitive vectors in
Verma modules of a higher rank $\cW$-algebra.
\smallskip

\bigskip
$$ n=2:\qquad
      \hbox{\lower 3pt \hbox{\epsffile{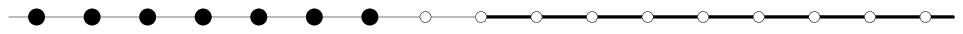} }} $$
$$ n=1:\qquad
      \hbox{\lower 3pt \hbox{\epsffile{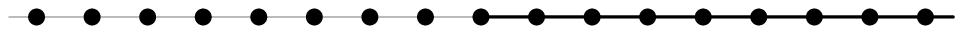} }} $$
$$ n=0:\qquad
      \hbox{\lower 3pt \hbox{\epsffile{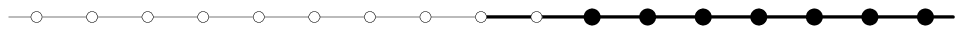} }} $$

\medskip
\noindent
\centerline{\vbox{\hsize 5.5in
\noindent
Figure \virreltbl. A graphical representation of
$H_{rel}(\cW_2,\fC)$. The points on the $\sltw$ weight lattice
correspond to shifted Liouville momenta $-i\La^L+2\rh$ and
\hbox{\lower 1pt \hbox{\epsffile{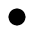}}} denotes
a single $\sltw\oplus\uone$ module. The fundamental Weyl chamber is
outlined by a thick line.}}
\bigskip
\nfig\virabstbl{}

The absolute cohomology is $H^\bullet(\cW_2,\fC)\cong
H_{rel}^{\bullet}(\cW_2,\fC)\oplus
H_{rel}^{\bullet-1}(\cW_2,\fC)$, \ie, it has a doublet structure with
the resulting pattern of $\sltw\oplus\uone$ modules as in Figure
\virabstbl.
\bigskip

$$ n=3:\qquad
      \hbox{\lower 3pt \hbox{\epsffile{vr2.eps} }} $$
$$ n=2:\qquad
      \hbox{\lower 3pt \hbox{\epsffile{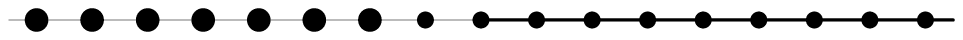} }} $$
$$ n=1:\qquad
      \hbox{\lower 3pt \hbox{\epsffile{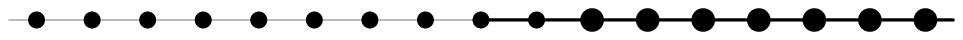} }} $$
$$ n=0:\qquad
      \hbox{\lower 3pt \hbox{\epsffile{vr0.eps} }} $$

\medskip
\centerline{\vbox{\hsize 5.5in
\noindent
Figure \virabstbl.  A graphical representation of
$H(\cW_2,\fC)$, adapting the conventions of Figure \virreltbl. The dots
\hbox{\lower 1pt \hbox{\epsffile{vdt1.eps}}} and
\hbox{\lower 1pt \hbox{\epsffile{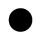}}} correspond,
to one and two $\sltw\oplus\uone$ modules, respectively.}}

\appsubsec{The BV-algebra and structure theorems}
\appsubseclab\AVbv

The ground ring of the BV-algebra $(\fH,\,\cdot\,,b_0)$, where
$\fH\equiv H(\cW_2,\fC)$ and $b_0\vps=b_0^{[2]}$, is isomorphic with
the algebra of polynomial functions on the complex plane,
\ie, $\fH^0\cong \cC_2$. Let us denote its generators by $\widehat x_i$,
$i=1,2$.  To make the analogy between this and the $\cW_3$ case more
evident, let us note that $\cC_2$ may also be identified with the ring
of regular functions on the base affine space of $\sltw$.
\smallskip

The geometric description of the higher ghost number cohomology, or at
least a part of it, is given by the following structure theorem.

\thm\strthsltw
\proclaim Theorem \strthsltw\ [\LZbv].
\item{i.} The map\foot{{\rm  In
[\LZbv] this map is denoted by $\psi$.}} $\pi:\fH\rightarrow
\cP(\cC_2)$ introduced in Section \SSpolyder\
is a BV-algebra homomorphism onto the BV-algebra of polyderivations
$(\cP(\cC_2),\,\cdot\,,\De_S)$.
\item{ii.} There exists an embedding $\imath:\cP(\cC_2)\rightarrow
\fH$ that preserves the dot product and satisfies $\pi\circ\imath={\rm
id}$.
\smallskip

\nfig\plembvir{}
In addition to $\imath(x_i)=\widehat x_i$, the embedding $\imath$ is
completely characterized by the image $\imath(\Om)=\widehat \Om$ of
the volume element $\Om$ in \eqJMvol, where $\widehat \Om$ is the
unique operator of ghost number two at the weight $(0,-2\rh)$, see
Figure \virabstbl.  Let us denote $\fH_+=\imath(\cP(\cC_2))$ and
$\fH_-={\rm ker}
\,\pi$. The decomposition $\fH\cong\fH_+\oplus \fH_-$ is shown in
Figure \plembvir.
\smallskip

One also finds that $b_0\widehat \Om\not\in\imath(\cP(\cC_2)$, and, as
easily seen from Figure \plembvir, $\widehat \Om$ is the only element
in $\fH_+$ with this property. By studying the action of $\fH$ on its
BV-ideal $\fH_-$, Lian and Zuckerman conclude that

\thm\genofvirbv
\proclaim Theorem \genofvirbv\ [\LZbv].
The BV-algebra $(\fH,\,,\cdot\,,b_0)$ is generated by $1$, the ground
ring generators $x_i$, and $\widehat \Om$.

\bigskip

$$ n=3:\qquad
      \hbox{\lower 3pt \hbox{\epsffile{vr2.eps} }} $$
$$ n=2:\qquad
      \hbox{\lower 3pt \hbox{\epsffile{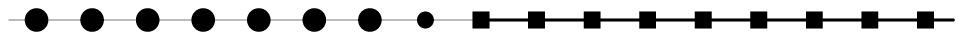} }} $$
$$ n=1:\qquad
      \hbox{\lower 3pt \hbox{\epsffile{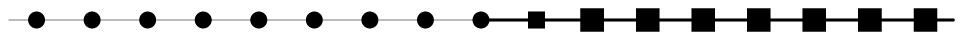} }} $$
$$ n=0:\qquad
      \hbox{\lower 3pt \hbox{\epsffile{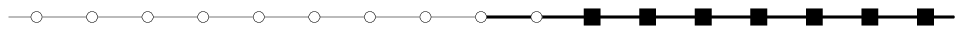} }} $$

\medskip
\centerline{\vbox{\hsize 5.5in
\noindent
Figure \plembvir. The decomposition $\fH\cong\fH_+\oplus \fH_-$. The
modules in $\fH_+$ are denoted by the squares while those in $\fH_-$ by
the dots.  Again, degeneracies are indicated by size.}}
\bigskip

Let us introduce on $\cC_2$ two structures of a ground ring ($\cC_2$)
module (see Section \twistedmodules): $M_1$, isomorphic to the ground
ring itself, and the twisted module $M_{-1}$ defined by the ``dual''
realization of the ground ring generators, $x_1\rightarrow -{\p\over\p
x_2}$ and $x_2\rightarrow -{\p\over\p x_1}$.
\smallskip

As shown in [\LZbv], the ghost number one cohomology in the negative
Weyl chamber, \ie, $\fH_-^1$, is isomorphic, as a ground ring module,
to $M_{-1}$.  Furthermore, $\fH_-$ has a natural structure of a
$G$-module of $\cP(\cC_2)$, with respect to which one may identify
it with the twisted polyderivations $\cP(\cC_2,M_{-1})$.
\smallskip

It may be worth emphasizing that the ``gluing'' of $\fH_+$ and $\fH_-$,
accomplished by the BV-operator $b_0$, is underlined by a simple
algebraic principle.

\thm\extenofvir
\proclaim Theorem \extenofvir. The BV-algebra $(\fH,\,\cdot\,,b_0)$
is the minimal extension of $(\cP(\cC_2),\,\cdot\,,\De_S)$, the
BV-algebra of polyvectors on the base affine space of $\sltw$,  in which
the cohomology of the BV-operator $b_0$ is trivial.

Let us illustrate the structure of $\fH$ on the example of
the BV-subalgebra $(\fH_0,\,\cdot\,,b_0)$ of $\sltw$ singlets. As read
off from Table \vircones, it consists of three doublets at the $\uone$
weights $0$, $-2\rh$ and $-4\rh$. As a BV-algebra it is generated by
$\widehat C$ and $\widehat \Om$, which have ghost number one and two,
respectively. Indeed, in terms of those two operators the three
doublets are explicitly given by, see [\LZbv],
\eqn\doublets{
(1, \widehat C)\,,\quad (b_0\widehat \Om,\widehat \Om)\,,\quad
(b_0(\widehat \Om\cdot b_0\widehat \Om),\widehat \Om\cdot b_0\widehat \Om)\,,}
where $b_0\widehat C=1$. The action of the BV operator in this basis
is obvious, and the only nonzero dot product, beyond the
ones listed in \doublets\ or involving the identity operator, is
\eqn\nonvanishpr{ \widehat C\cdot b_0(\widehat \Om\cdot b_0\widehat
\Om)=3\widehat \Om\cdot b_0\widehat \Om\,.}  Finally note that $1$,
$\widehat C$ and $\widehat \Om$ lie in $\fH_+$, while the remaining
ones, $b_0\widehat \Om$, $b_0(\widehat \Om\cdot b_0\widehat \Om)$ and
$\widehat \Om\cdot b_0\widehat \Om$, in $\fH_-$. Clearly, the latter three
span a BV-ideal of this finite-dimensional BV-algebra.

\vfill\eject

\footatend\vfill\eject\immediate\closeout\rfile
\baselineskip=14pt{{\bf  REFERENCES}}\bigskip{\frenchspacing%
\parindent=20pt\escapechar=` \input refs.tmp\vfill\eject}\nonfrenchspacing


\vfil\eject\end